\def\isarxiv{1} 

\ifdefined\isarxiv
\documentclass[11pt]{article}

\usepackage[numbers]{natbib}

\else
\documentclass[twoside,leqno]{article}
\usepackage{ltexpprt}
\fi
\usepackage{amsmath}
\usepackage{amssymb}
\usepackage{booktabs}
\usepackage{array}
\usepackage[shortlabels]{enumitem}
\usepackage{algpseudocode}
\usepackage{algorithm}
\ifdefined\isarxiv

\usepackage{amsthm}

\usepackage{algorithm}
\usepackage{subfig}
\usepackage{algpseudocode}
\usepackage{graphicx}
\usepackage{grffile}
\usepackage{wrapfig,epsfig}
\usepackage{url}
\usepackage{xcolor}
\usepackage{epstopdf}

\usepackage{bbm}
\usepackage{dsfont}
\fi

\ifdefined\isarxiv

\usepackage{tikz}
\usepackage{hyperref}  
\hypersetup{colorlinks=true,citecolor=blue,linkcolor=blue} 
\usetikzlibrary{arrows}
\usepackage[margin=1in]{geometry}

\else

\usepackage{microtype}
\usepackage{hyperref}
\usepackage{cleveref}

\fi

\ifdefined\isarxiv
\newtheorem{theorem}{Theorem}[section]
\newtheorem{lemma}[theorem]{Lemma}
\newtheorem{Definition}[theorem]{Definition}

\newtheorem{corollary}[theorem]{Corollary}

\newtheorem{fact}[theorem]{Fact}
\newtheorem{remark}[theorem]{Remark}

\else
\newtheorem{remark}{Remark}[section]
\makeatletter
\renewenvironment{proof}[1][Proof]{\@beginproof{#1}\ignorespaces}{\@endproof}
\makeatother
\fi
\newcommand{\wh}{\widehat}
\newcommand{\wt}{\widetilde}

\newcommand{\R}{\mathbb{R}}

\renewcommand{\d}{\mathrm{d}}

\renewcommand{\hat}{\wh}

\newcommand{\EO}{\mathsf{EO}}

\DeclareMathOperator*{\E}{{\mathbb{E}}}

\DeclareMathOperator{\Z}{\mathbb{Z}}

\DeclareMathOperator{\poly}{poly}

\DeclareMathOperator{\tr}{tr}

\DeclareMathOperator{\SO}{\mathsf{SO}}
\DeclareMathOperator{\vc}{vc}

\makeatletter
\newcommand*{\RN}[1]{\expandafter\@slowromancap\romannumeral #1@}
\makeatother

\makeatletter
\newcommand{\rthreecond}[2]{\phantomsection\def\@currentlabel{(#1)}\label{#2}}
\makeatother

\usepackage{lineno}

\begin{document}

\ifdefined\isarxiv

\date{}

\title{Convex Minimization with Integer Minima in $\widetilde O(n^4)$ Time\thanks{A preliminary version of this paper appeared at the 35th Annual ACM-SIAM Symposium on Discrete Algorithms (SODA 2024).}} 
\author{
Haotian Jiang\thanks{\texttt{jhtdavid96@gmail.com}. University of Washington. 
}\and Yin Tat Lee\thanks{\texttt{yintat@uw.edu}. University of Washington. 
}\and Zhao Song\thanks{\texttt{zsong@adobe.com}. Adobe Research.}\and Lichen Zhang\thanks{\texttt{lichenz@mit.edu}. Massachusetts Institute of Technology. Work partially done while visiting University of Washington. Supported in part by NSF grant No.\ 1955217 and NSF grant No.\ 2022448.}
}

\else
\newcommand\relatedversion{}
\renewcommand\relatedversion{\thanks{The full version of the paper can be accessed at \protect\url{https://arxiv.org/abs/2304.03426}}}
 \title{\Large Convex Minimization with Integer Minima in $\widetilde O(n^4)$ Time\relatedversion}
\author{Haotian Jiang\thanks{\texttt{jhtdavid96@gmail.com}. Microsoft Research.}\and Yin Tat Lee\thanks{\texttt{yintat@uw.edu}. University of Washington and Microsoft Research.}\and Zhao Song\thanks{\texttt{zsong@adobe.com}. Adobe Research.}\and Lichen Zhang\thanks{\texttt{lichenz@mit.edu}. Massachusetts Institute of Technology.}}
\fancyfoot[R]{\scriptsize{Copyright \textcopyright\ 2024 by SIAM\\
Unauthorized reproduction of this article is prohibited}}
\date{}
\fi

\ifdefined\isarxiv
\begin{titlepage}
  \maketitle
  \begin{abstract}
Given a convex function $f$ on $\mathbb{R}^n$ whose set of minimizers is nonempty and closed, lies in a box of radius $R$ with $\log R\le\mathrm{poly}(n)$, and has integral extreme points, we give a randomized algorithm that finds an integral minimizer of $f$ with high probability using $O(n^2\log n+n\log R)$ calls to a separation oracle, $O(n+\log R/\log n)$ calls to an approximate shortest vector routine with approximation factor $2^{O(n\log n)}$, such as the Lenstra-Lenstra-Lov\'asz (LLL) algorithm [Lenstra, Lenstra, Lov\'asz, Math. Ann. 1982], and $O(n^4 \log(nR)\log\log n)$ other arithmetic operations. The previous best polynomial time algorithm for this problem given in [Jiang, SODA 2021, JACM 2022] achieves $O(n(n\log\log n/\log n+\log R))$ oracle complexity. However, the overall runtime of Jiang's algorithm is at least $\widetilde{\Omega}(n^8)$, due to expensive sub-routines such as a shortest vector computation after every cutting plane step and the random walk based cutting plane method [Bertsimas, Vempala, JACM 2004]. We obtain our speedup by a nontrivial combination of a variant of the volumetric center cutting plane method (CPM) by [Vaidya, FOCS 1989] and its fast implementation given in [Jiang, Lee, Song, Wong, STOC 2020], run in blocks so that shortest vectors are computed only $O(n+\log R/\log n)$ times. To run the steps of the CPM with approximate leverage scores, we adapt the hybrid barrier of [Lee, Sidford, Wong, FOCS 2015], the volumetric barrier plus a small multiple of the logarithmic barrier, into the framework of [Jiang, Lee, Song, Wong, STOC 2020]. We compute the shortest vectors with a variant of the LLL algorithm that reduces segments of consecutive basis vectors recursively, adapted from [Schnorr, Inf. Comput. 2006] and [Zhou, Wang, Wang, ePrint 2026]. Started from the basis returned by the previous call, each call takes $O(n^3\log^3 n)$ arithmetic operations in exact arithmetic, so the total is $O(n^4\log^3 n+n^3\log^2 n\log R)$.  

For the special case of submodular function minimization (SFM), our result implies an algorithm for this problem using $O(n^3 \log n)$ calls to an evaluation oracle, $O(n)$ approximate shortest vector computations, and $O(n^4 \log n\log\log n)$ additional arithmetic operations. With the segment variant of LLL, it therefore uses $O(n^4\log^3 n)$ arithmetic operations in total. Its oracle complexity is better than that of the algorithms for this specific problem given in [Lee, Sidford, Wong, FOCS 2015] and [Dadush, V\'egh, Zambelli, SODA 2018, MOR 2021], which had the best running time before this work.

  \end{abstract}
  \thispagestyle{empty}
\end{titlepage}

\newpage

\else
\maketitle
\begin{abstract}
\small\baselineskip=9pt

\end{abstract}

\fi

\section{Introduction}

The problem of minimizing a convex function $f$ over $\mathbb{R}^n$, assuming access to a separation oracle $\SO$, has gained considerable interest since the seminal work of Gr\"{o}tschel, Lov\'{a}sz, and Schrijver \cite{gls81,gls88}.  
The separation oracle $\SO$ is such that when queried with a point $x\in \R^n$, it either returns ``YES'', which it may do only if $x$ minimizes $f$, or returns a vector $a\in\R^n$ with $a^\top y\geq a^\top x$ for every minimizer $y$ of $f$, where the inequality is strict when $x$ does not minimize $f$. So the hyperplane $\{y:a^\top y=a^\top x\}$ separates $x$ from the minimizers of $f$ whenever $x$ is not a minimizer. Every subgradient oracle of a convex function that is finite on $\R^n$ has this property (Section~\ref{sec:cpm:method}). The answer of $\SO$ may depend on the current and earlier queries and on random bits of its own. These bits, together with those of the shortest vector routine of Definition~\ref{def:approx_sv}, are independent of the random bits of our algorithm. For example, $\SO$ may be deterministic. Remark~\ref{rem:oracle_model} compares this definition with the one in the SODA 2024 version of this paper. One popular and successful approach for this problem is the cutting plane method, which dates back to the  center of gravity method, independently discovered by Levin \cite{l65} and Newman \cite{n65} in the 1960s. Since then, cutting plane methods have undergone substantial developments and improvements over the past six decades in terms of its oracle complexity\footnote{The oracle complexity of an algorithm is the number of oracle calls made by the algorithm in the worst case.} and runtime efficiency \cite{s77,yn76,k80,kte88,nn89,v89,av95,bv04,lsw15}. In particular, the cutting plane method with the fastest claimed running time is due to Jiang, Lee, Song, and Wong \cite{jlsw20} (see Remark~\ref{rem:jlsw20_floor} for an unproved step in its analysis).

Despite outstanding progress on the cutting plane method, direct application of these methods for minimizing a convex function $f$ on $\mathbb{R}^n$ via a separation oracle typically results in weakly-polynomial time algorithms, with the oracle complexity and runtime depending logarithmically on the accuracy parameter $\varepsilon > 0$ and the ``size''\footnote{Depending on the specific problem setting, the ``size'' of the function can be its range, Lipschitzness, condition number, length of binary representation, etc.} of the function $f$.  
 
A fundamental but extremely challenging question is to design a strongly-polynomial time algorithm that efficiently computes an exact minimizer of $f$ on $\mathbb{R}^n$, with its oracle complexity, number of arithmetic operations, and bit size all being polynomial only in the dimension $n$ of the problem and {\em not} dependent on the ``size'' of the function $f$. 
For example, it remains a major open problem to design a strongly-polynomial time algorithm for solving linear programs (LPs). This problem is widely known as Smale's 9th problem in his list of eighteen unsolved mathematical problems for the 21st century \cite{s98}. In particular, recent breakthroughs on LP solvers are all weakly-polynomial time algorithms, e.g., \cite{ls14, ls19, cls19, blss20,jswz21}. Despite this obstacle, strongly-polynomial time algorithms for LPs are known under additional structures, such as LPs with at most two non-zero entries per row \cite{m83,ac91,cm91} or per column \cite{v14,ov20} in the constraint matrix, LPs with bounded entries in the constraint matrix \cite{t86,vy96,dhnv20}, and LPs with 0-1 optimal solutions \cite{c12,c15}.

For minimizing a general convex function $f$ with access to a separation oracle, it is impossible to design a strongly-polynomial time algorithm unless the function $f$ satisfies additional combinatorial properties. One natural and general assumption, satisfied by many fundamental discrete optimization problems such  as submodular function minimization (SFM), maximum flow, maximum matching, and shortest path, is that the convex function $f$ has an integer minimizer.
Under the integer minimizer assumption, Jiang \cite{j21,j22}, improving upon the classical work of Gr\"{o}tschel, Lov\'{a}sz, and Schrijver \cite{gls81,gls88}, 
recently gave a strongly-polynomial time algorithm\footnote{In fact, Jiang \cite{j21,j22} gave a family of algorithms for this problem via an algorithmic reduction to the shortest vector problem. For instance, he gave an algorithm that achieves a nearly-optimal  oracle complexity of $O(n \log (nR))$ but using exponential time.} for finding an exact minimizer of $f$ using $O(n (n \log \log n / \log n + \log R))$ calls to $\SO$, where $R = 2^{\poly(n)}$ is the $\ell_\infty$-norm of the integer minimizer. 

In fact, Jiang's general result implies significant improvement to the oracle complexity of strongly-polynomial time algorithms even for the special problem of SFM (see Subsection~\ref{sec:sfm_application} for more details). 
Despite its favorable oracle complexity, Jiang's algorithm actually requires $\widetilde{\Omega}(n^8)$ additional arithmetic operations to implement, for the reasons that we recall after Theorem~\ref{thm:main_informal}. This additional part of the runtime is prohibitively large for its application to SFM. Before our work, the other state-of-the-art strongly-polynomial time algorithms for SFM \cite{lsw15,dvz21} had worse oracle complexity but used much fewer arithmetic operations. 

Unlike the special case of SFM, for the more general problem of minimizing a convex function with an integer minimizer, the $\widetilde{\Omega}(n^8)$ arithmetic operations in Jiang's algorithm is the state-of-the-art for strongly-polynomial time algorithms with near-optimal oracle complexity. 
The lack of a fast algorithm for the general problem, as well as the tradeoff between smaller oracle complexity and fewer additional arithmetic operations for the SFM algorithms that were the state of the art before our work, naturally lead to the following question:

\begin{center}
\emph{Can we minimize a convex function with integer minimizers using a separation oracle in a way that is both oracle-efficient and requires much fewer arithmetic operations?}
\end{center}

In this paper, we answer the above question affirmatively. We count arithmetic operations in exact arithmetic, without bounds on bit lengths. Specifically, we give a randomized algorithm that, when the set of minimizers is nonempty and closed, lies in $[-R,R]^n$ with $\log R\leq\poly(n)$, and has integral extreme points, finds an integral minimizer of a convex function with high probability using only $O(n^2\log n+n\log R)$ separation oracle calls, $O(n+\log R/\log n)$ calls to an approximate shortest vector routine with approximation factor $2^{O(n\log n)}$, and an additional $O(n^4 \log(nR)\log\log n)$ arithmetic operations. When applied to SFM, our result implies an algorithm that makes $O(n^3 \log n)$ evaluation oracle queries and $O(n)$ approximate shortest vector computations, and uses an additional $O(n^4 \log n\log\log n)$ arithmetic operations. This improves the oracle complexity of the strongly-polynomial time SFM algorithms in~\cite{lsw15,dvz21}, which had the best running time before our work. Compared with Jiang's algorithm~\cite{j22}, despite having a slightly bigger oracle complexity, our algorithm uses a $\widetilde{\Theta}(n^4)$ factor fewer arithmetic operations outside shortest vector computations when $R=\poly(n)$. It also computes far fewer approximate shortest vectors, $O(n+\log R/\log n)$ instead of one per cutting plane step. If the LLL algorithm starts each call from the basis it returned at the previous call, each shortest vector computation takes $O(n^4\log n)$ arithmetic operations (Lemma~\ref{lem:svp_total}). A variant of LLL reduces segments of consecutive basis vectors recursively, each in its own coordinates. Started from the previous basis as well, it takes $O(n^3\log^3n)$ operations per call, so the algorithm uses $O(n^4\log^3n+n^3\log^2n\log R)$ arithmetic operations in total (Lemma~\ref{lem:svp_segment}). In exact arithmetic, this total is far below that of Jiang's algorithm. For SFM, it has the form $n^4\log^{O(1)}n$ in exact arithmetic, like the totals of~\cite{lsw15,dvz21}. With exactly computed leverage scores, a deterministic variant uses $O(n^{\omega+2}\log n)$ arithmetic operations for SFM outside the shortest vector computations (Corollary~\ref{cor:exact_score}).

\subsection{Our Result}

The main result of this paper is a much more efficient algorithm for minimizing convex functions with integer minimizers:

\begin{theorem}[Main result, informal version of Theorem~\ref{thm:main_formal}]\label{thm:main_informal}
Given a separation oracle $\SO$ for a convex function $f:\R^n\to\R\cup\{+\infty\}$ and a number $R\geq1$ with $\log R\leq\poly(n)$. If the set of minimizers $K^*$ of $f$ is nonempty, closed and contained in the box $B_\infty(R)=[-R,R]^n$, and all extreme points of $K^*$ are integer vectors, then there exists a randomized algorithm that outputs an integer minimizer of $f$ with high probability using
\begin{itemize}
    \item ${O}(n^2\log n+n\log R)$ queries to $\SO$,
    \item ${O}(n+\log R/\log n)$ calls to an approximate shortest vector routine with a known approximation factor $2^{O(n\log n)}$, such as the LLL algorithm~\cite{lll82}, on lattices of rank at most $n$, and
    \item ${O}(n^4\log(nR)\log\log n)$ additional arithmetic operations outside these calls.
\end{itemize}
Let the routine be \textsc{\textup{SegmentLLL}} of Section~\ref{sec:segment_lll}, a variant of the LLL algorithm that reduces segments of consecutive basis vectors recursively. If each call starts from the basis returned by the previous call, these calls take $O(n^4\log^3n+n^3\log^2n\log R)$ arithmetic operations. The operations outside these calls number $O((n^4\log n+n^3\log R)\log\log n)$ (Lemma~\ref{lem:main_time_formal}), so the bound for the calls is also a bound on the total.
\end{theorem}

The strong assumption that all extreme points of $K^*$ are integer vectors guarantees that the algorithm outputs an {\em integer} minimizer of $f$. In fact, this assumption is necessary for the algorithm to efficiently compute an integer minimizer (see Remark 1.4 in \cite{j22} for an example). The set $K^*$ is closed whenever $f$ is finite on $\R^n$, since $f$ is then continuous, and also for the function of Corollary~\ref{cor:sfm_main} (proof of that corollary).

Previously, under the same assumptions on $f$ as in Theorem~\ref{thm:main_informal}, and for the oracle of Definition~1.1 of~\cite{j22}, Jiang~\cite{j22} gave an algorithm that computes an integer minimizer of $f$ using $O(n (n \log\log n/ \log n + \log R))$ separation oracle calls, which is better than the oracle complexity in Theorem~\ref{thm:main_informal}. However, his algorithm uses an enormous number of additional arithmetic operations, $\widetilde{\Omega}(n^8)$, for the following reasons: to obtain a subquadratic oracle complexity, \cite{j22} resorts to the random walk based cutting plane method by Bertsimas and Vempala~\cite{bv04} to work directly with the polytope $K$ formed by all the separating hyperplanes output by $\SO$. 
Unfortunately, the sampling approach in~\cite{bv04} requires $\Omega(n^6)$ arithmetic operations to perform one iteration of the cutting plane method. In addition, to guarantee that the sampling can be performed efficiently, \cite{j22} has to maintain two ``sandwiching'' ellipsoids, contained in and containing the polytope $K$ respectively, which requires additional computational cost to preserve after reducing the dimension of the problem. 

Another computational bottleneck in~\cite{j22} is that the algorithm approximates the length of the shortest vector after every iteration of the cutting plane method. To do so,~\cite{j22} uses the algorithm of~\cite{aks01}, which computes a $2^{n \log \log n/\log n}$-approximation to the shortest vector with a number of arithmetic operations polynomial in $n$ and in $\log D$, where $D$ bounds the squared lengths of the input basis vectors. Jiang's algorithm runs it in each of its $\wt O(n^2)$ iterations.  

To get around the aforementioned computational bottlenecks, 
we utilize an intricate combination of a computationally more efficient cutting plane method based on the volumetric barrier of Vaidya~\cite{v89} and a lazy schedule for the expensive shortest vector computations. 
Instead of working directly with the polytope $K$ and performing volume reduction in a step-by-step manner, we run a variant of Vaidya's cutting plane method~\cite{v89} in blocks of $O(n\log n)$ consecutive steps, and a measure of the volume decreases by a constant factor per step only in an amortized sense within each block. The analysis of the fast implementation of this method in~\cite{jlsw20} relies on an unproved lower bound on the leverage scores (Remark~\ref{rem:jlsw20_floor}). To run these steps with approximate leverage scores without such a bound, we adapt the hybrid barrier of~\cite{lsw15}, the volumetric barrier plus a small multiple of the logarithmic barrier, into the framework of~\cite{jlsw20}. We analyze the resulting method for a weight of order $1/\log n$ by a potential that measures every slack against that of a fixed integral minimizer, and we prove that the leverage score estimates of~\cite{jlsw20} stay accurate with high probability over any polynomial number of operations, which covers our algorithm when $\log R\leq\poly(n)$, even though our updates depend on earlier estimates. A block of $O(n\log n)$ steps then takes $O(n^3\log n\log\log n)$ arithmetic operations. Running the method in blocks also enables us to compute only $O(n+\log R/\log n)$ approximate shortest vectors in total, while still being able to avoid the appearance of extremely short vectors. 
Each of these approximate shortest vectors can be computed by any routine with approximation factor $2^{O(n\log n)}$, such as the LLL algorithm~\cite{lll82}. We count these computations separately from the other arithmetic operations and bound their cost for two routines. Suppose first that LLL starts each call from the basis it returned at the previous call. Then it performs no basis swap right after a dimension reduction. After a block of cutting plane steps it performs $O(n^3\log n)$ swaps, because within a block the Hessian changes by a factor of at most $2^{O(n\log n)}$ in either direction. Each step of LLL takes $O(n)$ operations, which gives $O(n^4\log n)$ operations per call and $O(n^5\log n+n^4\log R)$ operations in total (Lemma~\ref{lem:svp_total}).

Our second routine, \textsc{SegmentLLL} (Section~\ref{sec:segment_lll}), has approximation factor $2^{O(n\log n)}$. It uses $O(n^3\log^3 n)$ operations per call and $O(n^4\log^3 n+n^3\log^2 n\log R)$ in total (Lemmas~\ref{lem:segment_lll} and~\ref{lem:svp_segment}). It keeps the recursion of the algorithm SLLL$^+$ of Schnorr~\cite{sch06}, which reduces a segment of $2k$ consecutive basis vectors by reducing its first, middle and last $k$ vectors in the coordinates of the segment. It replaces the stopping test of SLLL$^+$ by a test of the kind used by Zhou, Wang and Wang~\cite{zww26}, with our own thresholds. A segment is then done when no Gram--Schmidt norm in it is smaller than an earlier one by more than a factor that depends only on the length of the segment. Their paper is a preprint that has not been peer reviewed, so we restate their argument in exact arithmetic and prove it. The output of \textsc{SegmentLLL} need not be reduced, but this weaker condition still gives approximation factor $2^{O(n\log n)}$, and it survives dimension reductions. The known LLL-type algorithms that use about $n^3$ arithmetic operations~\cite{s96,ns16} need integral inputs, and their bounds do not give this cost per call in our algorithm (Section~\ref{sec:preli:lll} and Remark~\ref{rem:svp_cost}).

\subsection{Applications to Submodular Function Minimization}
\label{sec:sfm_application}

Submodular function minimization (SFM) has been one of the cornerstone problems in combinatorial optimization since the seminal work of Edmonds \cite{e70}. Many classic combinatorial problems can be abstracted as optimization over a submodular function, such as graph cut function, set coverage function and economic utility function. For more comprehensive reviews of SFM, we refer readers to~\cite{m05,i08}.

\begin{table}[htp!]
    \centering
    \begin{tabular}{ | l | l | l | >{\raggedright\arraybackslash}p{3.3cm} | l | }
        \hline
        {\bf Papers} & {\bf Year} & {\bf Runtime} & {\bf Remarks} & {\bf General?} \\ \hline \hline
        \cite{gls81,gls88}    & 1981,88    & $\widetilde{O}(n^5 \cdot \EO + n^7)$~\cite{m05} & first strongly & $\checkmark$\\ \hline
        \cite{s00} & 2000  & $O(n^8 \cdot \EO + n^9)$ & first comb. strongly &\\ \hline
        \cite{iff01}  & 2000    & $ O(n^7 \log n \cdot \EO + \poly(n))$ & first comb. strongly & \\ \hline
       \cite{fi03}  & 2000    & $O(n^7 \cdot \EO + n^8)$  & & \\ \hline
        \cite{i03} & 2002 & $O(n^6 \log n \cdot \EO + n^7 \log n)$  & &\\ \hline
        \cite{v03}   & 2003    & $ O(n^7 \cdot \EO + n^8)$& &\\ \hline
        \cite{o09}   & 2007    & $O(n^5 \cdot \EO + n^6)$ &  &\\ \hline
        \cite{io09} & 2009 & $ O(n^5 \log n \cdot \EO + n^6 \log n) $& &\\ \hline
        \cite{lsw15} & 2015 & $O(n^3 \log^2 n \cdot \EO + n^4 \log^{O(1)} n)$ & best runtime before this work &\\ \hline
        \cite{lsw15} & 2015 & $O(n^3 \log n \cdot \EO + 2^{O(n)})$ & exponential time&\\ \hline
        \cite{dvz21} & 2018 & $O(n^3 \log^2 n \cdot \EO + n^4 \log^{O(1)} n)$& best runtime before this work &\\ \hline
        \cite{j21} & 2021 & $O(n^3 \frac{\log\log n}{\log n} \cdot \EO + n^8 \log^{O(1)} n)$ & best oracle complexity before~\cite{en26} & $\checkmark$\\ \hline 
        \cite{j21} & 2021 & $O(n^2 \log n \cdot \EO + 2^{O(n)})$ &  exponential time& $\checkmark$\\ \hline 
        {\bf This paper} & 2023 & $O(n^3 \log n \cdot \EO + n^4\log^3 n)$ & best runtime before~\cite{en26}, randomized & $\checkmark$\\
        \hline 
        \cite{en26} & 2026 & $O(n^2 \log n \cdot \EO + n^3\log^{O(1)} n)$ & after this work, randomized & \\ \hline 
    \end{tabular}
    \caption{Strongly-polynomial algorithms for submodular function minimization, and the algorithm of this paper. The oracle complexity measures the number of calls to the evaluation oracle $\mathsf{EO}$. In the case where a paper is published in both conference and journal, the year we provide is the earlier one. In the column ``General?'', $\checkmark$ means that the algorithm works for the more general problem of minimizing convex functions with integer minimizers studied in this paper. The algorithm of this paper is randomized and succeeds with high probability. The strongly polynomial algorithms of~\cite{lsw15} and~\cite{dvz21} with runtime $O(n^3\log^2 n\cdot\EO+n^4\log^{O(1)}n)$ are also randomized, since both rest on the cutting plane method of~\cite{lsw15}, and their bounds hold in expectation. The algorithms of~\cite{j22} are also randomized and succeed with high probability, and the exponential-time algorithm of~\cite{lsw15} uses the random-walk cutting plane method of~\cite{bv04}. The row of this paper counts arithmetic operations in exact arithmetic without bounding bit lengths, and with high probability its bound is strongly polynomial in the sense of~\cite{lsw15}. This count includes $O(n)$ calls to the routine \textsc{SegmentLLL} of Lemma~\ref{lem:segment_lll}, a variant of the LLL algorithm, on lattices of rank at most $n$. These calls take $O(n^4\log^3 n)$ operations, and the other operations number $O(n^4\log n\log\log n)$ (Corollary~\ref{cor:sfm_main}). A deterministic variant uses $O(n^{\omega+2}\log n)$ operations outside the lattice calls (Corollary~\ref{cor:exact_score}).}
    \label{tab:submodular}
\end{table}

We briefly recall the standard setting of SFM: we are given a submodular function $f: 2^V\rightarrow \Z$ over an $n$-element ground set $V$ where the function $f$ is accessed via an evaluation oracle, which returns the value $f(S)$ for a query set $S$ using time $\EO$. The goal is to find the minimizer of $f$ using queries to the evaluation oracle and additional arithmetic operations. 

An SFM algorithm is called strongly-polynomial time if its runtime depends polynomially only on $\EO$ and the dimension $n$, but not on the range of the function $f$. In their seminal work, Gr\"{o}tschel, Lov\'{a}sz, and Schrijver~\cite{gls84,gls88} gave the first strongly-polynomial time algorithm for SFM based on the ellipsoid method. Since then, there has been a long history of efforts in designing faster strongly-polynomial time algorithms for SFM (see Table~\ref{tab:submodular}). 

Before our work, the state-of-the-art strongly-polynomial time algorithms for SFM, in terms of the number of arithmetic operations used, were given by Lee, Sidford, and Wong \cite{lsw15} and Dadush, V\'egh, and Zambelli \cite{dvz21}. Both algorithms have runtime $O(n^3 \log^2 n\cdot \EO+n^4 \log^{O(1)}n)$. The oracle complexity of these algorithms were later improved to sub-cubic in terms of $n$ by Jiang~\cite{j21,j22}, where he gave a strongly-polynomial time algorithm with runtime $O(n^3 \log\log n/\log n \cdot \EO + n^8 \log^{O(1)} n)$ alongside an exponential time algorithm with a nearly-quadratic $O(n^2 \log n)$ oracle complexity.
When this work appeared, it was a major open problem in the area of SFM whether there exists a strongly-polynomial time algorithm with truly sub-cubic, i.e. $O(n^{3 - c})$ for some absolute constant $c > 0$, oracle complexity. Ene and Nguyen~\cite{en26} have since resolved it (see below). When the problem has certain structures, such as the minimizer is $k$-sparse, Graur, Jiang and Sidford~\cite{gjs23} have developed an algorithm with $\wt O(|V|\cdot \poly(k))$ queries. SFM has also been heavily studied in the parallel setting, see e.g.,~\cite{cgjs23}.

Unfortunately, while the algorithm in \cite{j22} answers major open questions in~\cite{lsw15} and makes significant progress on the oracle complexity for SFM, the number of additional arithmetic operations used by Jiang's algorithm is a factor of $\widetilde{\Theta}(n^4)$ larger than the algorithms in \cite{lsw15,dvz21}. 
In this paper, we complement Jiang's algorithm by giving the following SFM algorithm. It improves the oracle complexity of the algorithms in \cite{lsw15,dvz21}. Apart from $O(n)$ approximate shortest vector computations, it uses $O(n^4\log n\log\log n)$ arithmetic operations. These computations take $O(n^4\log^3 n)$ operations with the routine \textsc{SegmentLLL} of Lemma~\ref{lem:segment_lll} and $O(n^5\log n)$ with the LLL algorithm. 

\begin{corollary}[Submodular function minimization]
\label{cor:sfm_main}
Given an evaluation oracle $\EO$ for a submodular function $f$ defined over an $n$-element ground set, there exists a randomized algorithm that computes an exact minimizer of $f$ with high probability using
\begin{itemize}
    \item $O(n^3 \log n)$ queries to $\EO$,
    \item $O(n)$ calls to an approximate shortest vector routine (Definition~\ref{def:approx_sv}), given together with an upper bound $2^{O(n\log n)}$ on its approximation factor $\gamma$, on lattices of rank at most $n$, and
    \item $O(n^4 \log n\log\log n)$ additional arithmetic operations outside these calls.
\end{itemize}
If the routine is \textsc{\textup{SegmentLLL}}, run as in Lemma~\ref{lem:svp_segment}, these calls take $O(n^4\log^3 n)$ arithmetic operations, and this bound is also the total. If it is the LLL algorithm, run as in Lemma~\ref{lem:svp_total}, they take $O(n^5\log n)$.
\end{corollary}

Corollary~\ref{cor:sfm_main} almost immediately follows from Theorem~\ref{thm:main_informal} together with the standard fact that a separation oracle for the {\em Lov\'asz extension} of the submodular function $f$ can be implemented by making $n$ queries to the evaluation oracle of $f$ and $O(n \log n)$ additional arithmetic operations. Theorem~61 of~\cite{lsw15}, restated as Theorem~6.4 of~\cite{j22}, gives this construction and states its time as $O(n\cdot\EO+n^2)$.\footnote{We cite~\cite{lsw15} and~\cite{jlsw20} by the numbering of their arXiv versions 1508.04874v2 and 2004.04250v1.} Counting arithmetic operations only gives $O(n\log n)$, as we show in the proof below. The proof follows that of Theorem~1.7 of~\cite{j22}.

\begin{proof}[Proof of Corollary~\ref{cor:sfm_main}]
Let $\hat f$ be the Lov\'asz extension of $f$ on the cube $[0,1]^n$, with $\hat f=+\infty$ outside the cube. We define a separation oracle $\SO$ for $\hat f$ with the property of the paragraph \emph{Cut vectors} in Section~\ref{sec:cpm:method}. That is, $a^\top y\geq a^\top z$ for every minimizer $y$ of $\hat f$, with strict inequality when $z$ does not minimize $\hat f$. At a point $z$ outside the cube, let $i$ be the smallest index with $z_i\notin[0,1]$. The oracle returns $e_i$ if $z_i<0$ and $-e_i$ if $z_i>1$, and $a^\top y>a^\top z$ because every minimizer has $0\leq y_i\leq1$. At a point $z$ of the cube, it returns $a=-g$ for the greedy vector $g$. To compute $g$, it sorts the coordinates of $z$ in decreasing order, with ties broken by the smaller index, and takes differences of values of $f$. This $g$ is a subgradient at $z$ of the Lov\'asz extension on $\R^n$ (Theorem~61 of~\cite{lsw15}), so $a^\top(y-z)\geq\hat f(z)-\hat f(y)$, which is positive when $z$ does not minimize $\hat f$.

The set $K^*$ of minimizers of $\hat f$ is nonempty and closed, since $\hat f$ is continuous on the closed cube and $+\infty$ outside it. Also $K^*\subseteq[0,1]^n\subseteq B_\infty(1)$, and the extreme points of $K^*$ are integral by Theorem~6.3(d) of~\cite{j22}. We apply Theorem~\ref{thm:main_formal} with $R=1$ and the given upper bound $2^{O(n\log n)}$ on $\gamma$ in place of $\gamma$ (paragraph \emph{The good event} in Section~\ref{sec:minimization:alg}). With high probability, the algorithm outputs an integral minimizer $x^*$ of $\hat f$ with $O(n^2\log n)$ calls to $\SO$, $O(n)$ calls to \textsc{ApproxSV} and $O(n^4\log n\log\log n)$ other arithmetic operations. The point $x^*\in\{0,1\}^n$ is the indicator vector of a set $S$ that minimizes $f$, because $\hat f$ agrees with $f$ on $\{0,1\}^n$ and its minimum over the cube equals the minimum of $f$ (Theorem~6.3(a) and~(b) of~\cite{j22}). Each call of $\SO$ makes $O(n)$ queries to $\EO$ (Theorem~61 of~\cite{lsw15}, Theorem~6.4 of~\cite{j22}). Counting arithmetic operations only, a call takes $O(n\log n)$ of them inside the cube, for the sort and the differences, and $O(n)$ outside it. So the calls to $\SO$ make $O(n^3\log n)$ queries to $\EO$ and $O(n^3\log^2n)$ further operations, which is within $O(n^4\log n\log\log n)$. The bounds for \textsc{SegmentLLL} and for LLL are those of Theorem~\ref{thm:main_formal} with $R=1$, from Lemmas~\ref{lem:svp_segment} and~\ref{lem:svp_total}.
\end{proof}

Compared to the algorithms in~\cite{lsw15} and~\cite{dvz21}, which had the best running time before our work, our algorithm improves their oracle complexity from $O(n^3\log^2 n)$ to $O(n^3 \log n)$. Outside its $O(n)$ shortest vector computations, it uses $O(n^4\log n\log\log n)$ arithmetic operations, while their algorithms use $O(n^4 \log^{O(1)} n)$ operations in total. With \textsc{SegmentLLL}, our total is $O(n^4\log^3 n)$. So our algorithm makes fewer oracle queries than theirs, and its number of arithmetic operations has the same form $n^4\log^{O(1)}n$ as theirs. Like theirs, our count is in exact arithmetic without bounds on bit lengths. After this work, Ene and Nguyen~\cite{en26} improved our result with a randomized strongly polynomial algorithm that uses $O(n^2\log n)$ oracle queries and $O(n^3\log^{O(1)} n)$ additional arithmetic operations. A deterministic version uses the same number of queries and $O(n^{\omega+1}\log^{O(1)}n)$ additional operations.
We highlight that in \cite{lsw15}, the authors manage to achieve an $O(n^3 \log n)$ oracle complexity, but at the expense of  an exponential runtime.
It is also important to note that our improvements are achieved via a much more {\em general} algorithm, whereas the algorithms in \cite{lsw15,dvz21} work specifically for SFM. 

In comparison to~\cite{j22}, which had the best oracle complexity before~\cite{en26} and still has it for the general problem, our algorithm has a slightly worse oracle complexity,
but we improve the $\widetilde{\Omega}(n^8)$ additional arithmetic operations of his algorithm down to $O(n^4 \log n\log\log n)$ outside shortest vector computations and $O(n^4\log^3 n)$ in total. We also compute $O(n)$ approximate shortest vectors instead of one per cutting plane step.

Finally, note that the current best implementation of a separation oracle for the Lov\'asz extension requires $n$ queries to $\EO$, and the cutting plane method of~\cite{jlsw20} uses $O(n^2)$ amortized arithmetic operations per step by its main theorem (Theorem~1.1 of~\cite{jlsw20}), whose analysis has an unproved step (Remark~\ref{rem:jlsw20_floor}). So for a cutting plane algorithm for SFM that uses $T'$ iterations, the best runtime one can hope for with these techniques is $O(T'n \cdot \EO + T' n^2)$. Outside its $O(n)$ shortest vector computations, our algorithm matches this bound up to a factor $O(\log\log n)$ in the arithmetic operations. In particular, we use $O(n^2 \log n)$ iterations of the cutting plane method with a total runtime of $O(n^3 \log n\cdot \EO+n^4 \log n\log\log n)$ outside these computations. The factor $\log\log n$ comes from the $O(\log\log n)$ Newton steps per iteration that our analysis of the hybrid barrier needs (Section~\ref{sec:cpm:method}). The calls to \textsc{SegmentLLL} add a further $O(n^4\log^3 n)$ operations. In contrast, the algorithm in~\cite{j22} is far from this bound, with the $\widetilde{\Omega}(n^8)$ additional arithmetic operations being much larger than its oracle complexity. 

\paragraph{Subsequent work.} The algorithm of Ene and Nguyen~\cite{en26} discussed above saves a factor of $n$ in oracle queries over ours and a factor of $n$ up to polylogarithmic factors in arithmetic operations. Their randomized algorithm finds an exact minimizer with high probability, and their bounds, like ours, are in the evaluation oracle model with exact arithmetic. Their algorithm runs one cutting plane method with a single barrier function from start to finish, combined with the framework of Iwata, Fleischer and Fujishige~\cite{iff01}, and uses no lattice basis reduction. Their result is for SFM, not for the general problem of minimizing convex functions with integer minimizers studied in this paper. Moreover, their query bound matches that of the exponential time algorithm of~\cite{j22}.

\section{Technical Overview}\label{sec:tech}

In this section, we provide a brief overview of the techniques in prior work and in our work. In Section~\ref{sec:tech:prior}, we review the methods in~\cite{j22}. In Section~\ref{sec:tech:ours}, we present a preliminary overview of our approach. In Section~\ref{sec:tech:discuss}, we summarize current progress on the problem of minimizing convex functions with integer minimizers and discuss potential future directions.

\subsection{An Overview of Previous Work}
\label{sec:tech:prior}

Before discussing our main technical insights, we first review the approach of~\cite{j22} that achieves subquadratic oracle calls for minimizing convex functions with integer minimizers. The key ingredient in Jiang's algorithm is an interplay between the polytope, formed by the separating hyperplanes returned by the separation oracle, which contains the integer minimizer, and a lattice that captures the structures of integer points. 

In particular, in each iteration of the cutting plane method, Jiang's algorithm examines the length of the (approximate) shortest vector in the lattice under the ${\rm Cov}(K)$-norm, where ${\rm Cov}(K)$ is the covariance matrix of the polytope $K$. The length of this vector under ${\rm Cov}(K)$-norm provides a measurement for the width of the outer ellipsoid induced by ${\rm Cov}(K)^{-1}$. Thus, when this norm is small, it implies that the outer ellipsoid has a small width and we can safely proceed by reducing dimension and projecting the lattice onto the orthogonal complement of the shortest vector. On the other hand, if the shortest vector still has a relatively large norm, then the polytope $K$ can be further refined by using cutting plane method. To achieve the best possible oracle complexity, the algorithm in \cite{j22} performs the width measurement in a \emph{step-by-step} fashion, meaning that it actively checks the length of the shortest vector after every single cutting plane step. This ensures that the algorithm can enter the dimension reduction phase as soon as a short lattice vector can be obtained, which happens when the volume of $K$ is small enough. 

Unfortunately, such a step-by-step strategy and careful width measurement come at a price -- the approximate shortest vector subprocedure will be called in every iteration of the cutting plane method. Moreover, an expensive cutting plane method, such as the random walk based center of gravity method \cite{bv04}, that guarantees the volume shrinks in every step needs to be used.  

\subsection{Our Approach: Cutting Plane in Blocks and Lazy Width Measurement}
\label{sec:tech:ours}

Now we discuss our approach for overcoming the major computational bottleneck of~\cite{j22}, which is the step-by-step measurement of width mentioned above. For simplicity, we assume the integer minimizer $x^* \in \{0,1\}^n$ (i.e. $R = 1$) in the subsequent discussion. By delaying the width measurement, i.e. find an approximate shortest vector after a block of cutting plane steps, we can utilize a variant of the much more efficient cutting plane method due to Vaidya~\cite{v89}. However, this comes at the cost of a possibly much larger number of oracle calls, since the length of the shortest vector in the lattice might become very small when we actually perform the measurement, which will lead to a large increase in the volume of $K$ after dimension reduction. Nevertheless, we show that by carefully balancing the block size of cutting plane steps and the loss incurred during dimension reduction due to short lattice vectors, we can still achieve an $O(n^2 \log n)$ oracle complexity.

To explain our approach, we first provide a brief introduction to Vaidya's cutting plane method. Unlike classical cutting plane methods, such as the ellipsoid method and the center of gravity method, where each step is guaranteed to shrink the volume, Vaidya's algorithm and analysis rely on controlling the volume of the Dikin ellipsoid induced by the log barrier on polytope $K$. The algorithm iteratively finds the point $\vc(K)$ whose corresponding Dikin ellipsoid has the largest volume, which is called the volumetric center. Subsequently, the algorithm uses a separation oracle on the volumetric center $\vc(K)$. If $\vc(K)$ is a minimizer, then we are done. Otherwise, the algorithm computes the leverage score of each constraint with respect to the Hessian matrix, which measures the relative importance of each constraint. If all constraints are relatively important, then a new constraint is added based on the separating hyperplane returned by the separation oracle at $\vc(K)$. If one of the constraints has leverage score smaller than some tiny constant, then it is dropped to ensure that the polytope $K$ always has at most $O(n)$ constraints. Due to the increase in the volume of the polytope $K$ when constraints are dropped, the volume of $K$ only shrinks in an amortized sense. In particular, only after $O(n\log n)$ steps, the volume of $K$ is guaranteed to decrease by a multiplicative factor of $2^{-\Omega(n\log n)}$ from the initial volume. It is crucial to note that no guarantee on the volume shrinkage of $K$ is known if $o(n \log n)$ iterations of Vaidya's cutting plane method is run. 

Our idea is then to view the $O(n \log n)$ steps as a ``block'',  meaning that each time we measure the shortest vector and realize that it still has a large norm, we execute $O(n\log n)$ steps of our variant of Vaidya's cutting plane method and then re-examine the norm. This naturally induces a strategy that measures the width of the ellipsoid in a \emph{lazy} fashion. In our analysis, the log-volume of the Dikin ellipsoid at the current point plus a slack term (Definition~\ref{def:credit}) takes the place of the log-volume of $K$, and over a block it decreases by $\Omega(n\log n)$ (Step~1 of the proof of Lemma~\ref{lem:main_oracle_formal}), as the log-volume of $K$ does when the cutting plane method in~\cite{bv04} runs for $O(n\log n)$ steps. Therefore, the only issue is the decrease in the norm of the shortest vector during a block of cutting plane steps. Each step of our method changes the Hessian of the log barrier at the current point by at most a constant factor in the Loewner order. So over a block, the norm of any lattice vector, measured with the inverse Hessian, shrinks by a factor of at most $2^{O(n\log n)}$ (Lemma~\ref{lem:block_stability}). Fortunately, this shrinkage of the norm is acceptable and it incurs only an additional $O(\log n)$ factor in the number of oracle calls. 

One major advantage of our variant of Vaidya's cutting plane method is that its iterations can be performed very efficiently. Using the leverage score maintenance of~\cite{jlsw20} together with the hybrid barrier described below, a block of $O(n \log n)$ cutting plane steps can be done with $O(n^3 \log n\log\log n)$ arithmetic operations. Lazy width measurement also reduces the total number of approximate shortest vectors we need to compute from $\wt O(n^2)$ to $\wt O(n)$, so lattice reduction is run only $\wt O(n)$ times. We start LLL at each call from the basis it returned at the previous call. Within a block the Hessian changes by a factor of at most $2^{O(n\log n)}$ in either direction, so each call performs $O(n^3\log n)$ basis swaps. A call right after a dimension reduction performs none (Lemma~\ref{lem:svp_total}). Reducing segments of consecutive basis vectors recursively, each in its own coordinates, lowers the cost of a call to $O(n^3\log^3 n)$ operations (Lemma~\ref{lem:svp_segment}).

Although the efficiency issue of the cutting plane step and width measurement is addressed, we still need to carefully control the complexity of the dimension reduction step. This step can be particularly expensive due to the loss of structural information of the polytope $K$ after collapsing it onto a proper subspace $P$ on which the algorithm recurs, as discussed in~\cite{j22}. In the standard setting of Vaidya's cutting plane method, the polytope evolves in a ``slow-changing'' manner that makes it easy to maintain and update the center and its corresponding Hessian matrix. However, after reducing a dimension, we no longer have such a ``slow-changing'' property, and thus the new center and Hessian matrix have to be computed from scratch. 

A natural idea would be to formulate the problem of recomputing the center and Hessian as a convex optimization task constrained to the polytope $K \cap P$ we have access to. On the surface, this problem can be solved straightforwardly using the same cutting plane procedure efficiently. The caveat is that evaluating the volumetric function or its gradient requires $O(n^\omega)$ time\footnote{$\omega$ is the exponent of matrix multiplication~\cite{dwz23,vxxz24,lg24,adw25,OpenAIMM26}. The current best bound is $\omega<2.258$ over every field of characteristic zero~\cite[Theorem~1.1 and Corollaries~14.1--14.2]{OpenAIMM26}, which improves the earlier bound $\omega<2.371339$ of~\cite{adw25}. The running time of the deterministic variant in Corollary~\ref{cor:exact_score} is an explicit function of $\omega$. With $\omega<2.258$, it uses $O(n^{4.258}\log n)$ arithmetic operations for SFM outside the evaluation oracle queries and the calls to \textsc{ApproxSV}.}. As the cutting plane method evolves for $O(n\log n)$ iterations, this will lead to a total of $O(n^{\omega+1}\log n)$ arithmetic operations. To circumvent this issue, we start from a simpler convex body containing $K \cap P$ whose center is known exactly. Specifically, we choose the hyperrectangle centered at the center of the outer ellipsoid that contains $K \cap P$. We show that this method only affects the number of oracle calls by a factor of $O(\log n)$, despite causing the volume to blow up by a factor of $n^{O(n)}$. A similar approach is also taken in~\cite{j22} to ease the computation of centroid and covariance matrix of $K \cap P$. However, his method requires iteratively refining the hyperrectangle using constraints of $K \cap P$ until it coincides with $K \cap P$, at which point the algorithm relearns the collapsed polytope $K \cap P$. This approach can take as many as $\wt O(n^2)$ steps (without calling the separation oracle) with $\Omega(n^2)$ operations per step, and is mainly for achieving the best possible oracle complexity. Our approach is arguably simpler and much more efficient.

Another challenge is to run the steps of Vaidya's method with approximate leverage scores. The data structure of~\cite{jlsw20} maintains estimates of the leverage scores with $\ell_2$ error $1/\poly(\log n)$, with high probability by Lemma~\ref{lem:jlsw_adaptive}. Vaidya's Newton steps weigh each constraint by its leverage score, so such an error is harmless only if every leverage score stays bounded below by an absolute constant, which is not proved during runs of consecutive deletions (Remark~\ref{rem:jlsw20_floor}). To avoid any lower bound on the leverage scores, we adapt the hybrid barrier of~\cite{lsw15}, the volumetric barrier $F$ plus a small multiple $cL$ of the logarithmic barrier $L$, into the framework of~\cite{jlsw20}. Every constraint then has weight at least $c$ in the matrix that approximates the Hessian of $F+cL$, so an additive error of the estimates is a small relative error. As in~\cite{lsw15}, where the weight of the logarithmic barrier is of order $1/\log(nR/\epsilon)$ for accuracy $\epsilon$, our weight tends to zero. We take $c=\Theta(1/\log n)$, and our accounting cannot support a constant $c$ (see the discussion after Lemma~\ref{lem:hybrid_identity}). The analysis of~\cite{lsw15} pays for this weight with upper and lower bounds on all slacks. We pay for it differently. We fix an integral minimizer $x^*$, which lies in every polytope that the method maintains, and we measure the slack of every constraint at the minimizer $x_c(K)$ of $F+cL$ against its slack at $x^*$. These slack ratios enter a sum $D(K)\geq0$ of logarithms (Definition~\ref{def:credit}). An exact identity then shows that each added cut raises the potential $\mathcal{U}(K)=F(x_c(K))-cD(K)$ by a constant and that each deletion lowers it by at most a smaller constant (Lemmas~\ref{lem:hybrid_identity} and~\ref{lem:hybrid_step}). We also prove that the estimates of~\cite{jlsw20} stay accurate with high probability over any polynomial number of operations, and hence for our algorithm when $\log R\leq\poly(n)$, even though our updates depend on earlier estimates (Lemma~\ref{lem:jlsw_adaptive}). The analysis needs $O(\log\log n)$ Newton steps after each added or deleted constraint, which costs the factor $\log\log n$ in our count of arithmetic operations.

We also need the Dikin ellipsoid at the current point to sandwich the polytope. In Jiang's algorithm, the use of centroid and covariance matrix makes the analysis straightforward due to the standard nice geometric property that the polytope $K$ is sandwiched between ellipsoids induced by the covariance matrix at the centroid. For the minimizer $x_c$ of $F+cL$, optimality gives $\sum_i(\sigma_i(x_c)+c)\,s_i(y)/s_i(x_c)=\sum_i(\sigma_i(x_c)+c)$ for every point $y$ of the affine hull of $K$, where $s_i$ is the slack of constraint $i$ and $\sigma_i$ its leverage score. Since every weight $\sigma_i(x_c)+c$ is at least $c$, no slack exceeds its value at $x_c$ by more than a factor $O(n/c)$ when there are $O(n)$ constraints. This gives $E(x_c,H(x_c))\subseteq K\subseteq O(n/c)\cdot E(x_c,H(x_c))$, where $H$ is the Hessian of the log barrier and $E(x,H)$ is the ellipsoid that it defines. The second containment also follows from Lemma~19 of~\cite{lsw15} (see the paragraph after Lemma~\ref{lem:sandwich}). The same holds at the point $x_K$ returned by our method, with a factor $R'=O(n\log n)$ (Lemma~\ref{lem:sandwich}). The analogous statement at the volumetric center $\vc(K)$, the minimizer of $F$, with a factor $\poly(mn)$ (Lemma~\ref{lem:polytope_sandwich}), is used in our analysis only by the deterministic variant of Corollary~\ref{cor:exact_score}.

To carry out the analysis, we use the potential function developed in~\cite{j22}, which captures the volume of the polytope and the density of the lattice simultaneously. In place of the volume of $K$, we use the log-volume of the Dikin ellipsoid $E(x_K,H_K)$ plus $c$ times the sum $D(K)$ of Definition~\ref{def:credit}. We show that each block of cutting plane steps decreases this potential by $\Omega(n\log n)$. Unlike~\cite{j22}, the block cutting plane steps and Dikin ellipsoids cause extra losses on the volume of $K$ after dimension reduction, due to the possible appearance of very short vectors. We observe that such blowups are always at most $n^{O(n)}$ for each dimension reduction step. Thus, the potential increases by at most $O(n^2 \log n)$ in total, and we need to use a total of $O(n^2 \log n)$ oracle calls to counter this increment. To summarize, by trading for a slightly-worse number of oracle calls, we make more room for the algorithm to gain extra speedup through leverage score maintenance, fewer approximate shortest vector computations, and crude estimation of the convex body after reducing dimension.

\subsection{Discussion}
\label{sec:tech:discuss}

A natural question that arises from our result is whether it is possible to obtain a strongly-polynomial time algorithm with quadratic or subquadratic oracle complexity, matching the one in \cite{j22}, while achieving the same arithmetic cost as ours. Our algorithm has only one $\log n$ factor in the oracle complexity, but we contend that this is inherent for Vaidya's approach and for our variant of it. In our analysis, one dimension reduction may raise the potential by up to $O(n\log n)$ when the shortest vector routine has an approximation factor $\gamma\leq2^{O(n\log n)}$ (Step~4 of the proof of Lemma~\ref{lem:main_oracle_formal}), and both methods guarantee a decrease of their potential by $\Omega(n\log n)$ only after $\Theta(n\log n)$ cutting plane steps. Part of this raise comes from the hyperrectangle that we use in the dimension reduction phase to approximate the convex body, as this approximation increases the volume by a factor of $n^{O(n)}$. To resolve this issue, an algorithm that computes an approximate minimizer of our barrier on $K\cap P$ within the cost of one block of cutting plane steps would suffice. However, designing such an algorithm is an interesting and nontrivial data structure task, 
as discussed in the preceding subsection, since it requires maintaining and updating the log determinant of a Gram matrix under slow diagonal changes.

To achieve a subquadratic oracle complexity, a crucial ingredient in \cite{j22} is an approximate shortest vector algorithm with a sub-exponential approximation factor, first given in~\cite{aks01}. The main reason for the sub-exponential approximation ratio is a block-reduction scheme introduced in~\cite{s87} that computes a more general notion of reduced lattice basis. Recent improvements on basis reduction algorithms make use of this block-reduction idea to obtain more refined recursive structures. Therefore, it is of interest to design approximate shortest vector algorithms using $\wt O(n^3)$ arithmetic operations while achieving a sub-exponential approximation factor. Faster lattice reduction speeds up our algorithm even with an exponential approximation factor. With LLL started from the previous basis, each call uses $O(n^4\log n)$ arithmetic operations (Lemma~\ref{lem:svp_total}), and with \textsc{SegmentLLL} it uses $O(n^3\log^3n)$ (Lemma~\ref{lem:svp_segment}). With \textsc{SegmentLLL}, all $O(n+\log R/\log n)$ calls together cost $O(n^4\log^3n+n^3\log^2n\log R)$. Combined with Theorem~\ref{thm:main_formal}, the total is then $\wt O(n^4\log(nR))$, with high probability when $\log R\leq\poly(n)$. This is $\wt O(n^4)$ when $\log R=\poly(\log n)$, for instance for SFM, where $R=1$. Our bound on these calls still exceeds our bound $O((n^4\log n+n^3\log R)\log\log n)$ on the operations outside them by a factor $O(\log^2n/\log\log n)$. A routine with $O(n^3\log n\log\log n)$ operations per call on the lattices and norms produced by our algorithm would remove this gap. The algorithms of~\cite{s96,ns16} do not give such a routine. The bound of~\cite{s96} needs the potential of the swap argument to drop by a factor of at most $2^{O(n^2)}$, while for our inputs we can bound this drop only by $2^{O(n^3\log n)}$ per call. The algorithm of~\cite{ns16} needs an integral input and bounds the first vector only against the determinant of the lattice. Our analysis needs a bound against the shortest vector instead, although a factor $2^{O(n\log n)}$ is enough (Remark~\ref{rem:svp_cost}).

\section{Preliminary}\label{sec:preli}

In Section~\ref{sec:preli:basic}, we provide several basic notations, definitions and facts. In Section~\ref{sec:preli:lll}, we discuss LLL algorithm and shortest vector problem. In Section~\ref{sec:preli:convex_geometry}, we define and state several basic tools in convex geometry. In Section~\ref{sec:preli:lattice_projection}, we provide some related definitions about lattice projection. In Section~\ref{sec:preli:slicing_lemma}, we present the slicing lemma. In Section~\ref{sec:preli:dim_red}, we state a dimension reduction lemma. 

\subsection{Notations and Basic Facts}\label{sec:preli:basic}

\paragraph{Basic Notations.}
For an integer $n$, we use $[n]$ to denote the set $\{1,2,\cdots,n\}$. For any function $f$, we use $\wt{O}(f)$ to denote $f \cdot \poly(\log f)$.

\paragraph{Model of computation.} Throughout this paper, an arithmetic operation is an addition, subtraction, multiplication, division, comparison or square root of real numbers, or the rounding of a real number to a nearest integer, carried out exactly and at unit cost. We do not bound the bit lengths of the numbers. A randomized algorithm may also draw independent standard Gaussian real numbers at unit cost. The data structure of~\cite{jlsw20} draws its random numbers this way, and we use it as a black box in this model, with the nodes and weights of its quadrature formulas taken as given. The parameters of the paragraph \emph{Parameters} in Section~\ref{sec:cpm:hybrid} involve the natural logarithm, which is not among these operations. They depend only on $n$, and our algorithm takes them as given in the same way. The same holds for the bound $M$ of the operation test of our algorithm, except for one term that depends on $R$ and on the approximation factor $\gamma$. Both numbers are given to the algorithm, which computes this term by repeated doubling (paragraph \emph{The good event} in Section~\ref{sec:minimization:alg}). Following~\cite{lsw15}, we call an algorithm for SFM strongly polynomial if its numbers of oracle calls and arithmetic operations are bounded by a polynomial in $n$, independent of the range of the function values. Our SFM algorithm, with the routine \textsc{SegmentLLL} of Section~\ref{sec:segment_lll}, is strongly polynomial in this sense with high probability: for every constant $k$ and every large enough $n$, with probability at least $1-n^{-k}$, it makes $O(n^3\log n)$ oracle calls and $O(n^4\log^3n)$ arithmetic operations, bounds that do not depend on $k$. The definition of~\cite{gls88} also bounds the bit lengths of all numbers, which we do not do.

\paragraph{Matrices and Vectors.}
  For a vector $x$, we use $\| x \|_{2}$ to denote its $\ell_2$ norm. For a vector $x$, we use $x^\top$ to denote its transpose. For a matrix $A$, we use $A^\top$ to denote its transpose. We use ${\bf 0}_n$ to denote a length-$n$ vector where all the entries are zeros. We use ${\bf 1}_n$ to denote a length-$n$ vector where all the entries are ones.

We say a square matrix $A \in \R^{n \times n}$ is PSD ($A \succeq 0$) if for all vectors $x \in \R^n$, $x^\top A x \geq 0$. For a square matrix $A$, we use $\det(A)$ to denote the determinant of matrix $A$. For a square and invertible matrix $A$, we use $A^{-1}$ to denote the inverse of matrix $A$.

For a PSD matrix $A$, we define the induced matrix norm for any vector $x$ as $\| x \|_A := \sqrt{x^\top A x}$.

\paragraph{Ellipsoid.}

Given a point $x_0 \in \R^n$ and a PSD matrix $A \in \R^{n \times n}$, we use $E(x_0,A)$ to denote the (not necessarily full-rank) ellipsoid given by 
\begin{align*}
    E(x_0,A):=\{ x \in x_0 + W_A : (x-x_0)^\top A (x-x_0) \leq 1 \},
\end{align*}
where $W_A$ is the subspace spanned by eigenvectors corresponding to nonzero eigenvalues of $A$. When the ellipsoid is centered at ${\bf 0}_n$, we use the short-hand notation $E(A)$ to denote $E( {\bf 0}_n,A)$. 

\paragraph{Lattices.}
Let $b_1,\ldots,b_k\in \R^n$ be a set of linearly independent vectors, we use 
\begin{align*}
\Lambda(b_1,\ldots,b_k)=\{ \sum_{i=1}^k \lambda_i b_i, \lambda_i\in \Z\}
\end{align*}
denote the lattice generated by $b_1,\ldots, b_k$ and $k$ is the rank of the lattice. If $k=n$, then it's full rank. A basis of $\Lambda:=\Lambda(b_1,\ldots,b_k)$ is a set of $k$ linearly independent vectors that generates $\Lambda$ under integer combinations. Bases of $\Lambda$ are equivalent under unimodular transforms. We use $\lambda_1(\Lambda)$ to denote the $\ell_2$ norm of the shortest nonzero vector in $\Lambda$ and $\lambda_1(\Lambda,A)$ to denote the induced $A$-norm of the shortest nonzero vector in $\Lambda$.

Given a basis $B\in \R^{n\times k}$, the parallelotope associated to it is the polytope $P(B)=\{\sum_{i=1}^k \lambda_i b_i: \lambda_i\in [0,1),\forall i\in [k] \}$. The determinant of $\Lambda$ is the volume of $P(B)$, which is independent of basis.

The dual lattice of $\Lambda$ is defined as follows:
\begin{Definition}[Dual lattice]\label{def:dual_lattice}
Given a lattice $\Lambda \subseteq \R^n$, the dual lattice $\Lambda^*$ is the set of all vectors $x \in \mathrm{span}\{ \Lambda \}$ such that $\langle x, y \rangle \in \mathbb{Z}$ for all $y \in \Lambda$.
\end{Definition}
For more backgrounds about lattices, we refer the readers to lecture notes by Rothvoss \cite{r16}.

\paragraph{Leverage Score.}
We define leverage score, which is a standard concept in numerical linear algebra \cite{cw13,bwz16,swz17,swz19,jlsw20,syyz22,dsw22}. We remark that leverage score has multiple equivalent definitions, here we just present one of them.
\begin{Definition}[Leverage score]\label{def:leverage_score}
Given a matrix $A \in \R^{m \times n}$, we define the leverage score for matrix $A$ to be $\sigma \in \R^m$, i.e,
\begin{align*}
    \sigma_i = a_i^\top (A^\top A)^{-1} a_i, ~~~ \forall i \in [m]
\end{align*}
Note that $a_i^\top$ is the $i$-th row of $A$.
\end{Definition}

We state a useful fact here.
\begin{fact}[folklore]\label{fac:leverage_score_sum}
Given a matrix $A \in \R^{m \times n}$, we have the following identity of its leverage score:
\begin{align*} 
\sum_{i=1}^m \sigma_i=n.
\end{align*}
\end{fact}

\subsection{LLL Algorithm for Shortest Vector Problem}\label{sec:preli:lll}

Given a lattice $\Lambda$ and a corresponding basis $B\in \R^{n\times k}$, it is natural to seek an algorithm that finds the vector with norm $\lambda_1(L)$, or at least approximately finds it. The famous Lenstra-Lenstra-Lov\'asz algorithm serves such a purpose:

\begin{theorem}[LLL algorithm, \cite{lll82}]\label{thm:lll}
Let $b_1, \cdots, b_k \in \mathbb{Z}^n$ be a basis for lattice $\Lambda$ and $A \in \mathbb{Z}^{n \times n}$ be a PSD matrix that is full rank on $\mathrm{span}(\Lambda)$. Let $D \in \R$ such that $\| b_i \|_A^2 \leq D$ for any $i \in [k]$. Then there exists an algorithm that outputs in $O(n^4 \log(D))$ arithmetic operations a nonzero vector $b_1'$ such that
\begin{align*}
    \| b_1' \|_A^2 \leq 2^{k-1} \cdot \lambda_1^2( \Lambda, A )
\end{align*}
Moreover, the integers occurring in the algorithm have bit sizes at most $O(n \log (D))$.
\end{theorem}

Our algorithm uses lattice reduction only through the following interface.

\begin{Definition}[Approximate shortest vector routine]\label{def:approx_sv}
Let $\gamma\geq 1$. A \emph{$\gamma$-approximate shortest vector routine} \textsc{\textup{ApproxSV}} is a routine that takes as input a positive definite matrix $G\in\R^{d\times d}$ with $d\leq n$ and returns a unimodular matrix $U\in\Z^{d\times d}$ such that
\begin{align*}
    (U^\top GU)_{1,1}\leq\gamma^2\cdot\min_{x\in\Z^d\setminus\{{\bf 0}_d\}}x^\top Gx .
\end{align*}
Its output may depend on the current and earlier inputs and on random bits of its own. The random bits of $\SO$ and of \textsc{\textup{ApproxSV}}, taken together, are independent of the random bits of our algorithm. For example, both may be deterministic.
\end{Definition}

Let $B\in\R^{n\times d}$ be a basis of a lattice $\Lambda$ and let $M$ be positive definite on ${\rm span}(\Lambda)$. If \textsc{ApproxSV} is run on the Gram matrix $G=B^\top MB$, then $BU$ is a basis of $\Lambda$ whose first vector $v$ satisfies $\|v\|_M\leq\gamma\cdot\lambda_1(\Lambda,M)$. We write $v\gets\textsc{ApproxSV}(\Lambda,M)$ for this use. The LLL algorithm of Theorem~\ref{thm:lll} is such a routine with $\gamma=2^{(n-1)/2}$ (its factor on a lattice of rank $d\leq n$ is $2^{(d-1)/2}$). It only needs the Gram matrix, and it can return the unimodular matrix along with the reduced basis. Its approximation guarantee does not use integrality. Theorem~\ref{thm:lll} counts operations for integral input, while the Gram matrices in our algorithm are real. We therefore state the variant of LLL that we use, and its cost, in the model of computation of Section~\ref{sec:preli:basic}.

\paragraph{Reduced bases.} Let $G\in\R^{d\times d}$ be positive definite, and let $V\in\Z^{d\times d}$ be unimodular, with columns $u_1,\ldots,u_d$. Let $u_1^*,\ldots,u_d^*$ be the Gram--Schmidt orthogonalization of $u_1,\ldots,u_d$ in the inner product $\langle x,y\rangle_G=x^\top Gy$, and let $B_k=\langle u_k^*,u_k^*\rangle_G$ and $\mu_{k,j}=\langle u_k,u_j^*\rangle_G/B_j$ for $j<k$. For $i\leq d$, let $V_i$ consist of the first $i$ columns of $V$, and let $d_i(V)=\det(V_i^\top GV_i)=\prod_{k\leq i}B_k$. We call $V$ \emph{reduced for $G$} if $|\mu_{k,k-1}|\leq 1/2$ and $B_k\geq(3/4-\mu_{k,k-1}^2)B_{k-1}$ for $2\leq k\leq d$. This implies $B_k\geq B_{k-1}/2$ for every $k$. The matrix $V$ is reduced for $G$ if and only if $I_d$ is reduced for $V^\top GV$. Finally, let $P(V)=\prod_{i=1}^{d-1}d_i(V)$.

We use one standard fact. Let $\lambda_l$ be the $l$-th successive minimum of $\Z^d$ in the norm $x\mapsto(x^\top Gx)^{1/2}$. Take $l$ linearly independent $x\in\Z^d$ with $x^\top Gx\leq\lambda_l^2$. One of them is not in the span of $u_1,\ldots,u_{l-1}$. Write it as an integer combination of $u_1,\ldots,u_d$, and let $J\geq l$ be the largest index with a nonzero coefficient $c_J$. Its component along $u_J^*$ is $c_Ju_J^*$, so
\begin{align}\label{eq:successive_minima_gs}
    \lambda_l^2\geq c_J^2B_J\geq\min_{j\geq l}B_j .
\end{align}

\begin{lemma}[LLL in exact arithmetic]\label{lem:lll_exact}
There is a $2^{(n-1)/2}$-approximate shortest vector routine \textsc{\textup{LLL}} with the following properties. On input a positive definite $G\in\R^{d\times d}$, it returns a unimodular $U\in\Z^{d\times d}$ that is reduced for $G$. It uses $O(d^3+dS)$ arithmetic operations, where the number $S$ of swaps it performs satisfies $(4/3)^S\leq P(I_d)/P(U)$. In particular, if $I_d$ is reduced for $G$, then it performs no swap.
\end{lemma}

\begin{proof}
\textsc{LLL} is the algorithm of~\cite{lll82}, except that, as in the modification of Storjohann~\cite{s96}, each vector is size-reduced only against the vector just before it. Unlike~\cite{s96}, we omit the final size reduction of the whole basis, which the analysis below does not need. It computes the Gram--Schmidt data of $I_d$, that is, the factorization $G=LDL^\top$ with $L$ unit lower triangular and $D$ diagonal, with $O(d^3)$ operations. It sets $U=I_d$ and $k=2$, and while $k\leq d$ it repeats the following step. If $|\mu_{k,k-1}|>1/2$, it replaces $u_k$ by $u_k-ru_{k-1}$, where $r$ is an integer nearest to $\mu_{k,k-1}$. Then, if $B_k\geq(3/4-\mu_{k,k-1}^2)B_{k-1}$, it sets $k\gets k+1$. Otherwise it exchanges $u_{k-1}$ and $u_k$, which we call a swap, and sets $k\gets\max\{k-1,2\}$. Both changes affect only $u_{k-1}$, $u_k$, $B_{k-1}$, $B_k$ and the coefficients $\mu_{k-1,j}$, $\mu_{k,j}$, $\mu_{i,k-1}$ and $\mu_{i,k}$. The update formulas of~\cite{lll82} take $O(d)$ operations per step.

The replacement of $u_k$ does not change the span of $u_1,\ldots,u_i$ for any $i$, so it changes no $d_i$. A swap changes only $d_{k-1}$, which it multiplies by $(B_k+\mu_{k,k-1}^2B_{k-1})/B_{k-1}<3/4$. Every $d_i$ is the squared determinant of a lattice of rank $i$ whose nonzero vectors have squared length at least $\lambda_1^2$, so $d_i\geq(\lambda_1^2/i)^i>0$ by Minkowski's first theorem. Hence $S$ is finite and $(4/3)^S\leq P(I_d)/P(U)$. A step without a swap increases $k$ by one and a swap decreases it by at most one, so there are at most $d-1+2S$ steps, and $O(d^3+dS)$ operations in total. As in~\cite{lll82}, at the start of each step the two conditions hold at every index below $k$, because a step at index $k$ does not change $B_j$ or $\mu_{j,j-1}$ for $j\leq k-2$. So when the loop ends, $U$ is reduced for $G$. Finally, $B_k\geq B_{k-1}/2$ gives $(U^\top GU)_{1,1}=B_1\leq 2^{d-1}\min_kB_k$, and $\min_kB_k\leq\lambda_1^2$ by~\eqref{eq:successive_minima_gs}, so the guarantee of Definition~\ref{def:approx_sv} holds with $\gamma^2=2^{d-1}\leq 2^{n-1}$.
\end{proof}

\begin{lemma}[Reduced bases are nearly optimal]\label{lem:reduced_near_opt}
Let $G\in\R^{d\times d}$ be positive definite, let $V\in\Z^{d\times d}$ be unimodular, and let $\tau\geq 0$ be such that the Gram--Schmidt norms of $V$ for $G$ satisfy $B_j\geq 2^{-\tau}B_l$ for all $1\leq l<j\leq d$. Then $d_i(V)\leq 2^{i\tau}\,i^i\det(Y^\top GY)$ for $1\leq i\leq d$ and every $Y\in\Z^{d\times i}$ of rank $i$. If $V$ is reduced for $G$, this holds with $\tau=d$.
\end{lemma}

\begin{proof}
If $V$ is reduced for $G$, then $B_j\geq B_{j-1}/2$ for every $j$, so $B_j\geq 2^{-(j-l)}B_l\geq 2^{-d}B_l$ for $l<j$, which is the hypothesis with $\tau=d$. By~\eqref{eq:successive_minima_gs} and the hypothesis, $\lambda_l^2\geq\min_{j\geq l}B_j\geq 2^{-\tau}B_l$ for every $l$. The columns of $Y$ generate a lattice of rank $i$ contained in $\Z^d$. Its squared determinant is $\det(Y^\top GY)$, and its $l$-th successive minimum is at least $\lambda_l$. Minkowski's second theorem, together with the fact that the $i$-dimensional unit ball contains the cube $[-1/\sqrt i,1/\sqrt i]^i$ of volume $(2/\sqrt i)^i$, gives $\prod_{l\leq i}\lambda_l^2\leq i^i\det(Y^\top GY)$. Hence $d_i(V)=\prod_{l\leq i}B_l\leq 2^{i\tau}\prod_{l\leq i}\lambda_l^2\leq 2^{i\tau}i^i\det(Y^\top GY)$.
\end{proof}

\begin{lemma}[Restarting LLL]\label{lem:lll_restart}
Let $G_0,G_1\in\R^{d\times d}$ be positive definite with $e^{-a}G_0\preceq G_1\preceq e^aG_0$ for some $a\geq 0$, and let $I_d$ be reduced for $G_0$. Then \textsc{\textup{LLL}} on input $G_1$ performs $O(ad^2+d^3)$ swaps.
\end{lemma}

\begin{proof}
Let $U$ be the output, and write $d_i^{(0)}$ and $d_i^{(1)}$ for $d_i$ computed with $G_0$ and with $G_1$. Let $1\leq i\leq d-1$. Since $G_1\preceq e^aG_0$, we have $d_i^{(1)}(I_d)\leq e^{ai}d_i^{(0)}(I_d)$. Lemma~\ref{lem:reduced_near_opt} with $V=I_d$, the matrix $G_0$ and $Y=U_i$ gives $d_i^{(0)}(I_d)\leq 2^{id}i^id_i^{(0)}(U)$. Since $G_0\preceq e^aG_1$, we have $d_i^{(0)}(U)\leq e^{ai}d_i^{(1)}(U)$. Hence $\ln(d_i^{(1)}(I_d)/d_i^{(1)}(U))\leq 2ai+id\ln 2+i\ln i$. Summing over $i\leq d-1$ gives $\ln(P(I_d)/P(U))\leq ad^2+d^3$ for $P$ computed with $G_1$. By Lemma~\ref{lem:lll_exact}, the number of swaps is at most $(ad^2+d^3)/\ln(4/3)$.
\end{proof}

When \textsc{ApproxSV} is \textsc{LLL}, the lattice bookkeeping in the proof of Lemma~\ref{lem:main_time_formal} starts each call from the basis returned by the previous call. Lemma~\ref{lem:svp_total} uses Lemma~\ref{lem:lll_restart} to bound the cost of all calls.

Neumaier and Stehl\'e~\cite{ns16} give a faster LLL-type reduction algorithm for integral bases, based on a recursive blocking of the Gram-Schmidt orthogonalization of the basis.

\begin{theorem}[Theorems 1 and 2 of \cite{ns16}]\label{thm:ns16}
Let $B\in\Z^{n\times n}$ be a basis of a full-rank lattice $L\subseteq\R^n$, let $\beta=\log\max_i\|b_i\|_2$, and let $\zeta>0$ be fixed. There is an algorithm that returns a size-reduced basis $C$ of $L$ whose first vector satisfies
\begin{align*}
    \|c_1\|_2\leq(1+\zeta)\cdot(4/3)^{(n-1)/4}\cdot(\det L)^{1/n},
\end{align*}
within $O(n^{4+\varepsilon_0}\beta^{1+\varepsilon_0})$ bit operations, for every fixed $\varepsilon_0>0$.
\end{theorem}

Theorem~2 of~\cite{ns16} states the cost as $O(k\,n^3\,\mathsf{M}(n\beta))$ bit operations, where $\mathsf{M}(\ell)$ is the cost of multiplying two $\ell$-bit integers and $k=2^{O(\sqrt{\log n\cdot\log\log(n\beta)})}\log^2(n\beta)$. Its proof multiplies a count of arithmetic operations by the cost $O(\mathsf{M}(n\beta))$ of one operation on integers of $O(n\beta)$ bits. It counts the two-dimensional base cases of its recursion in bit operations. So, apart from these base cases, the proof of~\cite{ns16} counts about $kn^3$ arithmetic operations on integers of $O(n\beta)$ bits, for an integral basis. Here $k=(n\beta)^{o(1)}$, and for fixed $n$ it is polylogarithmic in $\beta$. Storjohann~\cite{s96} gives an earlier variant of LLL that returns the same basis as~\cite{lll82} for an integral basis with $\|b_i\|_2^2\leq B_0$ for all $i$, using $O(n^3\log B_0)$ arithmetic operations on integers of $O(n\log B_0)$ bits. If a number $B\geq 2^n$ bounds all $\|b_i\|_2^2$ and all Gram determinants $d_i$ of the leading basis vectors, the count is $O(n^2\log B)$ operations on integers of $O(\log B)$ bits, which is $O(n^3)$ only when $\log B=O(n)$. This count comes from the potential argument that we use in Lemma~\ref{lem:lll_exact}. Integrality gives $d_i\geq 1$, so the number of swaps is at most $\log_{4/3}\prod_{i<n}d_i$ for the input basis, which is $O(n\log B)$. Each step costs $O(n)$ operations.

We use neither bound. In exact arithmetic, our \textsc{LLL} is the algorithm of~\cite{s96} without its final size reduction. The analysis of~\cite{s96}, however, gives $O(n^3)$ operations only when the potential $P$ can drop by a factor of at most $2^{O(n^2)}$. For a call in our algorithm we can only bound this drop by $2^{O(n^3\log n)}$ (proof of Lemma~\ref{lem:lll_restart}), even for an integral input. The algorithm of~\cite{ns16} needs an integral basis, and its cost grows with the bit length of the input. The Gram matrices passed to \textsc{ApproxSV}, in contrast, are real, and this paper does not bound bit lengths. Theorem~\ref{thm:ns16} also compares $\|c_1\|_2$ with $(\det L)^{1/n}$ rather than with $\lambda_1(L)$, so as stated it is not an approximate shortest vector routine in the sense of Definition~\ref{def:approx_sv}. Instead, Lemma~\ref{lem:lll_restart} bounds the number of swaps of \textsc{LLL} in exact arithmetic, and Lemma~\ref{lem:svp_total} applies it to our algorithm. The routine \textsc{SegmentLLL} of Section~\ref{sec:segment_lll} is faster on our inputs (Lemma~\ref{lem:svp_segment}). Like \textsc{LLL}, it needs neither an integral input nor bounds on bit lengths. This is because its analysis compares the potential at the start of a call with the potential of the basis returned by the previous call, and not with the lower bound $d_i\geq 1$ that integrality gives.

\subsection{Convex Geometry}\label{sec:preli:convex_geometry}
In this section, we define notions about centroid and covariance. The work~\cite{j22} heavily exploits the structure of objects to design a subqudratic oracle complexity algorithm for minimizing convex functions. Our approach also relates to these notions.

For a convex body $K$, we use $\mathrm{vol}(K)$ to denote its volume, i.e., $\mathrm{vol}(K):=\int_{x \in K} \mathrm{d} x$. We first define the centroid of convex body,
\begin{Definition}
\label{def:cg}

Let $K\subseteq \R^n$ be a convex body. Let $g: K\rightarrow \R_+$ denote the uniform measure on convex body $K$. We define the \emph{centroid} of $K$ as 

\begin{align*}
    \mathrm{cg}(K) = & ~ \int_K g(x)\cdot x~\d x.
\end{align*}

Equivalently, we can write $\mathrm{cg}(K)$ as
\begin{align*}
    \mathrm{cg}(K) = & ~ \frac{1}{\mathrm{vol}(K)} \int_K x~\d x.
\end{align*}
\end{Definition}

Then, we define the covariance of convex body,
\begin{Definition}[Covariance of convex body, $\mathrm{Cov}(K)$]
\label{def:cov}

Let $K\subseteq \R^n$ be a convex body. We define the \emph{covariance matrix} of $K$ under uniform measure as

\begin{align*}
    \mathrm{Cov}(K) = & ~ \frac{1}{\mathrm{vol}(K)}\int_K (x-\mathrm{cg}(K))(x-\mathrm{cg}(K))^\top~\d x.
\end{align*}
\end{Definition}

It is well-known that any isotropic\footnote{A convex body is isotropic if the uniform distribution over the body has zero mean and covariance matrix being the identity.} convex body is enclosed by two balls.

\begin{lemma}[Ellipsoidal approximation of convex body, \cite{kls95}] 
\label{lem:kls_sandwich}
Let $K$ be an isotropic convex body in $\R^n$. Then, 
\begin{align*}
    \sqrt{\frac{n+2}{n}}\cdot B_2 \subseteq K \subseteq \sqrt{n(n+2)}\cdot B_2,
\end{align*}
where $B_2$ is the unit Euclidean ball in $\R^n$.

If $K$ is non-isotropic, then
\begin{align*}
    \sqrt{\frac{n+2}{n}}\cdot E({\rm cg}(K), {\rm Cov}(K)^{-1}) \subseteq K \subseteq \sqrt{n(n+2)}\cdot E({\rm cg}(K), {\rm Cov}(K)^{-1}).
\end{align*}
\end{lemma}
Only Lemma~\ref{lem:hessian_cov_close}, which our analysis does not use, applies this lemma.

\subsection{Lattice Projection}\label{sec:preli:lattice_projection}

We collect some standard facts of lattice projection that is directly implied by Gram-Schmidt process. We use $\Pi_W(\cdot)$ to denote the orthogonal projection onto the subspace $W$.

\begin{fact}[Lattice projection]\label{def:lattice_projection}
Let $\Lambda$ be a full rank lattice in $\R^n$ and $W$ be a linear subspace such that ${\rm dim}({\rm span}(\Lambda\cap W))={\rm dim}(W)$. Then
\begin{align*}
    \det(\Lambda) = & ~ \det(\Lambda\cap W)\cdot \det(\Pi_{W^\perp}(\Lambda)).
\end{align*}
\end{fact}

\begin{fact}[Dual of lattice projection]\label{def:dual_lattice_projection}
Let $\Lambda$ be a full rank lattice in $\R^n$ and $W$ be a linear subspace such that ${\rm dim}({\rm span}(\Lambda\cap W^\perp))={\rm dim}(W^\perp)$. Then
\begin{align*}
    (\Pi_W(\Lambda))^* = & ~ \Lambda^* \cap W.
\end{align*}
\end{fact}

\subsection{Slicing Lemma}\label{sec:preli:slicing_lemma}

We recall the high dimensional slicing lemma of~\cite{j21}, which Step~3 of the proof of Lemma~\ref{lem:main_oracle_formal} applies to an ellipsoid. The SODA 2024 version of this paper also proved a variant of it with the norm of $H_K^{-1}$ in place of that of $\mathrm{Cov}(K)$ (Lemma~\ref{lem:high_dimensional_slicing_lemma}). The analysis in this paper does not use that variant.

\begin{lemma}[Lemma 3.2 in \cite{j21}]
\label{lem:j21_slice}
Let $K$ be a convex body and $L$ be a full-rank lattice in $\R^n$. Let $W$ be an $(n-k)$-dimensional subspace of $\R^n$ such that ${\rm dim}(L\cap W)=n-k$. Then
\begin{align*}
    \frac{{\rm vol}(K\cap W)}{\det(L\cap W)} \leq & ~ \frac{{\rm vol}(K)}{\det(L)}\cdot \frac{k^{O(k)}}{\lambda_1(L^*, {\rm Cov}(K))^k}
\end{align*}
where $L^*$ is the dual lattice (see Definition~\ref{def:dual_lattice}) and $\lambda_1(L^*, {\rm Cov}(K))$ is the shortest nonzero vector in $L^*$ under the norm $\|\cdot \|_{{\rm Cov}(K)}$.
\end{lemma}

\subsection{Dimension Reduction Lemma}\label{sec:preli:dim_red}

The key building block of both~\cite{j21} and our algorithm is a lattice-dimension reduction step. The following result, the case of integral points of Lemma~3.1 of~\cite{j21}, shows that all integral points are preserved after a dimension reduction step. We include its short proof, which gives the hyperplane that our algorithm computes.

\begin{lemma}[Lemma 3.1 of~\cite{j21}]
\label{lem:dim_red}
Given an affine subspace $W=x_0+W_0$ where $W_0$ is a subspace of $\R^n$ and $x_0\in \R^n$ is some fixed point, and an ellipsoid $E=E(x_0, A)$ that has full rank on $W$. Given a vector $v\in \Pi_{W_0}(\Z^n)\setminus \{ {\bf 0}_n \}$ with $\|v\|_{A^{-1}}< 1/2$ then there exists a hyperplane $P\not \supseteq W$ such that $E\cap \Z^n\subseteq P\cap W$. In particular, if $z\in\Z^n$ satisfies $v=\Pi_{W_0}(z)$, then $P$ can be taken as
\begin{align*}
    P=\{y: v^\top y=(v-z)^\top x_0+[z^\top x_0]\},
\end{align*}
where $[t]$ denotes an integer nearest to $t$.
\end{lemma}

\begin{proof}
Here $A^{-1}$ is the inverse of $A$ on $W_0$. Since $v\in W_0\setminus\{{\bf 0}_n\}$, the function $y\mapsto v^\top y$ is not constant on $W$, so $P\not\supseteq W$. Also $E\cap\Z^n\subseteq W$, because $E\subseteq W$. Let $y\in E\cap\Z^n$, so that $y-x_0\in W_0$ and $\|y-x_0\|_A\leq1$. Since $z-v$ is orthogonal to $W_0$, we have $z^\top(y-x_0)=v^\top(y-x_0)$, and the Cauchy--Schwarz inequality on $W_0$ gives $|z^\top y-z^\top x_0|\leq\|v\|_{A^{-1}}\|y-x_0\|_A<1/2$. The number $z^\top y$ is an integer, so it is the only integer at distance less than $1/2$ from $z^\top x_0$, and $z^\top y=[z^\top x_0]$ for every choice of the nearest integer. Finally, $(z-v)^\top y=(z-v)^\top x_0$ for the same reason, so $v^\top y=z^\top y-(z-v)^\top x_0=(v-z)^\top x_0+[z^\top x_0]$, that is, $y\in P$.
\end{proof}
\section{Cutting Plane Method}\label{sec:cpm}

In Section~\ref{sec:cpm:log_barrier}, we introduce the definition of log-barrier and volumetric center. In Section~\ref{sec:cpm:leverage_score}, we introduce leverage score and related notations. In Section~\ref{sec:cpm:sandwiching_lemma}, we present the sandwiching lemma at the volumetric center, which our analysis uses only in the deterministic variant of Corollary~\ref{cor:exact_score}. Sections~\ref{sec:cpm:close_log_hessian_and_covariance} and~\ref{sec:cpm:slicing_lemma} keep two results of the SODA 2024 version of this paper, the closeness between the Hessian of the log barrier and the covariance matrix, and a slicing lemma in the norm of the inverse Hessian. The analysis in this paper does not use them. In Section~\ref{sec:cpm:hybrid}, we introduce the hybrid barrier, state its local properties, fix the parameters and bound the effect of adding or deleting one constraint. In Section~\ref{sec:cpm:method}, we present the leverage score maintenance and our cutting plane method. In Section~\ref{sec:cpm:hybrid_progress}, we analyze one block of the method. In Section~\ref{sec:cpm:main_lemma}, we present our main lemma. Section~\ref{sec:r3_proofs} contains the proofs.

\subsection{Log-barrier and Volumetric Center}\label{sec:cpm:log_barrier}

We start with defining the log-barrier, and it has been widely used in convex optimization \cite{r88,nn94,jklps20,hjstz22,jnw22,lsz+23,syz23,gsz23}.

\begin{Definition}[Log-barrier]\label{def:log_barrier}
Let $A\in \R^{m\times n}$ and $b\in \R^m$. Let $a_i^\top$ denote the $i$-th row of $A$. The \emph{log-barrier} is defined as
\begin{align*}
    \phi(x) = & ~ \sum_{i=1}^m -\ln (a_i^\top x-b_i)
\end{align*}
for $x\in \R^n$.
\end{Definition}

Let $K$ be the bounded full-dimensional polytope $K = \{ x: A x \geq b \}$ where $A \in \R^{m \times n}$, $b \in \R^m$ and $x \in \R^n$.

\begin{Definition}[Hessian and volumetric]\label{def:H}
Given $A \in \R^{m \times n}$ and $b \in \R^m$.
Let $H(x)$ be defined as
\begin{align*}
    H(x) = \sum_{i=1}^m \frac{ a_i a_i^\top }{ (a_i^\top x - b_i)^2 }
\end{align*}
where $a_i^\top$ denotes the $i$-th row of $A$.

Let (volumetric) function $F(x)$ be as
\begin{align*}
    F(x) := \frac{1}{2} \ln ( \det(H(x) ) )
\end{align*}
\end{Definition}

$H(x)$ is the Hessian of the logarithmic barrier function $\sum_{i=1}^m - \ln (a_i^\top x-b_i)$ and is positive definite for all $x$ in the interior of $K$.

\begin{Definition}[Volumetric center]\label{def:vc}
We define the volumetric center of $K$ as
\begin{align*}
    \mathrm{vc}(K) := & ~ \arg\min_{x\in {\rm int}(K)}~F(x).
\end{align*}
\end{Definition}
For $K$ in an affine subspace $W$, we take the interior relative to $W$, that is, we work in the coordinates of $W$.

Observe that $F$ is a convex function, hence one can run Newton-type algorithm to approximate $\vc(K)$ very fast.

\subsection{Leverage Score of Log Hessian}\label{sec:cpm:leverage_score}

We start with some definitions.

\begin{Definition}\label{def:sigma_x}
We define $\sigma_i(x)$ to be the $i$-th leverage score of matrix $H(x)$ as 

\begin{align*}
    \sigma_i(x) = & ~ \frac{a_i^\top H(x)^{-1} a_i}{(a_i^\top x-b_i)^2}, ~~~ \forall i \in [ m ].
\end{align*}
\end{Definition}

Using $\sigma_i$, we can write the gradient of $F$ in an convenient form:

\begin{align*}
    \nabla F(x) = & ~ -\sum_{i=1}^m \sigma_i(x)\frac{a_i}{a_i^\top x-b_i}.
\end{align*}

\begin{Definition}\label{def:Q}
We define $Q(x)$ as

\begin{align*}
    Q(x) = & ~ \sum_{i=1}^m \sigma_i(x)\frac{a_ia_i^\top}{(a_i^\top x-b_i)^2}.
\end{align*}
where $\sigma_i(x)$ is defined as Definition~\ref{def:sigma_x}.
\end{Definition}

It is well-known that $Q(x)$ is a good approximation of $\nabla^2 F(x)$:

\begin{lemma}[Lemma 3 of~\cite{v89}]
\label{lem:v89_3}
For any $x$ in the interior of $K$, we have
\begin{align*}
    Q(x) \preceq \nabla^2 F(x) \preceq 5Q(x).
\end{align*}
\end{lemma}

We use a sharper form of this bound. Lemma~\ref{lem:hybrid_hessian} in Section~\ref{sec:r3_proofs} computes
\begin{align*}
    \nabla^2F(x)=A^\top S_x^{-1}(3\Sigma_0(x)-2P(x)\circ P(x))S_x^{-1}A,
\end{align*}
where $S_x={\rm diag}(a_i^\top x-b_i)$, $\Sigma_0(x)={\rm diag}(\sigma(x))$, $P(x)$ is the $m\times m$ matrix with entries
\begin{align*}
    a_i^\top H(x)^{-1}a_j/((a_i^\top x-b_i)(a_j^\top x-b_j))
\end{align*}
and $\circ$ is the entrywise product. This matrix is unrelated to the hyperplane $P$ of a dimension reduction step. In Lemma~\ref{lem:hybrid_hessian}, in the proof of Lemma~\ref{lem:hybrid_im} and in the subsection on Vaidya's method in Section~\ref{sec:r3_proofs}, a bare $P$ denotes this matrix at the point under consideration. Lemma~\ref{lem:hybrid_hessian} also shows that $Q(x)\preceq\nabla^2F(x)\preceq3Q(x)$. This bound is due to Anstreicher~\cite[Theorem~A.4]{a96}. See also~\cite[Theorem~4.1]{a00} for its semidefinite generalization, and Lemma~6 of the arXiv version 1508.04874v2 of~\cite{lsw15}. The analysis of the hybrid barrier in Section~\ref{sec:cpm:hybrid} uses the constant $3$, not the constant $5$ of Lemma~\ref{lem:v89_3}.

Finally, we define $\mu(x)$ which quantifies:
\begin{Definition}
\label{def:mu_x}
Let $\mu(x)$ be the largest number $\lambda$ such that
\begin{align*}
    Q(x) \succeq & ~ \lambda H(x).
\end{align*}
\end{Definition}

The following lemma provides an upper bound on $\mu(x)$:

\begin{lemma}[Lemma 4 of~\cite{v89}]
\label{lem:v89_4}

For any $x$ in the interior of $K$, we have
\begin{align*}
    \frac{1}{4m} \leq \mu(x) \leq 1.
\end{align*}
Further, we have
\begin{align*}
    \mu(x) \geq & ~ \min_{i\in [m]}\{\sigma_i(x) \}.
\end{align*}
\end{lemma}

\subsection{Sandwiching Lemma}\label{sec:cpm:sandwiching_lemma}

We derive some sandwiching conditions regarding the ellipsoid induced by the Hessian of log barrier. 

\begin{lemma}[Sandwich convex body via log Hessian]
\label{lem:polytope_sandwich}
Let $K=\{x: Ax\geq b \}$ be a polytope where $A\in \R^{m\times n}$ and $b\in \R^m$. Let $H(x)$ be defined as Definition~\ref{def:H}. Then for any $x$ in the interior of $K$, we have that
\begin{align*}
    E(x, H(x)) \subseteq  & ~K ,
\end{align*}
and the following upper bound
\begin{align*}
    K\subseteq & ~ 90m^{3/4}n\cdot E({\rm vc}(K), H({\rm vc}(K))).
\end{align*}
\end{lemma}

\begin{proof}
Let us first consider the ellipsoid contained in $K$. By definition, we have that
\begin{align*}
    E(x, H(x)) = & ~ \{y: (y-x)^\top H(x) (y-x)\leq 1 \},
\end{align*}
without loss of generality assume $x=0$, then $y\in E(x, H(x))$ means

\begin{align*}
    \sum_{i=1}^m \frac{(a_i^\top y)^2}{b_i^2} \leq & ~ 1,
\end{align*}

which means that for any $i\in [m]$, it holds that
\begin{align*}
    (a_i^\top y)^2 \leq & ~ b_i^2.
\end{align*}

Since $x\in K$, we must have that $b_i\leq 0$ for all $i\in [m]$. Hence, the square condition only requires that $|a_i^\top y|\leq |b_i|$, so if $a_i^\top y\geq 0$, then clearly $y\in K$. Otherwise due to the absolute value constraint, it must be the case that $a_i^\top y\geq b_i$. This concludes the proof of $E(x, H(x))\subseteq K$.

For the ellipsoid that contains $K$, we note that $225\sqrt{m}F$ is a self-concordant barrier with complexity parameter $\nu=225\sqrt{m}n$~\cite{v89,av93,a97} and scaling does not change the minimizer. A $\nu$-self-concordant barrier $f$ with minimizer $x_f$ satisfies $K\subseteq (\nu+2\sqrt{\nu})\cdot E(x_f,\nabla^2 f(x_f))\subseteq 3\nu\cdot E(x_f,\nabla^2 f(x_f))$. Also, $E(x,cM)=c^{-1/2}\cdot E(x,M)$ for $c>0$. Therefore, we have
\begin{align*}
    K \subseteq & ~ 45m^{1/4}n\cdot E({\rm vc}(K), \nabla^2 F(\vc(K))),
\end{align*}
recall that $\nabla^2 F(\vc(K))\succeq \mu(\vc(K))\cdot H(\vc(K))$ (due to Lemma~\ref{lem:v89_3} and Definition~\ref{def:mu_x}) and $\frac{1}{4m}\leq \mu(\vc(K))\leq 1$ (See Lemma~\ref{lem:v89_4}), we conclude that
\begin{align*}
    \frac{1}{4m}\cdot H(\vc(K)) \preceq & ~ \nabla^2 F(\vc(K)),
\end{align*}
thus, $E(\vc(K),\nabla^2 F(\vc(K)))\subseteq 2\sqrt{m}\cdot E(\vc(K),H(\vc(K)))$, and we conclude that $K\subseteq 90m^{3/4}n\cdot E(\vc(K),H(\vc(K)))$.
\end{proof}

\subsection{Closeness of Log Hessian and Covariance}\label{sec:cpm:close_log_hessian_and_covariance}

This subsection and the next keep two results of the SODA 2024 version of this paper. The analysis in this paper does not use them. Lemma~\ref{lem:kls_sandwich} and Lemma~\ref{lem:polytope_sandwich} together imply the spectral approximation between the matrix $H({\rm vc}(K))$ (See Definition~\ref{def:H}) and ${\rm Cov}(K)$ (See Definition~\ref{def:cov}).

\begin{lemma}[Closeness of log Hessian and covariance]
\label{lem:hessian_cov_close}

Let $K=\{x: Ax\geq b\}$ be a polytope for $A\in \R^{m\times n}$ and $b\in \R^m$. Then
\begin{align*}
   \frac{1}{8100m^{1.5}n^2}\cdot H({\rm vc}(K))\preceq {\rm Cov}(K)^{-1} \preceq 4n^2\cdot H({\rm vc}(K)).
\end{align*}
\end{lemma}

\begin{proof}
The proof relies on two sandwiching lemmas. 

By Lemma~\ref{lem:kls_sandwich}, we have that
\begin{align*}
    E({\rm cg}(K), {\rm Cov}(K)^{-1}) \subseteq K \subseteq 2n\cdot E({\rm cg}(K), {\rm Cov}(K)^{-1}),
\end{align*}
by Lemma~\ref{lem:polytope_sandwich},
\begin{align*}
    E({\rm vc}(K), H({\rm vc}(K))) \subseteq K \subseteq 90m^{3/4}n\cdot E({\rm vc}(K), H({\rm vc}(K))).
\end{align*}
We thus have
\begin{align*}
    E({\rm cg}(K), {\rm Cov}(K)^{-1})\subseteq & ~ 90m^{3/4}n\cdot E(\vc(K), H(\vc(K))), \\
    E(\vc(K), H(\vc(K))) \subseteq & ~ 2n\cdot E({\rm cg}(K), {\rm Cov}(K)^{-1}).
\end{align*}
Without loss of generality, let's prove one side containment. Given
\begin{align*}
    E({\rm cg}(K), {\rm Cov}(K)^{-1}) \subseteq & ~ 90m^{3/4}n\cdot E({\rm vc}(K), H(\vc(K))),
\end{align*}
we note that re-centering and making their centers the same does not change the containment relation, therefore, we have the following:
\begin{align*}
    E({\rm Cov}(K)^{-1})\subseteq & ~ 90m^{3/4}n\cdot E(H(\vc(K))),
\end{align*}
this directly implies the Loewner ordering
\begin{align*}
    \frac{1}{8100m^{1.5}n^2} \cdot H(\vc(K)) \preceq & ~ {\rm Cov}(K)^{-1}
\end{align*}

Following a similar approach, we can show that 
\begin{align*}
    {\rm Cov}(K)^{-1} \preceq & ~ 4n^2 \cdot H(\vc(K))
\end{align*}
this completes the proof.
\end{proof}

\subsection{High Dimensional Slicing Lemma}\label{sec:cpm:slicing_lemma}

The SODA 2024 version of this paper proved the following variant of Lemma~\ref{lem:j21_slice}, which uses the Hessian of the log barrier instead of the covariance matrix. It incurs an extra $(mn)^{O(k)}$ factor. The analysis in this paper does not use it. Instead, it applies Lemma~\ref{lem:j21_slice} directly to an ellipsoid, whose covariance matrix is a multiple of the inverse Hessian.

\begin{lemma}[High dimensional slicing lemma,  volumetric version]\label{lem:high_dimensional_slicing_lemma}
Let $K=\{x:Ax\geq b\}$ be a bounded polytope with $m$ constraints, let $H_K$ denote the Hessian matrix of log barrier of $K$ at its volumetric center. Let $L$ be a full-rank lattice in $\R^n$. Let $W$ be an $(n-k)$-dimensional subspace of $\R^n$ such that $\mathrm{dim}( L \cap W ) = n -k$. Then we have
\begin{align*}
    \frac{ \mathrm{vol}(K\cap W) }{ \det(L \cap W) } \leq \frac{ \mathrm{vol}(K) }{ \det(L) } \cdot \frac{ k^{O(k)}\cdot (mn)^{O(k)} }{ \lambda_1 (L^*,H_K^{-1})^k }
\end{align*}
where $L^*$ is the dual lattice and $\lambda_1(L^*, H_K^{-1})$ is the shortest nonzero vector in $L^*$ under the norm $\|\cdot \|_{H_K^{-1}}$.
\end{lemma}
\begin{proof}

Let $v$ denote the shortest nonzero vector in $L^*$ under the norm $\|\cdot \|_{{\rm Cov}(K)}$ and $u$ be the shortest nonzero vector in $L^*$ under the norm $\|\cdot \|_{H_K^{-1}}$, by Lemma~\ref{lem:hessian_cov_close}, we know that 
\begin{align*} 
\frac{1}{4n^2}\cdot H_K^{-1}\preceq {\rm Cov}(K)\preceq 8100m^{1.5}n^2\cdot H_K^{-1}.
\end{align*}
We have
\begin{align*}
    \|v\|_{{\rm Cov}(K)} \geq & ~ \frac{1}{2m^{1.5}n}\cdot \|v\|_{H_K^{-1}} \\
    \geq & ~ \frac{1}{2m^{1.5}n} \cdot\|u\|_{H_K^{-1}},
\end{align*}
therefore, we have that 
\begin{align*}
    \frac{1}{\lambda_1(L^*, {\rm Cov}(K))} \leq & ~ \frac{2m^{1.5}n}{\lambda_1(L^*, H_K^{-1})}.
\end{align*}
Using Lemma~\ref{lem:j21_slice}, we have
\begin{align*}
    \frac{{\rm vol}(K\cap W)}{\det(L\cap W)} \leq & ~ \frac{{\rm vol}(K)}{\det(L)}\cdot \frac{k^{O(k)}}{\lambda_1(L^*, {\rm Cov}(K))^k}
\end{align*}
Finally, combining the above equations we conclude
\begin{align*}
    \frac{ \mathrm{vol}(K\cap W) }{ \det(L \cap W) } \leq \frac{ \mathrm{vol}(K) }{ \det(L) } \cdot \frac{ k^{O(k)}\cdot (mn)^{O(k)} }{ \lambda_1 (L^*,H_K^{-1})^k } .
\end{align*}
Thus, we complete the proof.
\end{proof}
\subsection{The Hybrid Barrier}\label{sec:cpm:hybrid}

The data structure of~\cite{jlsw20} estimates leverage scores with an additive $\ell_2$ error of $1/\poly(\log n)$. Vaidya's Newton steps weigh each row by its leverage score. So this error is harmless only if every score stays above an absolute constant, which is not proved during runs of deletions (Remark~\ref{rem:jlsw20_floor}). We therefore adapt the hybrid barrier of Section~5.2 of~\cite{lsw15} to the framework of~\cite{jlsw20}, without the weight corrections and the regularization of~\cite{lsw15}. The method minimizes $p=F+cL$, the volumetric barrier plus the logarithmic barrier with the weight $c=\Theta(1/\log n)$. Every row then has weight at least $c$ in the matrix $Q_c$ of Definition~\ref{def:hybrid}, so an additive error in the scores is a small relative error on every weight. The price is $N=O(\log\log n)$ damped Newton steps after each added or deleted row. Combinations of these two barriers go back to the path-following methods of~\cite{av93,a97}.

\paragraph{Setting.} From now on, $K=\{x\in W:a_i^\top x\geq b_i,\ i\in[m]\}$, where $W\subseteq\R^n$ is an affine subspace with linear part $W_0$ and $d=\dim(W)$. For a point $z$ in the relative interior of $K$, we write $s_i(z)=a_i^\top z-b_i$ and $S_z={\rm diag}(s(z))$, and we regard $H(z)=\sum_ia_ia_i^\top/s_i(z)^2$ as a quadratic form on $W_0$. We fix a basis of $W_0$ and identify linear and quadratic forms on $W_0$ with their coordinate vectors and matrices, so $a_i$ also denotes the restriction of $a_i^\top$ to $W_0$. When $H(z)$ is positive definite on $W_0$, we define $\sigma_i(z)$, $Q(z)$ and $\mu(z)$ as in Section~\ref{sec:cpm:leverage_score}, with the inverse of $H(z)$ on $W_0$. Then $\sum_i\sigma_i(z)=d$ by Fact~\ref{fac:leverage_score_sum}. We define $F(z)=\frac12\ln\det H(z)$ with respect to an orthonormal basis of $W_0$, so that
\begin{align}\label{eq:vol_dikin}
    {\rm vol}(E(z,H(z)))={\rm vol}(B_d)\,e^{-F(z)},
\end{align}
where $B_d$ is the unit ball of dimension $d$ and ${\rm vol}$ is the $d$-dimensional volume in $W$. Let $L(x)=-\sum_{i=1}^m\ln s_i(x)$ be the log-barrier of Definition~\ref{def:log_barrier}. When several polytopes occur, a subscript names the polytope, as in $p_K$, $F_K$, $\lambda_K$ and $H_K(z)$. From now on, $H_K$ without an argument denotes the matrix $H_K(x_K)$ that Algorithm~\ref{alg:ours} keeps.

\begin{Definition}[Hybrid barrier]\label{def:hybrid}
Let $c>0$. On the relative interior of $K$, let $p:=F+c\,L$. For $z$ such that $H(z)$ is positive definite on $W_0$, let $\Sigma_c(z):={\rm diag}(\sigma(z)+c{\bf 1}_m)$ and
\begin{align*}
    Q_c(z):=Q(z)+c\,H(z)=A^\top S_z^{-1}\Sigma_c(z)S_z^{-1}A=\sum_{i=1}^m(\sigma_i(z)+c)\frac{a_ia_i^\top}{s_i(z)^2}.
\end{align*}
Let $\mu_c(z):=\mu(z)+c$, which is the largest number $t$ with $Q_c(z)\succeq t\,H(z)$ on $W_0$. The \emph{decrement} of $p$ at $z$ is $\lambda(z):=\|\nabla p(z)\|_{Q_c(z)^{-1}}$, where $\nabla p(z)=-\sum_i(\sigma_i(z)+c)a_i/s_i(z)$ is the gradient along $W$. The \emph{hybrid center} $x_c(K)$ is the minimizer of $p$ over the relative interior of $K$, when it exists. For $t\in(0,1)$, let $\mathcal{S}(z,t):=\{y\in W:|s_i(y)/s_i(z)-1|\leq t\ \text{for all}\ i\in[m]\}$.
\end{Definition}

Since $Q(z)\succeq(\min_i\sigma_i(z))H(z)$, we have $\mu(z)\geq\min_i\sigma_i(z)\geq0$ and $\mu_c(z)\geq c$. From now on, we fix the radius $r:=1/10$. For $t\in(0,r]$, let
\begin{align*}
    \kappa(t):=\Big(\frac{1+t}{1-t}\Big)^2,\qquad\alpha(t):=\frac{(1-t)^2}{(1+t)^4}\qquad\text{and}\qquad\alpha'(t):=\frac{(1+t)^2}{(1-t)^4},
\end{align*}
and let $\alpha:=\alpha(r)$, $\alpha':=\alpha'(r)$ and $\beta:=3\alpha'$. On the set $\mathcal{S}(z,t)$, leverage scores change by at most the factor $\kappa(t)$, and $Q_c$ changes by factors between $\alpha(t)$ and $\alpha'(t)$ (Part~4 of the next lemma). The constants of the next two lemmas do not depend on $c$, and Section~\ref{sec:r3_local} proves the lemmas of this section.

\begin{lemma}[Local facts]\label{lem:hybrid_local}
Let $c>0$, and let $z$ be a relative interior point of $K$ such that $H(z)$ is positive definite on $W_0$. Write $s_i$, $H$, $\sigma_i$, $Q_c$ and $\mu_c$ for these quantities at $z$.
\begin{enumerate}
    \item $Q_c\preceq\nabla^2p(z)\preceq3Q(z)+cH\preceq3Q_c$. In particular, $p$ is strictly convex on the relative interior of $K$.
    \item For every $e\in\R^m$, $\|A^\top S_z^{-1}e\|_{Q_c^{-1}}^2\leq\sum_ie_i^2/(\sigma_i+c)\leq\|e\|_2^2/c$.
    \item For every $k\in[m]$, $s_k^{-2}\,a_k^\top Q_c^{-1}a_k\leq\min\{1/(\sigma_k+c),\sigma_k/\mu_c\}\leq\mu_c^{-1/2}$. Hence, for every $h\in W_0$, we have $|a_k^\top h|/s_k\leq(\sigma_k/\mu_c)^{1/2}\|h\|_{Q_c}$ and $|a_k^\top h|/s_k\leq\mu_c^{-1/4}\|h\|_{Q_c}$.
    \item Let $r'\in(0,r]$ and $y\in\mathcal{S}(z,r')$. Then $(1+r')^{-2}H\preceq H(y)\preceq(1-r')^{-2}H$, $\kappa(r')^{-1}\sigma_i\leq\sigma_i(y)\leq\kappa(r')\sigma_i$ for every $i$, $\alpha(r')Q_c\preceq Q_c(y)\preceq\alpha'(r')Q_c$, $\mu_c(y)\geq((1-r')/(1+r'))^4\mu_c$, $\mu(y)\geq((1-r')/(1+r'))^4\mu(z)$, and $\alpha(r')Q_c\preceq\nabla^2p(y)\preceq3\alpha'(r')Q_c$.
\end{enumerate}
\end{lemma}

The next lemma describes $p$ near a point $z_s$ of small decrement. Its constants $\gamma_0$, $c_0$, $r_{\rm N}$, $q_0$, $C_{\rm r}$, $C_{\rm x}$, $C_\lambda$ and $C_{\rm e}$ are absolute, and Section~\ref{sec:r3_params} fixes their values.

\begin{lemma}[Location, existence and recentering]\label{lem:hybrid_loc}
Let $c>0$ and let $z_s$ be a relative interior point of $K$ such that $H(z_s)$ is positive definite on $W_0$. We do not assume that $K$ is bounded. Write $\mu_s=\mu_c(z_s)$, $\lambda_s=\lambda(z_s)$ and $Q_0=Q_c(z_s)$.
\begin{enumerate}
    \item Suppose that $\lambda_s<(\alpha/2)\,r'\mu_s^{1/4}$ for some $r'\in(0,r]$. Then the hybrid center $x_c=x_c(K)$ exists, is unique and lies in $\mathcal{S}(z_s,r')$. Moreover, $\|x_c-z_s\|_{Q_0}\leq(2/\alpha)\lambda_s$ and $p(z_s)-p(x_c)\leq(2/\alpha)\lambda_s^2$. In particular, $x_c\in\mathcal{S}(z_s,C_{\rm r}\lambda_s\mu_s^{-1/4})$ whenever $C_{\rm r}\lambda_s\mu_s^{-1/4}\leq r$. Finally, $K$ is bounded.
    \item Let $\widehat Q$ be a positive definite matrix with $(1-\gamma_Q)Q_0\preceq\widehat Q\preceq(1+\gamma_Q)Q_0$ for some $\gamma_Q\in[0,\gamma_0]$. Let $y_0=z_s$ and $y_{k+1}=y_k-\beta^{-1}\widehat Q^{-1}(\nabla p(y_k)+e_k)$ for $k\geq0$, where the vectors $e_k$ satisfy $\|e_k\|_{Q_0^{-1}}\leq\epsilon_g$ whenever $y_k\in\mathcal{S}(z_s,r)$. If $\lambda_s\leq c_0\mu_s^{1/4}$ and $\epsilon_g\leq2c_0\mu_s^{1/4}$, then for every $k\geq0$ we have $y_k\in\mathcal{S}(z_s,r_{\rm N})$,
    \begin{align*}
        \|y_k-x_c\|_{Q_0}\leq q^k\cdot\frac{2}{\alpha}\lambda_s+C_{\rm x}\epsilon_g\qquad\text{and}\qquad\lambda(y_k)\leq C_\lambda q^k\lambda_s+C_{\rm e}\epsilon_g,
    \end{align*}
    where $q:=1-\alpha/\beta+\gamma_Q/(1-\gamma_Q)\leq q_0$.
    \item Let $y\in\mathcal{S}(z_s,r)$, let $\wt\sigma\in\R^m$ satisfy $\|\wt\sigma-\sigma(y)\|_2\leq\delta$, and let $\wt g=-\sum_i(\wt\sigma_i+c)a_i/s_i(y)$. Then $\|\wt g-\nabla p(y)\|_{Q_0^{-1}}\leq\alpha'^{1/2}\delta c^{-1/2}$.
    \item (Slacks at the hybrid center.) Suppose only that $x_c=x_c(K)$ exists. Then $K$ is bounded, and every $y\in K$ satisfies
    \begin{align*}
        s_i(y)\leq\Big(1+\frac{d+cm}{\sigma_i(x_c)+c}\Big)s_i(x_c)\leq\Big(1+\frac{n+cm}{c}\Big)s_i(x_c)\qquad\text{for every}\ i\in[m].
    \end{align*}
\end{enumerate}
\end{lemma}

In words, the hybrid center lies within distance $O(\lambda_s)$ of $z_s$ (Part~1). Damped Newton steps with a fixed matrix close to $Q_0$ shrink the distance to $x_c$ by the factor $q<1$ per step, up to an additive $O(\epsilon_g)$ (Part~2). Estimated scores cause gradient errors of size $O(\delta c^{-1/2})$ (Part~3), and Part~4 bounds every slack in $K$ by a multiple of its value at $x_c$.

\paragraph{Parameters.}\label{sec:cpm:params} The method uses three absolute constants, the leverage $w$ of an added cut, the threshold $\theta:=w/10$ below which it deletes a row, and the accuracy $\epsilon_{\rm im}$ of its inverse maintenance. The other parameters depend only on $n$:
\begin{itemize}
    \item $c:=w/(20\ln(n/w))=\Theta(1/\log n)$, the weight of the logarithmic barrier,
    \item $\lambda^*=\Theta(c^{3/4})$, a bound on the decrement at the points where the method calls $\SO$,
    \item $\delta_0=\Theta(c^{5/4})$, a bound on the $\ell_2$ error of the estimated leverage scores,
    \item $\bar m:=\lfloor n/(\theta-\delta_0)\rfloor+1=O(n)$, a bound on the number of rows, and $\bar\kappa:=1+(n+c\bar m)/c=O(n\log n)$,
    \item $\bar R$ and $R'$, both $O(n\log n)$, which bound the size of a polytope around its hybrid center and around the point that a block returns,
    \item $T=\Theta(n\log n)$, the number of steps of one block, and $N=O(\log(1/c))=O(\log\log n)$, the number of Newton steps after each added or deleted row.
\end{itemize}
The tests of Algorithm~\ref{alg:hcpm} use three more absolute constants, the thresholds $\tau_{\rm del}$ and $\tau_{\rm step}$ and the number $t_{\max}$ of pieces of a Newton step. Section~\ref{sec:r3_params} fixes all values and lists every inequality between them that the proofs use, as the conditions~\ref{cond:C1} to~\ref{cond:C7} and~\ref{cond:W1} to~\ref{cond:W6}. We often use
\begin{align}\label{eq:R4}
    c\ln(2\bar\kappa)\leq\frac{w}{10},
\end{align}
which holds for every $n\geq1$, because $2\bar\kappa\leq(n/w)^2$ and $c=w/(20\ln(n/w))$ (condition~\ref{cond:C3}). By Part~4 of Lemma~\ref{lem:hybrid_loc}, if $K$ has at most $\bar m$ rows and $x_c(K)$ exists, then $s_i(y)\leq\bar\kappa\,s_i(x_c(K))$ for every $y\in K$ and every row $i$.

The number $N$ is of order $\log(1/c)$ for the following reason. After an add or a delete, the decrement at the current point is $O(w)$ (Lemma~\ref{lem:hybrid_pert}), and the analysis needs it back at most $\lambda^*$ at the next current point. Lemma~\ref{lem:hybrid_loc} needs the decrement to be small compared with $\mu_c^{1/4}$, and we only know that $\mu_c\geq c$, so $\lambda^*$ is a power of $c$. Each damped Newton step lowers the bound on the decrement by the factor $q\leq q_0<1$ (Part~2 of Lemma~\ref{lem:hybrid_loc}). So $N=O(\log(w/\lambda^*))=O(\log(1/c))$ steps suffice, which is $O(\log\log n)$.

The next lemma bounds the effect of adding or deleting one row at a fixed point. Its constants are absolute and fixed in Section~\ref{sec:r3_params}. The constants $\theta_{\rm add}$ and $\theta'_{\rm add}$ are slightly below $\theta$, $w_{\rm add}$ is slightly above $w$, and $\epsilon_{\rm add}$, $\sigma_{\rm del}$, $\epsilon_{\rm del}$ and $\nu_{\rm del}$ are small. Later proofs use only Items~(i) and~(iii) of the add case, Items~(ii) and~(iii) of the delete case, and the positive definiteness of $H_{K_-}(z)$.

\begin{lemma}[Adding or deleting one constraint]\label{lem:hybrid_pert}
Let $z$ be a relative interior point of $K$ such that $H=H(z)$ is positive definite on $W_0$, and write $\sigma_i$, $\mu$, $Q_c$ and $\lambda$ for these quantities at $z$. Quantities of the new polytope carry a subscript $+$ or $-$.
\begin{enumerate}
    \item (Add.) Suppose $\mu\geq\theta_{\rm add}$. Let $K_+$ be $K$ with an added row $a^\top x\geq b'$ such that $s_a:=a^\top z-b'>0$ and $w':=s_a^{-2}a^\top H^{-1}a\leq w_{\rm add}$, and let $\sigma':=w'/(1+w')$. Then
    \begin{itemize}
        \item[(i)] $\mu_{c,+}(z)\geq(\mu+c)/(1+w')^2\geq\theta'_{\rm add}$,
        \item[(ii)] $\lambda_{K_+}(z)\leq(1+w')^{1/2}\lambda+w'+(\sigma'+c)(\sigma'/\mu_{c,+}(z))^{1/2}$, and
        \item[(iii)] $\lambda_{K_+}(z)\leq(1+\epsilon_{\rm add})\lambda+C_+w$.
    \end{itemize}
    \item (Delete.) Let $j\in[m]$ with $\sigma_j\leq\sigma_{\rm del}$, and suppose that $\mu\geq(1-\epsilon_{\rm del})\sigma_j-\nu$ for some $\nu\in[0,\nu_{\rm del}c]$. Let $K_-$ be $K$ without row $j$. Then $H_{K_-}(z)$ is positive definite on $W_0$, and
    \begin{itemize}
        \item[(i)] $Q_{c,-}(z)\succeq(1-C_Q\sigma_j)Q_c$,
        \item[(ii)] $\mu_{c,-}(z)\geq c_\mu(\sigma_j+c)$, and
        \item[(iii)] $\lambda_{K_-}(z)\leq(1-C_Q\sigma_j)^{-1/2}\lambda+J$, where $J:=C_-\sigma_j+C'_-(\sigma_jc)^{1/2}$.
    \end{itemize}
\end{enumerate}
\end{lemma}

In words, adding a row of leverage about $w$ at a point where $\mu$ is of order $\theta$ raises the decrement by $O(w)$ and keeps $\mu_c$ of order $\theta$. Deleting a row of leverage $\sigma_j$ raises the decrement by $O(\sigma_j+(\sigma_jc)^{1/2})$ and keeps $\mu_c$ of order $\sigma_j+c$.

\subsection{The Cutting Plane Method}\label{sec:cpm:method}

\paragraph{Leverage scores.} The method uses the data structure of~\cite{jlsw20} to estimate leverage scores. We restate its operations and costs as they are stated there.

\begin{lemma}[Theorem~5.1 of~\cite{jlsw20}]\label{lem:jlsw_main}
Given an initial matrix $A\in\R^{m\times n}$ with $m=O(n)$ and an initial weight vector $\eta\in\R^m_{>0}$, there is a randomized data structure that approximately maintains the leverage scores
\begin{align*}
    \sigma_i(\eta)=\big({\rm diag}(\eta)^{1/2}A(A^\top{\rm diag}(\eta)A)^{-1}A^\top{\rm diag}(\eta)^{1/2}\big)_{i,i},\qquad i\in[m],
\end{align*}
through the following operations.
\begin{itemize}
    \item \textsc{\textup{Init}}$(A,\eta)$ takes $O(n^{\omega+o(1)})$ arithmetic operations.
    \item \textsc{\textup{Update}} performs one of the following changes of the current pair $(A,\eta)$. It inserts a row $u$ with weight $\eta_u$, or it deletes a row $u$ with weight $\eta_u$, where in both cases $\eta_uuu^\top\preceq0.01A^\top{\rm diag}(\eta)A$ for the pair before the change. Or it replaces $\eta$ by $\eta'$, with the same rows, such that $\|\log\eta'-\log\eta\|_2\leq0.01$.
    \item \textsc{\textup{Query}} takes $O(n)$ arithmetic operations and outputs a vector $\wt\sigma\in\R^m$.
\end{itemize}
If the number of rows stays $O(n)$, each \textsc{\textup{Update}} takes amortized $O(n^2)$ arithmetic operations.
\end{lemma}

Theorem~5.1 of~\cite{jlsw20} also states that the outputs of \textsc{Query} can be made to have $\ell_2$ error $O(\log^{-\ell}n)$ for every constant $\ell>0$. Its proof treats $\ell=15$ and a fixed sequence of updates, and it bounds the error only in second moment. So it neither bounds every output nor covers updates that depend on earlier outputs, such as ours and those of the SODA 2024 version of this paper. Lemma~\ref{lem:jlsw_adaptive} in Section~\ref{sec:r3_adaptive} gives what we use. Let the operations be valid and depend on earlier outputs, and let $M$ be a deterministic number. Then for every constant $k'\geq1$ and every large enough $n$, with probability at least $1-3Mn^{-k'}$, every output among the first $M$ operations has $\ell_2$ error at most $\delta_{\rm J}(n)=O(\log^{-15}n)$. This is at most $\delta_0=\Theta(\log^{-5/4}n)$ for $n\geq n_0$, where $n_0$ is an absolute constant that exists but is not explicit (Section~\ref{sec:r3_params}, Step~5). For $n<n_0$, the method computes exact leverage scores, with $O(n^\omega)=O(1)$ operations per query.

\paragraph{Inverse maintenance.} The method also keeps two instances of the inverse maintenance of Theorem~B.4 of~\cite{jlsw20}, a variant of its Theorem~B.1 taken from~\cite{cls19}. An instance with accuracy $\epsilon_{\rm im}$ keeps weights $\wt\eta$ with $(1-\epsilon_{\rm im})\wt\eta_i\leq\eta_i\leq(1+\epsilon_{\rm im})\wt\eta_i$ for the current weights $\eta$, and the inverse of $A^\top{\rm diag}(\wt\eta)A$. If the number of rows stays $O(n)$, its initialization takes $O(n^{\omega+o(1)})$ operations and inserting or deleting a row takes $O(n^2)$. If every call of \textsc{Update} moves $\log\eta$ by at most $C$ in the $\ell_2$ norm, then for every constant $c''>0$, a call of \textsc{Update} takes amortized $O(n^2+Cn^2/(\epsilon_{\rm im}\log^{c''}n))$ operations. The instance \textsc{IM}$_H$ has the weights $s_i(z)^{-2}$ at the point $z$ where it sits, and $\widehat H^{-1}$ denotes the inverse that it keeps. Whatever the history of the instance, the bounds on $\wt\eta$ give
\begin{align}\label{eq:im_inverse}
    (1-\epsilon_{\rm im})H(z)^{-1}\preceq\widehat H^{-1}\preceq(1+\epsilon_{\rm im})H(z)^{-1}\quad\text{on }W_0.
\end{align}
The instance \textsc{IM}$_Q$ has the weights $(\wt\sigma_i+c)s_i(z)^{-2}$.

\paragraph{Coordinates and padding.} The subspace $W$ changes only between blocks. At the start of a block we fix a point $\bar x\in W$ and a matrix $Y\in\R^{n\times d}$ whose columns form a basis of $W_0$. The three data structures run on the rows $(Y^\top a_i)^\top t\geq b_i-a_i^\top\bar x$ in the coordinates $x=\bar x+Yt$, padded with $n-d$ dummy rows of weight $1$. So every instance has $n$ columns and at most $\bar m+n=O(n)$ rows, and Lemmas~\ref{lem:jlsw_main} and~\ref{lem:jlsw_adaptive} and the bounds on inverse maintenance apply with the parameter $n$, as Section~\ref{sec:r3_local} shows.

\paragraph{Cut vectors.} Recall from the introduction that the separation oracle either answers YES, only at a minimizer of $f$, or returns a vector $a$. This vector satisfies $a^\top y\geq a^\top z$ for every minimizer $y$, strictly when $z$ does not minimize $f$. A subgradient oracle of a finite $f$ has this property. Let $g$ be a subgradient at $z$ and $a=-g$. Then $a^\top(y-z)\geq f(z)-f(y)$, which is positive when $z$ does not minimize $f$. So does the oracle of Corollary~\ref{cor:sfm_main} (proof of that corollary). The set $K^*$ is closed and lies in $B_\infty(R)$, so it is compact.

We fix $x^*$ to be the lexicographically largest point of $K^*$. It is an extreme point of $K^*$, hence an integral minimizer of $f$ (Section~\ref{sec:r3_local}). Only the analysis uses $x^*$, and the method never computes it. Let $z\in W$, where $\dim(W)\geq1$. The \emph{cut vector} at $z$ is the answer $a$ of $\SO$ at $z$ if this answer is a vector with $Y^\top a\neq0$. Otherwise it is $e_i$ for the first coordinate $i$ that is not constant on $W$, so that $Y^\top e_i\neq0$. This is the device of~\cite[Section~1.3]{j22}, with the lexicographically largest point in place of the smallest. So every cut $a^\top x\geq a^\top z-\rho$ with $\rho>0$ is a proper halfspace inside $W$. If moreover $x^*\in W$, then the cut vector $a$ at $z$ satisfies
\begin{align}\label{eq:cut_xstar}
    a^\top x^*\geq a^\top z,
\end{align}
so the cut contains $x^*$ with slack at least $\rho$. Section~\ref{sec:r3_local} proves~\eqref{eq:cut_xstar}, using the strict inequality at points that do not minimize $f$ (Remark~\ref{rem:oracle_model}).

\paragraph{The method.} Algorithm~\ref{alg:hcpm} gives one block of the method. Each iteration of its outer loop starts at the current point $z$. It deletes the row with the smallest estimated score if this score is below $\theta$. Otherwise it calls $\SO$ once and adds the cut $a^\top x\geq a^\top z-\rho$ for the cut vector $a$ at $z$, with $\rho$ chosen so that the new row has leverage about $w$ at $z$. Then it takes $N$ damped Newton steps on $p$, with estimated scores in the gradient and a matrix $\widehat Q$ fixed at their start. Each step goes to \textsc{LS} in $t_{\max}$ equal pieces, so that every weight update is small enough for Lemma~\ref{lem:jlsw_main}.

\begin{algorithm}[!ht]
\caption{Cutting plane method with the hybrid barrier, one block}
\label{alg:hcpm}
\begin{algorithmic}[1]
\Procedure{HybridCPM}{$\SO,K,z$} \Comment{$K$ has $m\leq\bar m$ rows, and $\lambda_K(z)\leq\lambda^*$}
\State Fix $\bar x$ and $Y$, and form the padded instance (\emph{Coordinates and padding})
\State Initialize \textsc{LS} (Lemma~\ref{lem:jlsw_main}) and \textsc{IM}$_H$ with the weights $s_i(z)^{-2}$, and query $\wt\sigma$ from \textsc{LS}
\State \textbf{if} $\min_i(\wt\sigma_i+c)\leq0$ over the rows of $K$ \textbf{then} abort \Comment{keeps every weight of \textsc{IM}$_Q$ positive}
\State Initialize \textsc{IM}$_Q$ with the weights $(\wt\sigma_i+c)s_i(z)^{-2}$
\For{$k=1,\ldots,T$}
\State Query $\wt\sigma$ from \textsc{LS}
\If{$\min_i\wt\sigma_i<\theta$, the minimum taken over the rows of $K$}
\State $j\gets\arg\min_i\wt\sigma_i$
\State \textbf{if} $s_j(z)^{-2}a_j^\top\widehat H^{-1}a_j>\tau_{\rm del}$ \textbf{then} abort \Comment{keeps the deletion valid}
\State Delete row $j$ from $K$, \textsc{LS}, \textsc{IM}$_H$ and \textsc{IM}$_Q$
\Else
\State \textbf{if} $m+1>\bar m$ \textbf{then} abort \Comment{keeps at most $\bar m$ rows}
\State $a\gets$ the cut vector at $z$ (one call of $\SO$), $\wt h\gets a^\top\widehat H^{-1}a$ and $\rho\gets(\wt h/w)^{1/2}$
\State Add the row $a^\top x\geq a^\top z-\rho$ to $K$. Insert it into \textsc{LS} and \textsc{IM}$_H$ with weight $\rho^{-2}$, and into \textsc{IM}$_Q$ with weight $(w/(1+w)+c)\rho^{-2}$
\EndIf
\State Query $\wt\sigma$ from \textsc{LS}, and \textbf{if} $\min_i(\wt\sigma_i+c)\leq0$ over the rows of $K$ \textbf{then} abort
\State Update \textsc{IM}$_Q$ to the weights $(\wt\sigma_i+c)s_i(z)^{-2}$, and fix its matrix as $\widehat Q$
\State $y\gets z$
\For{$l=1,\ldots,N$}
\State $\wt g\gets-\sum_i(\wt\sigma_i+c)a_i/s_i(y)$ and $y'\gets y-\beta^{-1}\widehat Q^{-1}\wt g$
\State $y^{(t)}\gets y+\frac{t}{t_{\max}}(y'-y)$ for $t=0,1,\ldots,t_{\max}$
\State \textbf{if} $s_i(y^{(t)})\leq0$ for some row $i$ and some $t$, or $\sum_i\big(s_i(y^{(t+1)})/s_i(y^{(t)})-1\big)^2>\tau_{\rm step}^2$ for some $t<t_{\max}$, \textbf{then} abort \Comment{keeps every slack positive and every weight update valid}
\State Update the weights of \textsc{LS} to $s_i(y^{(t)})^{-2}$, for $t=1,\ldots,t_{\max}$ in turn
\State $y\gets y'$, and query $\wt\sigma$ from \textsc{LS}
\EndFor
\State $z\gets y$, and update \textsc{IM}$_H$ to the weights $s_i(z)^{-2}$
\EndFor
\State \Return $(K,z,H_K(z))$, where $H_K(z)$ is computed exactly
\EndProcedure
\end{algorithmic}
\end{algorithm}

Here \textsc{LS} runs with exponent $15$ and the weights $s_i(\cdot)^{-2}$ of the real rows, so its scores are the leverage scores $\sigma_i(\cdot)$ of Definition~\ref{def:sigma_x} at the current point, $z$ or $y$. The matrix $\widehat Q^{-1}$ is the inverse kept by \textsc{IM}$_Q$ after its update at $z$. A new row enters \textsc{IM}$_Q$ with the weight $(w/(1+w)+c)\rho^{-2}$, an estimate of its target weight $(\sigma'+c)s_a(z)^{-2}$, where $\sigma'$ is its leverage score in the new polytope. The tests make every operation on the three data structures valid, whatever the queries return (Section~\ref{sec:r3_local}). If a test fires, the block aborts and Algorithm~\ref{alg:ours} returns \textsc{fail}. Computing the output $H_K(z)$ takes $O(n^\omega)$ operations.

\begin{Definition}[Good block]\label{def:good_block}
We call one run of Algorithm~\ref{alg:hcpm} a \emph{block}. A block on $(K,z)$ is \emph{good} if the following hold.
\begin{itemize}
    \item $K$ has at most $\bar m$ rows.
    \item $z$ is a relative interior point of $K$ at which $H_K(z)$ is positive definite on $W_0$, so that $\lambda_K(z)$ is defined.
    \item $\lambda_K(z)\leq\lambda^*$.
    \item Every output of \textsc{\textup{LS}} during the block has $\ell_2$ error at most $\delta_0$.
\end{itemize}
\end{Definition}

Lemma~\ref{lem:hybrid_inv} in Section~\ref{sec:r3_invariant} shows, by induction over the steps, that no step of a good block aborts and that every current point has decrement at most $\lambda^*$. In a good block, each \textsc{Update} moves the logarithms of the weights by $O(1)$ in \textsc{IM}$_H$ and by $O(\log n)$ in \textsc{IM}$_Q$, in the $\ell_2$ norm. The bound on inverse maintenance with $c''=1$ then gives the next lemma, which Section~\ref{sec:r3_im} proves.

\begin{lemma}[Inverse maintenance]\label{lem:hybrid_im}
In a good block, every call of \textsc{\textup{Update}} of \textsc{\textup{IM}}$_H$ and of \textsc{\textup{IM}}$_Q$ takes amortized $O(n^2)$ arithmetic operations.
\end{lemma}

\begin{remark}[Leverage scores during runs of deletions]\label{rem:jlsw20_floor}
The fast implementation of Vaidya's method in~\cite{jlsw20}, which the SODA 2024 version of this paper uses, replaces the leverage scores by estimates with $\ell_2$ error $1/\poly(\log n)$. Its analysis in Appendix~A of~\cite{jlsw20} states that every leverage score stays at least $c_1$, the absolute constant below which the method deletes a row. The claim is in the paragraph on the running time per iteration, just before Section~A.1 of~\cite{jlsw20}. Read literally, this fails at the current point of every deletion, where the deleted row has score below $c_1$. The proofs of Lemmas~A.2 and~A.4 of~\cite{jlsw20} need only some absolute lower bound on all scores, which turns the additive error into a small relative error. But during a run of consecutive deletions, the rows with scores below $c_1$ stay in the polytope while the point moves, and no such bound is proved. Whether it holds is open. The hybrid barrier avoids the question, because every row keeps the weight $\sigma_i+c$ in $Q_c$ (Lemma~\ref{lem:hybrid_local}, Part~2). The price is $N=O(\log\log n)$ Newton steps per added or deleted row, and so a factor $\log\log n$ in the arithmetic operations.
\end{remark}

\subsection{Analysis of One Block}\label{sec:cpm:hybrid_progress}

The analysis of Section~\ref{sec:minimization} follows the potential argument of~\cite{j22}. It needs a quantity that bounds the volume of the polytope and rises by order $n\log n$ in every block. For Vaidya's method, the volumetric barrier at the volumetric center plays this role. Our method instead keeps its points near the hybrid center, the minimizer of $p=F+cL$. There the extra term $cL$ has no useful bound by itself. So we measure every slack against its value at the minimizer $x^*$ of Section~\ref{sec:cpm:method}. Lemma~\ref{lem:xstar} gives conditions under which $x^*$ lies in every polytope of a run, with positive slack in every row.

\begin{Definition}[A potential measured at a fixed minimizer]\label{def:credit}
Let $K$ be a polytope maintained by the algorithm whose hybrid center $x_c=x_c(K)$ exists and whose rows satisfy $s_i(x^*)>0$. We define
\begin{align*}
    D(K):=\sum_{i=1}^m\ln\frac{\bar\kappa\,s_i(x_c)}{s_i(x^*)}\qquad\text{and}\qquad \mathcal{U}(K):=F_K(x_c)-c\,D(K).
\end{align*}
\end{Definition}

The number $\bar\kappa$ is fixed for the whole run, so $D$ and $\mathcal{U}$ depend on the polytope alone. We call $\mathcal{U}$ the potential.

\paragraph{Proof overview.} \emph{The potential.} By~\eqref{eq:vol_dikin}, the term $F_K(x_c)$ is minus the logarithm of the volume of the Dikin ellipsoid at the hybrid center, up to an additive constant. It takes the place of the volume in the potential of~\cite{j22}. Since the gradient of $p$ vanishes at the hybrid center, Part~4 of Lemma~\ref{lem:hybrid_loc} gives $s_i(x^*)\leq\bar\kappa\,s_i(x_c)$ when $K$ has at most $\bar m$ rows. So $D(K)\geq0$ and $\mathcal{U}(K)\leq F_K(x_c)$, and a large potential still forces a small Dikin ellipsoid. At a box, $cD$ is a tiny fraction of the dimension (Lemma~\ref{lem:box_margin}), so a restart at a box costs little.

\emph{One step.} Lemma~\ref{lem:hybrid_identity} below writes the exact change of $\mathcal{U}$ under one added or deleted row as a sum of three terms. The first comes from the determinant of $H$. It is about $w/2$ for an add, since the new cut has leverage about $w$ at the hybrid center. For a delete it is $\frac12\ln(1-\sigma_j)$, at least about $-\theta/2=-w/20$, since the method deletes only rows with a small estimated score. The second term is $-{\rm rec}$, the decrease of $p$ from the old hybrid center to the new one. It is quadratic in the decrement of the new barrier at the old hybrid center, which is $O(w)$ by Lemma~\ref{lem:hybrid_pert}, so it is $O(w^2)$. The third term involves only the changed row. For an add it costs at most a little more than $c\ln\bar\kappa$, because the cut vector puts $x^*$ at least as far from the new cut as the current point. By~\eqref{eq:R4} this is at most $w/10$. For a delete the third term is nonnegative, again because $s_j(x^*)\leq\bar\kappa\,s_j(x_c)$. So an add raises $\mathcal{U}$ by a constant times $w$, and a delete lowers it by at most a smaller constant times $w$.

\emph{One block.} A delete removes a row, so during a block the deletes outnumber the adds by at most the number of rows at the start, which is $O(n)$. Hence a block of $T=\Theta(n\log n)$ steps raises $\mathcal{U}$ by $\Omega(wT)-O(n)$, which is at least $2n\ln n$ for our choice of $T$. Section~\ref{sec:minimization} uses $-\mathcal{U}$ in place of the logarithm of the volume.

\emph{Why $c$ may tend to zero.} Moving the hybrid center changes the slack of every row, and the term $cL$ of $p$ registers all these changes. The term $cD$ measures the same slacks against the fixed point $x^*$, so the changes of the old rows cancel exactly in $\mathcal{U}(K')-\mathcal{U}(K)$. Each step pays only for the row that it adds or deletes, and unlike~\cite{lsw15} we need no bound on the other slacks. The payment for an added row grows with $c$, which forces $c=O(w/\log n)$, as we explain after the identity.

\emph{Accuracy of the estimates.} Every row has weight at least $c$ in $Q_c$. So an additive $\ell_2$ error $\delta$ of the estimated scores changes $\widehat Q$ by a relative error $O(\delta/c)$, and the gradient by $O(\delta c^{-1/2})$ in the norm of $Q_c^{-1}$ (Lemma~\ref{lem:hybrid_loc}, Part~3). For $\delta\leq\delta_0=\Theta(c^{5/4})$, the Newton steps still bring the decrement back to $\lambda^*$ at every current point, and no step aborts (Lemma~\ref{lem:hybrid_inv}).

\emph{Near the hybrid center.} At decrement $\lambda^*=\Theta(c^{3/4})$, the hybrid center lies in $\mathcal{S}(z,r_{\rm f})$ for the current point $z$, where $r_{\rm f}=O(c^{1/2})$ (Lemma~\ref{lem:near_center}). So the slacks, the scores and the Hessians at the two points agree up to factors $1+O(c^{1/2})$, and we can prove the per-step bounds at the exact hybrid center.

\medskip
The next identity is the heart of the analysis.

\begin{lemma}[Exact step identity]\label{lem:hybrid_identity}
Let $K'$ arise from $K$ by adding a row $a^\top x\geq b'$ or by deleting row $j$. Suppose that $x:=x_c(K)$ and $x':=x_c(K')$ exist, that $x$ lies in the relative interior of $K'$, and that every row of $K$ and of $K'$ has positive slack at $x^*$. Let ${\rm rec}:=p_{K'}(x)-p_{K'}(x')\geq0$. For an add, let $w_x:=s_a(x)^{-2}a^\top H_K(x)^{-1}a$. Then
\begin{align*}
    \mathcal{U}(K')-\mathcal{U}(K)=\frac12\ln(1+w_x)-{\rm rec}-c\ln\frac{\bar\kappa\,s_a(x)}{s_a(x^*)}.
\end{align*}
For a deletion, let $\sigma_j(x)$ be the leverage score of row $j$ in $K$ at $x$. Then
\begin{align*}
    \mathcal{U}(K')-\mathcal{U}(K)=\frac12\ln(1-\sigma_j(x))-{\rm rec}+c\ln\frac{\bar\kappa\,s_j(x)}{s_j(x^*)}.
\end{align*}
\end{lemma}

\begin{proof}
By the matrix determinant lemma, $F_{K'}(x)-F_K(x)$ equals $\frac12\ln(1+w_x)$ for an add and $\frac12\ln(1-\sigma_j(x))$ for a deletion. Since $F_{K'}=p_{K'}-cL_{K'}$, we have
\begin{align*}
    F_{K'}(x')-F_{K'}(x)=-{\rm rec}+c\sum_{i\in K'}\big(\ln s_i(x')-\ln s_i(x)\big),
\end{align*}
where the sum runs over the rows of $K'$. By the definition of $D$, the difference $D(K')-D(K)$ equals the same sum, plus $\ln(\bar\kappa s_a(x)/s_a(x^*))$ for an add and minus $\ln(\bar\kappa s_j(x)/s_j(x^*))$ for a deletion. We subtract $c$ times this from $F_{K'}(x')-F_K(x)=(F_{K'}(x)-F_K(x))+(F_{K'}(x')-F_{K'}(x))$. The two sums cancel, which gives the identity. Finally ${\rm rec}\geq0$ because $x'$ minimizes $p_{K'}$.
\end{proof}

The identity uses no property of the estimated scores. For an add, its last term is the price of the logarithmic barrier. Suppose that $x^*$ lies on the hyperplane $\{y:a^\top y=a^\top z\}$ through the current point $z$, so that $s_a(x^*)=\rho$. For our parameters the slack of the new row at the hybrid center is at least $\rho/2$ (Lemma~\ref{lem:good_centers}), and heuristically the same holds whenever $\lambda^*c^{-1/4}$ is small. With $\bar\kappa\geq n/c$, this term then costs at least $c\ln(\bar\kappa/2)\geq c\ln(n/(2c))$. This lower bound increases with $c$ for $c<n/(2e)$, and it exceeds $w$ already at $c=w/\ln n$. The gain $\frac12\ln(1+w_x)$ of the cut is only about $w/2$, so this accounting needs $c=O(w/\log n)$. Whether the method itself makes enough progress with a constant $c$ is open.

The identity gives the following bounds for one step of a good block. Here $z_k$ and $K_k$ are the current point and the polytope at the start of iteration $k$ of the outer loop. The absolute constants $g_+>g_->0$, $w_+$ slightly above $w$ and $\theta_{\rm del}$ slightly above $\theta$ are fixed in Section~\ref{sec:r3_params}, like all constants below. Summing the bounds over the $T$ steps gives the progress of a block, with the absolute constants $g_{\rm b}=(g_+-g_-)/2$ and $C_{\rm b}$. Our choice of $T$ makes $g_{\rm b}wT-C_{\rm b}n$ at least $2n\ln n$. Section~\ref{sec:r3_block} proves both lemmas.

\begin{lemma}[Per-step bounds]\label{lem:hybrid_step}
Consider a good block on $(K,z)$ such that $K$ contains $x^*$ with positive slack in every row. Consider one iteration of its outer loop, with current point $z_k$, polytope $K_k$ before it and polytope $K_{k+1}$ after its add or delete. Then $x_k:=x_c(K_k)$ and $x_c(K_{k+1})$ exist, and $x_c(K_{k+1})\in\mathcal{S}_{K_{k+1}}(x_k,r)$.
\begin{enumerate}
    \item If a row $a$ is added, then $\mathcal{U}(K_{k+1})-\mathcal{U}(K_k)\geq g_+w$ and $w_x\leq w_+$, where $w_x$ is as in Lemma~\ref{lem:hybrid_identity} at $x=x_k$.
    \item If row $j$ is deleted, then $\mathcal{U}(K_{k+1})-\mathcal{U}(K_k)\geq-g_-w$ and $\sigma_j(x_k)\leq\theta_{\rm del}$.
\end{enumerate}
\end{lemma}

\begin{lemma}[Block progress]\label{lem:hybrid_block}
Consider a good block on $(K,x_K)$, where $K$ has $m_0$ rows and contains $x^*$ with positive slack in every row. Then no step of the block aborts, every polytope during the block has at most $\bar m$ rows, and the polytope $K'$ at its end satisfies
\begin{align*}
    \mathcal{U}(K')-\mathcal{U}(K)\geq g_{\rm b}wT-C_{\rm b}n\geq2n\ln n.
\end{align*}
\end{lemma}

\subsection{Main Lemma}\label{sec:cpm:main_lemma}

The next lemma collects what one block guarantees, and Section~\ref{sec:minimization} applies it to every block of Algorithm~\ref{alg:ours}. A block starts at the center of a box, where the decrement is $0$ (Lemma~\ref{lem:reduction_geometry}, Part~3), or at the output of the previous block, where it is at most $\lambda^*$ (Part~4 below). Part~2 rests on Lemma~\ref{lem:sandwich}. That lemma bounds the distance from the hybrid center to any point of a polytope with at most $\bar m$ rows by $\bar R=O(n/c)$, in the norm of the Hessian at the hybrid center. In Part~4, $\epsilon_{\rm f}$ is a small absolute constant.

\begingroup\emergencystretch=3em
\begin{lemma}[One block]\label{lem:hybrid_cpm}
Consider a good block on $(K,x_K)$, where $K$ contains $x^*$ with positive slack in every row. Then the block makes at most $T$ calls to $\SO$ and uses $O(n^3\log n\log\log n)$ arithmetic operations, and it outputs $(K',x_{K'},H_{K'})$ with $H_{K'}=H_{K'}(x_{K'})$ such that the following hold.
\begin{itemize}
    \item Part 1. $x^*\in K'$, every constraint of $K'$ is a constraint of $K$ or a cut $a^\top x\geq a^\top z-\rho$ built from the cut vector $a$ at a current point $z$, and $K'$ has at most $\bar m$ constraints.
    \item Part 2. $E(x_{K'},H_{K'})\subseteq K'\subseteq R'\cdot E(x_{K'},H_{K'})$.
    \item Part 3. $\mathcal{U}(K')\geq \mathcal{U}(K)+2n\ln n$.
    \item Part 4. $\lambda_{K'}(x_{K'})\leq\lambda^*$, $(1-\epsilon_{\rm f})H_{K'}(x_c(K'))\preceq H_{K'}\preceq(1+\epsilon_{\rm f})H_{K'}(x_c(K'))$ and $|F_{K'}(x_{K'})-F_{K'}(x_c(K'))|\leq\epsilon_{\rm f}d$.
\end{itemize}
Moreover, the block makes at most $T(3+(t_{\max}+1)N)+2=O(n\log n\log\log n)$ operations on \textsc{\textup{LS}}.
\end{lemma}\par\endgroup

The oracle count of Section~\ref{sec:minimization:oracle} needs a bound on how much the Hessian kept by Algorithm~\ref{alg:ours} can grow during one block. The cost bound of Lemma~\ref{lem:svp_total} needs a bound on how much it can shrink. The next lemma answers both, with a moderate absolute constant $C_H$. Section~\ref{sec:r3_block} proves it and Lemma~\ref{lem:hybrid_cpm}.

\begin{lemma}[Hessian stability within a block]\label{lem:block_stability}
Let $(K',x_{K'},H_{K'})$ be the output of a block on $(K,x_K)$ made by Algorithm~\ref{alg:ours}, let $W_0$ be the linear subspace parallel to the affine subspace that contains $K$, and let $H_K=H_K(x_K)$. Suppose that every output of \textsc{\textup{LS}} during this block and during the previous blocks has $\ell_2$ error at most $\delta_0$. Then, as quadratic forms on $W_0$,
\begin{align*}
    e^{-a}H_K\preceq H_{K'}\preceq e^{a}H_K,\qquad\text{where}\ a:=C_HT+10\,\epsilon_{\rm f}=O(n\log n).
\end{align*}
Consequently, $\lambda_1(\Lambda,H_{K'}^{-1})\geq e^{-a/2}\lambda_1(\Lambda,H_K^{-1})=2^{-O(n\log n)}\cdot\lambda_1(\Lambda,H_K^{-1})$ for every lattice $\Lambda$ that spans $W_0$.
\end{lemma}
\section{Shortest Vectors from the Previous Basis}\label{sec:segment_lll}

This section gives the routine \textsc{SegmentLLL}, a variant of the LLL algorithm that reduces segments of consecutive basis vectors recursively, and bounds its cost from a given starting basis (Lemma~\ref{lem:segment_lll}). Lemma~\ref{lem:svp_segment} in Section~\ref{sec:minimization:runtime} applies this bound to the calls of our algorithm. We use the notation of the paragraph \emph{Reduced bases} in Section~\ref{sec:preli:lll}. The analysis of \textsc{SegmentLLL} uses the following form of the argument of Lemma~\ref{lem:lll_restart}. It assumes less than reducedness of the starting basis, and it bounds each $d_i$ separately.

\begin{lemma}[Restarting from a nearly reduced basis]\label{lem:restart_bound}
Let $G_0,G_1\in\R^{d\times d}$ be positive definite with $e^{-a}G_0\preceq G_1\preceq e^aG_0$ for some $a\geq 0$, and let $\tau\geq 0$ be such that the Gram--Schmidt norms of $I_d$ for $G_0$ satisfy $B_j\geq 2^{-\tau}B_l$ for all $1\leq l<j\leq d$. Then, for every unimodular $U\in\Z^{d\times d}$ and $1\leq i\leq d-1$, we have $\log_2(d_i(I_d)/d_i(U))\leq i(2a\log_2e+\tau+\log_2i)$, and hence $\log_2(P(I_d)/P(U))\leq(3a+\tau+\log_2d)\,d^2/2$, where $d_i$ and $P$ are computed with $G_1$.
\end{lemma}

\begin{proof}
Write $d_i^{(0)}$ and $d_i^{(1)}$ for $d_i$ computed with $G_0$ and with $G_1$, and let $1\leq i\leq d-1$. As in the proof of Lemma~\ref{lem:lll_restart}, $d_i^{(1)}(I_d)\leq e^{ai}d_i^{(0)}(I_d)$ and $d_i^{(0)}(U)\leq e^{ai}d_i^{(1)}(U)$. Lemma~\ref{lem:reduced_near_opt} with $V=I_d$, the matrix $G_0$ and $Y=U_i$ gives $d_i^{(0)}(I_d)\leq 2^{i\tau}i^id_i^{(0)}(U)$. Hence $\log_2(d_i^{(1)}(I_d)/d_i^{(1)}(U))\leq i(2a\log_2e+\tau+\log_2i)$, which is the first claim. Summing over $i\leq d-1$, and using $2\log_2e<3$ and $\sum_{i<d}i\leq d^2/2$, gives the second claim.
\end{proof}

The routine below repeatedly multiplies a set of consecutive columns of the current basis by a unimodular matrix. The next lemma gives the resulting change of the Gram--Schmidt data. For a matrix $X$ and sets $I$ and $J$ of indices, $X_{I,J}$ denotes the submatrix of $X$ with rows in $I$ and columns in $J$.

\begin{lemma}[Updating the Gram--Schmidt data of consecutive columns]\label{lem:block_update}
Let $G\in\R^{d\times d}$ be positive definite, and let $V\in\Z^{d\times d}$ be unimodular, with $V^\top GV=LDL^\top$, where $L$ is unit lower triangular and $D={\rm diag}(B_1,\ldots,B_d)$. Let $\mathcal{J}=\{p,\ldots,q\}\subseteq[d]$ be a set of $k$ consecutive indices, let $W\in\Z^{k\times k}$ be unimodular, and let $V'$ be obtained from $V$ by multiplying the columns with indices in $\mathcal{J}$ by $W$. Let $L_{\mathcal{J}}=L_{\mathcal{J},\mathcal{J}}$ and $D_{\mathcal{J}}={\rm diag}(B_p,\ldots,B_q)$, and let $W^\top L_{\mathcal{J}}D_{\mathcal{J}}L_{\mathcal{J}}^\top W=L'_{\mathcal{J}}D'_{\mathcal{J}}L_{\mathcal{J}}'^\top$ with $L'_{\mathcal{J}}$ unit lower triangular and $D'_{\mathcal{J}}$ diagonal. Then $V'^\top GV'=L'D'L'^\top$, where $L'$ and $D'$ agree with $L$ and $D$ except that
\begin{gather*}
    L'_{\mathcal{J},\mathcal{J}}=L'_{\mathcal{J}},\qquad D'_{\mathcal{J},\mathcal{J}}=D'_{\mathcal{J}},\qquad L'_{\mathcal{J},[p-1]}=W^\top L_{\mathcal{J},[p-1]},\\
    L'_{[d]\setminus[q],\mathcal{J}}=L_{[d]\setminus[q],\mathcal{J}}L_{\mathcal{J}}^{-1}W^{-\top}L'_{\mathcal{J}}.
\end{gather*}
In particular, the Gram--Schmidt norms of $V'$ and $V$ for $G$ agree outside $\mathcal{J}$, and their products over $\mathcal{J}$ agree. Given $L$, $D$ and $W$, the new entries take $O(dk^2)$ arithmetic operations.
\end{lemma}

\begin{proof}
Let $v_1,\ldots,v_d$ be the columns of $V$, and let $v_l^*$, $\mu_{l,j}=L_{l,j}$ and $B_l$ be their Gram--Schmidt data in $\langle\cdot,\cdot\rangle_G$. Let $\pi$ be the projection, orthogonal in $\langle\cdot,\cdot\rangle_G$, onto the orthogonal complement of $v_1,\ldots,v_{p-1}$. For $l\in\mathcal{J}$ we have $\pi v_l=\sum_{j\in\mathcal{J},\,j\leq l}\mu_{l,j}v_j^*$. Write $v_{\mathcal{J}}^*$ for the row $(v_p^*,\ldots,v_q^*)$, so that $(\pi v_p,\ldots,\pi v_q)=v_{\mathcal{J}}^*L_{\mathcal{J}}^\top$, and $L_{\mathcal{J}}D_{\mathcal{J}}L_{\mathcal{J}}^\top$ is the Gram matrix of $\pi v_p,\ldots,\pi v_q$. Replacing $(v_p,\ldots,v_q)$ by $(v_p,\ldots,v_q)W$ replaces $(\pi v_p,\ldots,\pi v_q)$ by $v_{\mathcal{J}}^*L_{\mathcal{J}}^\top W$, whose Gram matrix is $W^\top L_{\mathcal{J}}D_{\mathcal{J}}L_{\mathcal{J}}^\top W$. As $W$ is unimodular, it changes neither the vectors $v_j^*$ with $j\notin\mathcal{J}$ nor the span of $v_p,\ldots,v_q$. So the new Gram--Schmidt data at the positions in $\mathcal{J}$ are $L'_{\mathcal{J}}$ and $D'_{\mathcal{J}}$, and the new vectors $v_p^*,\ldots,v_q^*$ form the row $v_{\mathcal{J}}^*L_{\mathcal{J}}^\top WL_{\mathcal{J}}'^{-\top}$. For $l\in\mathcal{J}$ and $j<p$, the coefficient $\mu_{l,j}=\langle v_l,v_j^*\rangle_G/B_j$ is linear in $v_l$, which gives the new $L_{\mathcal{J},[p-1]}$. For $l>q$, the component $v_{\mathcal{J}}^*(L_{\{l\},\mathcal{J}})^\top$ of $v_l$ in the span of $v_p^*,\ldots,v_q^*$ does not change, and its coefficients in the new vectors are the entries of $L_{\mathcal{J}}'^\top W^{-1}L_{\mathcal{J}}^{-\top}(L_{\{l\},\mathcal{J}})^\top$. This gives the new rows below $\mathcal{J}$. The other coefficients involve no $v_l$ and no $v_j^*$ with $l\in\mathcal{J}$ or $j\in\mathcal{J}$, so they do not change. The product of $B_p,\ldots,B_q$ is the determinant of the Gram matrix of $\pi v_p,\ldots,\pi v_q$, which does not change because $\det W=\pm 1$. Finally, the products, inverses and factorizations of $k\times k$ matrices take $O(k^3)$ operations, and the products with the two blocks of at most $d$ rows or columns take $O(dk^2)$ operations. As $k\leq d$, the total is $O(dk^2)$.
\end{proof}

\paragraph{Reduction by iterated subsegments.} A swap of \textsc{LLL} lowers $\log_2P$ by more than $\log_2(4/3)$, but it costs $O(d)$ operations, because it changes the Gram--Schmidt coefficients in two whole rows and two whole columns. Our bound on the drop of $\log_2P$ in a call of our algorithm is $O(n^3\log n)$, so this gives $O(n^4\log n)$ operations per call (Lemma~\ref{lem:svp_total}). The semi-reduction of Sch\"onhage~\cite{s84}, as presented in Section~5 of~\cite{s96}, runs \textsc{LLL} on $k$ consecutive basis vectors in their own coordinates, where a swap costs $O(k)$ operations. It then writes the result into the Gram--Schmidt data of the whole basis once, with $O(dk^2)$ operations (Lemma~\ref{lem:block_update}). These updates then become the main cost. Schnorr~\cite{sch06} avoids most of them by recursion. His algorithm SLLL$^+$ (Section~6 of~\cite{sch06}) reduces a segment of $2k$ consecutive basis vectors by reducing three segments of $k$ consecutive vectors inside it: its first half, its middle half and its last half. Each of them is reduced by the same procedure and in the coordinates of the enclosing segment. The transform found for a subsegment is written into the enclosing segment once, when the subsegment is done. So the result for $k$ vectors is written into a segment of $2k$ vectors, with $O(k^3)$ operations, and only the outermost segment is written into the whole basis. For an integral basis of $m$ vectors in $\Z^{m'}$, Theorem~9 of~\cite{sch06} states $O(m^2m'+m\log_2m\log_{1/\delta}M)$ arithmetic operations for SLLL$^+$, where $\delta\in[1/2,1)$ is the parameter of the reduction and $M$ is the largest of $2^m$ and the leading Gram determinants of the input. This is $O(m^3\log m)$ for bases whose vectors have length $2^{O(m)}$ in dimension $m'=O(m)$. We refer to results of~\cite{sch06} by their numbers in its preprint version of 25 October 2004, which is the version we checked.

We cannot use SLLL$^+$ as it stands, for three reasons. First, its output condition (Definition~4 of~\cite{sch06}) only compares the determinants of neighboring segments that are the two halves of one block of a fixed dyadic partition of the basis. The guarantee proved for its output (Theorem~8 of~\cite{sch06}) bounds the first vector against the determinant of the lattice instead of against $\lambda_1$. Definition~4 alone gives no bound against $\lambda_1$ either. For instance, for every integer $t\geq 1$, the basis $te_1,te_2,te_3,te_4,e_5,t^2e_6,t^2e_7,t^2e_8$ of $\Z^8$ satisfies Definition~4 of~\cite{sch06}, and its first vector is $t$ times as long as its fifth. Second, the count of operations of~\cite{sch06} uses that the potential of an integral basis is at least one. Third, the proof of Theorem~9 of~\cite{sch06} invokes the argument of its Theorem~4 for the decrease of the potential per reduction of a segment. This argument uses that a reduction ends below the bound that triggers it by a factor that grows with the size of the segment. For $j\geq 1$ and $k\geq 1$, let $D_{j,k}$ be the product of the squared Gram--Schmidt norms of the $j$-th block of $k$ consecutive basis vectors, and let $\alpha=1/(\delta-1/4)$. When the procedure of SLLL$^+$ ends on a segment of $2^{\sigma+1}$ vectors, its loop only guarantees $D_{j,2^{\sigma-1}}\leq(\alpha/\delta)^{4^{\sigma-1}}D_{j+1,2^{\sigma-1}}$ for its consecutive quarters. Counting the middle inequality twice, these give $D_{l,2^\sigma}\leq\delta(\alpha/\delta)^{4^\sigma}D_{l+1,2^\sigma}$ for its halves, where the factor $\delta$ comes from the two-dimensional base case. So, unless the bounds weaken from level to level, a reduction ends below the bound that triggers it only by the constant factor $\delta$, not by a factor that grows with the segment. The bounds of the iterated segments of Koy and Schnorr weaken in this way (Definition~3 of~\cite{ks01}).

We keep the recursion of SLLL$^+$ and change the stopping test, the order of the subsegments and the counting. A segment of $2^h$ vectors is done when none of its Gram--Schmidt norms is smaller than an earlier one by more than a factor $2^{h2^h}$. This test bounds every ratio $B_l/B_j$ with $l<j$ in the segment. Its bound weakens from each level to the next by more than the three subsegments can use up. A round processes the middle half first, then the first half and then the last half. Rounds are repeated until the test succeeds. Zhou, Wang and Wang~\cite{zww26} use a stopping test of this kind, which bounds the forward spread of the logarithms of the Gram--Schmidt norms. They also use this order and the argument that every round after which the test fails lowers the potential near the middle of the segment (Algorithm~2, Definition~3, Lemma~3 and Theorem~4 of~\cite{zww26}). They take the order of a round from the algorithm flatter of Ryan and Heninger~\cite{rh23}. Zhou, Wang and Wang work in finite precision on integral bases. Their paper is a recent preprint that has not been peer reviewed, and we do not use its results as stated. Instead, we restate their argument in exact arithmetic and prove every step that we use (Lemmas~\ref{lem:lll_step_monotone}, \ref{lem:centered_segment} and~\ref{lem:merge} below). Theorem~4 of~\cite{zww26} bounds the decrease of the whole potential. Its proof only sums over the indices near the middle of the segment. We state the bound for these indices, because our count needs it. Our thresholds $h2^h$ differ from theirs and weaken faster from level to level, which saves a logarithmic factor in the count. Our algorithm tolerates the faster weakening because it only needs the factor $2^{O(n\log n)}$. The count also needs a bound on the total decrease of the potential in a call. Schnorr uses that the potential of an integral basis is at least one. Zhou, Wang and Wang bound the decrease by the forward spread of the input (Lemma~2 of~\cite{zww26}). We use instead the per-index bound of Lemma~\ref{lem:restart_bound}, restricted to the indices near the middle of each segment. After a change of the Hessian, this bound is smaller. So \textsc{SegmentLLL} is a modification of SLLL$^+$ with its own proof. The restart bound and the count of operations in exact arithmetic are ours.

Let $\bar G\in\R^{\bar d\times\bar d}$ be positive definite, where $\bar d=2^{\bar h}$ for an integer $\bar h\geq 1$, and let $U\in\Z^{\bar d\times\bar d}$ be unimodular, with Gram--Schmidt norms $B_1,\ldots,B_{\bar d}$ for $\bar G$. Let $x_t=\log_2B_t$. For $y\in\R^m$ and a set $I\subseteq[m]$ of consecutive indices, let
\begin{align*}
    {\rm sp}_I(y)=\max\{y_i-y_j:\ i,j\in I,\ i\leq j\}\geq 0,
\end{align*}
and write ${\rm sp}(I)={\rm sp}_I(x)$. So ${\rm sp}(I)\leq\tau$ means that $B_j\geq 2^{-\tau}B_i$ for all $i\leq j$ in $I$. For $1\leq h\leq\bar h$, a \emph{segment of order $2^h$} is a set $J=\{p+1,\ldots,p+2^h\}\subseteq[\bar d]$, where $p$ is a multiple of $2^{h-1}$. Its \emph{halves} are $J_L=\{p+1,\ldots,p+2^{h-1}\}$ and $J_R=\{p+2^{h-1}+1,\ldots,p+2^h\}$, and for $h\geq 2$ its \emph{middle} is $J_M=\{p+2^{h-2}+1,\ldots,p+3\cdot 2^{h-2}\}$. These are segments of order $2^{h-1}$. In the notation of~\cite{sch06}, $J$ is the double segment $R_{l,2^{h-1}}$ with $l=p/2^{h-1}+1$, and $J_L$, $J_M$ and $J_R$ are $R_{l',2^{h-2}}$ for $l'=2l-1,2l,2l+1$. Let $\tau_h=h2^h$. A segment $J$ of order $2^h$ is \emph{flat} (for $U$ and $\bar G$) if ${\rm sp}(J)\leq\tau_h$ and, in case $h\geq 2$, both halves of $J$ are flat. So if $J$ is flat, then so are its halves, their halves, and so on. We say that $J$ has \emph{flat halves} if $h=1$ or both halves of $J$ are flat. If $U$ is replaced by $UW$, where $W\in\Z^{\bar d\times\bar d}$ is unimodular and differs from $I_{\bar d}$ only in rows and columns with indices in a set $J'$ of consecutive indices, then the Gram--Schmidt norms with indices outside $J'$ do not change, and neither does $\prod_{t\in J'}B_t$ (Lemma~\ref{lem:block_update}). In particular, flatness of a segment disjoint from $J'$ is preserved.

\begin{lemma}[Steps of \textsc{LLL} are monotone]\label{lem:lll_step_monotone}
Let $G\in\R^{d\times d}$ be positive definite and $V\in\Z^{d\times d}$ unimodular, and let $x_t=\log_2B_t$ for the Gram--Schmidt norms $B_t$ of $V$ for $G$. Let $V'$ be obtained from $V$ by one step of \textsc{\textup{LLL}} at an index $t+1\leq d$, as in the proof of Lemma~\ref{lem:lll_exact}, and let $x'_s=\log_2B'_s$ for its Gram--Schmidt norms $B'_s$. If the step makes no swap, then $x'=x$. If it makes a swap, then $x'_s=x_s$ for $s\notin\{t,t+1\}$, $x'_t+x'_{t+1}=x_t+x_{t+1}$ and $x_{t+1}\leq x'_t<x_t-\log_2(4/3)$. In both cases, for every set $J$ of consecutive indices with $t,t+1\in J$ and every $r\in J$,
\begin{align}\label{eq:lll_step_monotone}
    \sum_{s\in J,\,s\leq r}x'_s\leq\sum_{s\in J,\,s\leq r}x_s,\qquad\max_{s\in J,\,s\leq r}x'_s\leq\max_{s\in J,\,s\leq r}x_s,\qquad\min_{s\in J,\,s\geq r}x'_s\geq\min_{s\in J,\,s\geq r}x_s .
\end{align}
\end{lemma}

\begin{proof}
The size reduction in the step changes no $B_s$. Suppose that the step makes a swap, and let $\mu=\mu_{t+1,t}$ after the size reduction. The swap happens because $B_{t+1}<(3/4-\mu^2)B_t$, and it makes $B'_t=B_{t+1}+\mu^2B_t$, so $B_{t+1}\leq B'_t<(3/4)B_t$. By the proof of Lemma~\ref{lem:lll_exact} it changes only $d_t$, so $B'_s=B_s$ for $s\notin\{t,t+1\}$ and $B'_tB'_{t+1}=B_tB_{t+1}$. This proves the first claims, and it shows that $x'_t$ and $x'_{t+1}=x_t+x_{t+1}-x'_t$ both lie in $[x_{t+1},x_t]$. For~\eqref{eq:lll_step_monotone}, the sums agree for $r\neq t$ and drop by $x_t-x'_t\geq 0$ for $r=t$. The maxima agree for $r<t$ and do not increase for $r\geq t$, because $x'_t\leq x_t$ and $x'_{t+1}\leq x_t$. The minima agree for $r>t+1$ and do not decrease for $r\leq t+1$, because $x'_{t+1}\geq x_{t+1}$ and $x'_t\geq x_{t+1}$.
\end{proof}

The inequalities~\eqref{eq:lll_step_monotone} are the profile inequalities of Definition~3 of~\cite{zww26}. If they hold for $(x,x')$ and for $(x',x'')$, they hold for $(x,x'')$.

\begin{algorithm}[!ht]
\caption{Reduction by iterated subsegments}
\label{alg:segment_lll}
\begin{algorithmic}[1]
\Procedure{SegmentLLL}{$G$} \Comment{$G\in\R^{d\times d}$ positive definite, $d\leq n$}
\If{$d=1$} \State \Return $I_1$ \EndIf
\State Compute $G=LDL^\top$, let $B_{\max}$ be the largest entry of $D$, and let $\bar h=\lceil\log_2d\rceil$ and $\bar d=2^{\bar h}$
\State $\bar G\gets G\oplus B_{\max}I_{\bar d-d}$ and $U\gets I_{\bar d}$
\State \Call{Reduce}{$[\bar d]$}
\State \Return the leading $d\times d$ block of $U$
\EndProcedure
\Procedure{Reduce}{$J$} \Comment{$J$ a segment of order $2^h$}
\If{$h\geq 2$} \State \Call{Reduce}{$J_L$}, \Call{Reduce}{$J_R$} \EndIf
\State \Call{Merge}{$J$}
\EndProcedure
\Procedure{Merge}{$J$} \Comment{$J$ a segment of order $2^h$ with flat halves}
\If{$h=1$}
\State Run \textsc{LLL} on the Gram matrix of the projections of the two columns of $U$ with indices in $J$, and multiply these columns by its output
\Else
\Repeat
\State \Call{Merge}{$J_M$}, \Call{Merge}{$J_L$}, \Call{Merge}{$J_R$}
\Until{${\rm sp}(J)\leq\tau_h$}
\EndIf
\EndProcedure
\end{algorithmic}
\end{algorithm}

In Algorithm~\ref{alg:segment_lll}, every procedure changes the current matrix $U$ by multiplying some of its columns by a unimodular matrix. Gram--Schmidt norms, flatness and ${\rm sp}$ refer to the current $U$ and $\bar G$. One execution of the body of the repeat loop is a \emph{round}. For $J=\{t,t+1\}$, the projections in \textsc{Merge} are onto the orthogonal complement of $u_1,\ldots,u_{t-1}$ in $\langle\cdot,\cdot\rangle_{\bar G}$. So the Gram--Schmidt data of their Gram matrix are $B_t$, $B_{t+1}$ and $\mu_{t+1,t}$, and each step of \textsc{LLL} on it is a step of \textsc{LLL} at the index $t+1$ on $U$. So every change of $U$ is a step of \textsc{LLL} as in Lemma~\ref{lem:lll_step_monotone}, for $\bar G$. The last $\bar d-d$ columns of $U$ never change, and $B_t=B_{\max}$ for $t>d$ throughout. To see this, note that as long as this holds, these columns are orthogonal in $\langle\cdot,\cdot\rangle_{\bar G}$ to all other columns, so $\mu_{t+1,t}=0$ for $t\geq d$. Under the same condition, the largest of $B_1,\ldots,B_d$ is at most $B_{\max}$ by~\eqref{eq:lll_step_monotone} with $J=[\bar d]$ and $r=d$. So a step at an index $t+1>d$ makes neither a size reduction nor a swap. Hence the returned matrix is unimodular, and its Gram--Schmidt norms for $G$ are $B_1,\ldots,B_d$.

\begin{lemma}[Centered segment, after Lemma~3 of~\cite{zww26}]\label{lem:centered_segment}
Let $m_L,m_R\geq 1$, $m=m_L+m_R$, $I=[m_L]$, $I'=[m]\setminus I$, $1\leq b\leq\min\{m_L,m_R\}$ and $A=\{m_L-b+1,\ldots,m_L+b\}$. Let $x^0,x^1,x^2\in\R^m$ be such that the inequalities~\eqref{eq:lll_step_monotone} with $J=[m]$ hold for $(x,x')=(x^0,x^1)$ and for $(x,x')=(x^1,x^2)$, such that $x^1_s=x^0_s$ for $s\notin A$, and such that $\sum_{s\in A}x^1_s=\sum_{s\in A}x^0_s$. Let $\eta_I=\max\{{\rm sp}_I(x^0),{\rm sp}_I(x^2)\}$ and $\eta_{I'}=\max\{{\rm sp}_{I'}(x^0),{\rm sp}_{I'}(x^2)\}$. Then
\begin{align*}
    {\rm sp}_{[m]}(x^2)\leq\eta_I+{\rm sp}_A(x^1)+\eta_{I'}+\frac{2}{b^2}\sum_{r=m_L-b+1}^{m_L+b-1}\ \sum_{s\leq r}(x^0_s-x^2_s).
\end{align*}
\end{lemma}

\begin{proof}
Let $i_L=m_L-b+1$, $i_R=m_L+b$, $A_L=\{i_L,\ldots,m_L\}$ and $A_R=\{m_L+1,\ldots,i_R\}$. For $r\in[m]$ let $D_r=\sum_{s\leq r}(x^0_s-x^2_s)$ and $D^1_r=\sum_{s\leq r}(x^0_s-x^1_s)$. The first inequality of~\eqref{eq:lll_step_monotone} for $(x^1,x^2)$ gives $D_r\geq D^1_r$, and for $(x^0,x^1)$ it gives $D^1_r\geq 0$. Let
\begin{align*}
    g_L=\max\Big\{0,\ \min_{s\in A_L}x^0_s-\max_{s\in A_L}x^1_s\Big\},\qquad g_R=\max\Big\{0,\ \min_{s\in A_R}x^1_s-\max_{s\in A_R}x^0_s\Big\}.
\end{align*}
Let $1\leq q\leq b$. As $x^1_s=x^0_s$ for $s<i_L$, we have $D^1_{i_L+q-1}=\sum_{s=i_L}^{i_L+q-1}(x^0_s-x^1_s)$. This is at least $qg_L$, because it is nonnegative and, if $g_L>0$, each of its terms is at least $g_L$. As $x^1$ and $x^0$ agree outside $A$ and have the same sum over $A$, we have $D^1_{i_R-q}=\sum_{s=i_R-q+1}^{i_R}(x^1_s-x^0_s)$. In the same way, this is at least $qg_R$, because these indices lie in $A_R$. Hence $D_{m_L-b+q}\geq qg_L$ and $D_{m_L+b-q}\geq qg_R$. The indices $m_L-b+q$ and $m_L+b-q$ with $1\leq q<b$, together with $m_L$, are the $2b-1$ indices from $m_L-b+1$ to $m_L+b-1$, and $D_{m_L}\geq b\max\{g_L,g_R\}$. So
\begin{align*}
    \sum_{r=m_L-b+1}^{m_L+b-1}D_r\geq\sum_{q=1}^{b-1}q\,(g_L+g_R)+b\max\{g_L,g_R\}\geq\frac{b^2}{2}(g_L+g_R).
\end{align*}
Now let $y^+=\max_{s\in A_L}x^1_s$ and $y^-=\min_{s\in A_R}x^1_s$, so that $y^+-y^-\leq{\rm sp}_A(x^1)$. For $s\in I$ with $s<i_L$, we have $x^1_s=x^0_s\leq\min_{s'\in A_L}x^0_{s'}+\eta_I\leq y^++g_L+\eta_I$, and for $s\in A_L$ we have $x^1_s\leq y^+$. By the second inequality of~\eqref{eq:lll_step_monotone} for $(x^1,x^2)$ with $r=m_L$, we get $x^2_i\leq y^++g_L+\eta_I$ for all $i\in I$. In the same way, for $s\in I'$ with $s>i_R$ we have $x^1_s=x^0_s\geq\max_{s'\in A_R}x^0_{s'}-\eta_{I'}\geq y^--g_R-\eta_{I'}$, and for $s\in A_R$ we have $x^1_s\geq y^-$, so the third inequality with $r=m_L+1$ gives $x^2_j\geq y^--g_R-\eta_{I'}$ for all $j\in I'$. Hence $x^2_i-x^2_j\leq\eta_I+{\rm sp}_A(x^1)+\eta_{I'}+g_L+g_R$ for $i\in I$ and $j\in I'$. For $i\leq j$ both in $I$ or both in $I'$, the difference $x^2_i-x^2_j$ is at most $\eta_I$ or $\eta_{I'}$. Together with the bound on $g_L+g_R$, this proves the lemma.
\end{proof}

\begin{lemma}[Rounds of \textsc{Merge}, after Theorem~4 of~\cite{zww26}]\label{lem:merge}
For $h\geq 2$, let $k_h=h-1$ if $h\leq 3$ and $k_h=1+\lceil\log_2h\rceil$ if $h\geq 4$, let $b_h=2^{h-1-k_h}$, and for a segment $J=\{p+1,\ldots,p+2^h\}$ of order $2^h$ let $C(J)=\{p+2^{h-1}-b_h+1,\ldots,p+2^{h-1}+b_h-1\}$. Let $J$ be a segment of order $2^h$, where $1\leq h\leq\bar h$, and consider a call \textsc{\textup{Merge}}$(J)$ that starts when $J$ has flat halves. Then every call of \textsc{\textup{Merge}} made during it starts when its segment has flat halves, the call changes only columns of $U$ with indices in $J$, it terminates, and when it returns, $J$ is flat. Moreover, if $h\geq 2$ and ${\rm sp}(J)>\tau_h$ at the end of a round of the call, then
\begin{align*}
    \sum_{i\in C(J)}\log_2\frac{d_i^{\rm in}}{d_i^{\rm out}}>b_h^2\,2^{h-2},
\end{align*}
where $d_i^{\rm in}$ and $d_i^{\rm out}$ denote $d_i(U)$, computed with $\bar G$, at the start and at the end of the round.
\end{lemma}

\begin{proof}
For $h\leq 3$ we have $k_h=h-1$ and $b_h=1$. For $h\geq 4$ we have $2h\leq 2^{k_h}<4h$, and $k_h\leq h-1$ because $\lceil\log_2h\rceil\leq h-2$, so $2^{h-3}/h<b_h\leq 2^{h-2}/h$. In all cases $1\leq k_h\leq h-1$, $b_h\leq 2^{h-2}$, $2^h\leq 8hb_h$ and $2b_h-1\leq 2^h/h$. Moreover $(h-k_h)2^{h-k_h}\leq 2^{h-1}$, which reads $2\leq 2^{h-1}$ for $h\leq 3$ and follows from $2^{k_h}\geq 2h$ for $h\geq 4$. Hence
\begin{align}\label{eq:merge_gap}
    \tau_h-2\tau_{h-1}-\tau_{h-k_h}=2^h-(h-k_h)2^{h-k_h}\geq 2^{h-1}.
\end{align}

We use induction on $h$. For $h=1$, let $J=\{t,t+1\}$. The call makes no further call and changes only $u_t$ and $u_{t+1}$. By Lemma~\ref{lem:lll_exact}, it terminates with $B_{t+1}\geq B_t/2$, so ${\rm sp}(J)\leq 1<\tau_1$ and $J$ is flat. Let $h\geq 2$, and assume the lemma for smaller orders. Consider a round of the call that starts when $J_L$ and $J_R$ are flat. If $h\geq 3$, the halves of $J_M$ are the right half of $J_L$ and the left half of $J_R$, so they are flat. So, also for $h=2$, $J_M$ has flat halves, and by induction \textsc{Merge}$(J_M)$ terminates, changes only columns in $J_M$, and returns with $J_M$ flat. When \textsc{Merge}$(J_L)$ starts and $h\geq 3$, its halves are the left half of $J_L$, whose Gram--Schmidt norms have not changed in this round, and the left half of $J_M$. Both are flat, so in all cases $J_L$ has flat halves, and \textsc{Merge}$(J_L)$ returns with $J_L$ flat. When \textsc{Merge}$(J_R)$ starts and $h\geq 3$, its halves are the right half of $J_M$ and the right half of $J_R$, whose Gram--Schmidt norms have not changed in this round, because the earlier calls of the round change only columns with smaller indices. Both are flat, so in all cases $J_R$ has flat halves, \textsc{Merge}$(J_R)$ returns with $J_R$ flat, and $J_L$ stays flat. So every round terminates, every call of \textsc{Merge} made during it starts with flat halves, by induction for the calls made within the three calls, and every round ends with $J_L$ and $J_R$ flat. In particular every round starts with $J_L$ and $J_R$ flat, and when the test of a round succeeds, $J$ is flat.

Now consider a round after which ${\rm sp}(J)>\tau_h$. For $1\leq k\leq h-1$ let $J^{(k)}=\{p+2^{h-1}-2^{h-1-k}+1,\ldots,p+2^{h-1}+2^{h-1-k}\}$, so that $J^{(1)}=J_M$ and, for $k\geq 2$, $J^{(k)}$ is the middle of $J^{(k-1)}$, a segment of order $2^{h-k}$. The first call of the round is \textsc{Merge}$(J^{(1)})$, and for $2\leq k\leq h-1$ the first call of the first round of \textsc{Merge}$(J^{(k-1)})$ is \textsc{Merge}$(J^{(k)})$. Let $k=k_h$ and consider this first call \textsc{Merge}$(J^{(k)})$. Until it returns, the only changes of $U$ in the round are made within it. By the first part it starts with flat halves, so by induction it returns with $J^{(k)}$ flat and has changed only columns in $J^{(k)}$. Let $x^0,x^1,x^2\in\R^{2^h}$ be the vectors with entries $x_{p+r}$, $r\in[2^h]$, at the start of the round, when this call returns, and at the end of the round. Every change of $U$ in the round is a step of \textsc{LLL} at an index $t+1$ with $t,t+1\in J$. So Lemma~\ref{lem:lll_step_monotone}, applied with the set $J$ and with indices shifted by $p$, shows that the inequalities~\eqref{eq:lll_step_monotone} with $J=[2^h]$ hold for $(x^0,x^1)$ and for $(x^1,x^2)$. The vectors $x^0$ and $x^1$ agree outside the set $A$ of positions of $J^{(k)}$ and have the same sum over $A$. We apply Lemma~\ref{lem:centered_segment} with $m_L=m_R=2^{h-1}$ and $b=b_h=2^{h-1-k}$, for which $A$ is the set of that lemma. As $J_L$ and $J_R$ are flat at the start and at the end of the round, and $J^{(k)}$ is flat when the call returns, we get
\begin{align*}
    \tau_h<{\rm sp}_{[2^h]}(x^2)\leq 2\tau_{h-1}+\tau_{h-k}+\frac{2}{b_h^2}\sum_{r=2^{h-1}-b_h+1}^{2^{h-1}+b_h-1}\ \sum_{s\leq r}(x^0_s-x^2_s).
\end{align*}
Since $d_p(U)$ does not change during the round, $\sum_{s\leq r}(x^0_s-x^2_s)=\log_2(d_{p+r}^{\rm in}/d_{p+r}^{\rm out})$. Together with~\eqref{eq:merge_gap}, this gives the claimed bound.

Finally, the call terminates. By Lemma~\ref{lem:lll_step_monotone} no $d_i(U)$ increases, and $P(U)=\prod_{i<\bar d}d_i(U)\geq\prod_{i<\bar d}(\lambda_1^2/i)^i>0$ as in the proof of Lemma~\ref{lem:lll_exact}, where $\lambda_1$ is the length of a shortest nonzero vector of $\Z^{\bar d}$ in the norm $v\mapsto(v^\top\bar Gv)^{1/2}$. So a round after which the test fails lowers $\log_2P(U)$ by more than $b_h^2\,2^{h-2}\geq 1$. Hence there are finitely many such rounds, and the call returns after a round whose test succeeds, with $J$ flat.
\end{proof}

\begin{lemma}[Reduction by iterated subsegments]\label{lem:segment_lll}
Let $\kappa_n=\lceil\log_2n\rceil\,2^{\lceil\log_2n\rceil}$, so that $\kappa_n<2n\lceil\log_2n\rceil$. The routine \textsc{\textup{SegmentLLL}} of Algorithm~\ref{alg:segment_lll} has the following properties, where $G\in\R^{d\times d}$ with $d\leq n$ is positive definite.
\begin{itemize}
    \item[(a)] On input $G$, it returns a unimodular $U\in\Z^{d\times d}$ whose Gram--Schmidt norms for $G$ satisfy $B_j\geq 2^{-\kappa_n}B_l$ for all $1\leq l<j\leq d$. In particular, \textsc{\textup{SegmentLLL}} is a $\gamma$-approximate shortest vector routine with $\gamma=2^{\kappa_n/2}=2^{O(n\log n)}$.
    \item[(b)] On input $G$, it uses
    \begin{align*}
        O(d^3+d\log^2(2d)\cdot\Delta_{\max})
    \end{align*}
    arithmetic operations, where $\Delta_{\max}=\max_{1\leq i<d}\log_2(d_i(I_d)/d_i(U))$ for $d_i$ computed with $G$, and $\Delta_{\max}=0$ if $d=1$. Moreover, $\Delta_{\max}\geq 0$, and if $G$ is a multiple of $I_d$, then $U=I_d$.
\end{itemize}
\end{lemma}

\begin{proof}
For $d=1$ there is nothing to show, so let $d\geq 2$. For Part~(a), a call \textsc{Reduce}$(J)$ changes only columns in $J$ and returns with $J$ flat, by induction on the order of $J$. For order $2$ this is Lemma~\ref{lem:merge}. For order $2^h$ with $h\geq 2$, \textsc{Reduce}$(J_L)$ and \textsc{Reduce}$(J_R)$ return with $J_L$ and $J_R$ flat, the second without changing the Gram--Schmidt norms in $J_L$. Then \textsc{Merge}$(J)$ starts with flat halves and returns with $J$ flat by Lemma~\ref{lem:merge}. So at the end $[\bar d]$ is flat, and ${\rm sp}([d])\leq{\rm sp}([\bar d])\leq\tau_{\bar h}\leq\kappa_n$, as $\bar h\leq\lceil\log_2n\rceil$. As shown after Algorithm~\ref{alg:segment_lll}, the returned matrix is unimodular with Gram--Schmidt norms $B_1,\ldots,B_d$ for $G$. This proves the first claim. It gives $(U^\top GU)_{1,1}=B_1\leq 2^{\kappa_n}\min_kB_k$, and $\min_kB_k\leq\min_{x\in\Z^d\setminus\{{\bf 0}_d\}}x^\top Gx$ by~\eqref{eq:successive_minima_gs}, which is the guarantee of Definition~\ref{def:approx_sv} with $\gamma^2=2^{\kappa_n}$. Finally $\kappa_n<2n\lceil\log_2n\rceil$, because $2^{\lceil\log_2n\rceil}<2n$.

For Part~(b), no $d_i(U)$ increases during the run, by Lemma~\ref{lem:lll_step_monotone}. For $i<d$, $d_i(U)$ equals $d_i$ of the leading $d\times d$ block of $U$ computed with $G$, so $\Delta_i=\log_2(d_i(I_d)/d_i(U))\geq 0$ is the total decrease of $\log_2d_i(U)$ during the run. Its decrease during any part of the run is therefore at most $\Delta_i\leq\Delta_{\max}$. For $i\geq d$, $d_i(U)=\det(G)\,B_{\max}^{i-d}$ never changes. If $G$ is a multiple of $I_d$, then all $B_t$ are equal and all $\mu_{k,j}$ vanish, so no step of \textsc{LLL} changes $U$, and $U=I_d$.

We now describe how the procedures work and count their operations. Write $U^\top\bar GU=LDL^\top$ with $L$ unit lower triangular and $D={\rm diag}(B_1,\ldots,B_{\bar d})$. A call of \textsc{Reduce} or \textsc{Merge} on a segment $J$ works on the \emph{data} of $J$, which consist of the entries of $L$ with row and column in $J$ and the $B_t$ with $t\in J$. These are the Gram--Schmidt data of the projections of the columns of $U$ in $J$ onto the orthogonal complement of the earlier columns. The call keeps a unimodular matrix $W\in\Z^{J\times J}$, initially the identity, by which it has multiplied these columns. A call on a segment $J'\subseteq J$ that it makes receives a copy of the data of $J'$, with $O(|J'|^2)$ operations, and returns its matrix $W'$ and the final data of $J'$. The caller then multiplies the columns of $W$ with indices in $J'$ by $W'$ and updates its data by the formulas of Lemma~\ref{lem:block_update}, applied to the Gram matrix of these projections, to the set $J'$ and to the matrix $W'$. By that lemma, the result is the data of $J$ for the new $U$. These updates take $O(|J|\cdot|J'|^2)$ operations. The test ${\rm sp}(J)\leq\tau_h$ checks $2^{\tau_h}B_j\geq\max\{B_i:i\in J,\ i\leq j\}$ for all $j\in J$, which takes $O(2^h)$ operations with running maxima, once $2^{\tau_h}=(2^{2^h})^h$ is known. We compute these numbers for all $h\leq\bar h$ once, with $O(\bar h^2)$ operations. The outermost call works on the data of $\bar G$, which the factorization of $G$ gives with $O(d^3)$ operations. The matrix $W$ of the outermost call is the final $U$. So, apart from the calls that it makes, a round of a call \textsc{Merge}$(J)$ of order $2^h\geq 4$ and a call \textsc{Reduce}$(J)$ of order $2^h$ use $O(8^h)$ operations. A call \textsc{Merge}$(J)$ of order $2$ uses $O(1+N_J)$ operations, where $N_J$ is its number of swaps (Lemma~\ref{lem:lll_exact}).

For $2\leq h\leq\bar h$, let $R_h$ be the number of rounds of calls of \textsc{Merge} of order $2^h$, and let $F_h$ be the number of those rounds after which the test fails. For $1\leq h\leq\bar h$, let $M_h$ be the number of calls of \textsc{Merge} of order $2^h$. Each call has exactly one round whose test succeeds, so $R_h=M_h+F_h$. Calls of order $2^h$ are made three in each round of order $2^{h+1}$ and once by \textsc{Reduce} for each of the $2^{\bar h-h}$ segments $\{p+1,\ldots,p+2^h\}$ with $p$ a multiple of $2^h$. So $M_h=3R_{h+1}+2^{\bar h-h}$, where $R_{\bar h+1}=0$. Hence $R_h=\sum_{j=h}^{\bar h}3^{j-h}(F_j+2^{\bar h-j})$, and
\begin{align*}
    \sum_{h=2}^{\bar h}8^hR_h\leq\sum_{j=2}^{\bar h}8^j(F_j+2^{\bar h-j})\sum_{i\geq 0}(3/8)^i=O\Big(\sum_{j=2}^{\bar h}8^jF_j+\bar d^3\Big).
\end{align*}
Each swap lowers one $\log_2d_t(U)$ with $t<d$ by more than $\log_2(4/3)$ (Lemma~\ref{lem:lll_step_monotone}), so the total number of swaps is less than $\sum_{i<d}\Delta_i/\log_2(4/3)\leq d\Delta_{\max}/\log_2(4/3)$.

Let $2\leq h\leq\bar h$. By Lemma~\ref{lem:merge}, a round counted in $F_h$, of a call on a segment $J$, lowers $\sum_{i\in C(J)}\log_2d_i(U)$ by more than $b_h^2\,2^{h-2}$. The sets $C(J)$ of different segments $J$ of order $2^h$ are disjoint, because the points $p+2^{h-1}$ differ by multiples of $2^{h-1}$, and $C(J)$ lies within distance $b_h-1<2^{h-2}$ of its point. Two rounds of calls on the same segment do not overlap in time, because calls of the same order are never nested. As no $d_i(U)$ increases and $d_i(U)$ is constant for $i\geq d$, we get $F_h\,b_h^2\,2^{h-2}\leq|S_h|\,\Delta_{\max}$, where $S_h$ is the union of the sets $C(J)\cap[d-1]$. There are fewer than $2^{\bar h-h+1}$ segments of order $2^h$, and $|C(J)|=2b_h-1\leq 2^h/h$, so $|S_h|<2^{\bar h+1}/h$. With $2^h\leq 8hb_h$ this gives
\begin{align*}
    8^hF_h\leq 4\Big(\frac{2^h}{b_h}\Big)^2\frac{2^{\bar h+1}}{h}\,\Delta_{\max}\leq 2^{\bar h+9}\,h\,\Delta_{\max},
\end{align*}
and so $\sum_{h=2}^{\bar h}8^hF_h\leq 2^{\bar h+8}\,\bar h(\bar h+1)\,\Delta_{\max}$.

We add up the factorization, the numbers $2^{\tau_h}$, the rounds, the $\sum_h2^{\bar h-h}\cdot O(8^h)=O(\bar d^3)$ operations of the calls of \textsc{Reduce}, and the $M_1=3R_2+2^{\bar h-1}$ calls of order $2$ with their swaps. With $\bar d<2d$ and $\bar h+1\leq 2\log_2(2d)$, the total is $O(d^3+d\log^2(2d)\cdot\Delta_{\max})$.
\end{proof}
\section{Efficient Minimization via Fast Cutting and Lazy Width Measurement}\label{sec:minimization}

This section assembles our algorithm and proves Theorem~\ref{thm:main_formal}. Section~\ref{sec:minimization:alg} gives the algorithm, Section~\ref{sec:minimization:correct} shows that it finds an integral minimizer, Sections~\ref{sec:minimization:oracle} and~\ref{sec:minimization:runtime} count its oracle calls and its operations, and Section~\ref{sec:minimization:main} states the main result.

\subsection{Our Algorithm}\label{sec:minimization:alg}

Our algorithm simplifies the method of~\cite{j22}. It combines the cutting plane method of Section~\ref{sec:cpm:method} with an approximate shortest vector routine (Definition~\ref{def:approx_sv}). It maintains an affine subspace $W$, a lattice $\Lambda$, a polytope $K\subseteq W$, a point $x_K\in K$ and the Hessian $H_K=H_K(x_K)$ of the log barrier of $K$. Each iteration computes a short vector $v$ of $\Lambda$ in the norm $\|\cdot\|_{H_K^{-1}}$. If $v$ is long, a \emph{cutting iteration} runs one block of Algorithm~\ref{alg:hcpm}. Otherwise every integral point of the ellipsoid $2R'\cdot E(x_K,H_K)$, which contains $K$ in the analysis, lies on one hyperplane $P$, and a \emph{reduction iteration} intersects $W$ with it. The algorithm can fail in three ways. For $n\geq n_0$, the \emph{operation test} fails when earlier blocks have made at least $M$ operations on their data structures \textsc{LS}. A block can abort, and the \emph{slice test} fails when the slice of the ellipsoid by $P$ is empty or degenerate. The analysis rules out all three on a good event. The first line lets the algorithm accept every $R>0$, since $K^*\subseteq B_\infty(1)$ when $R<1$.

\begin{algorithm}[!ht]
\caption{Our Algorithm}
\label{alg:ours}
\begin{algorithmic}[1]
\Procedure{Main}{$\SO, R, \gamma$, \textsc{ApproxSV}} \Comment{Theorem~\ref{thm:main_informal}}
\State $R\gets\max\{R,1\}$
\State $W\gets \R^n$ be an affine subspace
\State $K\gets B_{\infty}(2R)$ be a polytope with $2n$ constraints
\Comment{$K$ can be parameterized by $A\in \R^{2n \times n}$ and $b\in \R^{2n}$}
\State $\Lambda\gets \Z^n$ be a lattice
\State $x_K\gets 0$, the center of $K$
\State $H_K\gets \sum_{i=1}^{2n} \frac{a_ia_i^\top}{(a_i^\top x_K-b_i)^2}=\frac{1}{2R^2}I_n$, the Hessian of the log barrier
\While{$\mathrm{dim}(W)>0$}
\State $v\gets \textsc{ApproxSV}(\Lambda, H_K^{-1})$ \Comment{Definition~\ref{def:approx_sv}}
\If{$\|v\|_{H_K^{-1}}\geq 2^{-100n\log n}$} \Comment{$\log$ to base $2$: the test is $\|v\|_{H_K^{-1}}^2\geq n^{-200n}$}
\State \textbf{if} $n\geq n_0$ and the earlier blocks made at least $M$ operations on their data structures \textsc{LS} \textbf{then} \Return \textsc{fail} \Comment{operation budget}
\State $(K', x_{K'}, H_{K'})\gets \textsc{HybridCPM}(\SO, K, x_K)$, and return \textsc{fail} if it aborts \Comment{Algorithm~\ref{alg:hcpm}, Lemma~\ref{lem:hybrid_cpm}}
\State $K\gets K', x_{K}\gets x_{K'}, H_K\gets H_{K'}$
\Else
\State Find $z\in \Z^n$ such that $v=\Pi_{W-x_K}(z)$
\State $P\gets \{y: v^\top y=(v-z)^\top x_K+ [z^\top x_K] \}$
\State $W\gets W\cap P$
\State Let $E(\bar z, A)$ be the ellipsoid $2R'\cdot E(x_K, H_K)\cap P$, with $R'$ as in Section~\ref{sec:cpm:params}, and return \textsc{fail} if this set is empty, or if it is a single point and $\dim(W)\geq1$ \Comment{never fails while $x^*\in K\subseteq R'\cdot E(x_K,H_K)$}
\State $K\gets \{\bar z+YJ^{-\top}t: t\in[-1,1]^{\dim(W)}\}$, where the columns of $Y$ form a basis of the linear subspace parallel to $W$ and $Y^\top AY=JJ^\top$ is a Cholesky factorization (Lemma~\ref{lem:reduction_geometry}). If $\dim(W)=0$, then $Y$ has no columns, $K=\{\bar z\}=W$, and the loop ends after this iteration
\State $x_K\gets \bar z, H_K\gets\sum_{i=1}^{2\dim(W)} \frac{a_ia_i^\top}{(a_i^\top x_K-b_i)^2}$
\State Construct hyperplane $P_0\gets \{y: v^\top y=0 \}$
\State $\Lambda \gets \Pi_{P_0}(\Lambda)$
\EndIf
\EndWhile
\State \Return the single point of $W$
\EndProcedure
\end{algorithmic}
\end{algorithm}

\paragraph{The lattice.} Let $W_0$ be the linear subspace parallel to $W$. We claim that $\Lambda=\Pi_{W_0}(\Z^n)$ is always a lattice of rank $\dim(W)$ that spans $W_0$. This holds at the start. A reduction iteration replaces $W_0$ by $W_0\cap P_0$ and $\Lambda$ by $\Pi_{P_0}(\Lambda)$, where $P_0$ is the orthogonal complement of $v\in\Lambda\subseteq W_0$. For $y\in\R^n$ and $u=\Pi_{W_0}(y)$, the point $\Pi_{P_0}(u)=u-(v^\top u/v^\top v)v$ lies in $W_0\cap P_0$, and $y-\Pi_{P_0}(u)=(y-u)+(v^\top u/v^\top v)v$ is orthogonal to $W_0\cap P_0$. So $\Pi_{P_0}\circ\Pi_{W_0}=\Pi_{W_0\cap P_0}$. Also, $v$ is the first vector of a basis of $\Lambda$ (Definition~\ref{def:approx_sv}), and the kernel of $\Pi_{P_0}$ is spanned by $v$, so $\Pi_{P_0}$ maps the other basis vectors to a basis of $\Pi_{P_0}(\Lambda)$. The claim follows by induction. It also shows that the integer vector $z$ with $v=\Pi_{W_0}(z)$ that a reduction iteration needs exists.

\paragraph{The good event.} For $n\geq n_0$, the operation test uses the budget
\begin{align*}
    M=\Theta((n^2\log n+n\log(\gamma nR))\log\log n)
\end{align*}
of Section~\ref{sec:r3_params}, with $R\geq1$ the radius after the first line. The algorithm takes $n_0$, $C_M$ and the factors of $M$ that depend only on $n$ as given. It computes the term $\lceil\log_2(\gamma R)\rceil$ by doubling $1$ until it reaches $\gamma R$, with $O(1+\log(\gamma R))$ operations. Any upper bound $\hat\gamma\geq\gamma$ may replace $\gamma$ in $M$ and in all bounds. Let $\mathcal{G}$ be the event that every output of \textsc{LS} among the first $M$ operations, counted over all blocks in order, has $\ell_2$ error at most $\delta_{\rm J}(n)$. For $n<n_0$ the scores are exact, and $\mathcal{G}$ is the sure event. On $\mathcal{G}$, every output that the algorithm uses has $\ell_2$ error at most $\delta_0$ (Lemma~\ref{lem:main_oracle_formal}), and $\mathcal{G}$ has high probability (proof of Theorem~\ref{thm:main_formal}).

Off $\mathcal{G}$, the algorithm may fail, return a wrong point or exceed the stated bounds, and the failure probability of Theorem~\ref{thm:main_formal} covers all of these. Under the hypotheses of that theorem, the algorithm always terminates for $n\geq2$, because at most $n$ iterations are dimension reductions and the number of blocks is bounded. For $n<n_0$, Lemma~\ref{lem:main_oracle_formal} bounds the number of blocks, since $\mathcal{G}$ is the sure event. For $n\geq n_0$, each block makes at least two operations on its \textsc{LS}, so the operation test allows at most $M/2+1$ blocks.

\subsection{Correctness of Our Algorithm}\label{sec:minimization:correct}

We show that the integral minimizer $x^*$ of Section~\ref{sec:cpm:method}, which the algorithm never computes, stays in every polytope with positive slack in every row. So no dimension reduction loses it.

\begin{lemma}[The fixed minimizer stays inside]\label{lem:xstar}
Let $R_0>0$ satisfy $4R_0\leq2^{100n\log n}$. Consider an initial part of a run of Algorithm~\ref{alg:ours}, or of the algorithm of Corollary~\ref{cor:exact_score}, whose reduction iterations use the ellipsoid $2R_0\cdot E(x_K,H_K)\cap P$. Suppose that every triple $(K,x_K,H_K)$ at which a reduction iteration of this part happens satisfies $K\subseteq R_0\cdot E(x_K,H_K)$. Then every polytope that the algorithm holds in this part contains $x^*$, with slack at least $\rho$ in a cut $a^\top x\geq a^\top z-\rho$, at least $R$ in a row of the initial box, and at least $1/2$ in a row of a reduction box. Every slice $2R_0\cdot E(x_K,H_K)\cap P$ of this part contains $x^*$ in its interior relative to $W\cap P$. Finally, $D(K)\geq0$ for every such polytope $K$ with at most $\bar m$ rows whose hybrid center exists.
\end{lemma}

\begin{proof}
We use induction over the iterations of the part. The initial box contains $x^*$ with slack in $[R,3R]$ in every row (Lemma~\ref{lem:box_margin}, Part~1). A cutting iteration keeps $W$, its deletions only enlarge the polytope, and each row that it adds is a cut $a^\top x\geq a^\top z-\rho$ at a point $z\in W$. Since $x^*\in K\subseteq W$, \eqref{eq:cut_xstar} gives this cut slack at least $\rho>0$ at $x^*$.

Consider a reduction iteration at $(K,x_K,H_K)$, with $x^*\in K\subseteq R_0\cdot E(x_K,H_K)$. The ellipsoid $E:=2R_0\cdot E(x_K,H_K)=E(x_K,H_K/(4R_0^2))$ has full dimension in $W$. In the norm of the inverse of its matrix, $v$ has length $2R_0\|v\|_{H_K^{-1}}<1/2$, because the iteration happens only when $\|v\|_{H_K^{-1}}<2^{-100n\log n}\leq1/(4R_0)$. As $v$ is a nonzero vector of $\Lambda=\Pi_{W_0}(\Z^n)$ (paragraph \emph{The lattice}), Lemma~\ref{lem:dim_red} shows that $P\not\supseteq W$ and that every integral point of $E$ lies on $P$. This includes $x^*\in K\subseteq E$. Since $x^*$ lies in $R_0\cdot E(x_K,H_K)$, inside the interior of $E$ relative to $W$, the slice $E\cap P$ has full dimension in $W\cap P$ and contains $x^*$ in its relative interior. Lemma~\ref{lem:box_margin} (Part~2) then puts $x^*$ in the new box, with slack at least $1/2$ in every row.

So every row is a row of the initial box, a cut or a row of a reduction box, and $x^*$ has positive slack in it. Hence every term of $D(K)$ is finite. If $K$ has $m\leq\bar m$ rows and $x_c=x_c(K)$ exists, Lemma~\ref{lem:hybrid_loc} (Part~4) with $y=x^*$ gives $s_i(x^*)\leq(1+(n+cm)/c)\,s_i(x_c)\leq\bar\kappa\,s_i(x_c)$ for every row $i$, so every term of $D(K)$ is nonnegative.
\end{proof}

The next lemma applies Lemma~\ref{lem:xstar} to our algorithm. While the estimated leverage scores have small errors, every block is good and no dimension reduction fails.

\begin{lemma}[Runs with small errors]\label{lem:small_error_run}
Let $n\geq2$, let $f$, $\SO$, $R$, $\gamma$ and \textsc{\textup{ApproxSV}} satisfy the assumptions of Theorem~\ref{thm:main_formal}, except that $\log\gamma$ may be arbitrary, and let $i\geq1$. Consider a run of Algorithm~\ref{alg:ours}. Suppose that every output of \textsc{\textup{LS}} used in the first $i$ blocks has $\ell_2$ error at most $\delta_0$. Then each of these blocks is a good block on a polytope that contains $x^*$ with positive slack in every row, and no step of these blocks aborts. Moreover, the following holds until the start of block $i+1$, or until the end of the run if there is no such block. Every triple $(K,x_K,H_K)$ at which a reduction iteration happens satisfies $K\subseteq R'\cdot E(x_K,H_K)$, every slice $2R'\cdot E(x_K,H_K)\cap P$ contains $x^*$ in its interior relative to $W\cap P$, so that no slice test fails, and $x^*$ lies in every polytope that the algorithm holds.
\end{lemma}

\begin{proof}
We prove all claims together, by induction over the iterations before block $i+1$. At the start of an iteration, $(K,x_K,H_K)$ is a box or the output of a block $j\leq i$, and $H_K=H_K(x_K)$. A box has $2\dim(W)\leq2n\leq\bar m$ rows by~\ref{cond:C4}, and $\lambda_K(x_K)=0$ at its center (Lemma~\ref{lem:box_margin}, Part~1, and Lemma~\ref{lem:reduction_geometry}, Part~3). The output of block $j$ has at most $\bar m$ rows and $\lambda_K(x_K)\leq\lambda^*$, by Lemma~\ref{lem:hybrid_cpm} (Parts~1 and~4), which applies to block $j$ by induction. So a cutting iteration runs a good block, on a polytope that contains $x^*$ with positive slack in every row by induction, and no step of the block aborts by Lemma~\ref{lem:hybrid_block}. At a reduction iteration, Lemma~\ref{lem:sandwich} (Part~3) gives $K\subseteq R'\cdot E(x_K,H_K)$. We apply Lemma~\ref{lem:xstar} with $R_0=R'$ to the run up to the current iteration (its hypothesis $4R'\leq2^{100n\log n}$ holds by~\ref{cond:W6}). So $x^*$ stays in every polytope with positive slack in every row, and every slice contains $x^*$ in its interior relative to $W\cap P$. Such a slice is not empty, and it is not a single point when $\dim(W\cap P)\geq1$, so the slice test does not fail.
\end{proof}

\begin{lemma}[Formal version of Theorem~\ref{thm:main_informal}, output guarantee part]\label{lem:main_output_formal}
Let $n\geq2$. Given a separation oracle $\SO$ for a convex function $f:\R^n\to\R\cup\{+\infty\}$ and a number $R\geq1$ such that the set of minimizers $K^*$ of $f$ is nonempty, closed and contained in $B_\infty(R)$ and all extreme points of $K^*$ are integral, and given the number $\gamma$ and a $\gamma$-approximate shortest vector routine \textsc{\textup{ApproxSV}} (Definition~\ref{def:approx_sv}), run Algorithm~\ref{alg:ours} with input $R$. If every output of \textsc{\textup{LS}} that it uses has $\ell_2$ error at most $\delta_0$, which holds on the event $\mathcal{G}$ (Lemma~\ref{lem:main_oracle_formal}), then the algorithm does not fail and finds an integral minimizer of $f$.
\end{lemma}

\begin{proof}
Lemma~\ref{lem:small_error_run}, applied to all blocks, shows that no block aborts, that no slice test fails and that $x^*$ lies in every polytope. Step~6 of the proof of Lemma~\ref{lem:main_oracle_formal}, which assumes only the error bound of this lemma, shows that the run has at most $\bar N_{\rm c}$ cutting iterations. The operation test is active only for $n\geq n_0$, and for such $n$ Step~7 shows that it does not fail. As each reduction iteration lowers $\dim(W)$ by one, the run ends with $W=K=\{\bar z\}$, and this point is $x^*$, an integral minimizer of $f$.
\end{proof}

\subsection{Oracle Complexity}\label{sec:minimization:oracle}

We bound the number of blocks with the argument of~\cite{j22}, applied to a quantity $\wt\Phi$ that equals $-\mathcal{U}(K)+\ln\det(\Lambda)$ up to the term $\ln{\rm vol}(B_d)$, which depends only on $d$, and a term $\vartheta_K$ of size at most $\epsilon_{\rm f}n$ (see~\eqref{eq:phi_tilde_U}). A block lowers $\wt\Phi$ by at least $n\ln n$, a run of dimension reductions raises it by a bounded amount, and a low value of $\wt\Phi$ forces a dimension reduction. We first record three geometric facts.

\begin{lemma}[Ellipsoids and boxes]\label{lem:reduction_geometry}
Let $W$ be an affine subspace of $\R^n$ with linear part $W_0$, and let $d'=\dim(W)$. For an ellipsoid $E\subseteq W$ with center $q$ and a scalar $\zeta\geq 0$, write $\zeta E:=q+\zeta(E-q)$.
\begin{enumerate}
    \item Let $E\subseteq W$ be an ellipsoid of full dimension in $W$. If $S\subseteq\zeta E$ for some $\zeta\geq 0$, $u\in S$ and $\xi\geq 0$, then $u+\xi(S-u)\subseteq(1+2\xi)\zeta E$.
    \item Let $E\subseteq W$ be an ellipsoid of full dimension in $W$ with center $q$, and let $V\subseteq W_0$ be a linear subspace. Then ${\rm vol}(E\cap(u+V))\leq{\rm vol}(E\cap(q+V))$ for every $u\in W$, where ${\rm vol}$ is the $\dim(V)$-dimensional volume.
    \item Let $\bar z\in W$, let $A$ be a PSD matrix whose range is $W_0$, let $Y\in\R^{n\times d'}$ be a matrix whose columns form a basis of $W_0$, let $Y^\top AY=JJ^\top$ be a Cholesky factorization, and let $G:=YJ^{-\top}$. Then the box $\mathcal{B}=\{\bar z+Gt:t\in[-1,1]^{d'}\}$ satisfies $E(\bar z,A)\subseteq \mathcal{B}\subseteq\sqrt{d'}\cdot E(\bar z,A)$, the Hessian of the log barrier of $\mathcal{B}$ at $\bar z$ equals $2A$ as a quadratic form on $W_0$, and $\bar z=\vc(\mathcal{B})$. Moreover, for every $c>0$, $\bar z$ is the hybrid center $x_c(\mathcal{B})$ of Definition~\ref{def:hybrid}, so the decrement of $p$ at $\bar z$ is $0$.
\end{enumerate}
\end{lemma}

\begin{proof}
For Part 1, let $\|\cdot\|$ be the norm on $W_0$ whose unit ball is $E-q$, so that $\zeta E=\{x\in W:\|x-q\|\leq\zeta\}$. For $s\in S$, the point $u+\xi(s-u)$ lies in $W$ and
\begin{align*}
    \|u+\xi(s-u)-q\|=\|(1-\xi)(u-q)+\xi(s-q)\|\leq|1-\xi|\zeta+\xi\zeta\leq(1+2\xi)\zeta .
\end{align*}

For Part 2, write $E=\{x\in W:(x-q)^\top M(x-q)\leq 1\}$, where $M$ is a PSD matrix whose range is $W_0$. The function $g(y)=(u+y-q)^\top M(u+y-q)$ on $V$ is a convex quadratic whose quadratic part $y^\top My$ is positive definite on $V$. Let $y_0\in V$ be its minimizer and let $\varrho^2=g(y_0)\geq 0$. The gradient of $g$ at $y_0$ vanishes on $V$, so $g(y)=\varrho^2+(y-y_0)^\top M(y-y_0)$ for $y\in V$. Since $E\cap(q+V)=q+\{y\in V:y^\top My\leq 1\}$, the slice $E\cap(u+V)$ is empty if $\varrho>1$, and otherwise it equals $u+y_0+\sqrt{1-\varrho^2}\cdot\big(E\cap(q+V)-q\big)$. In both cases its volume is at most ${\rm vol}(E\cap(q+V))$.

For Part 3, the matrix $G=YJ^{-\top}\in\R^{n\times d'}$ maps $\R^{d'}$ bijectively onto $W_0$, and for $y=Gt$ we have $y^\top Ay=t^\top J^{-1}(Y^\top AY)J^{-\top}t=\|t\|_2^2$. Hence $E(\bar z,A)=\{\bar z+Gt:\|t\|_2\leq 1\}$, and the two containments follow from $\|t\|_\infty\leq\|t\|_2\leq\sqrt{d'}\,\|t\|_\infty$. Let $g_1,\ldots,g_{d'}$ be the linear functions on $W_0$ with $G^{-1}y=(g_1(y),\ldots,g_{d'}(y))$. The $2d'$ constraints of $\mathcal{B}$ are $1\pm g_i(x-\bar z)\geq 0$, each with slack $1$ at $x=\bar z$. So the Hessian of the log barrier of $\mathcal{B}$ at $\bar z$ is the quadratic form $y\mapsto 2\sum_ig_i(y)^2=2\|G^{-1}y\|_2^2=2y^\top Ay$ on $W_0$. Finally, the constraint set of $\mathcal{B}$ is invariant under the reflection $x\mapsto 2\bar z-x$, and hence so are $F$, $L$ and $p=F+cL$. A convex function $g$ with $g(\bar z+y)=g(\bar z-y)$ satisfies $g(\bar z)\leq\frac{1}{2}(g(\bar z+y)+g(\bar z-y))=g(\bar z+y)$ for every $y$. So $\bar z$ minimizes $F$ and $p$. Hence $\bar z=\vc(\mathcal{B})$, and $\bar z=x_c(\mathcal{B})$ with $\nabla p(\bar z)=0$, as $p$ is strictly convex (Lemma~\ref{lem:hybrid_local}, Part 1).
\end{proof}

\begin{lemma}[Oracle complexity part of Theorem~\ref{thm:main_formal}]\label{lem:main_oracle_formal}
Let $n\geq2$. Given a separation oracle $\SO$ for a convex function $f:\R^n\to\R\cup\{+\infty\}$ and a number $R\geq1$ such that the set of minimizers $K^*$ of $f$ is nonempty, closed and contained in $B_\infty(R)$ and all extreme points of $K^*$ are integral, and given the number $\gamma$ and a $\gamma$-approximate shortest vector routine \textsc{\textup{ApproxSV}} (Definition~\ref{def:approx_sv}), the following holds on the event $\mathcal{G}$. If $n\geq n_0$, then Algorithm~\ref{alg:ours} makes at most $M$ operations on the data structures \textsc{\textup{LS}}. For every $n\geq2$, every output of \textsc{\textup{LS}} that it uses has $\ell_2$ error at most $\delta_0$, and it calls Algorithm~\ref{alg:hcpm} $O(n+\log(\gamma R)/\log n)$ times. Hence it outputs an integral minimizer of $f$ with at most $O(n^2 \log n+n\log(\gamma n R))$ calls to $\SO$.
\end{lemma}

\begin{proof}
Steps~1 to~6 assume that every output of \textsc{LS} used by the algorithm has $\ell_2$ error at most $\delta_0$, so that Lemmas~\ref{lem:small_error_run}, \ref{lem:hybrid_cpm} and~\ref{lem:block_stability} apply to every block. Step~7 shows that this holds on $\mathcal{G}$. At any time, let $W_0$ be the linear subspace parallel to $W$ and $d=\dim(W)$, so that $\Lambda=\Pi_{W_0}(\Z^n)$ is a lattice of rank $d$ that spans $W_0$ (paragraph \emph{The lattice} in Section~\ref{sec:minimization:alg}). We write ${\rm vol}$ for the $d$-dimensional volume in $W$ and $H_K^{-1}$ for the inverse of $H_K$ on $W_0$. At the start of every iteration, $K$ is a box with $x_K=x_c(K)$ (Lemma~\ref{lem:box_margin}, Part~1, and Lemma~\ref{lem:reduction_geometry}, Part~3), or the output of a block, where $x_c(K)$ exists (Lemma~\ref{lem:hybrid_cpm}, Part~4). Either way $K$ has at most $\bar m$ rows, so $D(K)\geq0$ by Lemmas~\ref{lem:small_error_run} and~\ref{lem:xstar}. We track the quantity
\begin{align}\label{eq:phi_tilde_U}
    \begin{split}
    \wt\Phi&:=\ln{\rm vol}(E(x_K,H_K))+c\,D(K)+\ln\det(\Lambda)\\
    &=-\mathcal{U}(K)+\ln{\rm vol}(B_d)+\ln\det(\Lambda)+\vartheta_K,
    \end{split}
\end{align}
where $\vartheta_K:=F_K(x_c(K))-F_K(x_K)$, by~\eqref{eq:vol_dikin} and the definition of $\mathcal{U}$. At a box $\vartheta_K=0$, and at the output of a block $|\vartheta_K|\leq\epsilon_{\rm f}d\leq\epsilon_{\rm f}n$ (Lemma~\ref{lem:hybrid_cpm}, Part~4).

\textbf{Step 1: the start and the cutting iterations.} At the start, $\Lambda=\Z^n$, $K=B_\infty(2R)$, $x_K=0$ and $H_K=\frac{1}{2R^2}I_n$. So $\ln{\rm vol}(E(x_K,H_K))=\ln{\rm vol}(B_n)+n\ln(\sqrt2R)\leq2+n\ln(2R)-\frac n2\ln2$, as ${\rm vol}(B_n)<e^2$ for every $n$. With $\det(\Lambda)=1$ and $c\,D(K)\leq\epsilon_{\rm f}n\leq\frac n2\ln2$ (Lemma~\ref{lem:box_margin}, Part~3, and~\ref{cond:C4}), we get $\wt\Phi\leq n\ln(2R)+2$. The first iteration is a cutting iteration, since every nonzero $v\in\Z^n$ has $\|v\|_{H_K^{-1}}=\sqrt2R\|v\|_2\geq\sqrt2>2^{-100n\log n}$. A cutting iteration keeps $W$ and $\Lambda$ and replaces $K$ by the output $K'$ of a block, with $\mathcal{U}(K')\geq\mathcal{U}(K)+2n\ln n$ (Lemma~\ref{lem:hybrid_cpm}, Part~3). By~\eqref{eq:phi_tilde_U} before and after the iteration, $\wt\Phi$ decreases by at least $2n\ln n-2\epsilon_{\rm f}n\geq n\ln n$.

\textbf{Step 2: the volume after a run of reduction iterations.} Fix a maximal run of $k\geq1$ consecutive reduction iterations. By Step~1 it follows a cutting iteration. The superscript $(0)$ marks quantities at the start of this cutting iteration, $(j)$ those at the start of the $j$-th reduction iteration of the run, and $(k+1)$ those right after the run. So $(K^{(1)},x^{(1)},H^{(1)})$ is the output of the block, $W^{(1)}=W^{(0)}$ and $\Lambda^{(1)}=\Lambda^{(0)}$. Let $d=\dim(W^{(1)})$, so that $\dim(W^{(k+1)})=d-k$, and let $E_j:=2R'\cdot E(x^{(j)},H^{(j)})$. The $j$-th reduction iteration forms the slice $S_j:=E(\bar z_{j+1},A_{j+1})=E_j\cap P_j$, of full dimension in $W^{(j+1)}$ by Lemma~\ref{lem:small_error_run}. Its box $K^{(j+1)}$ is that of Lemma~\ref{lem:reduction_geometry} (Part~3), with $x^{(j+1)}=\bar z_{j+1}$ and $H^{(j+1)}=2A_{j+1}$.

As in Lemma~\ref{lem:reduction_geometry}, let $\zeta E_1:=x^{(1)}+\zeta(E_1-x^{(1)})$ for $\zeta\geq0$. We claim that $E_j\subseteq\chi_jE_1$ for $1\leq j\leq k$, where $\chi_1=1$ and $\chi_{j+1}=(1+2\sqrt2R')\chi_j$. If $E_j\subseteq\chi_jE_1$, then $S_j$ and its center $\bar z_{j+1}$ lie in $\chi_jE_1$. Since $H^{(j+1)}=2A_{j+1}$, we have $E_{j+1}=\bar z_{j+1}+\sqrt2R'(S_j-\bar z_{j+1})$, and Part~1 of that lemma with $\xi=\sqrt2R'$ gives $E_{j+1}\subseteq\chi_{j+1}E_1$. Next, Part~3 gives $K^{(k+1)}\subseteq\bar z_{k+1}+\sqrt{d-k}\,(S_k-\bar z_{k+1})$, and Part~1 with $\xi=\sqrt{d-k}\leq\sqrt n$ gives $K^{(k+1)}\subseteq\chi E_1$, where
\begin{align*}
    \chi:=(1+2\sqrt n)\chi_k\leq(1+2\sqrt n)(1+2\sqrt2R')^{k-1}=n^{O(k)},
\end{align*}
as $R'=O(n\log n)$. Since $W^{(k+1)}$ is a translate of $W_0^{(k+1)}\subseteq W_0^{(1)}$, Part~2, applied to $\chi E_1$, gives
\begin{align}\label{eq:box_chain}
    {\rm vol}(K^{(k+1)})\leq{\rm vol}\big(\chi E_1\cap(x^{(1)}+W_0^{(k+1)})\big)=\chi^{d-k}\cdot{\rm vol}(E_1\cap W'),
\end{align}
where $W':=x^{(1)}+W_0^{(k+1)}$. As this bound uses a central slice of $E_1$, it does not need $K^{(1)}\subseteq K^{(0)}$.

\textbf{Step 3: the slicing bound.} By the paragraph \emph{The lattice}, $\Lambda^{(k+1)}=\Pi_{W_0^{(k+1)}}(\Lambda^{(1)})$. The orthogonal complement of $W_0^{(k+1)}$ in $W_0^{(1)}$ is spanned by the pairwise orthogonal vectors $v^{(1)},\ldots,v^{(k)}$ of the run. Let $z^{(j)}\in\Z^n$ be the integer vector with $v^{(j)}=\Pi_{W_0^{(j)}}(z^{(j)})$. Then $\Pi_{W_0^{(1)}}(z^{(j)})\in\Lambda^{(1)}$ is $v^{(j)}$ plus a vector in ${\rm span}(v^{(1)},\ldots,v^{(j-1)})$. So these $k$ vectors of $\Lambda^{(1)}$ lie in that complement and span it, and Fact~\ref{def:dual_lattice_projection}, applied inside $W_0^{(1)}$, gives $(\Lambda^{(k+1)})^*=L\cap W_0^{(k+1)}$ with $L:=(\Lambda^{(1)})^*$.

The box $K^{(k+1)}$ contains $E(x^{(k+1)},H^{(k+1)})$ (Lemma~\ref{lem:sandwich}, Part~1), and $c\,D(K^{(k+1)})\leq\epsilon_{\rm f}n$ (Lemma~\ref{lem:box_margin}, Part~3). With $\det(\Lambda^{(k+1)})=1/\det((\Lambda^{(k+1)})^*)$ and~\eqref{eq:box_chain}, this gives
\begin{align*}
    e^{\wt\Phi^{(k+1)}}\leq e^{\epsilon_{\rm f}n}\,{\rm vol}(K^{(k+1)})\det(\Lambda^{(k+1)})\leq e^{\epsilon_{\rm f}n}\cdot\chi^{d-k}\cdot\frac{{\rm vol}(E_1\cap W')}{\det(L\cap W_0^{(k+1)})}.
\end{align*}
We apply Lemma~\ref{lem:j21_slice} inside $W_0^{(1)}$, identified with $\R^d$, to the convex body $E_1-x^{(1)}$, the full-rank lattice $L$ with $L^*=\Lambda^{(1)}$, and the subspace $W_0^{(k+1)}$, in which $L\cap W_0^{(k+1)}$ has rank $d-k$. As volumes and covariance matrices do not change under translation,
\begin{align*}
    \frac{{\rm vol}(E_1\cap W')}{\det(L\cap W_0^{(k+1)})}\leq\frac{{\rm vol}(E_1)}{\det(L)}\cdot\frac{k^{O(k)}}{\lambda_1(\Lambda^{(1)},{\rm Cov}(E_1))^k}.
\end{align*}
The uniform distribution on $E_1=E(x^{(1)},H^{(1)}/(2R')^2)$ has the covariance matrix
\begin{align*}
    {\rm Cov}(E_1)=\frac{(2R')^2}{d+2}(H^{(1)})^{-1}\succeq(H^{(1)})^{-1},
\end{align*}
as $\sqrt{d+2}\leq2\sqrt{2n}\leq2R'$ by~\ref{cond:C4}.
Finally, $\det(L)=1/\det(\Lambda^{(1)})$, $\Lambda^{(1)}=\Lambda^{(0)}$, $D(K^{(1)})\geq0$ and $\wt\Phi^{(1)}\leq\wt\Phi^{(0)}$ (Step~1) give $\ln({\rm vol}(E_1)/\det(L))=d\ln(2R')+\wt\Phi^{(1)}-c\,D(K^{(1)})\leq d\ln(2R')+\wt\Phi^{(0)}$. Combining these bounds, we get
\begin{align}\label{eq:upper_bound_e}
    e^{\wt\Phi^{(k+1)}}\leq e^{\wt\Phi^{(0)}}\cdot\frac{e^{\epsilon_{\rm f}n}\cdot\chi^{d-k}\cdot(2R')^d\cdot k^{O(k)}}{\lambda_1(\Lambda^{(0)},(H^{(1)})^{-1})^k}.
\end{align}

\textbf{Step 4: the lower bound on $\lambda_1$.} In the cutting iteration $(0)$, \textsc{ApproxSV} returned a vector $v^{(0)}\in\Lambda^{(0)}$ with $2^{-100n\log n}\leq\|v^{(0)}\|_{(H^{(0)})^{-1}}\leq\gamma\cdot\lambda_1(\Lambda^{(0)},(H^{(0)})^{-1})$. Its block starts from $x^{(0)}$ with $H^{(0)}=H_{K^{(0)}}(x^{(0)})$ and returns $H^{(1)}$, so Lemma~\ref{lem:block_stability} gives
\begin{align}\label{eq:lower_bound_lambda}
    \lambda_1(\Lambda^{(0)},(H^{(1)})^{-1})\geq2^{-O(n\log n)}\cdot\lambda_1(\Lambda^{(0)},(H^{(0)})^{-1})\geq\frac{2^{-O(n\log n)}}{\gamma}.
\end{align}
In~\eqref{eq:upper_bound_e} we have $\chi^{d-k}\leq\chi^n=n^{O(nk)}$, $(2R')^d=n^{O(n)}$, $k^{O(k)}\leq n^{O(k)}$, and $e^{\epsilon_{\rm f}n}\leq e^n\leq n^{2nk}$ as $\epsilon_{\rm f}\leq(\ln2)/2\leq1$ by~\ref{cond:C4} and $n\geq2$. With~\eqref{eq:lower_bound_lambda}, this gives $e^{\wt\Phi^{(k+1)}}\leq e^{\wt\Phi^{(0)}}\cdot\gamma^k\cdot n^{O(nk)}$. So there is an absolute constant $C_1$ such that $\wt\Phi^{(k+1)}\leq\wt\Phi^{(0)}+k\ln\gamma+C_1nk\ln n$ for every maximal run of $k$ reduction iterations.

\textbf{Step 5: a small value of $\wt\Phi$ forces a reduction.} Consider the start of an iteration with $d\geq1$. Since $c\,D(K)\geq0$, we have $\ln({\rm vol}(E(x_K,H_K))\det(\Lambda))\leq\wt\Phi$. For $\varrho>0$, the set $\{y\in W_0:\|y\|_{H_K^{-1}}\leq \varrho\}$ has volume
\begin{align*}
    \varrho^d\,{\rm vol}(B_d)\det(H_K)^{1/2}=\frac{\varrho^d\,{\rm vol}(B_d)^2}{{\rm vol}(E(x_K,H_K))}\geq \varrho^d\,{\rm vol}(B_d)^2\det(\Lambda)\,e^{-\wt\Phi},
\end{align*}
with determinants taken on $W_0$. As $B_d$ contains the cube $[-1/\sqrt d,1/\sqrt d]^d$, we have ${\rm vol}(B_d)\geq d^{-d/2}$. By Minkowski's first theorem, the set contains a nonzero point of $\Lambda$ once its volume is at least $2^d\det(\Lambda)$, which holds for $\varrho=2d\,e^{\wt\Phi/d}$. So $\lambda_1(\Lambda,H_K^{-1})\leq2d\,e^{\wt\Phi/d}$. Let
\begin{align*}
    \Phi_{\min}:=-n\big(100n\ln n+\ln(2\gamma n)\big).
\end{align*}
If $\wt\Phi<\Phi_{\min}$, then $\wt\Phi<0$ and $d\leq n$ give $\wt\Phi/d\leq\wt\Phi/n$. So \textsc{ApproxSV} returns a vector of norm at most $\gamma\cdot2n\,e^{\wt\Phi/n}<n^{-100n}=2^{-100n\log n}$, and the iteration is a reduction iteration. Thus every cutting iteration starts with $\wt\Phi\geq\Phi_{\min}$.

\textbf{Step 6: counting.} Suppose that an $I$-th cutting iteration occurs. For $i\leq I$, let $\Psi_i$ be the value of $\wt\Phi$ at the start of the $i$-th one, so that $\Psi_1\leq n\ln(2R)+2$ by Step~1. For $i<I$, let $k_i\geq0$ be the number of reduction iterations between the $i$-th and the $(i+1)$-th cutting iteration. Step~1 if $k_i=0$, and Step~4 if $k_i\geq1$, give
\begin{align*}
    \Psi_{i+1}\leq\Psi_i-n\ln n+k_i\big(\ln\gamma+(C_1+1)n\ln n\big).
\end{align*}
Each reduction iteration lowers $\dim(W)$ by one, so $\sum_ik_i\leq n$. Summing over $i<I$ and using $\Psi_I\geq\Phi_{\min}$ from Step~5, we get
\begin{align*}
    \Phi_{\min}\leq n\ln(2R)+2-(I-1)\,n\ln n+n\ln\gamma+(C_1+1)n^2\ln n.
\end{align*}
Dividing by $n\ln n$ gives $I\leq\bar N_{\rm c}$ for some $\bar N_{\rm c}=O(n+\log(\gamma R)/\log n)$, with an absolute constant in the $O(\cdot)$. So the run has at most $\bar N_{\rm c}$ cutting iterations, and in particular finitely many.

\textbf{Step 7: the event $\mathcal{G}$.} For $n<n_0$, the operation test is inactive and every leverage score is exact, so Steps~1 to~6 apply. Let $n\geq n_0$. By the loops of Algorithm~\ref{alg:hcpm}, each block makes at most $T(3+(t_{\max}+1)N)+2=O(n\log n\log\log n)$ operations on its \textsc{LS}, whatever the outputs are. As the constant in $\bar N_{\rm c}$ is absolute, we can fix the absolute constant $C_M$ of Section~\ref{sec:r3_params} so that $\bar N_{\rm c}(T(3+(t_{\max}+1)N)+2)\leq M$. Consider a cutting iteration $i$ such that every output used in the blocks before it has $\ell_2$ error at most $\delta_0$. Steps~1 to~6 for the run up to the start of iteration $i$ use only these outputs (through Lemmas~\ref{lem:small_error_run} and~\ref{lem:block_stability}), so they give $i\leq\bar N_{\rm c}$. Hence fewer than $M$ operations on \textsc{LS} precede iteration $i$, the operation test does not fail there, and the operations of block $i$ are among the first $M$. Now let $\mathcal{G}$ hold, and suppose that some output used by the algorithm has $\ell_2$ error larger than $\delta_0$. The first such output lies in some block $i$, so it is among the first $M$ operations, and its error is at most $\delta_{\rm J}(n)\leq\delta_0$ by~\ref{cond:C7}, a contradiction. So on $\mathcal{G}$ every output used by the algorithm has $\ell_2$ error at most $\delta_0$, there are at most $\bar N_{\rm c}$ blocks, and there are at most $M$ operations on \textsc{LS}.

Finally, on $\mathcal{G}$ the hypothesis of Lemma~\ref{lem:main_output_formal} holds, and the proof of that lemma uses only Lemma~\ref{lem:small_error_run} and Steps~6 and~7. So the algorithm outputs an integral minimizer of $f$. It calls $\SO$ only inside blocks, at most $T=O(n\log n)$ times per block, so it makes at most $\bar N_{\rm c}T=O(n^2\log n+n\log(\gamma nR))$ calls to $\SO$.
\end{proof}

\subsection{Runtime Analysis}\label{sec:minimization:runtime}

We bound the operations outside the calls to \textsc{ApproxSV}, and then the cost of the calls for two choices of the routine.

\begin{lemma}[Runtime part of Theorem~\ref{thm:main_formal}]\label{lem:main_time_formal}
Under the assumptions of Lemma~\ref{lem:main_oracle_formal}, on the event $\mathcal{G}$, Algorithm~\ref{alg:ours} makes $O(n+\log(\gamma R)/\log n)$ calls to \textsc{\textup{ApproxSV}}, and outside these calls it uses $O((n^4\log n+n^3\log(\gamma R))\log\log n)$ arithmetic operations. In particular, for $\log\gamma=O(n\log n)$ these bounds are $O(n+\log R/\log n)$ and $O((n^4\log n+n^3\log R)\log\log n)$, which is $O(n^{4}\log(nR)\log\log n)$.
\end{lemma}

\begingroup\emergencystretch=3em
\begin{proof}
Each iteration of the while loop calls \textsc{ApproxSV} once. There are at most $n$ reduction iterations and, by Lemma~\ref{lem:main_oracle_formal}, $O(n+\log(\gamma R)/\log n)$ cutting iterations, which bounds the number of calls. We now count the other operations.

\textbf{Lattice bookkeeping.} We store a basis $B\in\R^{n\times d}$ of $\Lambda$, where $d=\dim(W)$, with an integer matrix $Z\in\Z^{n\times d}$ of preimages, so that $B=\Pi_{W_0}Z$ column by column. Initially $B=Z=I_n$. To compute $v$, we form the Gram matrix $G=(B^\top B)(B^\top H_KB)^{-1}(B^\top B)$ of $B$ in the norm $\|\cdot\|_{H_K^{-1}}$. Indeed, for $u=Bs$ we have $\|u\|_{H_K^{-1}}^2=\max\{(u^\top y)^2:y\in W_0,\ \|y\|_{H_K}\leq1\}=s^\top Gs$. We run \textsc{ApproxSV} on $G$, which returns a unimodular matrix $U$, and replace $B\gets BU$ and $Z\gets ZU$. Then $v$ is the first column of $B$, and in a reduction iteration $z$ is the first column of $Z$. The new basis consists of the projections onto $P_0$ of the other columns of $B$, with the other columns of $Z$ as preimages. By the paragraph \emph{The lattice}, it is a basis of $\Pi_{P_0}(\Lambda)$, and the preimages are correct since $\Pi_{W_0\cap P_0}=\Pi_{P_0}\circ\Pi_{W_0}$. These steps take $O(n^3)$ operations per iteration, so $O(n^4+n^3\log(\gamma R)/\log n)$ in total.

\textbf{Cutting iterations.} On $\mathcal{G}$, every output of \textsc{LS} that the algorithm uses has $\ell_2$ error at most $\delta_0$ (Lemma~\ref{lem:main_oracle_formal}). So by Lemmas~\ref{lem:small_error_run} and~\ref{lem:hybrid_cpm}, each block uses $O(n^3\log n\log\log n)$ operations. For $n<n_0$, each exact computation of the scores takes $O(n^\omega)=O(1)$ operations, and the same bound holds. So all blocks use $O((n^4\log n+n^3\log(\gamma R))\log\log n)$ operations.

\textbf{Reduction iterations.} Computing $P$ and the new $W$ takes $O(n^2)$ operations. The center $\bar z$ of the slice $E(\bar z,A)$ of $2R'\cdot E(x_K,H_K)$ by $P$ minimizes $(y-x_K)^\top H_K(y-x_K)$ over $y\in W\cap P$, and $A$ is a positive multiple of the restriction of $H_K$ to $W_0\cap P_0$. So both take $O(n^3)$ operations. The other steps compute a basis $Y$ of the new $W_0$, for instance the new lattice basis, and the Cholesky factorization $Y^\top AY=JJ^\top$. They also compute the $2\dim(W)$ rows of the new box, which are $\pm$ the rows of $J^\top(Y^\top Y)^{-1}Y^\top$, and its Hessian $H_K=2A$ (Lemma~\ref{lem:reduction_geometry}, Part~3). Counting square roots as arithmetic operations, they take $O(n^3)$ operations, so $O(n^4)$ over all reduction iterations.

\textbf{The operation test.} For $n\geq n_0$, computing $M$ takes $O(1+\log(\gamma R))$ operations (paragraph \emph{The good event} in Section~\ref{sec:minimization:alg}). Counting the operations on \textsc{LS} costs $O(1)$ per operation, so $O(M)$ in total by Lemma~\ref{lem:main_oracle_formal}. Both are within the bound below.

Adding up, the algorithm uses $O((n^4\log n+n^3\log(\gamma R))\log\log n)$ arithmetic operations outside the calls to \textsc{ApproxSV}. If $\log\gamma=O(n\log n)$, then $n^3\log\gamma=O(n^4\log n)$ and $\log(\gamma R)/\log n=O(n+\log R/\log n)$, which gives the remaining bounds of the lemma.
\end{proof}\par\endgroup

\begin{lemma}[Dimension reduction keeps the basis reduced]\label{lem:reduction_keeps_reduced}
Consider a reduction iteration of Algorithm~\ref{alg:ours}, with the lattice bookkeeping of the proof of Lemma~\ref{lem:main_time_formal}. Let $b_1,\ldots,b_d$ be the stored basis right after the call to \textsc{\textup{ApproxSV}}, so that $b_1=v$, and let $G$ be its Gram matrix in the norm $\|\cdot\|_{H_K^{-1}}$. Let $G'$ be the Gram matrix of the new stored basis in the norm $\|\cdot\|_{H_K^{-1}}$ of the next iteration. Then there is $s>0$ such that the Gram--Schmidt norms of $I_{d-1}$ for $G'$ are $sB_2,\ldots,sB_d$, where $B_1,\ldots,B_d$ are those of $I_d$ for $G$. If $I_d$ is reduced for $G$, then $I_{d-1}$ is reduced for $G'$.
\end{lemma}

\begin{proof}
Let $H=H_K$ and $W_0$ be as before the reduction, let $W'=W_0\cap P_0$, and let $\Pi$ be the orthogonal projection onto $P_0$, so that the new stored basis is $\Pi b_2,\ldots,\Pi b_d$. By the proof of Lemma~\ref{lem:main_time_formal}, the new Hessian is $H'=2A$, where $A$ is a positive multiple $\vartheta H$ of the restriction of $H$ to $W'$. Hence, for $h\in W'$,
\begin{align*}
    \|h\|_{H'^{-1}}^2=\max\{(h^\top y)^2:y\in W',\ y^\top H'y\leq 1\}=s\cdot\max\{(h^\top y)^2:y\in W',\ y^\top Hy\leq 1\},
\end{align*}
where $s=1/(2\vartheta)$. Write $\langle x,y\rangle=x^\top H^{-1}y$ for $x,y\in W_0$, with $H^{-1}$ the inverse of $H$ on $W_0$, and for $u\in W_0$ let $\pi u=u-(\langle u,v\rangle/\langle v,v\rangle)v$. We claim that $\|\Pi u\|_{H'^{-1}}^2=s\langle\pi u,\pi u\rangle$ for every $u\in W_0$. Every $y\in W'$ is orthogonal to $v$, so $y^\top\Pi u=y^\top u=y^\top\pi u$, and the Cauchy--Schwarz inequality gives $(y^\top\pi u)^2\leq(y^\top Hy)\langle\pi u,\pi u\rangle$. Conversely, $y=H^{-1}\pi u\in W_0$ satisfies $v^\top y=\langle v,\pi u\rangle=0$, so $y\in W'$, and $(y^\top\pi u)^2=\langle\pi u,\pi u\rangle^2=(y^\top Hy)\langle\pi u,\pi u\rangle$. This proves the claim. Both sides of the claim are quadratic forms in $u$, so $G'_{k,l}=s\langle\pi b_{k+1},\pi b_{l+1}\rangle$ for $1\leq k,l\leq d-1$.

Let $b_k^*$, $\mu_{k,j}$ and $B_k$ be the Gram--Schmidt data of $b_1,\ldots,b_d$ in $\langle\cdot,\cdot\rangle$, which are the data of $I_d$ for $G$. Since $b_1^*=v$, we have $\pi b_k=b_k^*+\sum_{2\leq j<k}\mu_{k,j}b_j^*$, so the Gram--Schmidt orthogonalization of $\pi b_2,\ldots,\pi b_d$ is $b_2^*,\ldots,b_d^*$, with the same coefficients $\mu_{k,j}$ for $2\leq j<k$. Hence the data of $I_{d-1}$ for $G'$ are $B'_k=sB_{k+1}$ and $\mu'_{k,j}=\mu_{k+1,j+1}$, and the conditions that define a reduced basis carry over from $G$ to $G'$.
\end{proof}

\begin{lemma}[Cost of the shortest vector computations with LLL]\label{lem:svp_total}
Under the assumptions of Lemma~\ref{lem:main_oracle_formal}, let \textsc{\textup{ApproxSV}} be the routine \textsc{\textup{LLL}} of Lemma~\ref{lem:lll_exact}, run on the Gram matrix of the stored basis as in the proof of Lemma~\ref{lem:main_time_formal}, so that $\gamma=2^{(n-1)/2}$. Then, on the event $\mathcal{G}$, the first call and every call that follows a reduction iteration perform no swap, every call that follows a cutting iteration performs $O(n^3\log n)$ swaps, and all calls together use $O(n^5\log n+n^4\log R)$ arithmetic operations.
\end{lemma}

\begin{proof}
We work on the event $\mathcal{G}$, where every output of \textsc{LS} that the algorithm uses has $\ell_2$ error at most $\delta_0$ (Lemma~\ref{lem:main_oracle_formal}), so Lemma~\ref{lem:block_stability} holds for every block. Say that the stored basis is reduced for a norm if $I_d$ is reduced for its Gram matrix in that norm. A call receives the Gram matrix $G$ of the stored basis $B$ and returns $U$ reduced for $G$. Then $B$ becomes $BU$, whose Gram matrix is $U^\top GU$. So after each call the stored basis is reduced for the norm $\|\cdot\|_{H_K^{-1}}$ of that call.

At the first call, $B=I_n$ and $H_K=\frac{1}{2R^2}I_n$, so $G=2R^2I_n$. All $\mu_{k,j}$ vanish and all $B_k$ are equal, so $I_n$ is reduced for $G$ and there is no swap. A call that follows a reduction iteration performs no swap by Lemmas~\ref{lem:reduction_keeps_reduced} and~\ref{lem:lll_exact}.

Consider a call that follows a cutting iteration. Let $H_0$ and $H_1$ be the values of $H_K$ at the start and at the end of that cutting iteration. Let $G_0$ and $G_1$ be the Gram matrices of the stored basis in the norms $\|\cdot\|_{H_0^{-1}}$ and $\|\cdot\|_{H_1^{-1}}$. The cutting iteration changes neither the stored basis nor $W$, so $I_d$ is reduced for $G_0$, and the call receives $G_1$. By Lemma~\ref{lem:block_stability}, $e^{-a}H_0\preceq H_1\preceq e^aH_0$ on $W_0$ for some $a=O(T)=O(n\log n)$. For $u\in W_0$ we have $\|u\|_{H^{-1}}^2=\max\{(u^\top y)^2/(y^\top Hy):y\in W_0\setminus\{{\bf 0}_n\}\}$, so $e^{-a}\|u\|_{H_0^{-1}}^2\leq\|u\|_{H_1^{-1}}^2\leq e^a\|u\|_{H_0^{-1}}^2$, and hence $e^{-a}G_0\preceq G_1\preceq e^aG_0$. By Lemma~\ref{lem:lll_restart}, the call performs $O(ad^2+d^3)=O(n^3\log n)$ swaps.

By Lemma~\ref{lem:lll_exact}, a call with $S$ swaps uses $O(n^3+nS)$ operations. The first call and the at most $n$ calls that follow reduction iterations use $O(n^4)$ operations together. By Lemma~\ref{lem:main_oracle_formal}, since $\log\gamma=O(n)$, there are $O(n+\log R/\log n)$ cutting iterations, so at most that many calls follow a cutting iteration. Each of these calls uses $O(n^4\log n)$ operations. The total is $O(n^5\log n+n^4\log R)$.
\end{proof}

\begin{lemma}[Cost of the shortest vector computations with \textsc{SegmentLLL}]\label{lem:svp_segment}
Under the assumptions of Lemma~\ref{lem:main_oracle_formal}, let \textsc{\textup{ApproxSV}} be the routine \textsc{\textup{SegmentLLL}} of Lemma~\ref{lem:segment_lll}, run on the Gram matrix of the stored basis as in the proof of Lemma~\ref{lem:main_time_formal}, so that $\log\gamma=O(n\log n)$. Then, on the event $\mathcal{G}$, the first call returns the identity matrix and uses $O(n^3)$ arithmetic operations, every other call uses $O(n^3\log^3n)$ operations, and all calls together use $O(n^4\log^3n+n^3\log^2n\log R)$ operations.
\end{lemma}

\begin{proof}
As in the proof of Lemma~\ref{lem:svp_total}, we work on the event $\mathcal{G}$, where Lemma~\ref{lem:block_stability} holds for every block. A call receives the Gram matrix $G$ of the stored basis $B$ and returns $U$, and $B$ becomes $BU$, whose Gram matrix is $U^\top GU$. The Gram--Schmidt norms of $I_d$ for $U^\top GU$ are those of $U$ for $G$. So by Lemma~\ref{lem:segment_lll}(a), after each call, the Gram--Schmidt norms of the stored basis in the norm $\|\cdot\|_{H_K^{-1}}$ of that call satisfy $B_j\geq 2^{-\kappa_n}B_l$ for all $l<j$.

At the first call, $G=2R^2I_n$, so the call returns $I_n$ and uses $O(n^3)$ operations by Lemma~\ref{lem:segment_lll}(b). Consider a call that follows a reduction iteration, and let $G'$ be the Gram matrix that it receives. By Lemma~\ref{lem:reduction_keeps_reduced}, the Gram--Schmidt norms of the identity for $G'$ are $sB_2,\ldots,sB_d$ for some $s>0$, where $B_1,\ldots,B_d$ are those of the stored basis right after the previous call, in the norm of that call. So these norms $B'_t=sB_{t+1}$ also satisfy $B'_j\geq 2^{-\kappa_n}B'_l$ for $l<j$. Lemma~\ref{lem:restart_bound}, applied with $G_0=G_1=G'$ and $a=0$ and with $\kappa_n$ as the exponent in its hypothesis $B_j\geq 2^{-\kappa_n}B_l$, gives $\log_2(d_i(I_{d-1})/d_i(U))\leq i(\kappa_n+\log_2i)\leq n(\kappa_n+\log_2n)$ for the output $U$ and every $1\leq i\leq d-2$. So $\Delta_{\max}=O(n^2\log n)$, and by Lemma~\ref{lem:segment_lll}(b) the call uses $O(n^3+n\log^2n\cdot n^2\log n)=O(n^3\log^3n)$ operations.

Consider a call that follows a cutting iteration, and let $H_0$, $H_1$, $G_0$, $G_1$ and $a=O(n\log n)$ be as in the proof of Lemma~\ref{lem:svp_total}, so that $e^{-a}G_0\preceq G_1\preceq e^aG_0$ and the call receives $G_1$. The stored basis was returned by the call at the start of the cutting iteration, so the Gram--Schmidt norms of $I_d$ for $G_0$ satisfy $B_j\geq 2^{-\kappa_n}B_l$ for $l<j$. Lemma~\ref{lem:restart_bound} gives $\log_2(d_i(I_d)/d_i(U))\leq i(2a\log_2e+\kappa_n+\log_2i)=O(n^2\log n)$ for the output $U$ and every $1\leq i\leq d-1$, with $d_i$ computed with $G_1$. So $\Delta_{\max}=O(n^2\log n)$, and the call uses $O(n^3\log^3n)$ operations.

There are at most $n$ calls that follow a reduction iteration. Since $\log\gamma=O(n\log n)$, Lemma~\ref{lem:main_oracle_formal} bounds the number of calls that follow a cutting iteration by $O(n+\log R/\log n)$. The total is $O(n^3)+O(n^3\log^3n)\cdot O(n+\log R/\log n)=O(n^4\log^3n+n^3\log^2n\log R)$.
\end{proof}

\subsection{Main Result}\label{sec:minimization:main}

We now add a bound on the probability of the good event to the lemmas above.

\begin{theorem}[Main result, formal version of Theorem~\ref{thm:main_informal}]\label{thm:main_formal}
Given a separation oracle $\SO$ for a convex function $f:\R^n\to\R\cup\{+\infty\}$ and a number $R\geq1$ such that the set of minimizers $K^*$ of $f$ is nonempty, closed and contained in $B_\infty(R)$ and all extreme points of $K^*$ are integral, and given the number $\gamma$ and a $\gamma$-approximate shortest vector routine \textsc{\textup{ApproxSV}} (Definition~\ref{def:approx_sv}) with $\log\gamma=O(n\log n)$, for instance the LLL algorithm of Lemma~\ref{lem:lll_exact} or the routine \textsc{\textup{SegmentLLL}} of Lemma~\ref{lem:segment_lll}, there exists a randomized algorithm (Algorithm~\ref{alg:ours}) with the following guarantee. For every constant $k\geq1$ and every large enough $n$, with probability at least $1-O((n^2\log n+n\log(nR))\log\log n)\cdot n^{-k}$, it outputs an integral minimizer of $f$ and uses
\begin{itemize}
    \item ${O}(n^2\log n+n\log R)$ calls to $\SO$,
    \item ${O}(n+\log R/\log n)$ calls to \textsc{\textup{ApproxSV}}, each on a lattice of rank at most $n$, and
    \item ${O}((n^4\log n+n^3\log R)\log\log n)$ additional arithmetic operations outside the calls to \textsc{\textup{ApproxSV}}, which is $O(n^4\log(nR)\log\log n)$.
\end{itemize}
In particular, if $\log R\leq n^{k_0}$ for an absolute constant $k_0$, then this holds with high probability, that is, with probability at least $1-n^{-k}$ for every constant $k$ and every large enough $n$. If \textsc{\textup{ApproxSV}} is \textsc{\textup{SegmentLLL}}, run as in Lemma~\ref{lem:svp_segment}, then the calls to \textsc{\textup{ApproxSV}} use $O(n^4\log^3n+n^3\log^2n\log R)$ arithmetic operations, and so does the whole algorithm. If \textsc{\textup{ApproxSV}} is the LLL algorithm, run as in Lemma~\ref{lem:svp_total}, then the calls use $O(n^5\log n+n^4\log R)$ arithmetic operations, and so does the whole algorithm.
\end{theorem}

\begin{proof}
We take $n\geq2$, since the theorem concerns large $n$. On the event $\mathcal{G}$, Lemma~\ref{lem:main_oracle_formal} shows that the algorithm outputs an integral minimizer of $f$ with $O(n^2\log n+n\log(\gamma nR))$ calls to $\SO$, which is $O(n^2\log n+n\log R)$ for $\log\gamma=O(n\log n)$. Lemma~\ref{lem:main_time_formal} gives the other two bounds on $\mathcal{G}$.

\textbf{The probability of $\mathcal{G}$.} For $n<n_0$, the event $\mathcal{G}$ is sure. Let $n\geq n_0$. We check the hypotheses of Lemma~\ref{lem:jlsw_adaptive} for the operations on \textsc{LS} of all blocks, counted in order. The budget $M$ is a deterministic function of $n$, $\gamma$ and $R$. Algorithm~\ref{alg:hcpm} issues only valid operations, whatever \textsc{LS} outputs, and its padded instance has at most $\bar m+n$ rows (paragraph \emph{Validity of the operations} in Section~\ref{sec:r3_local}). The instances \textsc{IM}$_H$ and \textsc{IM}$_Q$ run the data structure of Theorem~B.4 of~\cite{jlsw20}, which is deterministic (proof of Lemma~\ref{lem:hybrid_im}). So is every other step of Algorithms~\ref{alg:ours} and~\ref{alg:hcpm}, with ties broken by the smallest index, apart from the operations on \textsc{LS} and the calls of $\SO$ and \textsc{ApproxSV}. We fix the random bits of $\SO$ and \textsc{ApproxSV}, which are independent of those of \textsc{LS} (Definition~\ref{def:approx_sv}). Their answers are then functions of the current and earlier queries, so each operation on \textsc{LS} is a function of earlier outputs of \textsc{LS}. Conditionally on these bits, Lemma~\ref{lem:jlsw_adaptive} gives $\Pr[\mathcal{G}]\geq1-3Mn^{-k'}$ for every constant $k'\geq1$ and every $n\geq\max\{n_0,n_1(k')\}$. Averaging over the bits gives the same bound.

For $\log\gamma=O(n\log n)$ we have $M=O((n^2\log n+n\log(nR))\log\log n)$, so $k'=k$ gives the probability in the theorem. If $\log R\leq n^{k_0}$, then $M\leq n^{k_0+3}$ for large $n$, and $k'=k+k_0+4$ gives failure probability at most $3n^{k_0+3-k'}\leq n^{-k}$.

\textbf{Shortest vector routines.} By Lemma~\ref{lem:segment_lll}, \textsc{SegmentLLL} satisfies $\log\gamma=O(n\log n)$, and so does LLL, with $\gamma=2^{(n-1)/2}$. Lemmas~\ref{lem:svp_segment} and~\ref{lem:svp_total} bound the cost of the calls for \textsc{SegmentLLL} and for LLL on $\mathcal{G}$. Adding the operations outside the calls gives the totals, because $(n^4\log n+n^3\log R)\log\log n$ is at most $n^4\log^3n+n^3\log^2n\log R$.
\end{proof}

Our algorithm draws random bits only in the data structure of Lemma~\ref{lem:jlsw_main}. If the leverage scores are computed exactly and $\SO$ and \textsc{ApproxSV} are deterministic, we obtain a deterministic algorithm with a larger number of arithmetic operations. With exact scores we do not need the hybrid barrier, and we run Vaidya's method~\cite{v89} with the volumetric barrier $F$ in each block.

\begin{corollary}[Exact leverage scores]\label{cor:exact_score}
Let $n\geq2$, suppose that the assumptions of Theorem~\ref{thm:main_formal} hold, and suppose that $\SO$ and \textsc{\textup{ApproxSV}} are deterministic. Then there is a deterministic algorithm that outputs an integral minimizer of $f$ using $O(n^2\log n+n\log R)$ calls to $\SO$, $O(n+\log R/\log n)$ calls to \textsc{\textup{ApproxSV}}, each on a lattice of rank at most $n$, and $O(n^{\omega+2}\log n+n^{\omega+1}\log R)$ arithmetic operations outside these calls. For submodular function minimization, it makes $O(n^3\log n)$ evaluation oracle queries and $O(n)$ calls to \textsc{\textup{ApproxSV}}, and it uses $O(n^{\omega+2}\log n)$ arithmetic operations outside these calls.
\end{corollary}

With \textsc{SegmentLLL}, the total for submodular function minimization is $O(n^{\omega+2}\log n+n^4\log^3n)$. The proof is in Section~\ref{sec:r3_vaidya}. It uses Lemma~\ref{lem:sp_vaidya}, which analyzes one iteration of Vaidya's method with exact leverage scores, and it follows the proof of Theorem~\ref{thm:main_formal}.

\begin{remark}[Cost of the shortest vector computations]\label{rem:svp_cost}
The SODA 2024 version of this paper claimed $O(n^4\log(nR))$ arithmetic operations in total, by charging each call to \textsc{ApproxSV} $O(n^3)$ operations with a citation of~\cite{ns16}. No such bound is available for the inputs of our algorithm (Section~\ref{sec:preli:lll}). The algorithm of~\cite{ns16} needs integral bases, its cost grows with their bit length, and it bounds the first vector only against the determinant. The bound of~\cite{s96} needs the potential of the swap argument to drop by a factor of at most $2^{O(n^2)}$, while for a call in our algorithm we can only bound this drop by $2^{O(n^3\log n)}$. In place of the SODA 2024 version's bound, we prove Lemma~\ref{lem:svp_segment}. In exact arithmetic, \textsc{SegmentLLL} started from the previous basis uses $O(n^3\log^3n)$ operations per call, so the total is $O(n^4\log^3n+n^3\log^2n\log R)$. For $\log R=O(\log n)$ this bound is a factor $O(\log^2n)$ above the SODA 2024 version's claim. LLL started from the previous basis uses $O(n^4\log n)$ operations per call (Lemma~\ref{lem:svp_total}). The bound on the oracle complexity is unaffected by this change. The bound on the operations outside \textsc{ApproxSV} gains a factor $\log\log n$ for a different reason, explained in Remark~\ref{rem:jlsw20_floor}.

The cost of a call that follows a cutting iteration is governed by the drop of the potential $P$ in the call. We bound the drop of $\log_2d_i$ by $O(in\log n)$ for each $i$, and hence the drop of $\log_2P$ by $O(n^3\log n)$ (Lemma~\ref{lem:restart_bound}). These bounds have two sources of the same order. Within a block the Hessian changes by a factor of up to $2^{O(n\log n)}$ in either direction, which can change each $d_i$ by a factor $2^{O(in\log n)}$. A basis whose Gram--Schmidt norms satisfy $B_j\geq 2^{-\kappa_n}B_l$ for all $l<j$, as the output of \textsc{SegmentLLL} does, is also only guaranteed to be within a factor $2^{O(in\log n)}$ of an optimal one in $d_i$ (Lemma~\ref{lem:reduced_near_opt}). \textsc{LLL} pays $O(n)$ operations per unit of drop of $\log_2P$, which gives its $O(n^4\log n)$ operations per call.

Given this drop, a bound of $\widetilde O(n^3)$ per call needs a cost of $\poly(\log n)$ per unit of drop. \textsc{SegmentLLL} uses $O(n\log^2n)$ operations per unit of the largest drop $\Delta_{\max}=O(n^2\log n)$ of a single $\log_2d_i$. This amounts to $O(\log^2n)$ operations per unit of our bound $O(n^3\log n)$ on the drop of $\log_2P$ (Lemmas~\ref{lem:segment_lll} and~\ref{lem:svp_segment}). Of its three logarithmic factors, one is in $\Delta_{\max}$, for the two reasons above. The other two come from the recursion. A round on a segment of $2^h$ vectors costs $O(8^h)$ operations, but if its test fails, it is only guaranteed to lower the sum of the $\log_2d_i$ over $O(2^h/h)$ middle indices by $\Omega(8^h/h^2)$ (Lemma~\ref{lem:merge}). Summing over the $O(\log n)$ orders gives the third factor. Faster recursive algorithms do not give the lower bound on $\lambda_1$ that Step~4 of the proof of Lemma~\ref{lem:main_oracle_formal} needs whenever the returned vector is long. The algorithms of~\cite{ns16} and~\cite{sch06} bound the first vector only against the determinant (Section~\ref{sec:segment_lll}). For bases that satisfy its third condition, the segment reduction of~\cite{ks01} bounds every basis vector against the corresponding successive minimum (Theorem~2 of~\cite{ks01}), but its count of operations is for integral bases. Without that condition, and for its iterated segments (Definition~3 of~\cite{ks01}), it bounds the first vector only against the determinant (Theorems~5 and~6 of~\cite{ks01}). We do not know whether $O(n^3\log n)$ operations per call are possible.
\end{remark}

\begin{remark}[The oracle model]\label{rem:oracle_model}
The SODA 2024 version of this paper used the separation oracle of~\cite{gls81,gls88}, as formalized in Definition~1.1 of~\cite{j22}. It answers ``YES'' at every minimizer of $f$, and at every other point $x$ it returns a vector $a$ with $a^\top y>a^\top x$ for every minimizer $y$. If its random bits are independent of ours, such an oracle satisfies our definition, so our model contains that of the SODA 2024 version. Ours also allows a vector at a minimizer, as a subgradient oracle returns there. The oracle of Corollary~\ref{cor:sfm_main} is of this kind and never answers ``YES'', although~\cite{j22} introduces it in its Theorem~6.4 as an implementation of Definition~1.1. The SODA 2024 version's method did not say what to do after a ``YES''. Our cut vector handles this case with the lexicographic device of~\cite[Section~1.3]{j22}, which needs strict separation at points that do not minimize $f$. Otherwise the oracle could return a vector $a$ with $Y^\top a=0$ at such a point of $W$, and the resulting cut with $e_i$ need not contain $x^*$. The SODA 2024 version of this paper left the independence of the random bits implicit, and Lemma~\ref{lem:jlsw_adaptive} uses it.
\end{remark}
\section*{Acknowledgement}

We would like to thank Jonathan Kelner for many helpful discussions and for suggesting the title of the paper. Lichen Zhang is supported in part by NSF grant No.\ 1955217 and NSF grant No.\ 2022448.

The corrections in this version were found and written with the help of Claude Opus 5.5. In the 2026 revision of this paper, Claude Opus 5.5, a large language model running as an agent in Claude Code, mainly generated the hybrid-barrier cutting plane method and the analysis that replace the steps of the SODA 2024 version that relied on unproved claims about~\cite{jlsw20} (Remark~\ref{rem:jlsw20_floor} and Lemma~\ref{lem:jlsw_adaptive}), and independent Claude Opus 5.5 reviewers then checked them. This covers Sections~\ref{sec:cpm:hybrid}, \ref{sec:cpm:method} and~\ref{sec:cpm:hybrid_progress}, the proofs in Section~\ref{sec:r3_proofs}, and the changes to the analysis in Sections~\ref{sec:cpm:main_lemma} and~\ref{sec:minimization}. The main new statements are Lemmas~\ref{lem:hybrid_identity}, \ref{lem:hybrid_block} and~\ref{lem:jlsw_adaptive}, Corollary~\ref{cor:exact_score} and the revised Theorem~\ref{thm:main_formal}. The authors have verified the corrections and take full responsibility for the content of this paper.

\ifdefined\isarxiv
\bibliographystyle{alpha}
\bibliography{ref}
\else
\bibliography{ref}
\bibliographystyle{alpha}
\fi

\newpage
\appendix
\begingroup\emergencystretch=3em
\section{Proofs for the Hybrid Barrier Method}\label{sec:r3_proofs}

This section proves the lemmas behind the cutting plane method of Sections~\ref{sec:cpm:hybrid} and~\ref{sec:cpm:method} and its analysis in Section~\ref{sec:cpm:hybrid_progress}. Section~\ref{sec:r3_params} fixes every constant and lists the conditions that the proofs use. Section~\ref{sec:r3_local} proves the lemmas on the local behavior of $p$ and the facts on coordinates, on the validity of the operations and on the cut vector. Section~\ref{sec:r3_invariant} proves the invariant of a block, and Section~\ref{sec:r3_block} proves the lemmas on one block. Sections~\ref{sec:r3_adaptive} and~\ref{sec:r3_im} prove the accuracy of the leverage scores under adaptive updates and the cost of inverse maintenance. Section~\ref{sec:r3_vaidya} proves Corollary~\ref{cor:exact_score}, the deterministic variant with exact leverage scores.

\paragraph{Notation for one block.} We use the notation of Section~\ref{sec:cpm:hybrid}. We index the iterations of the outer loop of Algorithm~\ref{alg:hcpm} by $k=0,1,\ldots,T-1$. Let $K_k$ and $z_k$ be the polytope and the current point at the start of iteration $k$, so that $K_0=K$ and $z_0$ is the starting point, and let $x_k:=x_c(K_k)$ when it exists. Iteration $k$ adds or deletes a row at $z_k$, which gives $K_{k+1}$. An added row is $a^\top x\geq a^\top z_k-\rho$, where $a$ is the cut vector at $z_k$, $\wt h=a^\top\widehat H^{-1}a$ and $\rho=(\wt h/w)^{1/2}$. The iteration then takes the Newton steps $y_0:=z_k,y_1,\ldots,y_N$ in $K_{k+1}$, which end at the next current point $z_{k+1}:=y_N$. For these Newton steps we write $z_s:=z_k$, $\mu_s:=\mu_c(z_s)$, $\lambda_s:=\lambda_{K_{k+1}}(z_s)$ and $Q_0:=Q_c(z_s)$, all in $K_{k+1}$, as in Lemma~\ref{lem:hybrid_loc}. We write $e_l:=\wt g-\nabla p(y_l)$ for the error of the gradient $\wt g$ that the algorithm computes at $y_l$.

\subsection{Constants and Conditions}\label{sec:r3_params}

This section fixes every constant of the method and of its analysis, in the order in which we fix them. All other sections use these constants by name, apart from intermediate numbers derived inside a proof. The constants are not optimized. For example, at $n=10^3$ the hypothesis $\delta\leq\delta_0$ asks for $\delta\leq1.0\cdot10^{-17}$, which the data structure meets only for much larger $n$ (see the choice of $n_0$ in Step~5). The conditions at the end of the section collect every inequality between the constants that the proofs use, and the proofs cite them by their labels.

\paragraph{Step 1: the radius $r$.} Recall $r=1/10$ from Section~\ref{sec:cpm:hybrid}, the radius of the sets $\mathcal{S}(z,r)$ on which $p$ is well conditioned. With the functions $\kappa$, $\alpha(\cdot)$ and $\alpha'(\cdot)$ of that section, let
\begin{gather*}
    \alpha:=\alpha(r)=\frac{(1-r)^2}{(1+r)^4},\qquad \alpha':=\alpha'(r)=\frac{(1+r)^2}{(1-r)^4},\\
    \beta:=3\alpha'\qquad\text{and}\qquad \alpha_\mu:=\Big(\frac{1-r}{1+r}\Big)^4.
\end{gather*}
To five digits, $\alpha=0.55324$, $\alpha'=1.84423$, $\beta=5.53269$ and $\alpha_\mu=0.44813$. The factor $3$ in $\beta$ is the constant of Lemma~\ref{lem:hybrid_local} (Part~1).

\paragraph{Step 2: the Newton steps.} Table~\ref{tab:a1_newton} lists the constants of Lemma~\ref{lem:hybrid_loc}, which depend only on $r$.
\begin{table}[!htbp]
\centering\small
\begin{tabular}{@{}l l p{0.6\textwidth}@{}}
\toprule
Constant & Value & Role\\
\midrule
$\gamma_0$ & $1/100$ & largest error of the matrix $\widehat Q$ that Part~2 allows\\
$c_0$ & $1/160$ & constant in the two locality hypotheses of Part~2\\
$r_{\rm N}$ & $0.071$ & radius of the set $\mathcal{S}(z_s,r_{\rm N})$ that contains all Newton steps\\
$q_0$, $\ell_q$ & $0.9102$, $0.094$ & bound on the contraction factor $q$ of one Newton step, and a lower bound on $\ln(1/q_0)$\\
$C_{\rm r}$ & $3.62$ & factor in the radius of Part~1\\
$C_{\rm x}$, $C_\lambda$, $C_{\rm e}$ & $2.04$, $26.9$, $15.2$ & factors of the distance bound and of the decrement bound of Part~2\\
$C_{\rm g}$ & $20.65$ & links the error of the estimated scores to the decrement (Step~4)\\
\bottomrule
\end{tabular}
\caption{Constants of the Newton steps (Step~2 of Section~\ref{sec:r3_params}).}
\label{tab:a1_newton}
\end{table}

\paragraph{Step 3: the constant $w$ and the constants of one step.} We fix $w:=10^{-7}$, the leverage of an added cut, and $\theta:=w/10$, the threshold below which the method deletes a row. Tables~\ref{tab:a1_step} and~\ref{tab:a1_block} list the other constants of this step, which are absolute.
\begin{table}[!htbp]
\centering\small
\begin{tabular}{@{}l l p{0.5\textwidth}@{}}
\toprule
Constant & Value & Role\\
\midrule
$\epsilon_{\rm im}$ & $0.005$ & accuracy of both inverse maintenance instances\\
$\gamma_1$ & $0.0051$ & resulting bound on the error of $\widehat Q$ (Lemma~\ref{lem:hybrid_inv}, Part~4)\\
$\theta_{\rm add}$, $\theta'_{\rm add}$ & $0.94\,\theta$, $0.9399\,\theta$ & lower bounds on $\mu$ before and on $\mu_c$ after an add (Lemma~\ref{lem:hybrid_pert}, Part~1)\\
$w_{\rm add}$, $\epsilon_{\rm add}$ & $1.03\,w$, $10^{-4}$ & largest leverage of an added row, and the factor $1+\epsilon_{\rm add}$ on the old decrement\\
$C_+$ & $4.46$ & an add raises the decrement by at most $C_+w$\\
$\sigma_{\rm del}$, $\epsilon_{\rm del}$, $\nu_{\rm del}$ & $8\cdot10^{-5}$, $0.03$, $1/64$ & hypotheses of a deletion (Lemma~\ref{lem:hybrid_pert}, Part~2)\\
$C_Q$, $c_\mu$ & $1.4$, $0.7$ & change of $Q_c$ and lower bound on $\mu_c$ after a deletion\\
$C_-$, $C'_-$ & $2.22$, $1.21$ & factors of the bound on the decrement after a deletion\\
$\epsilon_{\rm cut}$ & $0.006$ & relative change of the slack of a new cut from $z_k$ to $x_k$ (Lemma~\ref{lem:good_centers})\\
$w_+$, $\theta_{\rm del}$ & $1.018\,w$, $1.001\,\theta$ & bounds on $w_x$ and on the score of a deleted row at $x_k$ (same lemma)\\
$g_+$, $g_-$ & $0.38$, $0.0532$ & gain of an add and loss of a delete, in units of $w$ (Lemma~\ref{lem:hybrid_step})\\
$g_{\rm b}$, $C_{\rm b}$, $g_T$ & $0.1634$, $2.3$, $0.163$ & progress of a block, with $g_{\rm b}=(g_+-g_-)/2$, and the choice of $T$ (Lemma~\ref{lem:hybrid_block})\\
$C_H$ & $0.22$ & change of the Hessian in one step (Lemma~\ref{lem:block_stability})\\
\bottomrule
\end{tabular}
\caption{Constants of one step (Step~3 of Section~\ref{sec:r3_params}).}
\label{tab:a1_step}
\end{table}
\begin{table}[!htbp]
\centering\small
\begin{tabular}{@{}l l p{0.5\textwidth}@{}}
\toprule
Constant & Value & Role\\
\midrule
$\epsilon_{\rm f}$, $\epsilon_D$ & $10^{-7}$, $2.1\cdot10^{-8}$ & accuracy of the point that a block returns, and the term $D$ of a box (Lemmas~\ref{lem:near_center}, \ref{lem:hybrid_cpm} and~\ref{lem:box_margin})\\
$C_{\rm s}$, $w_J$ & $5$, $0.26\,w$ & decrement at the start of the Newton steps (Lemma~\ref{lem:hybrid_inv})\\
$C_{\rm N}$ & $270$ & factor in the choice of $N$ (Step~4)\\
$h_{\max}$ & $0.068$ & length of one Newton step in the norm of $H$ (Lemma~\ref{lem:hybrid_inv}, Part~5)\\
$\tau_{\rm del}$, $\tau_{\rm step}$, $t_{\max}$ & $0.009$, $0.004$, $20$ & thresholds of the delete test and of the step test, and the number of equal pieces of a Newton step\\
$C_*$ & $2000$ & denominator of $\lambda^*$ (Step~4)\\
$C_R$, $C_\tau$, $C_{R'}$ & $1.034$, $1.033$, $1.01$ & size of the polytopes (Lemma~\ref{lem:sandwich})\\
\bottomrule
\end{tabular}
\caption{Constants of a block, of its tests and of the size of the polytopes (Step~3 of Section~\ref{sec:r3_params}).}
\label{tab:a1_block}
\end{table}

\paragraph{Step 4: the parameters that depend on $n$.} For $n\geq1$, let
\begin{gather*}
    c:=\frac{w}{20\ln(n/w)},\qquad \lambda^*:=\frac{c^{3/4}}{C_*},\qquad \delta_0:=\frac{\lambda^*c^{1/2}}{2C_{\rm g}}=\frac{c^{5/4}}{82600},\\
    \epsilon_g:=\alpha'^{1/2}\delta_0c^{-1/2}\qquad\text{and}\qquad r_{\rm f}:=C_{\rm r}\lambda^*c^{-1/4}.
\end{gather*}
Here $\epsilon_g$ bounds the error of the gradient that comes from the error $\delta_0$ of the scores, and $r_{\rm f}$ bounds the distance from a point with decrement at most $\lambda^*$ to the hybrid center. The paragraph \emph{Parameters} in Section~\ref{sec:cpm:hybrid} gives the meaning of the other parameters. For $\sigma\geq0$ we write $J(\sigma):=C_-\sigma+C'_-(\sigma c)^{1/2}$ for the bound of Lemma~\ref{lem:hybrid_pert} on the decrement after a deletion. We also let
\begin{gather*}
    \bar m:=\Big\lfloor\frac{n}{\theta-\delta_0}\Big\rfloor+1,\qquad \bar\kappa:=1+\frac{n+c\bar m}{c},\\
    \bar R:=\frac{C_Rn}{c}\qquad\text{and}\qquad R':=C_{R'}\big(\bar R+\bar m^{1/2}\big),
\end{gather*}
which bound the number of rows, the ratio of two slacks of one row, the size of a polytope around its hybrid center and its size around the point that a block returns. Finally, the number of steps of one block and the number of Newton steps after each added or deleted row are
\begin{align*}
    T:=\Big\lceil\frac{2n\ln n+C_{\rm b}n}{g_Tw}\Big\rceil\qquad\text{and}\qquad N:=\Big\lceil\frac{\ln(C_{\rm N}w/\lambda^*)}{\ell_q}\Big\rceil.
\end{align*}
So $c=\Theta(1/\log n)$, $\lambda^*=\Theta(c^{3/4})$, $\delta_0=\Theta(c^{5/4})$, $r_{\rm f}=\Theta(c^{1/2})$, $\bar m=O(n)$, $\bar\kappa=O(n\log n)$, $R'=O(n\log n)$, $T=\Theta(n\log n)$ and $N=O(\log(1/c))=O(\log\log n)$.

\paragraph{Step 5: constants that exist but are not explicit.} Let $C_{\rm J}$ be the absolute constant of the deterministic part of the data structure of~\cite{jlsw20}, defined in the proof of Lemma~\ref{lem:jlsw_adaptive}, and let $\delta_{\rm J}(n):=(C_{\rm J}+1)\log^{-15}n$ be the accuracy of that lemma. Since $\delta_0=\Theta(\log^{-5/4}n)$, there is a least integer $n_0$ such that $\delta_{\rm J}(n)\leq\delta_0(n)$ for every $n\geq n_0$. The constant $n_0$ is absolute, but it is not explicit, because $C_{\rm J}$ is not. For $n<n_0$ the method computes every leverage score exactly. We also write $\epsilon_{\rm J}:=(0.48-(\omega-2))/2$, which is positive because $\omega<2.48$. Last, $C_M$ is the absolute constant fixed in the proof of Lemma~\ref{lem:main_oracle_formal}, and for $n\geq n_0$ the operation budget of Algorithm~\ref{alg:ours} is
\begin{align*}
    M:=\big\lceil C_M\big(n^2\log n+n\lceil\log_2(\gamma R)\rceil\big)\log\log n\big\rceil .
\end{align*}

\paragraph{Step 6: Vaidya's method with exact leverage scores.} Only Corollary~\ref{cor:exact_score} and Lemma~\ref{lem:sp_vaidya} use these constants. We fix $\epsilon_{\rm V}:=4^{-20}$, $\sigma_\star:=8\epsilon_{\rm V}$, $c_{\rm V}:=10^{-7}$, $\pi_0:=10^{-4}$, $r_{\rm V}:=10^{-3}$, $q_{\rm V}:=93/100$, $\tau_{\rm V}:=2^{-7}$, $g^+_{\rm V}:=16/5$, $g^-_{\rm V}:=2$, $C_\pi:=720$, $C_{\rm V}:=128$ and $C_{\rm V,b}:=2.7$. For $n\geq1$, let
\begin{gather*}
    \bar m_{\rm V}:=\Big\lfloor\frac{n}{\epsilon_{\rm V}}\Big\rfloor+1,\qquad k_{\rm V}:=\Big\lceil\frac{\ln(c_{\rm V}/(2\epsilon_{\rm V}^4))}{\ln(1/q_{\rm V})}\Big\rceil,\qquad R'_{\rm V}:=C_{\rm V}\bar m_{\rm V}^{3/2}n,\\
    T_{\rm V}:=\Big\lceil\frac{2n\ln n+C_{\rm V,b}n}{g_{\rm V,b}\,\epsilon_{\rm V}}\Big\rceil\qquad\text{with}\qquad g_{\rm V,b}:=\frac{g^+_{\rm V}-g^-_{\rm V}}{2}=\frac35.
\end{gather*} So $T_{\rm V}=O(n\log n)$, $k_{\rm V}=O(1)$ and $R'_{\rm V}=O(n^{5/2})$.

\paragraph{Conditions.} The proofs use the values above only through the inequalities below. Conditions~(C1) to~(C7) follow by arithmetic from the definitions, with~(W1) and~(W5) for the size of $c$ and of $r_{\rm f}$. Conditions~(W1) to~(W5) need $w$ to be small, and~(W6) needs $w$ not to be too small. For each condition we give the numbers that show it. A condition stated for every $n$ holds for every $n\geq1$, and~(W6) for every $n\geq2$.

\begin{itemize}
    \item[(C1)]\rthreecond{C1}{cond:C1} \emph{Contraction of the Newton steps.} The constants of Steps~1 and~2 satisfy
    \begin{gather*}
        1-\frac{\alpha}{\beta}+\frac{\gamma_0}{1-\gamma_0}\leq q_0,\qquad \ln\frac1{q_0}\geq\ell_q,\qquad \beta(1-\gamma_0)(1-q_0)C_{\rm x}\geq1,\\
        \frac{2\beta}{\alpha^{3/2}}\leq C_\lambda\qquad\text{and}\qquad \frac{\beta C_{\rm x}}{\alpha^{1/2}}\leq C_{\rm e},
    \end{gather*}
    with left sides $0.91011$, $0.09409$, $1.0034$, $26.890$ and $15.174$. For every $n$, the error of the matrix $\widehat Q$ obeys
    \begin{align*}
        \Big(1+\frac{\epsilon_{\rm im}}{1-\epsilon_{\rm im}}\Big)\Big(1+\frac{\delta_0}{c}\Big)-1\leq\gamma_1\leq\gamma_0,
    \end{align*}
    because $\epsilon_{\rm im}/(1-\epsilon_{\rm im})\leq0.005026$ and $\delta_0/c\leq10^{-7}$ by~\ref{cond:C7}.
    \item[(C2)]\rthreecond{C2}{cond:C2} \emph{Locality and one row.} The locality constants of Lemma~\ref{lem:hybrid_loc} satisfy
    \begin{align*}
        \alpha C_{\rm r}>2,\qquad c_0<\frac{\alpha r}{2},\qquad \frac{4c_0}{\alpha}+2C_{\rm x}c_0\leq r_{\rm N}<r,
    \end{align*}
    with values $2.0027$, $0.0277$ and $0.0707$. The constants of an added row satisfy
    \begin{align*}
        \frac{w}{1-\epsilon_{\rm im}}\leq w_{\rm add}\qquad\text{and}\qquad w_{\rm add}+\Big(w_{\rm add}+\frac{w}{322}\Big)\Big(\frac{w_{\rm add}}{\theta'_{\rm add}}\Big)^{1/2}\leq C_+w,
    \end{align*}
    where the left side of the second inequality is $4.4500\,w$. With $\mu_{\rm del}:=3/4-\epsilon_{\rm del}=0.72$, the constants of a deleted row satisfy
    \begin{gather*}
        \max\Big\{\frac{1}{\mu_{\rm del}},\,1+4\nu_{\rm del}\Big\}\leq C_Q<\frac1{\sigma_{\rm del}},\qquad \mu_{\rm del}-\frac{(1-\mu_{\rm del})\sigma_{\rm del}}{1-\sigma_{\rm del}}\geq c_\mu,\\
        1-\frac{4\nu_{\rm del}\sigma_{\rm del}}{1-\sigma_{\rm del}}\geq c_\mu(1+4\nu_{\rm del}),\qquad (1-\sigma_{\rm del})^{-1/2}+\big(c_\mu(1-\sigma_{\rm del})\big)^{-1/2}\leq C_-,\\
        \big(c_\mu(1-\sigma_{\rm del})\big)^{-1/2}\leq C'_-,
    \end{gather*}
    with left sides $1.3889$, $0.71998$, $0.99999$, $2.1953$ and $1.1953$, against $c_\mu(1+4\nu_{\rm del})=0.74375$ in the third. Also $w_+\leq w_{\rm add}$.
    \item[(C3)]\rthreecond{C3}{cond:C3} \emph{Progress.} For every $n$,
    \begin{align*}
        c\ln(2\bar\kappa)\leq\frac{w}{10}.
    \end{align*}
    Indeed $\bar m\leq10.6\,n/w$ by~\ref{cond:C7}, so $\bar\kappa=1+n/c+\bar m\leq21(n/w)\ln(n/w)$, because $\ln(n/w)\geq16.1$ by~\ref{cond:W1}. Hence $\ln(2\bar\kappa)\leq\ln(n/w)+\ln(42\ln(n/w))\leq2\ln(n/w)$, as $42x\leq e^x$ for $x\geq16.1$, and we multiply by $c=w/(20\ln(n/w))$. In particular $c\ln((1+\epsilon_{\rm cut})\bar\kappa)\leq w/10$ and $2c\ln(2\bar\kappa)\leq w/5\leq\epsilon_D\leq\epsilon_{\rm f}$.
    \item[(C4)]\rthreecond{C4}{cond:C4} \emph{Block length and size.} For every $n$,
    \begin{gather*}
        \frac{g_++g_-}{2}\,w\bar m\leq C_{\rm b}n,\qquad 2n\leq\bar m,\qquad c\bar m\leq(C_\tau-1)n,\\
        C_\tau(1+c)^{1/2}\leq C_R\qquad\text{and}\qquad \sqrt{2n}\leq R',
    \end{gather*}
    because $\bar m\leq10.6\,n/w$ and $c\leq w/322$ give left sides at most $2.296\,n$ and $0.033\,n$ in the first and third inequality. Also $(1+r_{\rm f})\max\{1,r_{\rm f}/(1-r_{\rm f})\}\leq C_{R'}$, since $r_{\rm f}\leq3.2\cdot10^{-8}$ by~\ref{cond:W5}. The constants of the block satisfy $g_T\leq g_{\rm b}$, $2\ln(1/(1-\epsilon_{\rm f}))\leq10\,\epsilon_{\rm f}$, $\epsilon_{\rm f}\leq(\ln2)/2$ and
    \begin{align*}
        \ln\frac{1}{(1-r)^2}+w_+\leq C_H\qquad\text{and}\qquad \ln\big((1+r)^2\big)+\ln\frac{1}{1-\theta_{\rm del}}\leq C_H,
    \end{align*}
    with left sides at most $0.2108$ and $0.1907$.
    \item[(C5)]\rthreecond{C5}{cond:C5} \emph{The decrement after the Newton steps.} We have $2C_\lambda C_{\rm s}=269\leq C_{\rm N}$ and $\alpha'^{1/2}C_{\rm e}=20.642\leq C_{\rm g}$. For every $n$,
    \begin{gather*}
        C_+w+(1+\epsilon_{\rm add})\lambda^*\leq C_{\rm s}w,\qquad J(\theta+\delta_0)\leq w_J\\
        \text{and}\qquad(1+\epsilon_{\rm add})\lambda^*+w_J\leq C_{\rm s}w,
    \end{gather*}
    as $\lambda^*\leq3.7\cdot10^{-11}$ by~\ref{cond:W1} and $J(\theta+\delta_0)\leq0.2434\,w$. The choice of $\delta_0$ gives $C_{\rm g}\delta_0c^{-1/2}=\lambda^*/2$ exactly. Since $\ln(1/q_0)\geq\ell_q$, the choice of $N$ gives $q_0^N\leq\lambda^*/(C_{\rm N}w)$. Together,
    \begin{align*}
        C_\lambda C_{\rm s}w\,q_0^N\leq\frac{\lambda^*}{2}\qquad\text{and}\qquad C_{\rm e}\,\epsilon_g\leq C_{\rm g}\delta_0c^{-1/2}=\frac{\lambda^*}{2}.
    \end{align*}
    \item[(C6)]\rthreecond{C6}{cond:C6} \emph{Validity of every operation.} The numbers $0.01$ below are those of Lemma~\ref{lem:jlsw_main}. We have
    \begin{gather*}
        \frac{w}{1-\epsilon_{\rm im}}\leq0.01,\qquad\frac{\tau_{\rm del}}{1-\epsilon_{\rm im}}<0.01,\qquad 2\cdot1.01\cdot\tau_{\rm step}<0.01\\
        \text{and}\qquad \frac{1+r_{\rm N}}{1-r_{\rm N}}\cdot\frac{h_{\max}}{t_{\max}}\leq\tau_{\rm step},
    \end{gather*}
    with left sides at most $1.0051\cdot10^{-7}$, $0.009046$, $0.00808$ and $0.00392$, and $|\ln(1+x)|\leq1.01|x|$ for $|x|\leq\tau_{\rm step}$. For every $n$, $(1+\epsilon_{\rm im})(\theta+\delta_0)\leq\tau_{\rm del}$ and $\delta_0<c$.
    \item[(C7)]\rthreecond{C7}{cond:C7} \emph{Accuracy.} For every $n$, $\delta_0\leq10^{-7}c$, because $\delta_0/c=c^{1/4}/82600$ and $c\leq w/322$ by~\ref{cond:W1}. Hence $2\delta_0\leq\nu_{\rm del}c$, $\delta_0\leq\theta/20$, $\theta-\delta_0\geq\theta_{\rm add}$ and $\bar m\leq n/(0.95\,\theta)+1\leq10.6\,n/w$. For every $n\geq n_0$, $\delta_{\rm J}(n)\leq\delta_0(n)$, by the choice of $n_0$.
\end{itemize}

\begin{itemize}
    \item[(W1)]\rthreecond{W1}{cond:W1} \emph{The size of $c$, and locality after one step.} We have $\ln(1/w)=16.118\geq16.1$, so $c\leq w/322$ for every $n$. Hence $c^{1/2}\leq1.77\cdot10^{-5}$ and $\lambda^*\leq3.7\cdot10^{-11}$. After an add, the decrement stays within the locality radius of Lemma~\ref{lem:hybrid_loc} (Part~2), as
    \begin{align*}
        C_{\rm s}w=5\cdot10^{-7}\leq6.1\cdot10^{-5}\leq c_0\,(\theta'_{\rm add})^{1/4}.
    \end{align*}
    After a delete, for every $n$ and every $\sigma\in[0,\theta_{\rm del}]$,
    \begin{align*}
        (1+\epsilon_{\rm add})\lambda^*+J(\sigma)\leq c_0\big(c_\mu(\sigma+c)\big)^{1/4}.
    \end{align*}
    Indeed $(\sigma c)^{1/2}\leq\sigma+c$ and $\lambda^*\leq C_*^{-1}c^{1/2}(\sigma+c)^{1/4}$, so the left side is at most $\big((C_-+C'_-)(\theta_{\rm del}+w/322)^{3/4}+(1+\epsilon_{\rm add})C_*^{-1}(w/322)^{1/2}\big)(\sigma+c)^{1/4}\leq3.6\cdot10^{-6}(\sigma+c)^{1/4}$, while $c_0c_\mu^{1/4}\geq0.0057$. Finally $\epsilon_g\leq2c_0c^{1/4}$ and $r_{\rm f}\leq r$ for every $n$.
    \item[(W2)]\rthreecond{W2}{cond:W2} \emph{Gain and loss of one step.} For every $n$,
    \begin{gather*}
        \frac12\ln\Big(1+\frac{(1-r_{\rm f})^2w}{(1+\epsilon_{\rm im})(1+\epsilon_{\rm cut})^2}\Big)-\frac{2}{\alpha}(C_+w)^2-\frac{w}{10}\geq g_+w,\\
        \frac12\ln\frac{1}{1-\theta_{\rm del}}+\frac{2}{\alpha}J(\theta_{\rm del})^2\leq g_-w.
    \end{gather*}
    The left sides are at least $0.3915\,w$ and at most $0.05006\,w$. The middle term of the first, at most $7.2\cdot10^{-6}w$, is the bound on ${\rm rec}$ after an add (Lemma~\ref{lem:good_centers}).
    \item[(W3)]\rthreecond{W3}{cond:W3} \emph{Short Newton steps.} For every $n$, the step lengths of Lemma~\ref{lem:hybrid_inv} (Part~5) after an add and after the deletion of a row with $\sigma\in[0,\theta+\delta_0]$ satisfy
    \begin{gather*}
        2\Big(\frac{\alpha'}{\alpha_\mu\theta'_{\rm add}}\Big)^{1/2}\Big(\frac{2}{\alpha}\big(C_+w+(1+\epsilon_{\rm add})\lambda^*\big)+C_{\rm x}\epsilon_g\Big)\leq h_{\max},\\
        2\Big(\frac{\alpha'}{\alpha_\mu c_\mu(\sigma+c)}\Big)^{1/2}\Big(\frac{2}{\alpha}\big((1+\epsilon_{\rm add})\lambda^*+J(\sigma)\big)+C_{\rm x}\epsilon_g\Big)\leq h_{\max}.
    \end{gather*}
    The left sides are at most $0.0675$ and $0.0043$. The first is the condition that keeps $w$ as small as $10^{-7}$.
    \item[(W4)]\rthreecond{W4}{cond:W4} \emph{One row.} We have $(1+w_{\rm add})^2\theta'_{\rm add}\leq\theta_{\rm add}$, $(1+w_{\rm add})^{1/2}\leq1+\epsilon_{\rm add}$, $(1-C_Q\theta_{\rm del})^{-1/2}\leq1+\epsilon_{\rm add}$ and $\theta_{\rm del}\leq\sigma_{\rm del}$.
    \item[(W5)]\rthreecond{W5}{cond:W5} \emph{Near the hybrid center.} For every $n$, $r_{\rm f}=C_{\rm r}C_*^{-1}c^{1/2}\leq3.2\cdot10^{-8}$ by~\ref{cond:W1}. This gives
    \begin{gather*}
        (1+r_{\rm f})^2\leq1+\epsilon_{\rm f},\qquad(1-r_{\rm f})^2\geq1-\epsilon_{\rm f},\qquad\ln\frac{1}{1-r_{\rm f}}\leq\epsilon_{\rm f},\qquad\kappa(r_{\rm f})^{-2}\geq1-\epsilon_{\rm del},\\
        \kappa(r_{\rm f})(\theta+\delta_0)\leq\theta_{\rm del},\qquad\Big(\frac{1-r_{\rm f}}{1+r_{\rm f}}\Big)^4(\theta-\delta_0)\geq\theta_{\rm add}\qquad\text{and}\\
        \frac{(1+r_{\rm f})^2w}{(1-\epsilon_{\rm im})(1-\epsilon_{\rm cut})^2}\leq w_+,
    \end{gather*}
    where the last left side is $1.0172\,w$. For a cut added at a point $z$ with $\lambda(z)\leq\lambda^*$, the slack at the hybrid center is within the factor $1\pm\epsilon_{\rm cut}$ of its slack at $z$, because
    \begin{align*}
        \frac{2\lambda^*}{\alpha}\Big(\frac{w}{(1-\epsilon_{\rm im})(\theta-\delta_0)}\Big)^{1/2}\leq4.3\cdot10^{-10}\leq\epsilon_{\rm cut}.
    \end{align*}
    \item[(W6)]\rthreecond{W6}{cond:W6} \emph{The threshold of the dimension reduction.} For every $n\geq2$,
    \begin{align*}
        4R'\leq2^{100n\log n}=n^{100n},
    \end{align*}
    where $\log$ is to base $2$. Indeed $R'\leq21\,n\ln(n/w)/w$ by the definitions and~\ref{cond:C7}, so $\ln(4R')\leq21+\ln n+\ln\ln(10^7n)$, which is at most $100\,n\ln n$ for $n\geq2$.
\end{itemize}

\begin{itemize}
    \item[(V1)]\rthreecond{V1}{cond:V1} \emph{Vaidya's Newton step.} $\epsilon_{\rm V}\leq10^{-12}$, $\epsilon_{\rm V}^4\leq c_{\rm V}/2$, $C_\pi c_{\rm V}\leq\pi_0$ and $(C_\pi c_{\rm V})^{1/2}/10\leq r_{\rm V}$, where the last two left sides are at most $7.2\cdot10^{-5}$ and $8.5\cdot10^{-4}$. Also $\pi_0<r^2/16$, $c_{\rm V}<r^2/4$, so that $\epsilon_{\rm V}^4<r^2/4$, and $r_{\rm V}\leq r$. The values of Step~1 give $\alpha\geq1/2$, $3\alpha'\leq6$ and $\alpha_\mu\geq2/5$.
    \item[(V2)]\rthreecond{V2}{cond:V2} \emph{One add or delete in Vaidya's method.} $3600\,\epsilon_{\rm V}\leq\pi_0$, $9000\,\epsilon_{\rm V}\leq c_{\rm V}/2$, $4(3600\,\epsilon_{\rm V})^{1/2}\leq r_{\rm V}$ and $16\,\epsilon_{\rm V}\leq r_{\rm V}$, with left sides at most $3.3\cdot10^{-9}$, $8.2\cdot10^{-9}$, $2.3\cdot10^{-4}$ and $1.5\cdot10^{-11}$.
    \item[(V3)]\rthreecond{V3}{cond:V3} \emph{Counting in Vaidya's method.} $g^+_{\rm V}>g^-_{\rm V}$, and for every $n$, $\frac{g^+_{\rm V}+g^-_{\rm V}}{2}(n+\epsilon_{\rm V})\leq C_{\rm V,b}n$. For every $n\geq2$, $4R'_{\rm V}\leq2^{100n\log n}$, because $\bar m_{\rm V}\leq1.01\cdot4^{20}n$ and $4C_{\rm V}=512$ give $\ln(4R'_{\rm V})\leq\ln512+1.5\ln(1.01\cdot4^{20}n)+\ln n\leq48+2.5\ln n\leq100\,n\ln n$.
\end{itemize}

\subsection{Proofs for the Hybrid Barrier and the Method}\label{sec:r3_local}

This section proves Lemmas~\ref{lem:hybrid_local}, \ref{lem:hybrid_loc} and~\ref{lem:hybrid_pert}, and the facts on coordinates, on the validity of the operations and on the cut vector that Section~\ref{sec:cpm:method} states.

\paragraph{Coordinates and padding.} In the coordinates $x=\bar x+Yt$ of Section~\ref{sec:cpm:method}, a row $a_i^\top x\geq b_i$ becomes $(Y^\top a_i)^\top t\geq b_i-a_i^\top\bar x$, with the same slack at every point. A quadratic form $M$ on $W_0$ has the matrix $Y^\top MY$ in the basis $Y$, and another basis changes $F$ only by an additive constant. So slacks, leverage scores, decrements, Newton steps and the sets $\mathcal{S}(z,r')$ do not depend on the choice of $\bar x$ and $Y$, and any choice works, for instance $\bar x=0$ and $Y=I_n$ for the initial box. The three data structures run on the padded matrix
\begin{align*}
    A_{\rm pad}=\begin{bmatrix}AY&0\\0&I_{n-d}\end{bmatrix}\in\R^{(m+n-d)\times n},
\end{align*}
whose first $m$ rows carry the weights of the method and whose last $n-d$ rows carry the fixed weight $1$. For all weights, $A_{\rm pad}^\top{\rm diag}(\eta)A_{\rm pad}$ is block diagonal, with the blocks $Y^\top A^\top{\rm diag}(\eta_{[m]})AY$ and $I_{n-d}$. Hence every real row, inserted and deleted ones included, has the same leverage score in the padded instance as in $K$. The top-left block of an inverse kept by an inverse maintenance instance is the inverse of $Y^\top A^\top{\rm diag}(\wt\eta_{[m]})AY$ for the kept weights $\wt\eta$ of the real rows, so~\eqref{eq:im_inverse} holds for it. The dummy rows never change their weights, so they add nothing to $\|\Delta\log\eta\|_2$.

The case $c=0$ of Part~3 of the next lemma is due to Anstreicher~\cite[Theorem~A.4]{a96} (see the discussion after Lemma~\ref{lem:v89_3}). We prove it to keep the paper self-contained.

\begin{lemma}[Derivatives of the hybrid barrier]\label{lem:hybrid_hessian}
Let $K=\{x\in W:a_i^\top x\geq b_i,\ i\in[m]\}$, where $W\subseteq\R^n$ is an affine subspace with linear part $W_0$. Let $L(x)=-\sum_{i=1}^m\ln(a_i^\top x-b_i)$, let $c\geq0$ and let $p=F+c\,L$. Let $x$ be a relative interior point of $K$ such that $H(x)$ is positive definite on $W_0$. Write $s_i=a_i^\top x-b_i$, $S={\rm diag}(s)$, $P_{ij}=a_i^\top H(x)^{-1}a_j/(s_is_j)$ for $i,j\in[m]$, $\sigma_i=P_{ii}$, $\Sigma_c={\rm diag}(\sigma+c\mathbf{1})$ and $Q_c=A^\top S^{-1}\Sigma_cS^{-1}A$. Thus $\sigma_i=\sigma_i(x)$, $\Sigma_0={\rm diag}(\sigma)$ and $Q_c=Q(x)+c\,H(x)$. Then the following hold.
\begin{enumerate}
    \item $\nabla F(x)=-A^\top S^{-1}\sigma$, $\nabla L(x)=-A^\top S^{-1}\mathbf{1}$ and $\nabla p(x)=-A^\top S^{-1}(\sigma+c\mathbf{1})$.
    \item $\nabla^2F(x)=A^\top S^{-1}(3\Sigma_0-2P\circ P)S^{-1}A$ and $\nabla^2L(x)=H(x)$, where $\circ$ denotes the entrywise product.
    \item $Q_c\preceq\nabla^2p(x)\preceq3Q(x)+c\,H(x)\preceq3Q_c$. In particular, $Q(x)\preceq\nabla^2F(x)\preceq3Q(x)$.
\end{enumerate}
If $W\neq\R^n$, we regard $H(x)$ as a quadratic form on $W_0$ and take its inverse and its determinant with respect to a fixed basis of $W_0$ (another basis changes $F$ only by an additive constant). Derivatives are then taken along $W$, and the identities and inequalities above hold for linear and quadratic forms on $W_0$.
\end{lemma}

\begin{proof}
By the paragraph \emph{Coordinates and padding}, we may assume that $W=\R^d$, where $d=\dim(W)$, and that $H(x)=A^\top S^{-2}A$ is positive definite. Then $A$ has rank $d$, so $F$ and $L$ are smooth wherever all slacks are positive. With $B=S^{-1}A$, the matrix $P=BH(x)^{-1}B^\top$ is the orthogonal projection onto the column space of $B$, so $P\succeq0$ and $I-P\succeq0$.

Fix $h\in\R^d$. Along $x(\tau)=x+\tau h$, let $s_i=a_i^\top x(\tau)-b_i$, $v_i=a_i/s_i$, $u_i=a_i^\top h/s_i$ and $H=\sum_jv_jv_j^\top$, and write a dot for the derivative in $\tau$. Since $\dot s_i=a_i^\top h$, we have $\dot v_i=-u_iv_i$ and $\dot u_i=-u_i^2$, so
\begin{align*}
    \dot H=-2\sum_ju_jv_jv_j^\top\qquad\text{and}\qquad\frac{\d}{\d\tau}H^{-1}=-H^{-1}\dot HH^{-1}=2\sum_ju_jH^{-1}v_jv_j^\top H^{-1}.
\end{align*}
With $\sigma_i=v_i^\top H^{-1}v_i$ and $P_{ij}=v_i^\top H^{-1}v_j$ along $x(\tau)$, Jacobi's formula $\frac{\d}{\d\tau}\ln\det H=\tr(H^{-1}\dot H)$ and the product rule give
\begin{align}\label{eq:sigma_dot}
\begin{split}
    \frac{\d}{\d\tau}F&=\frac12\tr(H^{-1}\dot H)=-\sum_i\sigma_iu_i\qquad\text{and}\\
    \dot\sigma_i&=2\dot v_i^\top H^{-1}v_i+v_i^\top\Big(\frac{\d}{\d\tau}H^{-1}\Big)v_i=-2\sigma_iu_i+2\sum_jP_{ij}^2u_j.
\end{split}
\end{align}
Differentiating once more, we get
\begin{align*}
    \frac{\d^2}{\d\tau^2}F=-\sum_i\big(\dot\sigma_iu_i+\sigma_i\dot u_i\big)=3\sum_i\sigma_iu_i^2-2\sum_{i,j}P_{ij}^2u_iu_j=u^\top(3\Sigma_0-2P\circ P)u.
\end{align*}
At $\tau=0$ we have $u=S^{-1}Ah$. As $h$ is arbitrary and both sides are symmetric, this gives $\nabla F(x)$ and $\nabla^2F(x)$. For $L$, the first two derivatives in $\tau$ are $-\sum_iu_i$ and $\sum_iu_i^2$. This proves Parts~1 and~2, and it gives $\nabla^2p(x)=A^\top S^{-1}(3\Sigma_0-2P\circ P+cI)S^{-1}A$.

For Part~3, the Schur product theorem states that the entrywise product of two positive semidefinite matrices is positive semidefinite. It gives $P\circ P\succeq0$ and $\Sigma_0-P\circ P=(I-P)\circ P\succeq0$, where the identity holds because the diagonal of $P$ is $\sigma$. Hence
\begin{align*}
    \Sigma_0+cI\preceq3\Sigma_0-2P\circ P+cI\preceq3\Sigma_0+cI\preceq3(\Sigma_0+cI),
\end{align*}
using $c\geq0$ in the last step. The map $M\mapsto A^\top S^{-1}MS^{-1}A$ preserves the order $\preceq$, which gives Part~3.
\end{proof}

Both constants in Part~3 are attained for $c=0$, so the lemma cannot be improved. For the box $[-1,1]^d$ at its center, every $\sigma_i$ equals $1/2$ and the opposite constraints $x_k\leq1$ and $x_k\geq-1$ cancel in $u^\top(P\circ P)u$ for every $u=S^{-1}Ah$, so $\nabla^2F=3Q$. If $b=0$ and $h=x$, then $u=\mathbf{1}$ and $(\Sigma_0-P\circ P)\mathbf{1}=0$, because $\sum_jP_{ij}^2=(P^2)_{ii}=\sigma_i$, so $h^\top\nabla^2F(x)h=h^\top Q(x)h$.

\begin{proof}[Proof of Lemma~\ref{lem:hybrid_local}]
Part~1 is Part~3 of Lemma~\ref{lem:hybrid_hessian}. Since $H(z)$ is positive definite on $W_0$, the forms $a_i$ span the dual of $W_0$, so $H$ is positive definite at every relative interior point. As $Q_c\succeq cH$, the function $p$ is strictly convex.

For Parts~2 and~3, we use that $g^\top M^{-1}g=\max_h(2g^\top h-h^\top Mh)$, over $h\in W_0$, for a linear form $g$ and a positive definite form $M$ on $W_0$. Let $B:=S_z^{-1}A$, so that $Q_c=B^\top\Sigma_cB$ with $\Sigma_c=\Sigma_c(z)$. Replacing $Bh$ by an arbitrary $y\in\R^m$ gives
\begin{align*}
    \|B^\top e\|_{Q_c^{-1}}^2=\max_h\big(2e^\top Bh-(Bh)^\top\Sigma_cBh\big)\leq\max_{y\in\R^m}\big(2e^\top y-y^\top\Sigma_cy\big)=\sum_i\frac{e_i^2}{\sigma_i+c},
\end{align*}
which proves Part~2. For Part~3, $Q_c\succeq(\sigma_k+c)a_ka_k^\top/s_k^2$, so the same formula with $\tau:=a_k^\top h/s_k$ gives $s_k^{-2}a_k^\top Q_c^{-1}a_k\leq\max_\tau(2\tau-(\sigma_k+c)\tau^2)=1/(\sigma_k+c)$. Also $Q_c\succeq\mu_cH$, so $s_k^{-2}a_k^\top Q_c^{-1}a_k\leq s_k^{-2}a_k^\top H^{-1}a_k/\mu_c=\sigma_k/\mu_c$. The minimum of two numbers is at most their geometric mean, and $\sigma_k/((\sigma_k+c)\mu_c)\leq1/\mu_c$. Finally, $|a_k^\top h|\leq\|a_k\|_{Q_c^{-1}}\|h\|_{Q_c}$ for $h\in W_0$, which gives both bounds.

For Part~4, every ratio $s_i(y)/s_i(z)$ lies in $[1-r',1+r']$, so every term $a_ia_i^\top/s_i(y)^2$ lies between $(1+r')^{-2}$ and $(1-r')^{-2}$ times $a_ia_i^\top/s_i^2$. This gives the bounds on $H(y)$. Inverting them shows that $\sigma_i(y)=s_i(y)^{-2}a_i^\top H(y)^{-1}a_i$ lies within the factor $\kappa(r')$ of $\sigma_i$. As $\kappa(r')\geq1$ and $c\geq0$, also $\sigma_i(y)+c$ lies within the factor $\kappa(r')$ of $\sigma_i+c$. So every weight $(\sigma_i+c)/s_i^2$ changes by a factor in $[\kappa(r')^{-1}(1+r')^{-2},\kappa(r')(1-r')^{-2}]=[\alpha(r'),\alpha'(r')]$, which gives the bounds on $Q_c(y)$. Next, $Q_c(y)\succeq\alpha(r')Q_c\succeq\alpha(r')\mu_cH\succeq\alpha(r')(1-r')^2\mu_cH(y)$, and $\alpha(r')(1-r')^2=((1-r')/(1+r'))^4$, which bounds $\mu_c(y)$. The same argument with $c=0$ bounds $\mu(y)$. The last claim follows from Part~1 at $y$ and the bounds on $Q_c(y)$.
\end{proof}

\begin{proof}[Proof of Lemma~\ref{lem:hybrid_loc}]
We prove Part~4 first, since Part~1 uses it. Since $p(x_c)$ is finite, $\det H(x_c)>0$, so the matrix $H(x_c)$ is positive definite on $W_0$. So no nonzero $h\in W_0$ has $a_i^\top h=0$ for every $i$, that is, the forms $a_i$ span the dual of $W_0$. Let $v_i:=\sigma_i(x_c)+c>0$, so that $\sum_iv_i=d+cm$. By Lemma~\ref{lem:hybrid_hessian} (Part~1), $\nabla p(x_c)=-\sum_iv_ia_i/s_i(x_c)$, and this form vanishes on $W_0$ because $x_c$ minimizes $p$ over the relative interior of $K$. For $y\in K$, let $u_i:=s_i(y)/s_i(x_c)-1\geq-1$. Since $y-x_c\in W_0$,
\begin{align}\label{eq:center_balance}
    \sum_iv_iu_i=\sum_iv_i\frac{a_i^\top(y-x_c)}{s_i(x_c)}=-\nabla p(x_c)^\top(y-x_c)=0.
\end{align}
Hence $\sum_{i:\,u_i>0}v_iu_i=\sum_{i:\,u_i<0}v_i|u_i|\leq\sum_iv_i=d+cm$, so $v_iu_i\leq d+cm$ for every $i$. This is the first bound of Part~4, and the second follows from $d\leq n$ and $\sigma_i(x_c)\geq0$. So every $a_i^\top(y-x_c)$ lies between $-s_i(x_c)$ and $(d+cm)s_i(x_c)/v_i$, and $K$ is bounded because the forms $a_i$ span the dual of $W_0$. This is the identity~(6.12) in the proof of Lemma~19 of~\cite{lsw15}, without regularization and weight corrections.

For Part~1, let $\mathcal{S}':=\mathcal{S}(z_s,r')$. It is convex, and every slack is at least $(1-r')s_i(z_s)>0$ on it, so it lies in the relative interior of $K$. It is compact, as $H(z_s)$ is positive definite on $W_0$ and $\|y-z_s\|_{H(z_s)}^2\leq mr'^2$ for $y\in\mathcal{S}'$. Let $y$ lie on the relative boundary of $\mathcal{S}'$. Some $k$ has $|a_k^\top(y-z_s)|=r's_k(z_s)$, so $\ell:=\|y-z_s\|_{Q_0}\geq r'\mu_s^{1/4}$ by Lemma~\ref{lem:hybrid_local} (Part~3). On $[z_s,y]\subseteq\mathcal{S}'$, Part~4 of that lemma gives $\nabla^2p\succeq\alpha(r')Q_0\succeq\alpha Q_0$, as $\alpha(\cdot)$ is decreasing. Hence
\begin{align*}
    p(y)\geq p(z_s)-\lambda_s\ell+\frac{\alpha}{2}\ell^2=p(z_s)+\ell\Big(\frac{\alpha}{2}\ell-\lambda_s\Big)>p(z_s),
\end{align*}
because $(\alpha/2)\ell\geq(\alpha/2)r'\mu_s^{1/4}>\lambda_s$. So $p$ attains its minimum over $\mathcal{S}'$ at a relative interior point $x_c$ of $\mathcal{S}'$. This is a local, hence global, minimum of the strictly convex function $p$ on the relative interior of $K$, so it is unique. As $p(x_c)\leq p(z_s)$, the same bound gives $\|x_c-z_s\|_{Q_0}\leq(2/\alpha)\lambda_s$, and convexity gives $p(z_s)-p(x_c)\leq\langle\nabla p(z_s),z_s-x_c\rangle\leq\lambda_s\|z_s-x_c\|_{Q_0}\leq(2/\alpha)\lambda_s^2$. If $\lambda_s=0$, then $x_c=z_s$. Otherwise the claim with $r'=C_{\rm r}\lambda_s\mu_s^{-1/4}\leq r$ applies, as $\alpha C_{\rm r}>2$ by~\ref{cond:C2}, and it gives the statement after ``in particular''. Finally, $K$ is bounded by Part~4.

For Part~2, Part~1 applies with $r'=r$, because $\lambda_s\leq c_0\mu_s^{1/4}$ and $c_0<\alpha r/2$ by~\ref{cond:C2}. So $x_c$ exists, lies in $\mathcal{S}(z_s,r)$ and has $\|x_c-z_s\|_{Q_0}\leq(2/\alpha)\lambda_s$. We show by induction on $k$ that $y_k\in\mathcal{S}(z_s,r_{\rm N})$ and $\|y_k-x_c\|_{Q_0}\leq q^k\|z_s-x_c\|_{Q_0}+C_{\rm x}\epsilon_g$, which holds for $k=0$. Suppose that it holds for $k$. As the segment $[x_c,y_k]$ lies in the convex set $\mathcal{S}(z_s,r)$, we have $\nabla p(y_k)=M_k(y_k-x_c)$ with $M_k=\int_0^1\nabla^2p(x_c+\tau(y_k-x_c))\,{\rm d}\tau$, and $\alpha Q_0\preceq M_k\preceq\beta Q_0$ by Lemma~\ref{lem:hybrid_local} (Part~4). In the coordinates $\xi_k:=Q_0^{1/2}(y_k-x_c)$, the step reads
\begin{align*}
    \xi_{k+1}=(I-\beta^{-1}XM')\xi_k-\beta^{-1}Xe'_k.
\end{align*}
Here $X=Q_0^{1/2}\widehat Q^{-1}Q_0^{1/2}=I+E$ has spectrum in $[1/(1+\gamma_Q),1/(1-\gamma_Q)]$, so $\|E\|\leq\gamma_Q/(1-\gamma_Q)$. The matrix $M'=Q_0^{-1/2}M_kQ_0^{-1/2}$ has spectrum in $[\alpha,\beta]$, and $\|e'_k\|_2=\|e_k\|_{Q_0^{-1}}\leq\epsilon_g$ because $y_k\in\mathcal{S}(z_s,r)$. So $\|I-\beta^{-1}XM'\|\leq\|I-\beta^{-1}M'\|+\|\beta^{-1}EM'\|\leq(1-\alpha/\beta)+\gamma_Q/(1-\gamma_Q)=q$, and the error term has norm at most $\epsilon_g/(\beta(1-\gamma_Q))$. Since $\gamma_Q\leq\gamma_0$, condition~\ref{cond:C1} gives $q\leq q_0<1$ and $\beta(1-\gamma_Q)(1-q)C_{\rm x}\geq1$. Hence
\begin{align*}
    \|\xi_{k+1}\|_2\leq q\big(q^k\|\xi_0\|_2+C_{\rm x}\epsilon_g\big)+\frac{\epsilon_g}{\beta(1-\gamma_Q)}\leq q^{k+1}\|\xi_0\|_2+C_{\rm x}\epsilon_g,
\end{align*}
which is the distance bound for $k+1$. Adding the bound on $\|x_c-z_s\|_{Q_0}$ gives $\|y_{k+1}-z_s\|_{Q_0}\leq(4/\alpha)\lambda_s+C_{\rm x}\epsilon_g$. So by Lemma~\ref{lem:hybrid_local} (Part~3), the two hypotheses and~\ref{cond:C2}, every ratio $s_i(y_{k+1})/s_i(z_s)$ is within $\mu_s^{-1/4}((4/\alpha)\lambda_s+C_{\rm x}\epsilon_g)\leq4c_0/\alpha+2C_{\rm x}c_0\leq r_{\rm N}$ of $1$. This completes the induction. Finally, $Q_c(y_k)\succeq\alpha Q_0$ and $0\preceq M_k\preceq\beta Q_0$ give
\begin{align*}
    \lambda(y_k)=\|M_k(y_k-x_c)\|_{Q_c(y_k)^{-1}}\leq\alpha^{-1/2}\beta\,\|y_k-x_c\|_{Q_0}\leq\frac{2\beta}{\alpha^{3/2}}\,q^k\lambda_s+\frac{\beta C_{\rm x}}{\alpha^{1/2}}\,\epsilon_g,
\end{align*}
which is at most $C_\lambda q^k\lambda_s+C_{\rm e}\epsilon_g$ by~\ref{cond:C1}.

For Part~3, $\wt g-\nabla p(y)=-A^\top S_y^{-1}(\wt\sigma-\sigma(y))$. Part~2 of Lemma~\ref{lem:hybrid_local} at $y$ bounds its norm in $Q_c(y)^{-1}$ by $\delta c^{-1/2}$. Since $y\in\mathcal{S}(z_s,r)$, Part~4 of that lemma gives $Q_c(y)\preceq\alpha'Q_0$, so $Q_0^{-1}\preceq\alpha'Q_c(y)^{-1}$.
\end{proof}

\begin{proof}[Proof of Lemma~\ref{lem:hybrid_pert}]
Let $u_i:=H^{-1/2}a_i/s_i(z)$, so that $\sum_iu_iu_i^\top=I_d$ and $\sigma_i=\|u_i\|_2^2$.

\emph{Part~1.} Let $u:=H^{-1/2}a/s_a$, so that $\|u\|_2^2=w'$ and $H_+=H^{1/2}(I+uu^\top)H^{1/2}\preceq(1+w')H$. By the Sherman--Morrison formula, the new row has leverage score $\sigma'$ in $K_+$, and every old row has the score $\sigma_i-\zeta_i$, where $\zeta_i:=(u_i^\top u)^2/(1+w')$. The Cauchy--Schwarz inequality gives $\zeta_i\leq\sigma_iw'/(1+w')$, so $\sigma_i-\zeta_i\geq\sigma_i/(1+w')$ and $\zeta_i\leq w'(\sigma_i-\zeta_i)$. Also $\sum_i\zeta_i=\|u\|_2^2/(1+w')=\sigma'$. Hence
\begin{align*}
    Q_{c,+}(z)\succeq\frac{Q_c}{1+w'}\succeq\frac{(\mu+c)H}{1+w'}\succeq\frac{(\mu+c)H_+}{(1+w')^2},
\end{align*}
which gives the first bound of~(i). The second follows from $\mu\geq\theta_{\rm add}$, $w'\leq w_{\rm add}$ and~\ref{cond:W4}. For~(ii), $\nabla p_+(z)-\nabla p(z)=\sum_i\zeta_ia_i/s_i-(\sigma'+c)a/s_a$. In the norm of $Q_{c,+}(z)^{-1}$, Lemma~\ref{lem:hybrid_local} (Part~2) in $K_+$ bounds the sum by $(\sum_i\zeta_i^2/(\sigma_i-\zeta_i+c))^{1/2}\leq(w'\sigma')^{1/2}\leq w'$. Part~3 of that lemma in $K_+$, for the new row, bounds the last term by $(\sigma'+c)(\sigma'/\mu_{c,+}(z))^{1/2}$. Finally, $Q_{c,+}(z)^{-1}\preceq(1+w')Q_c^{-1}$ gives $\|\nabla p(z)\|_{Q_{c,+}(z)^{-1}}\leq(1+w')^{1/2}\lambda$. For~(iii), $(1+w')^{1/2}\leq1+\epsilon_{\rm add}$ by~\ref{cond:W4}. As $\sigma'\leq w'\leq w_{\rm add}$, $c\leq w/322$ by~\ref{cond:W1} and~(i) holds, the last term of~(ii) is at most $(w_{\rm add}+w/322)(w_{\rm add}/\theta'_{\rm add})^{1/2}$. So the two additive terms of~(ii) sum to at most $C_+w$, by~\ref{cond:C2}.

\emph{Part~2.} The matrix $H_-=H^{1/2}(I-u_ju_j^\top)H^{1/2}$ is positive definite, because $\sigma_j\leq\sigma_{\rm del}<1/C_Q\leq1$ by~\ref{cond:C2}. By the Sherman--Morrison formula, every other row has the score $\sigma_i+\zeta_i$ in $K_-$, where $\zeta_i:=(u_i^\top u_j)^2/(1-\sigma_j)\leq\sigma_i\sigma_j/(1-\sigma_j)$, and $\sum_{i\neq j}\zeta_i=\sigma_j$. Let $\pi_j:=a_ja_j^\top/s_j^2$ with $s_j=s_j(z)$. Then $s_j^{-2}a_j^\top H_-^{-1}a_j=\sigma_j/(1-\sigma_j)$, and
\begin{align*}
    Q_{c,-}(z)\succeq Q_c-(\sigma_j+c)\pi_j,\qquad\pi_j\preceq\sigma_jH,\qquad\pi_j\preceq\frac{\sigma_j}{1-\sigma_j}H_-\qquad\text{and}\qquad Q_c\succeq(\mu+c)H.
\end{align*}
In the first case, $\sigma_j\geq4\nu$ and $\mu\geq(1-\epsilon_{\rm del})\sigma_j-\sigma_j/4=\mu_{\rm del}\sigma_j$, with $\mu_{\rm del}=3/4-\epsilon_{\rm del}$. In the second case, $\sigma_j<4\nu\leq4\nu_{\rm del}c$, and we use $\mu\geq0$. Let $\sigma^\circ:=\sigma_{\rm del}/(1-\sigma_{\rm del})\geq\sigma_j/(1-\sigma_j)$. Below, every inequality between constants is from~\ref{cond:C2}.

(i) We have $(\sigma_j+c)\pi_j\preceq\eta\,Q_c$ with $\eta:=(\sigma_j+c)\sigma_j/(\mu+c)$. Here $\eta\leq\sigma_j/\mu_{\rm del}$ in the first case, as $\mu_{\rm del}\leq1$, and $\eta\leq(1+\sigma_j/c)\sigma_j\leq(1+4\nu_{\rm del})\sigma_j$ in the second. Both are at most $C_Q\sigma_j<1$. Hence $Q_{c,-}(z)\succeq(1-\eta)Q_c\succeq(1-C_Q\sigma_j)Q_c$.

(ii) Let $\varphi(\mu):=\mu+c-(\sigma_j-\mu)_+\sigma_j/(1-\sigma_j)$, which is increasing in $\mu$. By the first and last of the four bounds above,
\begin{align*}
    Q_{c,-}(z)\succeq(\mu+c)H-(\sigma_j+c)\pi_j=(\mu+c)H_--(\sigma_j-\mu)\pi_j\succeq\varphi(\mu)H_-,
\end{align*}
so $\mu_{c,-}(z)\geq\varphi(\mu)$. In the first case, $\varphi(\mu)\geq\varphi(\mu_{\rm del}\sigma_j)\geq(\mu_{\rm del}-(1-\mu_{\rm del})\sigma^\circ)\sigma_j+c\geq c_\mu\sigma_j+c$. In the second case, $\varphi(\mu)\geq\varphi(0)=c-\sigma_j^2/(1-\sigma_j)\geq(1-4\nu_{\rm del}\sigma^\circ)c\geq c_\mu(1+4\nu_{\rm del})c$. Both are at least $c_\mu(\sigma_j+c)$, as $c_\mu\leq1$ and, in the second case, $\sigma_j<4\nu_{\rm del}c$.

(iii) We have $\nabla p_-(z)-\nabla p(z)=(\sigma_j+c)a_j/s_j-\sum_{i\neq j}\zeta_ia_i/s_i$. Since $\zeta_i\leq(\sigma_i+\zeta_i+c)\sigma_j/(1-\sigma_j)$, Lemma~\ref{lem:hybrid_local} (Part~2) in $K_-$ bounds the norm of the sum in $Q_{c,-}(z)^{-1}$ by $(\sum_{i\neq j}\zeta_i^2/(\sigma_i+\zeta_i+c))^{1/2}\leq\sigma_j(1-\sigma_j)^{-1/2}$. Since $Q_{c,-}(z)\succeq\mu_{c,-}(z)H_-$, the first term has norm at most $(\sigma_j+c)(\sigma_j/((1-\sigma_j)\mu_{c,-}(z)))^{1/2}$. By~(ii) this is at most $(c_\mu(1-\sigma_{\rm del}))^{-1/2}(\sigma_j(\sigma_j+c))^{1/2}\leq(c_\mu(1-\sigma_{\rm del}))^{-1/2}(\sigma_j+(\sigma_jc)^{1/2})$. So the two terms sum to at most $J$, by the last two inequalities of~\ref{cond:C2}. Finally,~(i) gives $\|\nabla p(z)\|_{Q_{c,-}(z)^{-1}}\leq(1-C_Q\sigma_j)^{-1/2}\lambda$.
\end{proof}

\paragraph{Validity of the operations.} We show that the tests of Algorithm~\ref{alg:hcpm} make every operation on \textsc{LS}, \textsc{IM}$_H$ and \textsc{IM}$_Q$ well defined and valid, whatever the queries return. Every block starts with $H_K(z)$ positive definite on $W_0$. This holds for boxes, and for the output of a block by what follows. Since $Y^\top a\neq0$, the form $a$ is nonzero on $W_0$, so~\eqref{eq:im_inverse} gives $\wt h>0$, and $\rho$ is defined. The added row has slack $\rho>0$ at $z$, and by~\eqref{eq:im_inverse} its leverage $w':=\rho^{-2}a^\top H_K(z)^{-1}a=w\,a^\top H_K(z)^{-1}a/\wt h$ lies in $[w/(1+\epsilon_{\rm im}),w/(1-\epsilon_{\rm im})]$. So $w'\leq0.01$, by~\ref{cond:C6}. By the deletion test and~\eqref{eq:im_inverse}, a deleted row has leverage at most $\tau_{\rm del}/(1-\epsilon_{\rm im})<0.01$, by~\ref{cond:C6}. So every insertion and deletion meets the condition of Lemma~\ref{lem:jlsw_main}, and $H_K(z)$ stays positive definite on $W_0$. The step test keeps every slack positive. When it passes, the vector $\xi$ with entries $\xi_i:=s_i(y^{(t+1)})/s_i(y^{(t)})-1$ has $\ell_2$ norm at most $\tau_{\rm step}$. The weights of \textsc{LS} are $s_i^{-2}$, so every entry of $\Delta\log\eta$ is $-2\ln(1+\xi_i)$, and the dummy rows keep weight $1$. As $|\ln(1+x)|\leq1.01|x|$ for $|x|\leq\tau_{\rm step}$, each of the $t_{\max}$ weight updates has $\|\Delta\log\eta\|_2\leq2\cdot1.01\cdot\tau_{\rm step}<0.01$, by~\ref{cond:C6}. The two weight tests keep every weight of \textsc{IM}$_Q$ positive, so $\widehat Q$ is positive definite. The row test keeps at most $\bar m$ rows, so the padded instance has at most $\bar m+n=O(n)$ rows.

\paragraph{The fixed minimizer and the cut vector.} We prove the two facts of the paragraph \emph{Cut vectors} in Section~\ref{sec:cpm:method}. First, $x^*$ is an extreme point of $K^*$. Indeed, suppose that $x^*=(u+v)/2$ with $u\neq v$ in $K^*$, and let $j$ be the first coordinate where $u$ and $v$ differ. Both agree with $x^*$ before $j$, and one of $u_j$ and $v_j$ exceeds $x^*_j$. That point of $K^*$ is lexicographically larger than $x^*$, a contradiction. As the extreme points of $K^*$ are integral, $x^*$ is an integral minimizer of $f$, and $x^*\in K^*\subseteq B_\infty(R)$.

Second, let $x^*\in W$, let $z\in W$, and let $a$ be the cut vector at $z$. If $a$ is the answer of $\SO$, then~\eqref{eq:cut_xstar} is the property of the oracle. Otherwise $z$ minimizes $f$. Indeed, either $\SO$ answered YES, or it returned a vector $a'$ with $Y^\top a'=0$, and then $a'^\top x^*=a'^\top z$ because $x^*-z\in W_0$, which rules out strict inequality. Now $a=e_i$, and every coordinate $j<i$ is constant on $W$, so $z_j=x^*_j$. Since $z\in K^*$ and $x^*$ is the lexicographically largest point of $K^*$, we get $x^*_i\geq z_i$, which is~\eqref{eq:cut_xstar}. Finally, the cut $a^\top x\geq a^\top z-\rho$ has slack $a^\top(x^*-z)+\rho\geq\rho$ at $x^*$.

\subsection{The Invariant of One Block}\label{sec:r3_invariant}

This section proves Lemma~\ref{lem:hybrid_inv}, the invariant of a good block, in the notation for one block from the start of Section~\ref{sec:r3_proofs}. Its key step bounds the error of the matrix $\widehat Q$ of the Newton steps, using that every weight of $Q_c$ is at least $c$.

\begin{lemma}[Invariant of the hybrid method]\label{lem:hybrid_inv}
Consider a good block on a polytope $K$, started at a point $z_0$, and let $\delta\leq\delta_0$ be the largest $\ell_2$ error of the outputs of \textsc{\textup{LS}} during the block. Then the following hold for every step $k$ of the block.
\begin{enumerate}
    \item For every $0\leq k\leq T$, the point $z_k$ satisfies $\lambda_{K_k}(z_k)\leq\lambda^*$, and $K_k$ has at most $\bar m$ rows. In particular this holds for the output point $z_T$.
    \item If step $k$ adds a row, then $K_k$ has at most $\bar m-1$ rows, $\sigma_i(z_k)\geq\theta-\delta_0$ for every row $i$, the new row has leverage $w':=a^\top H_{K_k}(z_k)^{-1}a/\rho^2\in[w/(1+\epsilon_{\rm im}),w/(1-\epsilon_{\rm im})]$ at $z_k$, and the Newton steps start with $\lambda_s\leq C_+w+(1+\epsilon_{\rm add})\lambda^*$ and $\mu_s\geq\theta'_{\rm add}$.
    \item If step $k$ deletes row $j$, then $\sigma_j:=\sigma_j(z_k)<\theta+\delta_0$, $\sigma_i(z_k)\geq\sigma_j-2\delta$ for every row $i$, and the Newton steps start with $\lambda_s\leq(1+\epsilon_{\rm add})\lambda^*+J(\sigma_j)$ and $\mu_s\geq c_\mu(\sigma_j+c)$, where $J(\sigma_j)=C_-\sigma_j+C'_-(\sigma_jc)^{1/2}\leq w_J$.
    \item In both cases $\lambda_s\leq C_{\rm s}w$ and $\mu_s\geq c$. The matrix $\widehat Q$ fixed at the start of the Newton steps satisfies $(1-\gamma_Q)Q_0\preceq\widehat Q\preceq(1+\gamma_Q)Q_0$ with $\gamma_Q\leq\gamma_1$, and the Newton steps satisfy the hypotheses of Lemma~\ref{lem:hybrid_loc} (Part~2), with the bound $\epsilon_g$ on the vectors $e_l$ that Part~3 of that lemma provides.
    \item No step of the block aborts. Every add is preceded by at most $\bar m-1$ rows, the estimate $s_j(z_k)^{-2}a_j^\top\widehat H^{-1}a_j$ of the leverage score of a deleted row is at most $(1+\epsilon_{\rm im})(\theta+\delta_0)$, and every output of a query has $\min_i(\wt\sigma_i+c)\geq c-\delta>0$. Every Newton step $h=y_{l+1}-y_l$ has $\|h\|_{H(y_l)}\leq h_{\max}$ and passes the test of Algorithm~\ref{alg:hcpm} on its $t_{\max}$ pieces, so \textsc{\textup{LS}} receives $t_{\max}$ weight updates $\eta\mapsto\eta'$ with $\|\log\eta'-\log\eta\|_2\leq0.01$ each.
\end{enumerate}
The hypothesis $\lambda_K(z_0)\leq\lambda^*$ holds when $z_0$ is the output of a previous block (Part~1) and when $z_0$ is the center of a box, where $\lambda=0$ (Lemma~\ref{lem:reduction_geometry}, Part~3, and Lemma~\ref{lem:box_margin}).
\end{lemma}

\begin{proof}
We use induction over the steps. Suppose that Part~1 holds at step $k$, that $H_{K_k}$ is positive definite on $W_0$ at the relative interior point $z_k$ of $K_k$, and that \textsc{IM}$_H$ sits at $z_k$ with the rows of $K_k$. For $k=0$ this holds by the hypothesis and the initialization. The paragraph \emph{No step aborts} below shows, from facts about step $k$ alone, that no test of step $k$ fires. The queries at $z_k$ have $|\wt\sigma_i-\sigma_i(z_k)|\leq\delta$ for every row $i$.

\emph{Add.} Every estimate is at least $\theta$, so every $\sigma_i(z_k)\geq\theta-\delta_0$. The scores at $z_k$ sum to $d\leq n$, so $K_k$ has $m\leq n/(\theta-\delta_0)<\bar m$ rows, that is, at most $\bar m-1$. By the paragraph \emph{Validity of the operations} in Section~\ref{sec:r3_local}, the new row has slack $\rho>0$ at $z_k$ and leverage $w'\in[w/(1+\epsilon_{\rm im}),w/(1-\epsilon_{\rm im})]$. So $z_k$ is a relative interior point of $K_{k+1}$, and $H_{K_{k+1}}(z_k)\succeq H_{K_k}(z_k)$ is positive definite. As $Q\succeq(\min_i\sigma_i)H$ term by term, $\mu(z_k)\geq\theta-\delta_0\geq\theta_{\rm add}$ by~\ref{cond:C7}, and $w'\leq w_{\rm add}$ by~\ref{cond:C2}. So Lemma~\ref{lem:hybrid_pert} (Part~1) gives the bounds on $\mu_s$ and $\lambda_s$, which proves Part~2.

\emph{Delete.} Let $j$ be the deleted row. Every row $i$ has $\sigma_i(z_k)\geq\wt\sigma_i-\delta\geq\wt\sigma_j-\delta\geq\sigma_j-2\delta$. So $\mu(z_k)\geq\sigma_j-2\delta\geq(1-\epsilon_{\rm del})\sigma_j-\nu$ with $\nu:=2\delta\leq\nu_{\rm del}c$ by~\ref{cond:C7}, and $\sigma_j\leq\wt\sigma_j+\delta<\theta+\delta_0\leq\theta_{\rm del}\leq\sigma_{\rm del}$ by~\ref{cond:W5}, as $\kappa(r_{\rm f})\geq1$, and~\ref{cond:W4}. So Lemma~\ref{lem:hybrid_pert} (Part~2) applies. It shows that $H_{K_{k+1}}(z_k)$ is positive definite on $W_0$, and it gives $\mu_s\geq c_\mu(\sigma_j+c)$ and $\lambda_s\leq(1-C_Q\sigma_j)^{-1/2}\lambda^*+J(\sigma_j)$, where $(1-C_Q\sigma_j)^{-1/2}\leq1+\epsilon_{\rm add}$ by~\ref{cond:W4}. As $J$ is increasing, $J(\sigma_j)\leq J(\theta+\delta_0)\leq w_J$ by~\ref{cond:C5}. This proves Part~3.

\emph{The Newton steps.} In both cases $\lambda_s\leq C_{\rm s}w$ by~\ref{cond:C5}, and $\mu_s=\mu(z_s)+c\geq c$. The estimates $\wt\sigma$ queried at $z_s$ have error at most $\delta<c$ by~\ref{cond:C6}, so every $\wt\sigma_i+c$ is positive. Let $\wt Q:=\sum_i(\wt\sigma_i+c)a_ia_i^\top/s_i(z_s)^2$. As $Q_0\succeq cH(z_s)$, every $v\in W_0$ has
\begin{align*}
    |v^\top(\wt Q-Q_0)v|=\Big|\sum_i\big(\wt\sigma_i-\sigma_i(z_s)\big)\frac{(a_i^\top v)^2}{s_i(z_s)^2}\Big|\leq\delta\,v^\top H(z_s)v\leq\frac{\delta}{c}\,v^\top Q_0v .
\end{align*}
\textsc{IM}$_Q$ keeps weights $\wt\eta_i$ with $(1-\epsilon_{\rm im})\wt\eta_i\leq(\wt\sigma_i+c)/s_i(z_s)^2\leq(1+\epsilon_{\rm im})\wt\eta_i$ (paragraph \emph{Inverse maintenance} in Section~\ref{sec:cpm:method}). So $(1+\epsilon_{\rm im})^{-1}\wt Q\preceq\widehat Q\preceq(1-\epsilon_{\rm im})^{-1}\wt Q$ on the block of the real rows. Let
\begin{align*}
    \gamma_Q:=\Big(1+\frac{\epsilon_{\rm im}}{1-\epsilon_{\rm im}}\Big)\Big(1+\frac{\delta}{c}\Big)-1=\frac{1+\delta/c}{1-\epsilon_{\rm im}}-1,
\end{align*}
so that $\gamma_Q\geq\epsilon_{\rm im}+\delta/c$, and $\gamma_Q\leq\gamma_1\leq\gamma_0$ by~\ref{cond:C1}, as $\delta\leq\delta_0$. With the bound $(1-\delta/c)Q_0\preceq\widetilde Q\preceq(1+\delta/c)Q_0$ above, the upper factor is $(1-\epsilon_{\rm im})^{-1}(1+\delta/c)=1+\gamma_Q$. The lower one is $(1+\epsilon_{\rm im})^{-1}(1-\delta/c)\geq1-\epsilon_{\rm im}-\delta/c\geq1-\gamma_Q$. So $(1-\gamma_Q)Q_0\preceq\widehat Q\preceq(1+\gamma_Q)Q_0$.

The estimates used at $y_l$ are queried after the weights of \textsc{LS} are set to $s_i(y_l)^{-2}$, so they have error at most $\delta\leq\delta_0$ at $y_l$. So Lemma~\ref{lem:hybrid_loc} (Part~3) gives $\|e_l\|_{Q_0^{-1}}\leq\epsilon_g$ whenever $y_l\in\mathcal{S}(z_s,r)$. After an add, $\lambda_s\leq C_{\rm s}w\leq c_0(\theta'_{\rm add})^{1/4}\leq c_0\mu_s^{1/4}$ by~\ref{cond:W1}. After a delete, $\sigma_j\leq\theta_{\rm del}$, and~\ref{cond:W1} gives $\lambda_s\leq c_0(c_\mu(\sigma_j+c))^{1/4}\leq c_0\mu_s^{1/4}$. Also $\epsilon_g\leq2c_0c^{1/4}\leq2c_0\mu_s^{1/4}$ by~\ref{cond:W1}. So Lemma~\ref{lem:hybrid_loc} (Part~2) applies. It keeps every $y_l$ in $\mathcal{S}(z_s,r_{\rm N})$, and with~\ref{cond:C5} it gives
\begin{align*}
    \lambda_{K_{k+1}}(z_{k+1})\leq C_\lambda q_0^N\lambda_s+C_{\rm e}\epsilon_g\leq C_\lambda C_{\rm s}w\,q_0^N+C_{\rm e}\epsilon_g\leq\frac{\lambda^*}{2}+\frac{\lambda^*}{2}=\lambda^* .
\end{align*}
So $z_{k+1}$ is a relative interior point of $K_{k+1}$, at which $H_{K_{k+1}}$ is positive definite by Lemma~\ref{lem:hybrid_local} (Part~4). The iteration ends by moving \textsc{IM}$_H$ to $z_{k+1}$, and $K_{k+1}$ has at most $\bar m$ rows. This proves Part~4 and completes the induction for Parts~1 to~4.

\emph{No step aborts.} The add case shows that the row test passes. Every estimate has $\wt\sigma_i+c\geq c-\delta>0$, so the weight tests pass. A deleted row has $\sigma_j<\theta+\delta_0$, so by~\eqref{eq:im_inverse} and~\ref{cond:C6} its estimate $s_j(z_k)^{-2}a_j^\top\widehat H^{-1}a_j$ is at most $(1+\epsilon_{\rm im})(\theta+\delta_0)\leq\tau_{\rm del}$. For a Newton step $h=y_{l+1}-y_l$, Part~2 of Lemma~\ref{lem:hybrid_loc}, with $q^l\leq1$ and $x_c:=x_c(K_{k+1})$, gives
\begin{align*}
    \|h\|_{Q_0}\leq\|y_{l+1}-x_c\|_{Q_0}+\|y_l-x_c\|_{Q_0}\leq2\Big(\frac{2}{\alpha}\lambda_s+C_{\rm x}\epsilon_g\Big).
\end{align*}
As $y_l\in\mathcal{S}(z_s,r)$, Lemma~\ref{lem:hybrid_local} (Part~4) gives $Q_c(y_l)\preceq\alpha'Q_0$ and $\mu_c(y_l)\geq\alpha_\mu\mu_s$. With $Q_c(y_l)\succeq\mu_c(y_l)H(y_l)$, this yields
\begin{align*}
    \|h\|_{H(y_l)}\leq\Big(\frac{\alpha'}{\alpha_\mu\mu_s}\Big)^{1/2}\|h\|_{Q_0}\leq2\Big(\frac{\alpha'}{\alpha_\mu\mu_s}\Big)^{1/2}\Big(\frac{2}{\alpha}\lambda_s+C_{\rm x}\epsilon_g\Big)\leq h_{\max},
\end{align*}
by~\ref{cond:W3} and the bounds of Parts~2 and~3. The segment $[y_l,y_{l+1}]$ lies in the convex set $\mathcal{S}(z_s,r_{\rm N})$, so its slacks are positive. Each of its points $y$ has $s_i(y)\geq\frac{1-r_{\rm N}}{1+r_{\rm N}}s_i(y_l)$, hence $\|v\|_{H(y)}\leq\frac{1+r_{\rm N}}{1-r_{\rm N}}\|v\|_{H(y_l)}$ for $v\in W_0$. So for each of the $t_{\max}$ pieces of the step, the vector with entries $s_i(y^{(t+1)})/s_i(y^{(t)})-1=a_i^\top(y^{(t+1)}-y^{(t)})/s_i(y^{(t)})$ has $\ell_2$ norm
\begin{align*}
    \|y^{(t+1)}-y^{(t)}\|_{H(y^{(t)})}\leq\frac{1+r_{\rm N}}{1-r_{\rm N}}\cdot\frac{h_{\max}}{t_{\max}}\leq\tau_{\rm step},
\end{align*}
by~\ref{cond:C6}. So the step passes the test, and the paragraph \emph{Validity of the operations} bounds each weight update by $0.01$.
\end{proof}

\subsection{Proofs for One Block}\label{sec:r3_block}

This section proves the lemmas of Sections~\ref{sec:cpm:hybrid_progress} and~\ref{sec:cpm:main_lemma}, after four lemmas that they use.

\begin{lemma}[Near the hybrid center]\label{lem:near_center}
Let $z$ be a relative interior point of $K$ at which $H(z)$ is positive definite on $W_0$, and suppose that $\lambda_K(z)\leq\lambda^*$. Then $x_c:=x_c(K)$ exists and lies in $\mathcal{S}(z,r_{\rm f})$, where $r_{\rm f}=C_{\rm r}\lambda^*c^{-1/4}$. Moreover, $(1-\epsilon_{\rm f})H(x_c)\preceq H(z)\preceq(1+\epsilon_{\rm f})H(x_c)$ on $W_0$, and $|F_K(z)-F_K(x_c)|\leq\epsilon_{\rm f}d$.
\end{lemma}

\begin{proof}
Since $\mu_c(z)\geq c$, we have $(\alpha/2)\,r_{\rm f}\,\mu_c(z)^{1/4}\geq(\alpha C_{\rm r}/2)\lambda^*>\lambda^*\geq\lambda_K(z)$ by~\ref{cond:C2}, and $r_{\rm f}\leq r$ by~\ref{cond:W1}. So Lemma~\ref{lem:hybrid_loc} (Part~1) with $r'=r_{\rm f}$ shows that $x_c$ exists and lies in $\mathcal{S}(z,r_{\rm f})$. So every ratio $s_i(z)/s_i(x_c)$ lies in $[(1+r_{\rm f})^{-1},(1-r_{\rm f})^{-1}]$, and $(1-r_{\rm f})^2H(x_c)\preceq H(z)\preceq(1+r_{\rm f})^2H(x_c)$, which gives the bounds on $H(z)$ by~\ref{cond:W5}. Taking determinants on the $d$-dimensional space $W_0$ gives $|F_K(z)-F_K(x_c)|\leq d\ln(1/(1-r_{\rm f}))\leq\epsilon_{\rm f}d$, again by~\ref{cond:W5}.
\end{proof}

The next lemma collects the facts about one step of a good block that do not involve the point $x^*$. Lemmas~\ref{lem:hybrid_step} and~\ref{lem:block_stability} use them.

\begin{lemma}[Hybrid centers in a good block]\label{lem:good_centers}
Consider a good block, with the notation for one block, and a step $k$ of it. Then $x_k=x_c(K_k)$ exists and lies in $\mathcal{S}(z_k,r_{\rm f})$, the hybrid center $x_c(K_{k+1})$ exists and lies in $\mathcal{S}_{K_{k+1}}(x_k,r)$, and the decrease ${\rm rec}:=p_{K_{k+1}}(x_k)-p_{K_{k+1}}(x_c(K_{k+1}))$ obeys the following bounds.
\begin{enumerate}
    \item If step $k$ adds the row $a^\top x\geq a^\top z_k-\rho$, then $|s_a(x_k)-\rho|\leq\epsilon_{\rm cut}\rho$, so $x_k$ lies in the relative interior of $K_{k+1}$. Moreover ${\rm rec}\leq(2/\alpha)(C_+w)^2$, and $w_x:=s_a(x_k)^{-2}a^\top H_{K_k}(x_k)^{-1}a$ satisfies
    \begin{align*}
        \frac{(1-r_{\rm f})^2w}{(1+\epsilon_{\rm im})(1+\epsilon_{\rm cut})^2}\leq w_x\leq w_+.
    \end{align*}
    \item If step $k$ deletes row $j$, then $\sigma_j(x_k)\leq\theta_{\rm del}$ and ${\rm rec}\leq(2/\alpha)J(\sigma_j(x_k))^2$.
\end{enumerate}
\end{lemma}

\begin{proof}
By Lemma~\ref{lem:hybrid_inv} (Part~1), $\lambda_{K_k}(z_k)\leq\lambda^*$. So Lemma~\ref{lem:near_center} shows that $x_k$ exists and lies in $\mathcal{S}(z_k,r_{\rm f})$, where $r_{\rm f}\leq r$ by~\ref{cond:W1}, and Lemma~\ref{lem:hybrid_local} (Part~4) with $r'=r_{\rm f}$ compares the scores, $H$ and $\mu$ at $x_k$ with those at $z_k$. Also $\lambda_{K_k}(x_k)=0$. In both cases below, Lemma~\ref{lem:hybrid_pert} at $x_k$ bounds the decrement of $x_k$ in $K_{k+1}$ by some $\lambda'$, and $\mu_c(x_k)$ in $K_{k+1}$ from below by some $\mu'$, with $\lambda'\leq c_0\mu'^{1/4}<(\alpha r/2)\mu'^{1/4}$ by~\ref{cond:C2}. Then Lemma~\ref{lem:hybrid_loc} (Part~1) at $x_k$ in $K_{k+1}$, with $r'=r$, puts $x_c(K_{k+1})$ in $\mathcal{S}_{K_{k+1}}(x_k,r)$ and gives ${\rm rec}\leq(2/\alpha)\lambda'^2$.

\emph{Add.} As in the proof of Lemma~\ref{lem:near_center}, Lemma~\ref{lem:hybrid_loc} (Part~1) at $z_k$ in $K_k$ gives $\|x_k-z_k\|_{Q_c(z_k)}\leq(2/\alpha)\lambda^*$. By Lemma~\ref{lem:hybrid_inv} (Part~2), $Q_c(z_k)\succeq(\theta-\delta_0)H(z_k)$ and $a^\top H(z_k)^{-1}a=w'\rho^2\leq w\rho^2/(1-\epsilon_{\rm im})$. As $s_a(z_k)=\rho$, the Cauchy--Schwarz inequality on $W_0$ and~\ref{cond:W5} give
\begin{align*}
    |s_a(x_k)-\rho|=|a^\top(x_k-z_k)|&\leq\|a\|_{H(z_k)^{-1}}\|x_k-z_k\|_{H(z_k)}\\
    &\leq\frac{2\lambda^*}{\alpha}\Big(\frac{w}{(1-\epsilon_{\rm im})(\theta-\delta_0)}\Big)^{1/2}\rho\leq\epsilon_{\rm cut}\,\rho .
\end{align*}
Lemma~\ref{lem:hybrid_local} (Part~4) also gives $(1-r_{\rm f})^2H(z_k)^{-1}\preceq H(x_k)^{-1}\preceq(1+r_{\rm f})^2H(z_k)^{-1}$. With the bounds on $w'$ and on $s_a(x_k)$, this gives the bounds on $w_x$, by~\ref{cond:W5}. Also $\mu(x_k)\geq((1-r_{\rm f})/(1+r_{\rm f}))^4(\theta-\delta_0)\geq\theta_{\rm add}$ by~\ref{cond:W5}, and $w_x\leq w_+\leq w_{\rm add}$ by~\ref{cond:C2}. So Lemma~\ref{lem:hybrid_pert} (Part~1) applies at $x_k$, with $\mu'=\theta'_{\rm add}$ and $\lambda'=C_+w\leq C_{\rm s}w\leq c_0(\theta'_{\rm add})^{1/4}$, by~\ref{cond:C5} and~\ref{cond:W1}.

\emph{Delete.} By Lemma~\ref{lem:hybrid_inv} (Part~3), $\sigma_j(z_k)<\theta+\delta_0$ and $\sigma_i(z_k)\geq\sigma_j(z_k)-2\delta_0$ for every row $i$. The scores at $x_k$ are within the factor $\kappa(r_{\rm f})$ of those at $z_k$, and $\mu\geq\min_i\sigma_i$. So~\ref{cond:W5} gives
\begin{align*}
    \mu(x_k)\geq\kappa(r_{\rm f})^{-1}\big(\sigma_j(z_k)-2\delta_0\big)\geq\kappa(r_{\rm f})^{-2}\sigma_j(x_k)-2\delta_0\geq(1-\epsilon_{\rm del})\,\sigma_j(x_k)-2\delta_0,
\end{align*}
where $2\delta_0\leq\nu_{\rm del}c$ by~\ref{cond:C7}. Also $\sigma_j:=\sigma_j(x_k)\leq\kappa(r_{\rm f})(\theta+\delta_0)\leq\theta_{\rm del}\leq\sigma_{\rm del}$, by~\ref{cond:W5} and~\ref{cond:W4}. So Lemma~\ref{lem:hybrid_pert} (Part~2) applies at $x_k$ with $\nu=2\delta_0$, with $\mu'=c_\mu(\sigma_j+c)$ and $\lambda'=J(\sigma_j)\leq c_0\mu'^{1/4}$ by~\ref{cond:W1}.
\end{proof}

The next lemma treats the initial box and the boxes of reduction iterations. It allows any radius $R_0$ of the reduction ellipsoid, since Algorithm~\ref{alg:ours} uses $R_0=R'$ and Corollary~\ref{cor:exact_score} uses $R_0=R'_{\rm V}$.

\begin{lemma}[Box margins]\label{lem:box_margin}
Let $x^*\in B_\infty(R)$ be the fixed integral minimizer, and let $R_0>0$.
\begin{enumerate}
    \item (Initial box.) Let $K_0=B_\infty(2R)=\{x\in\R^n:-2R\leq x_k\leq2R,\ k\in[n]\}$. Then $0=x_c(K_0)$, $\lambda_{K_0}(0)=0$ and $H_{K_0}(0)=I/(2R^2)$. Every row of $K_0$ has slack $2R$ at $0$ and slack in $[R,3R]$ at $x^*$.
    \item (Reduction box.) Let $(K,x_K,H_K)$ with $H_K=H_K(x_K)$ satisfy $x^*\in K\subseteq R_0\cdot E(x_K,H_K)$. Let $P$ be a hyperplane with $x^*\in P$ and $W\not\subseteq P$, and let $E(\bar z,A):=2R_0\cdot E(x_K,H_K)\cap P$, an ellipsoid of full dimension in $W\cap P$ with center $\bar z$. Let $\mathcal{B}=\{\bar z+Gt:t\in[-1,1]^{d'}\}$, with $d'=\dim(W\cap P)$ and $G=YJ^{-\top}$, be the box of Lemma~\ref{lem:reduction_geometry} (Part~3) for $(\bar z,A)$. Then $x^*=\bar z+Gt^*$ with $\|t^*\|_2\leq1/2$. Hence $x^*\in \mathcal{B}$, and every row of $\mathcal{B}$ has slack $1$ at $\bar z$ and slack in $[1/2,3/2]$ at $x^*$. Moreover $\bar z=x_c(\mathcal{B})$ and $\lambda_{\mathcal{B}}(\bar z)=0$. If $d'=0$, then $G$ has no columns and $\mathcal{B}=\{\bar z\}=W\cap P=\{x^*\}$, and these claims hold with the usual conventions for empty matrices and sums.
    \item (The term $D$ of a box.) For both kinds of boxes, with $d'$ the dimension of the box, $0\leq D\leq2d'\ln(2\bar\kappa)$ and $c\,D\leq\epsilon_Dd'\leq\epsilon_{\rm f}n$.
\end{enumerate}
\end{lemma}

\begin{proof}
For Part~1, the rows $x_k+2R\geq0$ and $2R-x_k\geq0$ of $K_0$ have slack $2R$ at $0$ and slack in $[R,3R]$ at $x^*\in B_\infty(R)$. The box $K_0$ is the box of Lemma~\ref{lem:reduction_geometry} (Part~3) with $\bar z=0$, $A=I/(4R^2)$ and $Y=I$, so that $J=I/(2R)$ and $G=2R\,I$. Its rows are $2R$ times those of that lemma, and $H$, the scores, $p$ up to an additive constant, $x_c$ and $\lambda$ do not change under positive scaling of rows. That lemma gives $0=x_c(K_0)$, $\lambda_{K_0}(0)=0$ and $H_{K_0}(0)=2A=I/(2R^2)$.

For Part~2, let $\bar z$ be the point of $W\cap P$ that minimizes $\|y-x_K\|_{H_K}$, and let $h:=\|\bar z-x_K\|_{H_K}$. For $y\in W\cap P$, the vector $y-\bar z$ is $H_K$-orthogonal to $\bar z-x_K$, so $\|y-x_K\|_{H_K}^2=h^2+\|y-\bar z\|_{H_K}^2$. As $x^*\in K\cap P\subseteq R_0\cdot E(x_K,H_K)$, this gives $\|x^*-\bar z\|_{H_K}^2\leq R_0^2-h^2$, and
\begin{align*}
    2R_0\cdot E(x_K,H_K)\cap P=\{y\in W\cap P:\|y-\bar z\|_{H_K}^2\leq4R_0^2-h^2\}.
\end{align*}
So this ellipsoid has center $\bar z$, and $A$ is the restriction of $H_K/(4R_0^2-h^2)$ to the linear part of $W\cap P$. Hence
\begin{align*}
    (x^*-\bar z)^\top A(x^*-\bar z)=\frac{\|x^*-\bar z\|_{H_K}^2}{4R_0^2-h^2}\leq\frac{R_0^2-h^2}{4R_0^2-h^2}\leq\frac14 .
\end{align*}
By the proof of Lemma~\ref{lem:reduction_geometry} (Part~3), $y^\top Ay=\|t\|_2^2$ for $y=Gt$, so $x^*=\bar z+Gt^*$ with $\|t^*\|_2\leq1/2$. The rows of $\mathcal{B}$ are $1\pm g_k(x-\bar z)\geq0$, where $(g_1,\ldots,g_{d'})=G^{-1}$, so their slacks are $1$ at $\bar z$ and $1\pm t^*_k\in[1/2,3/2]$ at $x^*$. Lemma~\ref{lem:reduction_geometry} (Part~3) also gives $\bar z=x_c(\mathcal{B})$ and $\lambda_{\mathcal{B}}(\bar z)=0$.

For Part~3, the box has $2d'$ rows, and by Parts~1 and~2 the slack of every row at $x^*$ is between $1/2$ and $3/2$ times its slack at the hybrid center. So every term of $D$ lies in $[\ln(2\bar\kappa/3),\ln(2\bar\kappa)]$, where $\bar\kappa\geq3/2$. Hence $0\leq D\leq2d'\ln(2\bar\kappa)$, and $c\,D\leq d'\cdot2c\ln(2\bar\kappa)\leq\epsilon_Dd'\leq\epsilon_{\rm f}n$ by~\ref{cond:C3}, since $d'\leq n$.
\end{proof}

\begin{lemma}[Sandwich for the hybrid method]\label{lem:sandwich}
Let $K$ have $m$ rows.
\begin{enumerate}
    \item For every relative interior point $x$ of $K$, $E(x,H(x))\subseteq K$.
    \item If $x_c=x_c(K)$ exists, then every $y\in K$ satisfies $\|y-x_c\|_{H(x_c)}\leq(\tau^2+c\tau)^{1/2}/c$, where $\tau:=d+cm$. If $m\leq\bar m$, then $\tau\leq C_\tau n$ and this radius is at most $\bar R$.
    \item If $m\leq\bar m$ and $x_K$ is a relative interior point of $K$ with $\lambda_K(x_K)\leq\lambda^*$, then $x_c(K)$ exists, $K\subseteq R'\cdot E(x_K,H(x_K))$, $(1-\epsilon_{\rm f})H(x_c(K))\preceq H(x_K)\preceq(1+\epsilon_{\rm f})H(x_c(K))$ and $|F_K(x_K)-F_K(x_c(K))|\leq\epsilon_{\rm f}d$. If $K$ is the initial box with $x_K=0$, or a box made by a reduction iteration with $x_K=\bar z$ as in Lemma~\ref{lem:box_margin}, and $H_K=H(x_K)$, then $K\subseteq\sqrt{2d}\cdot E(x_K,H_K)\subseteq R'\cdot E(x_K,H_K)$.
\end{enumerate}
Consequently, every triple $(K,x_K,H_K)$ on which Algorithm~\ref{alg:ours} calls \textsc{\textup{ApproxSV}} satisfies $K\subseteq R'\cdot E(x_K,H_K)$, provided that every block outputs a point with $\lambda\leq\lambda^*$. The row bound $m\leq\bar m$ holds in every state, because Algorithm~\ref{alg:hcpm} aborts instead of adding a row to a polytope with $\bar m$ rows.
\end{lemma}

At the hybrid center and without regularization, inequality~(6.7) in Lemma~19 of~\cite{lsw15} gives $\|y-x_c\|_{Q_c(x_c)}\leq(12cm+6n)/\sqrt c$. Since $Q_c(x_c)\succeq cH(x_c)$, this is the radius $O(n/c)$ of Part~2 when $m\leq\bar m$. We prove it in our affine setting, which gives the constant of $\bar R$.

\begin{proof}
For Part~1, let $y\in E(x,H(x))$. Then $y-x\in W_0$ and $\sum_i(a_i^\top(y-x))^2/s_i(x)^2\leq1$, so $|a_i^\top(y-x)|\leq s_i(x)$ and $s_i(y)\geq0$ for every $i$.

For Part~2, let $y\in K$, and let $v_i:=\sigma_i(x_c)+c\geq c$ and $u_i:=s_i(y)/s_i(x_c)-1\geq-1$ as in the proof of Lemma~\ref{lem:hybrid_loc} (Part~4). By~\eqref{eq:center_balance}, $\sum_{u_i>0}v_iu_i=\sum_{u_i<0}v_i|u_i|\leq\sum_iv_i=\tau$, so $u_i\leq\tau/c$ whenever $u_i>0$. Hence
\begin{align*}
    \|y-x_c\|_{H(x_c)}^2=\sum_iu_i^2\leq\frac1c\sum_iv_iu_i^2\leq\frac1c\Big(\frac{\tau}{c}\sum_{u_i>0}v_iu_i+\sum_{u_i<0}v_i\Big)\leq\frac{\tau^2+c\tau}{c^2}.
\end{align*}
If $m\leq\bar m$, then $\tau\leq n+c\bar m\leq C_\tau n$ by~\ref{cond:C4}. We may assume $d\geq1$, since otherwise $K=\{x_c\}$. Then $\tau\geq1$, and the radius $(\tau/c)(1+c/\tau)^{1/2}$ is at most $C_\tau(1+c)^{1/2}n/c\leq C_Rn/c=\bar R$, by~\ref{cond:C4}.

For Part~3, Lemma~\ref{lem:near_center} at $x_K$ gives $x_c=x_c(K)\in\mathcal{S}(x_K,r_{\rm f})$ and the bounds on $H(x_K)$ and on $F_K(x_K)$. Every row has $s_i(x_c)/s_i(x_K)\leq1+r_{\rm f}$, so $\|v\|_{H(x_K)}\leq(1+r_{\rm f})\|v\|_{H(x_c)}$ for $v\in W_0$. Every row also has $|s_i(x_K)/s_i(x_c)-1|\leq r_{\rm f}/(1-r_{\rm f})$, so $\|x_K-x_c\|_{H(x_c)}\leq\bar m^{1/2}r_{\rm f}/(1-r_{\rm f})$. For $y\in K$, Part~2 and~\ref{cond:C4} then give
\begin{align*}
    \|y-x_K\|_{H(x_K)}\leq(1+r_{\rm f})\Big(\bar R+\frac{\bar m^{1/2}r_{\rm f}}{1-r_{\rm f}}\Big)\leq C_{R'}\big(\bar R+\bar m^{1/2}\big)=R' .
\end{align*}
For a box made by a reduction iteration, with center $\bar z$ and matrix $A$, Lemma~\ref{lem:reduction_geometry} (Part~3) gives $\mathcal{B}\subseteq\sqrt{d'}\cdot E(\bar z,A)$ and $H_{\mathcal{B}}(\bar z)=2A$, so $\mathcal{B}\subseteq\sqrt{2d'}\cdot E(\bar z,H_{\mathcal{B}}(\bar z))$. The initial box is the case $\bar z=0$ and $A=I/(4R^2)$. Finally $\sqrt{2d}\leq\sqrt{2n}\leq R'$ by~\ref{cond:C4}.

For the consequence, \textsc{ApproxSV} is called at a box, which has $2d\leq\bar m$ rows by~\ref{cond:C4}, or at the output of a block. The output of a block has at most $\bar m$ rows by the abort rule of Algorithm~\ref{alg:hcpm}, and its point has $\lambda\leq\lambda^*$ by assumption. So Part~3 covers both cases.
\end{proof}

\begin{proof}[Proof of Lemma~\ref{lem:hybrid_step}]
As $x^*\in K\subseteq W$, every cut $a^\top x\geq a^\top z-\rho$ added during the block has slack at least $\rho>0$ at $x^*$, by~\eqref{eq:cut_xstar}. Deletions only remove rows, so every $K_k$ contains $x^*$ with positive slack in every row, as Lemma~\ref{lem:hybrid_identity} needs. By Lemma~\ref{lem:hybrid_inv} (Part~1), $K_k$ has at most $\bar m$ rows. Lemma~\ref{lem:good_centers} gives the claims on the hybrid centers, on $w_x$ and on $\sigma_j(x_k)$, and we insert its bounds on ${\rm rec}$ into the identity of Lemma~\ref{lem:hybrid_identity}.

\emph{Add.} The last term of the identity is at least $-c\ln((1+\epsilon_{\rm cut})\bar\kappa)\geq-w/10$, because $s_a(x_k)\leq(1+\epsilon_{\rm cut})\rho\leq(1+\epsilon_{\rm cut})s_a(x^*)$ and by~\ref{cond:C3}. With the lower bound on $w_x$ and the bound on ${\rm rec}$, condition~\ref{cond:W2} gives
\begin{align*}
    \mathcal{U}(K_{k+1})-\mathcal{U}(K_k)\geq\frac12\ln\Big(1+\frac{(1-r_{\rm f})^2w}{(1+\epsilon_{\rm im})(1+\epsilon_{\rm cut})^2}\Big)-\frac{2}{\alpha}(C_+w)^2-\frac{w}{10}\geq g_+w .
\end{align*}

\emph{Delete.} The last term of the identity is nonnegative, because $s_j(x^*)\leq\bar\kappa\,s_j(x_k)$ by Lemma~\ref{lem:hybrid_loc} (Part~4). As $\sigma_j(x_k)\leq\theta_{\rm del}$ and $J$ is increasing, ${\rm rec}\leq(2/\alpha)J(\theta_{\rm del})^2$, and the identity and~\ref{cond:W2} give
\begin{align*}
    \mathcal{U}(K_{k+1})-\mathcal{U}(K_k)\geq-\frac12\ln\frac{1}{1-\theta_{\rm del}}-\frac{2}{\alpha}J(\theta_{\rm del})^2\geq-g_-w .
\end{align*}
\end{proof}

\begin{proof}[Proof of Lemma~\ref{lem:hybrid_block}]
By Lemma~\ref{lem:hybrid_inv} (Parts~1 and~5), no step of the block aborts and every polytope of the block has at most $\bar m$ rows. Let $m_{\rm add}$ and $m_{\rm del}=T-m_{\rm add}$ be the numbers of adds and deletes. The final polytope has $m_0+m_{\rm add}-m_{\rm del}\geq0$ rows, so $m_{\rm add}\geq(T-m_0)/2$. Summing Lemma~\ref{lem:hybrid_step} over the $T$ steps gives
\begin{align*}
    \mathcal{U}(K')-\mathcal{U}(K)\geq g_+w\,m_{\rm add}-g_-w\,(T-m_{\rm add})\geq g_{\rm b}wT-\frac{g_++g_-}{2}\,w\,m_0,
\end{align*}
where the second step uses $m_{\rm add}\geq(T-m_0)/2$ and $g_{\rm b}=(g_+-g_-)/2$. As $m_0\leq\bar m$, condition~\ref{cond:C4} bounds the last term by $C_{\rm b}n$. Finally, $g_T\leq g_{\rm b}$ by~\ref{cond:C4}, so the choice of $T$ gives $g_{\rm b}wT\geq g_TwT\geq2n\ln n+C_{\rm b}n$.
\end{proof}

\begin{proof}[Proof of Lemma~\ref{lem:hybrid_cpm}]
By Lemma~\ref{lem:hybrid_block}, no step aborts, $K'$ has at most $\bar m$ rows, and Part~3 holds. Every row of $K'$ is a row of $K$ or a cut at a current point, which contains $x^*$ by~\eqref{eq:cut_xstar}, as $x^*\in K\subseteq W$. This proves Part~1. The output point $x_{K'}$ is the last current point, so $\lambda_{K'}(x_{K'})\leq\lambda^*$ by Lemma~\ref{lem:hybrid_inv} (Part~1), and Lemma~\ref{lem:near_center} gives the rest of Part~4. Part~2 follows from Parts~1 and~3 of Lemma~\ref{lem:sandwich}, since $H_{K'}=H_{K'}(x_{K'})$.

Each iteration of the outer loop calls $\SO$ at most once. On \textsc{LS} it makes one query, at most one insertion or deletion, one more query, and $N$ Newton steps with $t_{\max}$ weight updates and one query each. With \textsc{Init} and the first query, the block makes at most $T(3+(t_{\max}+1)N)+2=O(n\log n\log\log n)$ operations on \textsc{LS}, whatever \textsc{LS} outputs.

For the arithmetic operations, consider one Newton step. Computing $\wt g$ and $\widehat Q^{-1}\wt g$ takes $O(mn)=O(n^2)$ operations. The step test computes $s_i(y)$ and $u_i:=a_i^\top(y'-y)$ for every row once, with $O(n^2)$ operations, and then tests each piece with $O(n)$ operations, as $s_i(y^{(t)})=s_i(y)+\frac{t}{t_{\max}}u_i$. The $t_{\max}$ updates and the query of \textsc{LS} take amortized $O(n^2)$ operations each, by Lemma~\ref{lem:jlsw_main}. The rest of an iteration also takes amortized $O(n^2)$ operations. It updates \textsc{IM}$_H$ and \textsc{IM}$_Q$ once each (Lemma~\ref{lem:hybrid_im}) and changes at most one row in each structure. It also computes $\wt h$ or the estimate of the deleted score, converts a cut vector $a$ to $Y^\top a$ and runs the weight tests. So the $T$ iterations take $O(Tn^2N)=O(n^3\log n\log\log n)$ operations. The three initializations take $O(n^{\omega+o(1)})$ operations, and forming $AY$, converting the output Hessian to the coordinates of $\R^n$ and computing $H_{K'}(x_{K'})$ take $O(n^3)$.
\end{proof}

\begin{proof}[Proof of Lemma~\ref{lem:block_stability}]
By induction over the blocks, every block of the run up to this one is good. Indeed, it starts at the center of a box, where $\lambda=0$ by Lemma~\ref{lem:reduction_geometry} (Part~3), or at the output of a previous good block, where $\lambda\leq\lambda^*$ by Lemma~\ref{lem:hybrid_inv} (Part~1). Its polytope has at most $\bar m$ rows, by~\ref{cond:C4} for a box and by the same lemma otherwise.

Let $K_0=K,K_1,\ldots,K_T=K'$ be the polytopes of this block, and let $x_k:=x_c(K_k)$. Lemma~\ref{lem:good_centers} shows that $x_k$ exists, that $x_{k+1}\in\mathcal{S}_{K_{k+1}}(x_k,r)$, that $w_x\leq w_+$ for an add, and that $\sigma_j(x_k)\leq\theta_{\rm del}$ for a delete. Lemma~\ref{lem:hybrid_local} (Part~4) in $K_{k+1}$ gives $(1+r)^{-2}H_{K_{k+1}}(x_k)\preceq H_{K_{k+1}}(x_{k+1})\preceq(1-r)^{-2}H_{K_{k+1}}(x_k)$. By the Cauchy--Schwarz inequality, $aa^\top/s_a(x_k)^2\preceq w_xH_{K_k}(x_k)$ and $a_ja_j^\top/s_j(x_k)^2\preceq\sigma_j(x_k)H_{K_k}(x_k)$. So an add gives $H_{K_k}(x_k)\preceq H_{K_{k+1}}(x_k)\preceq(1+w_x)H_{K_k}(x_k)$, and a delete gives $(1-\sigma_j(x_k))H_{K_k}(x_k)\preceq H_{K_{k+1}}(x_k)\preceq H_{K_k}(x_k)$. As $\ln(1+w_x)\leq w_+$, condition~\ref{cond:C4} gives $e^{-C_H}H_{K_k}(x_k)\preceq H_{K_{k+1}}(x_{k+1})\preceq e^{C_H}H_{K_k}(x_k)$ in both cases. Hence
\begin{align*}
    e^{-C_HT}H_K(x_0)\preceq H_{K'}(x_T)\preceq e^{C_HT}H_K(x_0).
\end{align*}
At the start, $H_K=H_K(x_0)$ if $K$ is a box, by Lemma~\ref{lem:reduction_geometry} (Part~3). Otherwise $x_K$ is the output of the previous block, and $H_K$ is within the factors $1\pm\epsilon_{\rm f}$ of $H_K(x_0)$ by Lemma~\ref{lem:near_center}. At the end, the same lemma puts $H_{K'}$ within these factors of $H_{K'}(x_T)$. These two factors add at most $2\ln(1/(1-\epsilon_{\rm f}))\leq10\,\epsilon_{\rm f}$ to the exponent, by~\ref{cond:C4}. This proves the first claim.

Finally, $H_{K'}\preceq e^aH_K$ gives $\|u\|_{H_{K'}^{-1}}\geq e^{-a/2}\|u\|_{H_K^{-1}}$ for every $u\in W_0$. Applying this to a shortest nonzero vector $u$ of $\Lambda$ in the norm $\|\cdot\|_{H_{K'}^{-1}}$ gives the last claim, as $a=O(T)=O(n\log n)$.
\end{proof}

\subsection{Leverage Scores under Adaptive Updates}\label{sec:r3_adaptive}

Our algorithm chooses each operation on \textsc{LS} as a function of earlier outputs, while Theorem~5.1 of~\cite{jlsw20} bounds the error of a query only in second moment and for operations fixed in advance. The next lemma restores the error bound with high probability. Its proof follows the error analysis of~\cite{jlsw20} and replaces the last, second-moment step by a tail bound for martingales in Hilbert spaces.

\begin{lemma}[Leverage scores under adaptive updates]\label{lem:jlsw_adaptive}
Run the data structure of Lemma~\ref{lem:jlsw_main}, with the parameters of Table~5 of~\cite{jlsw20} and error exponent $15$, on the padded instance used by Algorithm~\ref{alg:hcpm}, which has $n$ columns and at most $\bar m+n$ rows. Suppose that a client chooses each operation as a function of the earlier query outputs, and that it aborts instead of issuing an operation that violates the row bound or the conditions of Lemma~\ref{lem:jlsw_main} on insertions, deletions and weight updates. The client may also call \textsc{\textup{Init}} again on a new padded instance, as at the start of every block of Algorithm~\ref{alg:ours}, and we count the operations on all these instances together, in order. Let $\delta_{\rm J}(n):=(C_{\rm J}+1)\log^{-15}n$, where $C_{\rm J}$ is the absolute constant of the deterministic phases of the data structure (defined in the proof). Let $M$ be a deterministic positive integer and $k'\geq1$ a constant. Then there is $n_1(k')$ such that for every $n\geq n_1(k')$, with probability at least $1-3Mn^{-k'}$, every query among the first $M$ operations returns $\wt\sigma$ with $\|\wt\sigma-\sigma(\eta)\|_2\leq\delta_{\rm J}(n)$, where $\eta$ is the current weight vector. In particular, if $M\leq n^{10}$, then with $k'=20$ this probability is at least $1-3n^{-10}$.
\end{lemma}

\begin{proof}
We write $N_{\rm J}=100\log^2n$ for the sampling parameter of Table~5 of~\cite{jlsw20} (numbering of the arXiv version 2004.04250v1, as throughout), which is unrelated to the number $N$ of Newton steps. As the client aborts instead of issuing an invalid operation, every batch of operations satisfies Assumption~7.1 of~\cite{jlsw20} for $n\geq n_1(k')$, whatever the earlier outputs were.

\emph{The deterministic phases.} The estimate combines an inner, a middle and an outer phase, and the inner and middle phases are deterministic (Theorems~6.3 and~7.2 of~\cite{jlsw20}). Under Assumption~7.1, the accounting in the proof of Theorem~5.1 of~\cite{jlsw20} for the exponent $15$ bounds their contribution to the error of a query by $C_{\rm J}\log^{-15}n+n^{-0.06}$, for an absolute constant $C_{\rm J}$ that this defines. The error of a query is at most this bound plus the norm of the current error of the outer phase.

\emph{Epochs.} The only randomness is the Gaussian matrices that each outer update samples (steps 29 and 32 of Algorithm~5 of~\cite{jlsw20}). An initialization of the outer phase sets its estimate to the exact scores and samples nothing. It happens at the start, at every restart of Algorithm~1 of~\cite{jlsw20}, and at every call of \textsc{Init} by the client. An epoch, the operations between two initializations, contains at most $T_{\rm out}=n^{\omega-2}$ outer updates. We view score vectors as elements of the Hilbert space $\ell^2$ of real sequences indexed by row identities, where a row inserted later gets a new index. The coordinate of a deleted row is ignored, so the error on the current rows is at most the norm of the error vector.

\emph{Filtration.} We number the outer updates over all epochs, in order, by $t=1,2,\ldots$, and the epochs by $e=1,2,\ldots$. The first $M$ operations contain at most $M$ of each. Let $\mathcal{F}_0$ be trivial, and let $\mathcal{F}_t$ be generated by the Gaussians of the first $t$ outer updates. The client's operations are functions of earlier outputs, and the deterministic parts of the data structure are functions of the operations and of earlier Gaussians. So the following are $\mathcal{F}_{t-1}$-measurable: whether the $t$-th outer update occurs among the first $M$ operations, its epoch and position there, its batch and the state just before it. Its Gaussians are independent of $\mathcal{F}_{t-1}$. Let $\Delta\wt\sigma_t$ be the change of the outer estimate in the $t$-th outer update, and $\Delta\sigma_t$ the change of the true scores over its batch. Conditionally on $\mathcal{F}_{t-1}$, each coordinate of $\Delta\wt\sigma_t$ is a polynomial of degree at most $2$ in independent standard Gaussians (Lemma~8.10 of~\cite{jlsw20}, after rescaling the variances to $1$). Let
\begin{align*}
    \xi_t:=\E[\Delta\wt\sigma_t\mid\mathcal{F}_{t-1}]-\Delta\sigma_t\qquad\text{and}\qquad X_t:=\Delta\wt\sigma_t-\E[\Delta\wt\sigma_t\mid\mathcal{F}_{t-1}],
\end{align*}
with $X_t=\xi_t=0$ if the $t$-th outer update is not among the first $M$ operations. Theorem~8.1 of~\cite{jlsw20}, for the batch and the state fixed by $\mathcal{F}_{t-1}$, gives $\|\xi_t\|_2\leq n^{-100}+\poly(n)2^{-2N_{\rm J}}$ and, with the parameters of Table~5,
\begin{align*}
    \E[\|X_t\|_2^2\mid\mathcal{F}_{t-1}]\leq\E[\|\Delta\wt\sigma_t-\Delta\sigma_t\|_2^2\mid\mathcal{F}_{t-1}]\leq v:=C_vn^{-0.48}
\end{align*}
for an absolute constant $C_v$. With $\epsilon_{\rm J}$ from Section~\ref{sec:r3_params}, $vT_{\rm out}=C_vn^{-2\epsilon_{\rm J}}$. Here $\epsilon_{\rm J}>0$, as $\omega<2.48$ follows from the best known bound $\omega<2.258$ over every field of characteristic zero~\cite[Theorem~1.1 and Corollaries~14.1--14.2]{OpenAIMM26}. After the first $j$ outer updates of epoch $e$, the outer estimate has error $E^{(e)}_j=\sum_t(X_t+\xi_t)$, where $t$ runs over these $j$ updates.

\emph{Step 1: the tail of one update.} If $X$ is a polynomial of degree at most $2$ in independent standard Gaussians and $q_*\geq2$, then $\|X\|_{L^{q_*}}\leq(q_*-1)\|X\|_{L^2}$. This follows from the hypercontractivity of the Ornstein--Uhlenbeck operator $U_\varrho$, which states $\|U_\varrho g\|_{L^{q_*}}\leq\|g\|_{L^2}$ for $0\leq\varrho\leq(q_*-1)^{-1/2}$ (see Chapter~11 of~\cite{o14}). Indeed, write $X=X_0+X_1+X_2$ with $X_j$ in the span of the Hermite polynomials of degree $j$. The $X_j$ are orthogonal and $U_\varrho X_j=\varrho^jX_j$, so with $\varrho=(q_*-1)^{-1/2}$,
\begin{align*}
    \|X\|_{L^{q_*}}=\Big\|U_\varrho\Big(\sum_j\varrho^{-j}X_j\Big)\Big\|_{L^{q_*}}\leq\Big(\sum_j\varrho^{-2j}\|X_j\|_{L^2}^2\Big)^{1/2}\leq\varrho^{-2}\|X\|_{L^2}.
\end{align*}
We apply this conditionally on $\mathcal{F}_{t-1}$ to each coordinate $X_{t,i}$. With conditional norms, Minkowski's inequality in $L^{q_*/2}$ gives
\begin{align*}
    \E[\|X_t\|_2^{q_*}\mid\mathcal{F}_{t-1}]=\Big\|\sum_iX_{t,i}^2\Big\|_{L^{q_*/2}}^{q_*/2}\leq\Big(\sum_i\|X_{t,i}\|_{L^{q_*}}^2\Big)^{q_*/2}\leq\big((q_*-1)^2v\big)^{q_*/2}.
\end{align*}
Let $q_*:=k'\ln n$, which is at least $2$ for $n\geq n_1(k')$, and $\beta_*:=e\,(q_*-1)v^{1/2}=O(k'v^{1/2}\log n)$. Markov's inequality gives $\Pr[\|X_t\|_2>\beta_*\mid\mathcal{F}_{t-1}]\leq e^{-q_*}=n^{-k'}$.

\emph{Step 2: truncation.} Let $\mathcal{E}_t:=\{\|X_t\|_2\leq\beta_*\}$ and $Y_t:=X_t\mathbf{1}_{\mathcal{E}_t}-\E[X_t\mathbf{1}_{\mathcal{E}_t}\mid\mathcal{F}_{t-1}]$. Then $(Y_t)$ is a martingale difference sequence in $\ell^2$ with respect to $(\mathcal{F}_t)$, and $\|Y_t\|_2\leq2\beta_*$. As $\E[X_t\mid\mathcal{F}_{t-1}]=0$, on the event $\mathcal{E}:=\bigcap_t\mathcal{E}_t$ we have $X_t=Y_t-\E[X_t\mathbf{1}_{\mathcal{E}_t^c}\mid\mathcal{F}_{t-1}]$, and by the Cauchy--Schwarz inequality
\begin{align*}
    \big\|\E[X_t\mathbf{1}_{\mathcal{E}_t^c}\mid\mathcal{F}_{t-1}]\big\|_2\leq\E[\|X_t\|_2^2\mid\mathcal{F}_{t-1}]^{1/2}\Pr[\mathcal{E}_t^c\mid\mathcal{F}_{t-1}]^{1/2}\leq v^{1/2}n^{-k'/2}.
\end{align*}

\emph{Step 3: one epoch.} By Theorem~3.5 of~\cite{p94}, a martingale $(f_j)$ in a Hilbert space whose differences $d_j$ satisfy $\sum_j\|d_j\|_{L^\infty}^2\leq b^2$ has $\Pr[\max_j\|f_j\|\geq u]\leq2\exp(-u^2/(2b^2))$ for every $u\geq0$, since a Hilbert space is $(2,1)$-smooth in the sense of~\cite{p94}. Fix an epoch $e$. For $j\geq1$, let $\tau_j$ be the number $t$ of the $j$-th outer update of epoch $e$, with $\tau_j=\infty$ if there is no such update among the first $M$ operations, and let $\tau_0=0$. As the epoch and position of the $t$-th update are $\mathcal{F}_{t-1}$-measurable, $\{\tau_j=t\}\in\mathcal{F}_{t-1}$ for every $t$. Let $\mathcal{H}_j$ be the $\sigma$-field of the events $A$ with $A\cap\{\tau_j\leq t\}\in\mathcal{F}_t$ for every $t$, and let $Z_j:=Y_{\tau_j}$ if $\tau_j<\infty$ and $Z_j:=0$ otherwise. Then $Z_j$ is $\mathcal{H}_j$-measurable, $\mathcal{H}_{j-1}\subseteq\mathcal{H}_j$ and $\|Z_j\|_2\leq2\beta_*$. Moreover $\E[Z_j\mid\mathcal{H}_{j-1}]=0$. Indeed, let $A\in\mathcal{H}_{j-1}$. As $\tau_{j-1}\leq t-1$ on $\{\tau_j=t\}$, the set $A\cap\{\tau_j=t\}=(A\cap\{\tau_{j-1}\leq t-1\})\cap\{\tau_j=t\}$ lies in $\mathcal{F}_{t-1}$, so $\E[Z_j\mathbf{1}_A]=\sum_t\E[Y_t\mathbf{1}_{A\cap\{\tau_j=t\}}]=0$. So $(Z_j)$ is a martingale difference sequence with respect to $(\mathcal{H}_j)$, and $Z_j=0$ for $j>T_{\rm out}$. The theorem with $b^2=4\beta_*^2T_{\rm out}$ and $u_*:=\beta_*(8k'T_{\rm out}\ln n)^{1/2}$ shows that, outside an event of probability at most $2n^{-k'}$, every $j$ has $\|\sum_{i\leq j}Z_i\|_2<u_*$, where
\begin{align*}
    u_*=O\big(k'^{3/2}(vT_{\rm out})^{1/2}\log^{3/2}n\big)=O\big(k'^{3/2}n^{-\epsilon_{\rm J}}\log^{3/2}n\big).
\end{align*}
On $\mathcal{E}$, Step~2 gives $E^{(e)}_j=\sum_{i\leq j}\big(Z_i+\xi_{\tau_i}-\E[X_{\tau_i}\mathbf{1}_{\mathcal{E}_{\tau_i}^c}\mid\mathcal{F}_{\tau_i-1}]\big)$ for $j\leq T_{\rm out}$. So on $\mathcal{E}$ and outside this event, every error of epoch $e$ satisfies
\begin{align*}
    \|E^{(e)}_j\|_2\leq u_*+T_{\rm out}\big(v^{1/2}n^{-k'/2}+n^{-100}+\poly(n)2^{-2N_{\rm J}}\big)=O\big(k'^{3/2}n^{-\epsilon_{\rm J}}\log^{3/2}n\big),
\end{align*}
because $T_{\rm out}\leq n\leq n^{k'}$ gives $T_{\rm out}v^{1/2}n^{-k'/2}\leq(vT_{\rm out})^{1/2}$, and $2^{-2N_{\rm J}}=2^{-200\log^2n}$ is below every inverse polynomial in $n$.

\emph{Step 4: union bound.} The event $\mathcal{E}_t$ holds trivially when there is no $t$-th outer update among the first $M$ operations. So some $\mathcal{E}_t$ fails with probability at most $\sum_{t\leq M}\E[\Pr[\mathcal{E}_t^c\mid\mathcal{F}_{t-1}]]\leq Mn^{-k'}$, and the event of Step~3 fails for some epoch $e\leq M$ with probability at most $2Mn^{-k'}$. With probability at least $1-3Mn^{-k'}$ neither happens, and then every query among the first $M$ operations has error at most
\begin{align*}
    C_{\rm J}\log^{-15}n+n^{-0.06}+O\big(k'^{3/2}n^{-\epsilon_{\rm J}}\log^{3/2}n\big)\leq(C_{\rm J}+1)\log^{-15}n
\end{align*}
for every $n\geq n_1(k')$. We choose $n_1(k')$ large enough for this inequality, for $k'\ln n\geq2$, and for the conditions in~\cite{jlsw20} that hold only for large $n$. These are the length bound $n^{0.08}$ on batches in Assumption~7.1, Theorem~8.1 with the parameters of Table~5, and the bounds $n^{-0.07}$ and $n^{-0.06}$ in the proof of Theorem~5.1. In the padded instance every real row has the same leverage score as in $K$, so the bound holds on the real rows. The last claim is the case $M\leq n^{10}$ and $k'=20$.
\end{proof}

\subsection{Inverse Maintenance}\label{sec:r3_im}

We use the bound of~\cite{jlsw20} quoted in the paragraph \emph{Inverse maintenance} of Section~\ref{sec:cpm:method}, with $c''=1$. So it suffices to show that the number of rows stays $O(n)$ and that every \textsc{Update} moves the vector of the logarithms of the weights by $C=O(\log n)$ in $\ell_2$.

\begin{proof}[Proof of Lemma~\ref{lem:hybrid_im}]
Consider a good block. By Lemma~\ref{lem:hybrid_inv} (Part~1), every polytope of the block has at most $\bar m$ rows, so the padded instance has $O(n)$ rows, and its dummy rows keep their weights. Step $k$ updates \textsc{IM}$_Q$ at $z_k$ and \textsc{IM}$_H$ at $z_{k+1}$, both with the rows of $K_{k+1}$.

\emph{The point moves little.} By Lemma~\ref{lem:hybrid_inv} (Part~4), the Newton steps of step $k$ satisfy the hypotheses of Lemma~\ref{lem:hybrid_loc} (Part~2), with $z_s=z_k$ and $Q_0=Q_c(z_k)\succeq\mu_sH(z_k)$. That part gives $z_{k+1}\in\mathcal{S}(z_k,r_{\rm N})$. It also gives the first inequality below, through $\|z_{k+1}-z_k\|_{Q_0}\leq\|y_N-x_c\|_{Q_0}+\|x_c-z_k\|_{Q_0}$, the bounds of Lemma~\ref{lem:hybrid_loc} (Parts~1 and~2) on both terms, and $Q_0\succeq\mu_sH(z_k)$. The second follows from the bounds on $\lambda_s$ and $\mu_s$ in Parts~2 and~3 of Lemma~\ref{lem:hybrid_inv}, from~\ref{cond:W3} and from $\alpha'\geq1\geq\alpha_\mu$.
\begin{align}\label{eq:im_move}
    \|z_{k+1}-z_k\|_{H(z_k)}\leq\frac{2}{\mu_s^{1/2}}\Big(\frac2\alpha\lambda_s+C_{\rm x}\epsilon_g\Big)\leq h_{\max}.
\end{align}
So the numbers $s_i(z_{k+1})/s_i(z_k)-1$ form a vector of $\ell_2$ norm at most $h_{\max}$, with entries in $[-r_{\rm N},r_{\rm N}]$, where $r_{\rm N}<r$ by~\ref{cond:C2}. As $|\ln(1+x)|\leq2|x|$ for $|x|\leq1/2$, the numbers $2\ln(s_i(z_{k+1})/s_i(z_k))$ form a vector of $\ell_2$ norm at most $4h_{\max}$.

\emph{The instance \textsc{IM}$_H$.} Between its \textsc{Update}s at $z_k$ and at $z_{k+1}$, step $k$ deletes a row, or it inserts one with the weight $\rho^{-2}=s_a(z_k)^{-2}$ that the formula of the other rows gives. So every log-weight changes by $-2\ln(s_i(z_{k+1})/s_i(z_k))$, and $C\leq4h_{\max}=O(1)$.

\emph{The instance \textsc{IM}$_Q$.} Its \textsc{Update}s happen at $z_k$ with the rows of $K_{k+1}$, and at $z_{k+1}$ with the rows of $K_{k+2}$. A row $i$ present at both changes its log-weight by
\begin{align*}
    \ln\frac{\wt\sigma'_i+c}{\wt\sigma_i+c}-2\ln\frac{s_i(z_{k+1})}{s_i(z_k)},
\end{align*}
where $\wt\sigma$ and $\wt\sigma'$ are the queries of the two \textsc{Update}s. Their errors are at most $\delta_0\leq c/2$, as~\ref{cond:C7} gives $\delta_0\leq10^{-7}c$, so $\wt\sigma_i+c$ and $\wt\sigma'_i+c$ are at least $c/2$, and the first term has $\ell_2$ norm at most $(2/c)\|\wt\sigma'-\wt\sigma\|_2$ over these rows. Let $\sigma$ be the true scores of $K_{k+1}$ along the segment $y(\tau)=z_k+\tau(z_{k+1}-z_k)$, which lies in $\mathcal{S}(z_k,r_{\rm N})$. By~\eqref{eq:sigma_dot} with $h=z_{k+1}-z_k$, we have $\dot\sigma_i=2\sum_jP_{ij}^2\ell_j-2\sigma_i\ell_i$, where $\ell_i=a_i^\top h/s_i(y)$. As $P$ is a projection with diagonal $\sigma$, the Cauchy--Schwarz inequality gives
\begin{align*}
    \sum_i\Big(\sum_jP_{ij}^2\ell_j\Big)^2\leq\sum_i\sigma_i\sum_jP_{ij}^2\ell_j^2\leq\sum_j\sigma_j\ell_j^2\leq\|\ell\|_2^2,
\end{align*}
so $\|\dot\sigma\|_2\leq4\|\ell\|_2$. By Lemma~\ref{lem:hybrid_local} (Part~4) and~\eqref{eq:im_move}, $\|\ell\|_2=\|h\|_{H(y)}\leq2h_{\max}$, so the scores move by at most $8h_{\max}$ in $\ell_2$. By the Sherman--Morrison formula, the row change at $z_{k+1}$ moves the old scores by amounts $e_i\geq0$ with $\sum_ie_i=w'/(1+w')$ for an add of leverage $w'$, and $\sum_ie_i=\sigma_j$ for the deletion of row $j$. Both sums are at most $2w$, by Parts~2 and~3 of Lemma~\ref{lem:hybrid_inv}, since $w/(1-\epsilon_{\rm im})\leq2w$ and $\theta+\delta_0\leq2w$. Hence $\|\wt\sigma'-\wt\sigma\|_2\leq2\delta_0+8h_{\max}+2w$. A row added at $z_{k+1}$ has its weight changed by the factor $(\wt\sigma'_a+c)/(w/(1+w)+c)$, and both numbers lie in $[c/2,2]$. So
\begin{align*}
    C\leq4h_{\max}+\frac2c\big(2\delta_0+8h_{\max}+2w\big)+\ln\frac4c=O(c^{-1})=O(\log n).
\end{align*}
For both instances, the first \textsc{Update} compares with the initialization at $z_0$ in the same way.

Last, the bound of~\cite{jlsw20} is stated as an amortized expected time, but the data structure is deterministic (footnote~5 to Theorem~5.1 in the arXiv version 1810.07896v2 of~\cite{cls19}). So the bound holds for every sequence of calls with the property above, however it was generated.
\end{proof}

\subsection{Vaidya's Method with Exact Leverage Scores}\label{sec:r3_vaidya}

This section proves Corollary~\ref{cor:exact_score}. With exact leverage scores we need no hybrid barrier. So here $c=0$, and each block runs Vaidya's method~\cite{v89} with the volumetric barrier $F$ (Algorithm~\ref{alg:exact_vaidya}).

\paragraph{Setting.} Let $K=\{t\in\R^d:a_i^\top t\geq b_i,\ i\in[m]\}$ be a polyhedron with nonempty interior and nonzero rows, whose constraint matrix $A$ has rank $d\leq n$. For $z\in{\rm int}(K)$ we write $s_i=s_i(z)$ and $v_i=a_i/s_i$. Then $H_K(z)=\sum_iv_iv_i^\top$, $\sigma_i(z)=v_i^\top H_K(z)^{-1}v_i$, $Q_K(z)=\sum_i\sigma_i(z)v_iv_i^\top$ and $\nabla F_K(z)=-\sum_i\sigma_i(z)v_i$. Let $\mu_K(z)$ be the largest $\mu$ with $Q_K(z)\succeq\mu H_K(z)$ (Definition~\ref{def:mu_x}). The matrix $P$ with entries $P_{ij}=v_i^\top H_K(z)^{-1}v_j$ is the orthogonal projection onto the column space of $S^{-1}A$, and $\sigma_i=P_{ii}$. Let
\begin{align*}
    \lambda_K(z):=\|\nabla F_K(z)\|_{Q_K(z)^{-1}}\qquad\text{and}\qquad\pi_K(z):=\lambda_K(z)^2/\mu_K(z).
\end{align*}
A minimizer of $F_K$ on ${\rm int}(K)$ is unique by (S1) below, and it is then $\vc(K)$. In this case let $g_K(z):=F_K(z)-F_K(\vc(K))$ and $\Gamma_K(z):=g_K(z)/\mu_K(\vc(K))$. We drop the subscript $K$ when only one polyhedron occurs. The constants are fixed in Section~\ref{sec:r3_params} (Step~6), and they are not optimized.

\paragraph{The algorithm.} Algorithm~\ref{alg:exact_vaidya} runs one block of Vaidya's method in the coordinates of the current affine subspace, as Algorithm~\ref{alg:hcpm} does. It adds a cut when every leverage score is at least $\epsilon_{\rm V}$, and otherwise it deletes a row of smallest score. The algorithm of Corollary~\ref{cor:exact_score} is Algorithm~\ref{alg:ours} with three changes. Every call of Algorithm~\ref{alg:hcpm} becomes a call of Algorithm~\ref{alg:exact_vaidya} from $x_K$, the reduction ellipsoid is $2R'_{\rm V}\cdot E(x_K,H_K)\cap P$, and the operation budget is dropped. Lemma~\ref{lem:sp_vaidya} analyzes one iteration. Its Part~(c) allows the deleted row to be smallest only up to a tolerance $\eta_m$, and the algorithm has $\eta_m=0$.

\begin{algorithm}[!ht]
\caption{Vaidya's method with exact leverage scores, one block}
\label{alg:exact_vaidya}
\begin{algorithmic}[1]
\Procedure{ExactVaidya}{$\SO,K,z$} \Comment{$K$ has at most $\bar m_{\rm V}$ rows, and $\Gamma_K(z)\leq\epsilon_{\rm V}^4$}
\For{$k=1,\ldots,T_{\rm V}$}
\State Compute $H(z)$, $H(z)^{-1}$ and every leverage score $\sigma_i(z)$ exactly
\If{$\min_i\sigma_i(z)<\epsilon_{\rm V}$}
\State Delete from $K$ a row $i$ with the smallest $\sigma_i(z)$
\Else
\State $a\gets$ the cut vector at $z$ (one call of $\SO$), and $\rho\gets\big(a^\top H(z)^{-1}a\,(1-\sigma_\star)/\sigma_\star\big)^{1/2}$
\State Add the row $a^\top x\geq a^\top z-\rho$ to $K$
\EndIf
\For{$l=1,\ldots,k_{\rm V}$}
\State $z\gets z-\frac1{10}Q(z)^{-1}\nabla F(z)$, for the current $K$
\EndFor
\EndFor
\State \Return $(K,z,H(z))$
\EndProcedure
\end{algorithmic}
\end{algorithm}

\begin{lemma}[Vaidya's iteration]\label{lem:sp_vaidya}
Let $K$ be a bounded polytope as above with $m$ constraints, let $z\in{\rm int}(K)$, and let $x_{\rm V}=\vc(K)$.
\begin{enumerate}[label=(\alph*)]
    \item (Newton step.) If $\Gamma_K(z)\leq c_{\rm V}$, then $z_+=z-\frac1{10}Q_K(z)^{-1}\nabla F_K(z)$ satisfies $\|z_+-z\|_{H_K(z)}\leq r_{\rm V}$ and $g_K(z_+)\leq q_{\rm V}\,g_K(z)$.
    \item (Adding a constraint.) Let $\Gamma_K(z)\leq\epsilon_{\rm V}^4$ and $\sigma_i(z)\geq\epsilon_{\rm V}/2$ for every constraint $i$. Let $K_+=K\cap\{x:a^\top x\geq a^\top z-\rho\}$ with $a\neq0$ and $\rho>0$ such that the leverage score at $z$ of the new constraint in $K_+$ lies in $[(1-\tau_{\rm V})\sigma_\star,\sigma_\star]$. Then $\vc(K_+)$ exists, $F_{K_+}(\vc(K_+))-F_K(\vc(K))\geq g^+_{\rm V}\epsilon_{\rm V}$ and $\Gamma_{K_+}(z)\leq c_{\rm V}/2$.
    \item (Deleting a constraint.) Let $\Gamma_K(z)\leq\epsilon_{\rm V}^4$, and let $i$ be a constraint with $\sigma_i(z)<2\epsilon_{\rm V}$ and $\sigma_i(z)\leq\sigma_j(z)+\eta_m$ for every constraint $j$, where $0\leq\eta_m\leq\epsilon_{\rm V}/(16m)$. Let $K_-$ be $K$ without constraint $i$. Then $K_-$ is bounded, $F_K(\vc(K))-F_{K_-}(\vc(K_-))\leq g^-_{\rm V}\epsilon_{\rm V}$ and $\Gamma_{K_-}(z)\leq c_{\rm V}/2$.
    \item (The invariant.) If $\Gamma_K(z)\leq\epsilon_{\rm V}^4$, then $\|z-x_{\rm V}\|_{H_K(x_{\rm V})}\leq2\epsilon_{\rm V}^2$, $(1-\epsilon_{\rm V})H_K(x_{\rm V})\preceq H_K(z)\preceq(1+\epsilon_{\rm V})H_K(x_{\rm V})$, $\pi_K(z)\leq C_\pi\epsilon_{\rm V}^4$ and $K\subseteq C_{\rm V}m^{3/2}n\cdot E(z,H_K(z))$.
\end{enumerate}
\end{lemma}

\begin{proof}
We first prove three facts for a polyhedron $K$ as above, which need not be bounded, and then (d), (a), (b) and (c).

\emph{(S1) Basic facts.} Let $z\in{\rm int}(K)$. (i) $0<\sigma_i\leq1$, $\sum_i\sigma_i=d$ and $v_iv_i^\top\preceq\sigma_iH$. (ii) $\min_i\sigma_i\cdot H\preceq Q\preceq H$ and $\max\{\min_i\sigma_i,1/(4m)\}\leq\mu\leq1$. (iii) $Q\preceq\nabla^2F\preceq3Q$, and $F$ is strictly convex. For (i), $P$ is a projection of rank $d$ with diagonal $\sigma$, $\sigma_i>0$ as $a_i\neq0$, and $(v_i^\top h)^2\leq\sigma_i\|h\|_H^2$ by the Cauchy--Schwarz inequality. In (ii), the Loewner inequalities hold term by term. With $\tau=1/(2m)$, (i) gives $\sum_{\sigma_i\leq\tau}(v_i^\top h)^2\leq m\tau\|h\|_H^2$, so $h^\top Qh\geq\tau\sum_{\sigma_i>\tau}(v_i^\top h)^2\geq\tau(1-m\tau)\|h\|_H^2=\|h\|_H^2/(4m)$. Part (iii) is Lemma~\ref{lem:hybrid_hessian} with $c=0$, together with $Q\succeq\mu H\succ0$.

\emph{(S2) Moving the point.} Let $z\in{\rm int}(K)$ and $\|z'-z\|_{H(z)}\leq\eta\leq r$. By (S1)(i), $|s_i(z')-s_i(z)|\leq s_i(z)\eta$, so $z'\in\mathcal{S}(z,\eta)$. Lemma~\ref{lem:hybrid_local} (Part~4) with $r'=\eta$ has factors that do not depend on $c>0$ and only improve as $r'$ decreases. Its Loewner inequalities survive the limit $c\to0$, where $Q_c\to Q$ and $\nabla^2p\to\nabla^2F$. With $\alpha\geq1/2$, $3\alpha'\leq6$ and $\alpha_\mu\geq2/5$ from~\ref{cond:V1}, this gives $\mu(z')\geq\frac25\mu(z)$ and
\begin{gather*}
    (1+\eta)^{-2}H(z)\preceq H(z')\preceq(1-\eta)^{-2}H(z),\qquad\frac12Q(z)\preceq Q(z')\\
    \text{and}\qquad\frac12Q(z)\preceq\nabla^2F(z')\preceq6Q(z).
\end{gather*}

\emph{(S3) Local analysis.} Let $z\in{\rm int}(K)$, $\lambda=\lambda(z)$ and $B=\{y:\|y-z\|_{H(z)}\leq r\}$, which lies in ${\rm int}(K)$ by (S2). For $y\in B$, Taylor's formula and (S2) give
\begin{align}\label{eq:vaidya_taylor}
    \frac14\|y-z\|_{Q(z)}^2\leq F(y)-F(z)-\nabla F(z)^\top(y-z)\leq3\|y-z\|_{Q(z)}^2.
\end{align}
Also $|\nabla F(z)^\top(y-z)|\leq\lambda\|y-z\|_{Q(z)}$ and $\|y-z\|_{H(z)}\leq\mu(z)^{-1/2}\|y-z\|_{Q(z)}$.

\emph{(S3a)} Suppose that $\pi(z)\leq\pi_0$. Let $\|y-z\|_{H(z)}=r$ and $\ell=\|y-z\|_{Q(z)}\geq\mu(z)^{1/2}r$. As $\pi_0<r^2/16$ by~\ref{cond:V1}, we have $\ell>4\lambda$, and~\eqref{eq:vaidya_taylor} gives $F(y)-F(z)\geq\ell(\ell/4-\lambda)>0$. So $F$ attains its minimum over $B$ at an interior point $x_{\rm V}$ of $B$, which minimizes the convex $F$ over ${\rm int}(K)$. If $K$ were unbounded, some $h\neq0$ would have $Ah\geq0$. No slack decreases along $h$, so $H(x_{\rm V}+h)\preceq H(x_{\rm V})$, and $x_{\rm V}+h$ would be a second minimizer, against (S1)(iii). So $K$ is bounded and $x_{\rm V}=\vc(K)$. Finally, $F(x_{\rm V})\leq F(z)$ and~\eqref{eq:vaidya_taylor} give $\ell_{\rm V}:=\|x_{\rm V}-z\|_{Q(z)}\leq4\lambda$, so $\|x_{\rm V}-z\|_{H(z)}\leq4\pi(z)^{1/2}\leq4\pi_0^{1/2}<r$.

\emph{(S3b)} For such $z$,~\eqref{eq:vaidya_taylor} at $y=x_{\rm V}$ gives $g_K(z)\leq\lambda\ell_{\rm V}-\ell_{\rm V}^2/4\leq\lambda^2$. Also $\mu(x_{\rm V})\geq\frac25\mu(z)$ by (S2), so $\Gamma_K(z)\leq\lambda^2/\mu(x_{\rm V})\leq\frac52\pi(z)$.

\emph{(S3c)} For such $z$, the point $z_+=z-\frac1{10}Q(z)^{-1}\nabla F(z)$ has $\|z_+-z\|_{Q(z)}=\lambda/10$, so $\|z_+-z\|_{H(z)}\leq\pi(z)^{1/2}/10$. By~\eqref{eq:vaidya_taylor}, $F(z_+)\leq F(z)-\lambda^2/10+3\lambda^2/100$. As $\lambda^2\geq g_K(z)$, we get $g_K(z_+)\leq(1-\frac7{100})g_K(z)=q_{\rm V}\,g_K(z)$.

\emph{(S3d)} Let $K$ be bounded, and let $x\in{\rm int}(K)$ satisfy $\Gamma_K(x)<r^2/4$. We show that $\|x-x_{\rm V}\|_{H(x_{\rm V})}\leq2\Gamma_K(x)^{1/2}$ and $\pi(x)\leq C_\pi\Gamma_K(x)$. Let $u=x-x_{\rm V}$. As $\nabla F(x_{\rm V})=0$,~\eqref{eq:vaidya_taylor} at $x_{\rm V}$ gives $F(y)-F(x_{\rm V})\geq\frac14\mu(x_{\rm V})\|y-x_{\rm V}\|_{H(x_{\rm V})}^2$ whenever $\|y-x_{\rm V}\|_{H(x_{\rm V})}\leq r$. If $\|u\|_{H(x_{\rm V})}>r$, convexity at the point of $[x_{\rm V},x]$ at distance $r$ would give $\Gamma_K(x)\geq r^2/4$. So $\|u\|_{H(x_{\rm V})}\leq r$, and the first claim follows. Next, $\nabla F(x)=Mu$ with $M=\int_0^1\nabla^2F(x_{\rm V}+\tau u)\,\d\tau$, and $\frac12Q(x_{\rm V})\preceq M\preceq6Q(x_{\rm V})$ by (S2). So $MQ(x_{\rm V})^{-1}M\preceq36\,Q(x_{\rm V})$ and $\|\nabla F(x)\|_{Q(x_{\rm V})^{-1}}^2\leq36\|u\|_{Q(x_{\rm V})}^2\leq144\,g_K(x)$. With $Q(x)\succeq\frac12Q(x_{\rm V})$ and $\mu(x)\geq\frac25\mu(x_{\rm V})$, we get $\pi(x)\leq288\,g_K(x)/(\frac25\mu(x_{\rm V}))=C_\pi\Gamma_K(x)$.

Now we prove the lemma, using $\epsilon_{\rm V}\leq10^{-12}$ from~\ref{cond:V1}.

(d) As $\epsilon_{\rm V}^4<r^2/4$ by~\ref{cond:V1}, the first and third claims are (S3d). By (S2), $H(z)$ lies between $(1+2\epsilon_{\rm V}^2)^{-2}H(x_{\rm V})$ and $(1-2\epsilon_{\rm V}^2)^{-2}H(x_{\rm V})$, which gives the second. Lemma~\ref{lem:polytope_sandwich}, applied in $\R^d$, gives $\|x-x_{\rm V}\|_{H(x_{\rm V})}\leq90m^{3/4}d$ for $x\in K$. So $\|x-z\|_{H(z)}\leq(1+\epsilon_{\rm V})^{1/2}(90m^{3/4}d+2\epsilon_{\rm V}^2)\leq C_{\rm V}m^{3/2}n$.

(a) As $c_{\rm V}<r^2/4$, (S3d) and~\ref{cond:V1} give $\pi_K(z)\leq C_\pi c_{\rm V}\leq\pi_0$. Then (S3c) and $(C_\pi c_{\rm V})^{1/2}/10\leq r_{\rm V}$, also from~\ref{cond:V1}, give the claim.

(b) Write $H,\sigma_i,v_i,Q,\mu,\lambda$ for $K$ at $z$, and let $v=a/\rho$, $\psi=v^\top H^{-1}v$ and $p_i=v_i^\top H^{-1}v$. By the Sherman--Morrison formula, the new constraint has score $\sigma_{\rm new}=\psi/(1+\psi)$ in $K_+$ at $z$, and an old one has $\sigma_i^+=\sigma_i-p_i^2/(1+\psi)\geq(1-\sigma_{\rm new})\sigma_i$, since $p_i^2\leq\sigma_i\psi$. So $Q_{K_+}(z)\succeq(1-\sigma_{\rm new})Q$, and (S1)(ii) for $K_+$ gives $\mu_{K_+}(z)\geq\min\{\sigma_{\rm new},(1-\sigma_{\rm new})\epsilon_{\rm V}/2\}\geq0.49\,\epsilon_{\rm V}$, since $(1-\tau_{\rm V})\sigma_\star\leq\sigma_{\rm new}\leq\sigma_\star$. Also $\mu\geq\epsilon_{\rm V}/2$, $\psi=\sigma_{\rm new}/(1-\sigma_{\rm new})\leq8.08\,\epsilon_{\rm V}$, and $\lambda\leq(C_\pi\epsilon_{\rm V}^4)^{1/2}\leq27\epsilon_{\rm V}^2$ by (d), as $\mu\leq1$. The gradient of $F_{K_+}$ at $z$ is
\begin{align*}
    \nabla F_{K_+}(z)=-\sum_i\sigma_i^+v_i-\sigma_{\rm new}v=\nabla F_K(z)+\frac1{1+\psi}\Big(\sum_ip_i^2v_i-\psi v\Big).
\end{align*}
Here $\|v\|_{Q^{-1}}\leq(\psi/\mu)^{1/2}$. By the Cauchy--Schwarz inequality, with $p_i^4/\sigma_i\leq\psi p_i^2$ and $\sum_ip_i^2=\psi$,
\begin{align*}
    \Big(\sum_ip_i^2v_i^\top h\Big)^2\leq\Big(\sum_i\frac{p_i^4}{\sigma_i}\Big)\Big(\sum_i\sigma_i(v_i^\top h)^2\Big)\leq\psi^2h^\top Qh\qquad\text{for every }h,
\end{align*}
so $\|\sum_ip_i^2v_i\|_{Q^{-1}}\leq\psi$. Hence $\lambda_{K_+}(z)\leq(1-\sigma_{\rm new})^{-1/2}(\lambda+\psi^{3/2}(2/\epsilon_{\rm V})^{1/2}+\psi)\leq42\,\epsilon_{\rm V}$, using $\psi^{3/2}(2/\epsilon_{\rm V})^{1/2}\leq32.5\,\epsilon_{\rm V}$, and $\pi_{K_+}(z)\leq(42\epsilon_{\rm V})^2/(0.49\epsilon_{\rm V})=3600\,\epsilon_{\rm V}$. By~\ref{cond:V2}, (S3a) and (S3b), $x_{\rm V}^+=\vc(K_+)$ exists, $\Gamma_{K_+}(z)\leq9000\,\epsilon_{\rm V}\leq c_{\rm V}/2$, and $r_+:=\|x_{\rm V}^+-z\|_{H_{K_+}(z)}\leq4(3600\,\epsilon_{\rm V})^{1/2}\leq r_{\rm V}$.

For the progress, let $s(x)$ be the new slack and $\psi(x)=\|a\|_{H_K(x)^{-1}}^2/s(x)^2$. Then $\det H_{K_+}=(1+\psi)\det H_K$, so $F_{K_+}=F_K+\frac12\ln(1+\psi)$. As $H_{K_+}(z)\succeq H$, an old slack changes from $z$ to $x_{\rm V}^+$ by at most $\|a_j\|_{H^{-1}}r_+\leq s_j(z)r_+$, and the new one by at most $\sigma_{\rm new}^{1/2}\rho\,r_+$. So $\psi(x_{\rm V}^+)\geq((1-r_+)/(1+r_+))^2\psi\geq0.98\,\sigma_\star$. With $F_K(x_{\rm V}^+)\geq F_K(x_{\rm V})$, we get $F_{K_+}(x_{\rm V}^+)-F_K(x_{\rm V})\geq\frac12\ln(1+0.98\,\sigma_\star)\geq0.48\,\sigma_\star\geq g^+_{\rm V}\epsilon_{\rm V}$.

(c) Write $H,\sigma_j,v_j,Q,\mu,\lambda,P$ for $K$ at $z$. By (S1)(i), $H_-:=H-v_iv_i^\top\succeq(1-\sigma_i)H$, so the constraint matrix of $K_-$ has rank $d$. By the Sherman--Morrison formula, $\|v_i\|_{H_-^{-1}}^2=\sigma_i/(1-\sigma_i)$, and the scores in $K_-$ at $z$ are $\sigma_j^-=\sigma_j+P_{ij}^2/(1-\sigma_i)\geq\sigma_j$. By (S1)(ii) for $K_-$, its $\mu_-:=\mu_{K_-}(z)$ satisfies $\mu_-\geq\sigma_i-\eta_m$ and $\mu_-\geq1/(4m)\geq4\eta_m/\epsilon_{\rm V}$. So $\mu_-\geq\sigma_i-\epsilon_{\rm V}\mu_-/4$, which gives $\mu_-\geq\sigma_i/2$. Let $\xi=\sigma_i^2/((1-\sigma_i)\mu_-)$. As $\sigma_i<2\epsilon_{\rm V}$,
\begin{align*}
    \xi\leq5\epsilon_{\rm V},\qquad\sigma_i\xi/\mu_-\leq9\epsilon_{\rm V}\qquad\text{and}\qquad\sigma_i^2/\mu_-\leq4\epsilon_{\rm V}.
\end{align*}
Next, $\sigma_iv_iv_i^\top\preceq\sigma_i^2(1-\sigma_i)^{-1}H_-\preceq\xi Q_{K_-}(z)$ and $Q-\sigma_iv_iv_i^\top\preceq Q_{K_-}(z)$. So $Q\preceq(1+\xi)Q_{K_-}(z)$, which gives $\|x\|_{Q_{K_-}(z)^{-1}}\leq(1+\xi)^{1/2}\|x\|_{Q^{-1}}$, and $\mu_-\geq\mu/(1+\xi)$ as $H_-\preceq H$. The gradient of $F_{K_-}$ at $z$ is
\begin{align*}
    \nabla F_{K_-}(z)=-\sum_{j\neq i}\sigma_j^-v_j=\nabla F_K(z)+\sigma_iv_i-\frac1{1-\sigma_i}\sum_{j\neq i}P_{ij}^2v_j.
\end{align*}
Here $\|\sigma_iv_i\|_{Q_{K_-}(z)^{-1}}^2\leq\sigma_i^2\|v_i\|_{H_-^{-1}}^2/\mu_-=\sigma_i\xi$. As $P_{ij}^2\leq\sigma_i\sigma_j^-$ and $\sum_jP_{ij}^2=\sigma_i$, the Cauchy--Schwarz argument of (b) gives $\|\sum_{j\neq i}P_{ij}^2v_j\|_{Q_{K_-}(z)^{-1}}\leq\sigma_i$. So $\lambda_{K_-}(z)\leq(1+\xi)^{1/2}\lambda+(\sigma_i\xi)^{1/2}+\sigma_i/(1-\sigma_i)$, and
\begin{align*}
    \pi_{K_-}(z)\leq3\Big((1+\xi)^2\pi_K(z)+\frac{\sigma_i\xi}{\mu_-}+\frac{1.1\,\sigma_i^2}{\mu_-}\Big)\leq3\big(1.1\,C_\pi\epsilon_{\rm V}^4+9\epsilon_{\rm V}+4.4\epsilon_{\rm V}\big)\leq41\,\epsilon_{\rm V}.
\end{align*}
By~\ref{cond:V2}, (S3a) and (S3b), $K_-$ is bounded, $x_{\rm V}^-=\vc(K_-)$ exists, $\Gamma_{K_-}(z)\leq103\,\epsilon_{\rm V}\leq c_{\rm V}/2$, and $r_-:=\|x_{\rm V}^--z\|_{H_-}\leq4(41\,\epsilon_{\rm V})^{1/2}\leq r_{\rm V}$.

For the progress, let $\psi_i(x)=\|a_i\|_{H_{K_-}(x)^{-1}}^2/s_i(x)^2$, so that row $i$ has the score $\sigma_{K,i}(x)=\psi_i(x)/(1+\psi_i(x))$ in $K$ at $x$, and $\psi_i(z)\leq2.1\,\epsilon_{\rm V}$. The slack of row $i$ changes from $z$ to $x_{\rm V}^-$ by at most $\|a_i\|_{H_-^{-1}}r_-=s_i(z)\psi_i(z)^{1/2}r_-$, so $x_{\rm V}^-\in{\rm int}(K)$. With (S2) for $K_-$, this gives $\psi_i(x_{\rm V}^-)\leq((1+r_-)/(1-r_-))^2\psi_i(z)$, so $\sigma_{K,i}(x_{\rm V}^-)\leq2.2\,\epsilon_{\rm V}$. Since $\det H_{K_-}=(1-\sigma_{K,i})\det H_K$ and $F_K(x_{\rm V}^-)\geq F_K(x_{\rm V})$, we get $F_{K_-}(x_{\rm V}^-)\geq F_K(x_{\rm V})+\frac12\ln(1-2.2\,\epsilon_{\rm V})\geq F_K(x_{\rm V})-g^-_{\rm V}\epsilon_{\rm V}$.
\end{proof}

\begin{proof}[Proof of Corollary~\ref{cor:exact_score}]
We work under the assumptions of Theorem~\ref{thm:main_formal}, so $\log\gamma=O(n\log n)$. The algorithm uses no data structure and no random bits, and $\SO$ and \textsc{ApproxSV} are deterministic, so the bounds below hold for every run.

\emph{One block.} Consider a call of Algorithm~\ref{alg:exact_vaidya} on $(K,x_K)$, where $K$ is bounded with at most $\bar m_{\rm V}$ rows and $\Gamma_K(x_K)\leq\epsilon_{\rm V}^4$. It makes at most $T_{\rm V}$ calls to $\SO$. Each iteration and each Newton step takes $O(n^\omega)$ operations, as $m=O(n)$, so the block takes $O(T_{\rm V}k_{\rm V}n^\omega)=O(n^{\omega+1}\log n)$ operations. We claim that its output $(K',x_{K'},H_{K'})$ satisfies the following.
\begin{enumerate}
    \item Every row of $K'$ is a row of $K$ or a cut $a^\top x\geq a^\top z-\rho$ built from the cut vector $a$ at a current point $z$, and $K'$ is bounded with at most $\bar m_{\rm V}$ rows.
    \item $E(x_{K'},H_{K'})\subseteq K'\subseteq R'_{\rm V}\cdot E(x_{K'},H_{K'})$.
    \item $F_{K'}(\vc(K'))\geq F_K(\vc(K))+2n\ln n$.
    \item $\Gamma_{K'}(x_{K'})\leq\epsilon_{\rm V}^4$ and $0\leq F_{K'}(x_{K'})-F_{K'}(\vc(K'))\leq\epsilon_{\rm V}^4$.
    \item $e^{-a_{\rm V}}H_K\preceq H_{K'}\preceq e^{a_{\rm V}}H_K$, where $H_K=H(x_K)$ and $a_{\rm V}:=2(k_{\rm V}+1)T_{\rm V}\ln\frac1{1-r_{\rm V}}=O(n\log n)$.
\end{enumerate}
We show by induction that at the start of every iteration the polytope is bounded with at most $\bar m_{\rm V}$ rows and $\Gamma(z)\leq\epsilon_{\rm V}^4$. A deleted row has $\sigma_i(z)<\epsilon_{\rm V}$ and the smallest score, so Lemma~\ref{lem:sp_vaidya}~(c) applies with $\eta_m=0$. Before an addition every score is at least $\epsilon_{\rm V}$, and the cut vector is nonzero on the linear part of the subspace. By the Sherman--Morrison formula, the new row has the score $\psi/(1+\psi)=\sigma_\star$ at $z$, where $\psi=a^\top H(z)^{-1}a/\rho^2=\sigma_\star/(1-\sigma_\star)$, so Lemma~\ref{lem:sp_vaidya}~(b) applies. In both cases the new polytope is bounded, $\Gamma(z)\leq c_{\rm V}/2$, and $F(\vc)$ rises by at least $g^+_{\rm V}\epsilon_{\rm V}$ or drops by at most $g^-_{\rm V}\epsilon_{\rm V}$. The scores sum to $d\leq n$, so an addition starts with at most $n/\epsilon_{\rm V}$ rows and ends with at most $\bar m_{\rm V}$. Each Newton step keeps the polytope and multiplies $\Gamma$ by at most $q_{\rm V}$, by Lemma~\ref{lem:sp_vaidya}~(a), so the choice of $k_{\rm V}$ gives $\Gamma(z)\leq q_{\rm V}^{k_{\rm V}}c_{\rm V}/2\leq\epsilon_{\rm V}^4$ at the end.

Part~1 follows from the induction, and Part~2 from Lemma~\ref{lem:sandwich} (Part~1) and Lemma~\ref{lem:sp_vaidya}~(d). For Part~3, let $m_+$ and $m_-$ be the numbers of additions and deletions, and $m_0\leq\bar m_{\rm V}$ the number of rows of $K$. Then $m_++m_-=T_{\rm V}$ and $m_-\leq m_++m_0$, so $m_+\geq(T_{\rm V}-m_0)/2$, and
\begin{align*}
    F_{K'}(\vc(K'))-F_K(\vc(K))\geq\epsilon_{\rm V}\big(g^+_{\rm V}m_+-g^-_{\rm V}m_-\big)\geq\epsilon_{\rm V}\Big(g_{\rm V,b}T_{\rm V}-\frac{g^+_{\rm V}+g^-_{\rm V}}{2}\,m_0\Big)\geq2n\ln n,
\end{align*}
by the choice of $T_{\rm V}$, $\epsilon_{\rm V}m_0\leq n+\epsilon_{\rm V}$ and~\ref{cond:V3}. Part~4 follows from the induction and $\mu\leq1$. For Part~5, a deletion multiplies $H(z)$ by at least $1-\epsilon_{\rm V}$, and an addition by at most $1+2\sigma_\star$. Here $1-\epsilon_{\rm V}\geq(1+r_{\rm V})^{-2}$ and $1+2\sigma_\star=1+16\epsilon_{\rm V}\leq(1-r_{\rm V})^{-2}$, as $16\epsilon_{\rm V}\leq r_{\rm V}$ by~\ref{cond:V2}. A Newton step moves $z$ by at most $r_{\rm V}\leq r$ in the norm of $H(z)$, by Lemma~\ref{lem:sp_vaidya}~(a) and~\ref{cond:V1}, so (S2) in the proof of that lemma applies. So each of the $(k_{\rm V}+1)T_{\rm V}$ changes multiplies $H(z)$ by a factor between $(1+r_{\rm V})^{-2}$ and $(1-r_{\rm V})^{-2}$.

\emph{The rest of the proof.} We run the proofs of Lemmas~\ref{lem:main_output_formal}, \ref{lem:main_oracle_formal} and~\ref{lem:main_time_formal} with $c=0$. There $F_K(\vc(K))$ takes the place of $\mathcal{U}(K)$, the terms $c\,D(K)$ are absent, and $R'_{\rm V}=O(n^{5/2})$ replaces $R'$. Parts~1 to~4 replace Lemmas~\ref{lem:hybrid_block} and~\ref{lem:hybrid_cpm}, and Part~5 replaces Lemma~\ref{lem:block_stability}. The uses of the hybrid method change as follows.
\begin{itemize}
    \item \emph{Starting points.} A block starts at the output of the previous block, or at the center $\bar z$ of a box with $2d\leq\bar m_{\rm V}$ rows, where $\Gamma=0$ because $\bar z$ is the volumetric center (Lemma~\ref{lem:reduction_geometry}, Part~3).
    \item \emph{Correctness.} Every triple at a reduction iteration has $K\subseteq R'_{\rm V}\cdot E(x_K,H_K)$, by Part~2 for the output of a block and by Lemma~\ref{lem:sandwich} (Part~3) for a box, as $\sqrt{2d}\leq R'_{\rm V}$. So Lemma~\ref{lem:xstar} with $R_0=R'_{\rm V}$, whose hypothesis holds by~\ref{cond:V3}, keeps $x^*$ in every polytope, and the slice test does not fail, as in the proof of Lemma~\ref{lem:small_error_run}. The rest of the proof of Lemma~\ref{lem:main_output_formal} applies, with no aborts and no operation budget.
    \item \emph{The quantity $\wt\Phi$.} By~\eqref{eq:vol_dikin} and Part~4, $\wt\Phi:=\ln{\rm vol}(E(x_K,H_K))+\ln\det(\Lambda)$ equals $-F_K(\vc(K))+\ln{\rm vol}(B_d)+\ln\det(\Lambda)+\vartheta_K$, where $\vartheta_K:=F_K(\vc(K))-F_K(x_K)\in[-\epsilon_{\rm V}^4,0]$. This is~\eqref{eq:phi_tilde_U} with $c=0$. By Parts~3 and~4, a cutting iteration lowers $\wt\Phi$ by at least $2n\ln n-\epsilon_{\rm V}^4\geq n\ln n$.
    \item \emph{Runs of reductions.} The proof of Lemma~\ref{lem:main_oracle_formal} bounds a maximal run of reduction iterations using only $K\subseteq R'\cdot E(x_K,H_K)$, $E(x_K,H_K)\subseteq K$, $\sqrt{d+2}\leq2R'=n^{O(1)}$ and Lemma~\ref{lem:block_stability}. These hold with $R'_{\rm V}$, by Part~2, Lemma~\ref{lem:sandwich} and Part~5.
    \item \emph{Counting and cost.} The counting argument of that proof uses only the bound $\ln{\rm vol}(E(x_K,H_K))+\ln\det(\Lambda)\leq\wt\Phi$, so there are $O(n+\log(\gamma R)/\log n)$ cutting iterations and at most $n$ reduction iterations. With $T_{\rm V}=O(n\log n)$ calls to $\SO$ and $O(n^{\omega+1}\log n)$ operations per block, this gives the bounds of the corollary. The remaining operations are $O(n^4+n^3\log(\gamma R)/\log n)$, as in the proof of Lemma~\ref{lem:main_time_formal}, which is within the bound since $\omega\geq2$.
\end{itemize}
Lemmas~\ref{lem:svp_total} and~\ref{lem:svp_segment} use only the number of calls to \textsc{ApproxSV} and the factor $2^{\pm O(n\log n)}$ of Part~5, so they hold here as well.

For submodular function minimization, a separation oracle for the Lov\'asz extension makes $n$ queries to $\EO$ and $O(n\log n)$ other operations, as in the proof of Corollary~\ref{cor:sfm_main}. This oracle is deterministic, so the hypothesis on $\SO$ holds. Its minimizers lie in $B_\infty(1)$. So the algorithm makes $O(n^2\log n)$ oracle calls, hence $O(n^3\log n)$ queries to $\EO$, and $O(n)$ calls to \textsc{ApproxSV}. It uses $O(n^{\omega+2}\log n+n^3\log^2n)=O(n^{\omega+2}\log n)$ other operations. With \textsc{SegmentLLL}, Lemma~\ref{lem:svp_segment} adds $O(n^4\log^3n)$ operations for the calls.
\end{proof}
\par\endgroup

\end{document}